\documentclass[usenames,prd,showpacs,letterpaper,twocolumn,floatfix,nofootinbib]{revtex4}%
\usepackage{graphicx}
\usepackage{bm}
\usepackage{epsf}
\usepackage{rotating}
\usepackage{epsfig,graphics,rotate}
\usepackage{wrapfig}
\usepackage{amssymb}
\usepackage{amsmath}
\usepackage{amsfonts}
\usepackage{array,hhline,dcolumn}%
\usepackage{gensymb}
\usepackage{xcolor}
\usepackage{textcomp}
\usepackage[colorlinks=true,allcolors=blue]{hyperref}
\usepackage{tikz}
\usepackage{tikz-feynman}
\usepackage{multirow}
\usepackage{float}
\usepackage{subfloat}
\usepackage{caption}
\usepackage{subcaption}
\usetikzlibrary{trees}
\usetikzlibrary{decorations.pathmorphing}
\usetikzlibrary{decorations.markings}

\definecolor{kyelloworange}   {RGB}{255, 210,  110}
\definecolor{jblue}  {RGB}{20,50,100}
\definecolor{npurple}  {RGB} {153, 51, 204}
\definecolor{wred}   {RGB}{217,0,56}
\definecolor{white}   {RGB}{255,255,255}
  
\definecolor{korange}   {RGB}{235, 80,  43}
\definecolor{korange2}   {RGB}{245, 100,  63}
\definecolor{kyelloworange}   {RGB}{255, 210,  110}
\definecolor{kyelloworange2}   {RGB}{240, 170,  90}
\definecolor{kred}   {RGB}{204,  102, 153}
\definecolor{kpurple}   {RGB}{153,  61, 190}
\definecolor{kpurplelight}   {RGB}{213,  161, 230}

\tikzset{
	    gaugeboson/.style={decorate,decoration={snake, amplitude = 6pt, post length = 1 pt, pre length = 4 pt},draw=magenta},
	    fermion/.style={draw=black,postaction={decorate},decoration={markings,mark=at position .55}},
	    fermionin/.style={draw=black,postaction={decorate},decoration={markings,mark=at position .55 with {\arrow[draw=black]{<}}}},
	    fermionout/.style={draw=black,postaction={decorate},decoration={markings,mark=at position .55 with {\arrow[draw=black]{>}}}},
	    gluon/.style={decorate,draw=magenta,decoration={coil,amplitude = 6pt,segment length=8pt}},
	    connect/.style={draw=black,postaction={decorate},decoration={markings}}, %,mark=at position .55 with {\arrow[draw=red]{<}}}}
	    gluon2/.style={decorate,draw=magenta,decoration={coil,amplitude = 3pt,segment length=4pt}},
	    gaugeboson2/.style={decorate,decoration={snake, amplitude = 3pt, segment length=6pt},draw=black}
	}
	
\tikzset{
	  photon/.style={decorate, decoration={snake}, draw=npurple,very thick},
	  boson/.style={decorate, decoration={snake}, draw=npurple,very thick},
	  electron/.style={draw=jblue,very thick, postaction={decorate},
	           decoration={markings,mark=at position .55 with {\arrow[draw=jblue]{>}}}
	  },
	  electron2/.style={draw=jblue,very thick, postaction={decorate},
	           decoration={markings,mark=at position .55 with {\arrow[draw=jblue]{<}}}
	  },
	  fermion/.style={draw=jblue,very thick, postaction={decorate},
	            decoration={markings,mark=at position .55 with {\arrow[draw=jblue]{}}}
	  },
	  gluon/.style={decorate, draw=korange,very thick, %kred
	    decoration={coil,amplitude=4pt, segment length=6pt}},
	  higgs/.style={draw=wred,very thick, postaction={decorate},
	           decoration={markings,mark=at position .55 with {\arrow[draw=wred]{>}}}
	  },
	  nothing/.style={draw=white,very thick}
}

\providecommand{\U}[1]{\protect\rule{.1in}{.1in}}
\newcommand{\mat}[1]{\boldsymbol{#1}}

\begin{document}
\title{Where Are We With Light Sterile Neutrinos?}
\author{A. Diaz$^{1}$, C.A. Arg\"uelles$^{1}$, G.H. Collin$^{1}$, J.M. Conrad$^{1}$, M.H. Shaevitz$^{2}$}
\affiliation{$^{1}$ Massachusetts Institute of Technology, Cambridge, MA 02139, USA}
\affiliation{$^{2}$ Columbia University, New York, NY 10027, USA}

\begin{abstract}
We review the status of searches for sterile neutrinos in the $\sim 1$ eV range, with an emphasis on the latest results from short baseline oscillation experiments and how they fit within sterile neutrino oscillation models.    We present global fit results to a three-active-flavor plus one-sterile-flavor model (3+1), where we find an improvement of $\Delta \chi^2=35$ for 3 additional parameters compared to a model with no sterile neutrino.  This is a 5$\sigma$ improvement, indicating that an effect that is like that of a sterile neutrino is highly preferred by the data.  However we note that separate fits to the appearance and disappearance oscillation data sets within a 3+1 model do not show the expected overlapping allowed regions in parameter space.   This ``tension'' leads us to explore two options: 3+2, where a second additional mass state is introduced, and a 3+1+decay model, where the $\nu_4$ state can decay to invisible particles.   The 3+1+decay model, which is also motivated by improving compatibility with cosmological observations,  yields the larger improvement, with a  $\Delta \chi^2=8$ for 1 additional parameter beyond the 3+1 model, which is a $2.6\sigma$ improvement.  Moreover the tension between appearance and disappearance experiments is reduced compared to 3+1, although disagreement remains.    In these studies, we use a frequentist approach and also a Bayesean method of finding credible regions. 

With respect to this tension, we review possible problems with the global fitting method.   We note multiple issues, including problems with reproducing the experimental results, especially in the case of experiments that do not provide adequate data releases.     We discuss an unexpected 5 MeV excess, observed in the reactor flux energy spectrum, that may be affecting the oscillation interpretation of the short baseline reactor data.    We emphasize the care that must be taken in mapping to the true neutrino energy in the case of oscillation experiments that are subject to multiple interaction modes and nuclear effects.   We point to problems with the ``Parameter-Goodness-of-Fit test'' that is used to quantify the tension.  Lastly, we point out that analyses presenting limits often receive less scrutiny that signals.

While we provide a snapshot of the status of sterile neutrino searches today and global fits to their interpretation, we emphasize that this is a fast-moving field.    We briefly review experiments that are expected to report new data in the immediate future.   Lastly, we consider the 5-year horizon, where we propose that decay-at-rest neutrino sources are the best method of finally resolving the confusing situation.

\end{abstract}

\pacs{14.60.Pq,14.60.St}
\maketitle
\tableofcontents

\section{Introduction}

From the beginning, neutrino physics has been propelled forward by
pursuit of anomalies.   Of these, some eventually developed into
decisive signals, laying the
ground work for today's ``neutrino Standard Model'' ($\nu$SM).   Others were
disproven, and forced us to improve our understanding of neutrino
models, sources, and detectors in the process.    In keeping with this cycle, anomalies have been observed in
short-baseline (SBL) oscillation experiments since the
1990's.  These potentially
point to the existence of a new kind of neutrino, called a sterile neutrino, although other experiments have
substantially limited exotic neutrino interpretations.   Resolving
the question of whether these results point to new physics is a priority of our field.  However,
in the past year alone, the confusion has only mounted.   

In this review, we consider the present status of the short baseline anomalies and their interpretations.  We explain the motivation for, and phenomenology of, sterile neutrinos.  We provide updated global fits to relevant data sets in the simplest single sterile neutrino model along with discussions of their frequentist and Bayesian interpretations.  Since the global fits point to data discrepancies with this simple model, we also consider more complex explanations.  Finally, we discuss how future measurements could impact our understanding.

\section{The Road to Oscillations is Paved with Interesting Anomalies}

For the sake of this discussion, we will
define an ``anomalous signal'' as a 2$\sigma$ effect with no clear
Standard Model (SM) explanation.    We freely admit that this is an
arbitrary line that reflects the personal taste of the authors on
the point where a signal reaches a significance making it worthy of further
exploration.   Using this definition, several anomalies appear in short baseline $\nu_\mu
\rightarrow \nu_e$ appearance and $\nu_e \rightarrow \nu_e$ disappearance 
oscillation experiments.  

However, before leaping in to these relatively recent anomalies, it is useful to consider the past history of neutrino physics.  
Let us begin this story thirty years ago, as the development of 
the three-neutrino oscillation model provides a valuable context for the four-or-more neutrino questions we are asking today.

\subsection{The State of Oscillation Physics in the 1990's}

Looking back to the 1990's, we were in situation regarding three-neutrino oscillations that was remarkably similar to where we are today with the sterile neutrino question.

\subsubsection{Anomalous Signals Existed}

In the mid-to-late 1990's, there were two classes of neutrino anomalies.  The
first set belonged to solar neutrino experiments.   Deficits of $\nu_e$
interactions were observed -- compared to
prediction -- in the SAGE~\cite{SAGEsolar} and Gallex~\cite{Gallexsolar} experiments, with a threshold of 0.23
MeV; the Homestake experiment~\cite{Homestake}, with a threshold of 0.8 MeV; and the
Kamioka~\cite{Kamsolar} and early-Super-K data sets~\cite{SKsolarearly}, with a threshold of 7 MeV.
The second set belonged to experiments that used neutrinos produced in the atmosphere.   To minimize the systematic uncertainties, these
experiments looked at the ratio-of-ratios: $(\nu_\mu/\nu_e)_{data}/(\nu_\mu/\nu_e)_{prediction}$.   Anomalies in these ratio-of-ratios
were observed in the Kamioka sub-GeV~\cite{Kamsub} and multi-GeV~\cite{Kammulti} data sets, and in the IMB sub-GeV data set~\cite{IMBsub}.     Both the solar and atmospheric observations sparked  the consideration of
neutrino oscillation models.

\subsubsection{Limits Contradicted Some of the Anomalies}

The situation was particularly confusing in the atmospheric sector,
as the Frejus~\cite{Frejus} and NUSEX~\cite{NUSEX} experiments
%(yes, that was its real name)
released data-to-simulation results for the ratio-of-ratios that were consistent with
unity---directly contradicting the anomalous signals.    The IMB
multi-GeV sample also agreed with unity~\cite{IMBmulti},
although the uncertainties were large enough to be additionally consistent with the measured
atmospheric anomalies.     The existence of these limits caused many to question whether the atmospheric anomaly, in particular, was
a new-physics effect, or was just due to an unidentified systematic uncertainty associated with
water-based Cherenkov detectors.

\subsubsection{In Retrospect, Unidentified Systematic Uncertainties Were Leading to Discrepancies}

In 1998, Super-K published the first very-high-statistics atmospheric
result~\cite{SKatmos1998}.    If interpreted within an oscillation scenario, this indicated
a mass splitting, $\Delta m^2$, that was a factor of about five times below
the one implied by the Kamiokande atmospheric result.    All atmospheric experiments that
have followed have found results consistent with Super-K.   This
indicates that Kamiokande had a source of systematic uncertainty associated with their measurements or analysis that was
never identified.    This should not be surprising---when we discover
anomalies, it is almost always at the edge of a detector's
capability.    Thus, it is likely that a true signal will end up
marginally distorted by unknown issues with event reconstruction or
backgrounds.  This makes the initial interpretation of
anomalies difficult.

\subsubsection{Theoretical Models Needed to be Expanded}

The solar neutrino data set was particularly confusing because the
data did not fit well to a vacuum oscillation solution~\cite{justso, nojustso}.   A new explanation was required, which took the form of the
MSW hypothesis~\cite{MSofMSW, WofMSW}.    This explanation noted that the $\nu_e$ flux produced by the
Sun has traveled through a highly dense environment of electrons that
produces a weak-force potential.   When this potential is added to the
Hamiltonian for vacuum oscillations, the oscillatory behavior is
destroyed.   The neutrino flavor change then occurs either through a resonance, leading to a small
mixing angle solution, or to an adiabatic transition, leading to a
large mixing angle solution.    Surprisingly, in either case the neutrinos exit the
Sun as a pure mass eigenstate, and so do not oscillate as they travel to the
Earth.  The MSW solutions implied a mass
splitting of $\sim 10^{-5}$ eV$^2$, which was five orders of magnitude above the 
vacuum oscillation solution.    If we had not thought broadly about the source of
the solar neutrino anomalies and how matter could effect flavor change  -- which led to the
development of the MSW solution --  we probably would have never proposed the
KamLAND experiment~\cite{Kamlandex}, and would remain greatly confused
about the three-neutrino oscillation physics today!     

An important point here is that anomalies should not necessarily be attributed to physics beyond the SM, but rather one should consider previously neglected SM effects.

\subsubsection{The Theoretical Bias Was Against an Oscillation Explanation}

By the mid-1990's, it was clear that if neutrinos were to have masses,
they would be many orders of magnitude smaller than the masses of the charged
particles.   Small-but-non-zero mass was not regarded as particularly
appealing then, nor is it appealing today.  Although most physicists were skeptical of massive neutrinos on theoretical grounds, many would consider them if their masses were of order 10~eV to 100~eV because they could explain dark matter~\cite{DMmass}.
Among the smaller set of theorists who accepted
neutrino masses well below 10~eV, there
was a strong prejudice that the ``correct answer'' had to be the small
mixing angle MSW solution.  This was based on an analogy to the quark
sector where the mixing angles are also small.     This scenario is a nice
illustration of how nature's taste may not match our taste in ``beautiful theories.''     

\subsubsection{The Resolution of the Anomalies}

The resolution to the debate arose from truly adventurous thinking in
detector technology.    Today, we simply accept that detectors like SNO~\cite{SNOdet} and Super-K~\cite{SKdet} can
be constructed.   Yet these were extraordinary achievements and represented massive
steps forward in sensitivity.   These giant leaps in detector
technology  were essential
to our understanding.    If we had continued to just make incremental
steps in sensitivity, the confusion surrounding the three-neutrino anomalies would have continued for decades.

Instead, we have now developed a consistent, highly predictive picture
that we call the ``$\nu$SM.''     This incorporates neutrino mass and
mixing into the phenomenological picture of the Standard Model,
without any reference to the underlying theory of the source of mass and mixing, which is still not understood.

\section{Two, Three, Four, and More}

The resolution of the anomalies described above came about from introducing neutrino mass and mixing into the picture, which leads directly to an effect called vacuum neutrino oscillations.   Before considering the $\nu$SM, which involves three neutrino flavors, it is instructive to introduce the phenomenology of neutrino oscillations in a two-neutrino picture.    This picture will also be useful as we consider the new set of anomalies that lead to the potential introduction of sterile neutrinos as additional flavors.

\subsection{Two Neutrino Oscillations}

Neutrino mass eigenstates need not correspond to pure neutrino flavor eigenstates, but, instead, may be rotated to form a linear combination.   In a two neutrino picture, if we call $\nu_\alpha$ and $\nu_\beta$ the flavor eigenstates and $\nu_1$ and $\nu_2$ the mass eigenstates, then they are related by
\begin{align}  \label{twonu}
\left(
\begin{array}
[c]{c}%
\nu_{\alpha}\\
\nu_{\beta}
\end{array}
\right)   &  {=}\left(
\begin{array}
[c]{cc}%
\cos\theta & \sin\theta \\
-\sin\theta & \cos\theta \\
\end{array}
\right) \left(
\begin{array}
[c]{c}%
\nu_{1}\\
\nu_{2}
\end{array}
\right).
\end{align}

If this is the physical case, then a neutrino born as $\nu_\alpha$ can transform to $\nu_\beta$ as it propagates.   In a vacuum, or within material of minimal density, the probability that this occurs is given by the vacuum neutrino oscillation formula:
\begin{equation}
P_{\nu_{\alpha}\rightarrow\nu_{\beta}}=\sin^{2}2\theta \sin^{2}\left(  1.27\ \Delta
m_{ij}^{2}\left(  \text{eV}^{2}\right)  \frac{L(\text{m})}{E(\text{MeV}%
)}\right), \label{vacosc}
\end{equation}
where the $\theta$ is the mixing angle and $\Delta m_{ij}^{2}=m_{i}%
^{2}-m_{j}^{2}$ for the two mass eigenstates. The value of $L/E$ for an experimental
setup sets the scale of the $\Delta m^{2}$ sensitivity with the first
oscillation maximum at a distance of
\begin{equation}
L^{Max}\left(  \text{m}\right)  =
\frac{\pi}{2} \frac{1}{1.27} \ \frac{E\left(  \text{MeV}\right) }{ \Delta m^{2}\left(  \text{eV}^{2}\right)}. 
\end{equation}

There are two experimental methods for searching for indications of neutrino
oscillations, the ``disappearance" method and the ``appearance" method. In a
disappearance search, one looks for a change in event rate of a given type of
neutrino over distance and energy. The power of this method is most impacted by the knowledge of
the neutrino flux and interaction cross sections.  
The appearance
method involves looking for neutrinos of type $\nu_\beta$ not present in an initially nearly-pure beam of $\nu_\alpha$. The sensitivity for an appearance search is
dependent on knowing the initial (``intrinisic'') contamination of $\nu_\beta$ in the beam before oscillations and
knowing the backgrounds associated with the primary neutrinos that mimic a $\nu_\beta$ event.

\begin{figure}[t]
\begin{center}
{\includegraphics[height=2.8in]{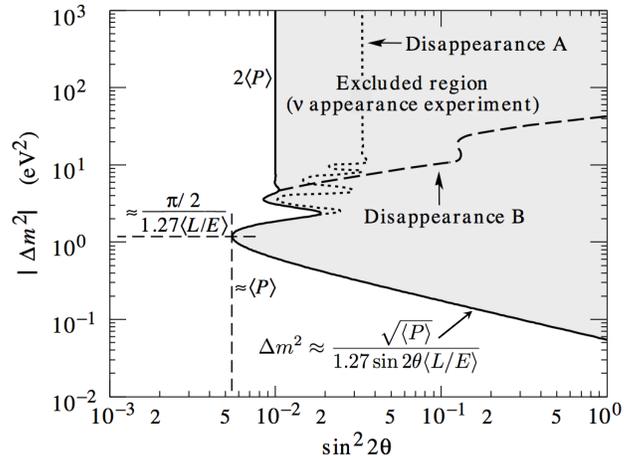}}
\end{center}
\vspace{-0.1in}
\caption{Neutrino oscillation parameter ranges excluded by toy appearance or disappearance experiments.
For the appearance solid curve, the average probability of appearance is assumed to be $\langle P \rangle =0.5\%$.  For disappearance, the regions are set by the measured limit on the allowed disappearance probability $P_{dis}$.  At high $\Delta m^2$ for disappearance, the $\sin^2 2\theta$ limit is either set by the normalization uncertainty associated with knowledge of the beam flux and neutrino cross section (Disappearance type A = dotted curve) or for a two detector or $L/E$ shape experiment goes to $\sin^2 2\theta = 1$  (Disappearance type B - dashed curve). (For this plot, the average $\langle L/E \rangle = 1$ km/GeV and the $L/E$ values are assumed to be normally distributed with a $\sigma$ of 20\%.) Plot from 1996 PDG \cite{Barnett:1996hr}.  
\label{app_disapp_cartoon}}
\end{figure}

The oscillation sensitivity limit for disappearance, at any $\Delta m^2$ and $\sin^2 2\theta$, is related to the experimental error by
\begin{equation}
P_{dis}  = \sin^22\theta \sin^{2}\left(  1.27\ \Delta
m^{2}  \frac{L}{E}\right)  < C \frac{\delta N_\alpha }{ N_\alpha }, \label{dissurv}
\end{equation}
where $P_{dis}$ is the disappearance probability, which is related to the survival probability by $1-P_{dis}$. In Eq.~\ref{dissurv}, $N_\alpha$ is the number of observed events of the given flavor $\alpha$,  $\delta N_\alpha$ is the combined systematic and statistical error on measuring the $\alpha$ flavor events, and $C$ is a factor that depends on the confidence level of the limit.  
For a 90\% C.L. limit that is set using a single-sided normal distribution, $C=1.28$. 
At high $\Delta m^2$ compared to $L/E$, the $\sin^{2}\left(  1.27\ \Delta
m^{2}  \frac{L}{E}\right)$ factor averages to 0.5 due to the experimental resolution on $E$ and $L$ and the limit becomes
\begin{equation}\label{eq:hi_dm2_limit}
\sin^22\theta < 2C \frac{\delta N_\alpha }{ N_\alpha }.
\end{equation}
Thus, one can use the high $\Delta m^2$ limit to determine the fractional error that an experiment is claiming. For example, in Fig.~\ref{app_disapp_cartoon}, the large $\Delta m^2$ limit is $\sin^22\theta < 0.031$ at 90\% C.L., which corresponds to $\delta N_\alpha/N_\alpha = 1.2\%$.

Many disappearance experiments use the shape of the observed energy spectrum at a single or multiple $L$ distances to look for indications of oscillations by observing a change with $L$ and $E$.  This can help remove the dependence on knowing the overall neutrino flux normalization but this method becomes insensitive at high $\Delta m^2$ where the rapid oscillations are not observable due to detector energy and position resolution and the limit goes to $\sin^22\theta = 1$.  Using measurements at several different $L$-values can also eliminate the uncertainty in knowing the shape and normalization of the neutrino energy spectrum.  For example, comparing the rates in a near and far detector can be used in an oscillation search if the relative systematic uncertainties can be kept small.  On the other hand, for a two-detector comparison measurement at high $\Delta m^2$, the sensitivity degrades markedly since the rapid oscillations removes any spectral difference in the two detectors.  Thus, for multi-detector measurements, the high $\Delta m^2$ limit becomes the same as a single detector disappearance measurement where the sensitivity is determined by the combined systematic and statistical error on measuring the $\alpha$ flavor events, $\delta N_\alpha$, as described in Eq.~\ref{eq:hi_dm2_limit}. Also, for a shape-only, single-detector measurement, the rapid oscillation at high $\Delta m^2$ effectively removes any oscillation sensitivity as shown in Fig.~\ref{app_disapp_cartoon} for the ``Disappearance B'' curve but some $\Delta m^2$ sensitivity can be recovered treating the measurement as a counting experiment as described in Eq.~\ref{eq:hi_dm2_limit}. 

Fig.~\ref{app_disapp_cartoon} illustrates this
in showing separately the two types of disappearance measurements, typically referred to as counting experiments (Disappearance A) and shape experiments (Disappearance B).  Of course, experiments usually perform both types of analyses and combine the results to maximize coverage of their oscillation search.  In order to achieve this, the correlated systematic uncertainties must be included in the shape plus normalization analysis, which typically leads to reduced sensitivity as outlined above.

For appearance, 
\begin{equation}
P_{app} = \sin^22\theta \sin^{2}\left(  1.27\ \Delta
m^{2}  \frac{L}{E}\right) < C \frac{\delta N_\beta }{ N_{\text{full~osc}} } 
\end{equation}
where $C$ is, again, related to the confidence level; $\delta N_\beta$ is the statistical and systematic error on the appearance signal, and $N_{\text{full~osc}}$ is the number of events one would have if all neutrinos of the original flavor, $\alpha$, converted to flavor $\beta$.

The two neutrino vacuum oscillation formula has several specific features that are helpful to consider when designing an experiment or interpreting an oscillation plot.  For an appearance measurement, the sensitivity of an experiment is set by the number of ``right-flavor'' events and the uncertainty in background rates.  At low $\Delta m^2$, the sensitivity boundary curve is given by:
\begin{equation}
\Delta m^2 \approx \frac{\sqrt{\langle P \rangle}}{1.27\sin 2\theta \langle L/E \rangle},
\end{equation}
where $\langle P \rangle$ is the average probability. The limit of an experiment's sensitivity to $\Delta m^2$ at $\sin^2 2\theta=1$ is then given by:
\begin{equation}
\Delta m^2_{min} = \sqrt{\langle P\rangle}/(1.27 \langle L/E \rangle).
\end{equation}
From this one sees that it is difficult to increase sensitivity to low $\Delta m^2$ through extended running, since the improvement due to statistical error will go as the fourth-root of $N$.    To go down  in $\Delta m^2$ reach, one therefore needs to adjust the $L$ and $E$ in the design to access lower values. On a log-log plot, the sensitivity to low $\sin^2 2\theta$ will increase with $\Delta m^2$ with a slope of $-1/2$.  The maximum sensitivity to $\sin^2 2\theta$ is reached at the $\Delta m^2$, $L$ and $E$ that satisfy
\begin{equation}
1.27 \Delta m^2 L/E = \pi/2,
\end{equation}
which is often approximated in general discussions about design as $L/E \sim 1/\Delta m^2$.

In this region, the oscillation limit depends on the deviation of the energy or $L/E$ dependence of an experiment from the expectation.
Variations in $\sin^2 2\theta$ sensitivity will be seen just above the maximum sensitivity due to points where stochastic and systematic fluctuations of the data points match the oscillation prediction.   At high $\Delta m^2$, the $\sin^2(1.27 \Delta m^2 L/E)$ term will oscillate rapidly and average to 1/2, and the experiment loses sensitivity to this parameter.   The sensitivity curve plot of $\Delta m^2$ {\it vs.} $\sin^2 2\theta$ will then extend straight up with a line that is equal to 
$\sin^2 2\theta_{high~\Delta m^2} = 2 \langle P\rangle$.

\subsection{Oscillations and the $\nu$SM}

\begin{figure}[t]
\begin{center}
{\includegraphics[height=2.in]{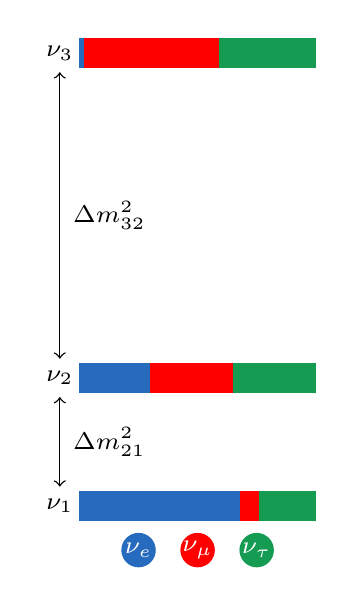}}~{\includegraphics[height=2.in]{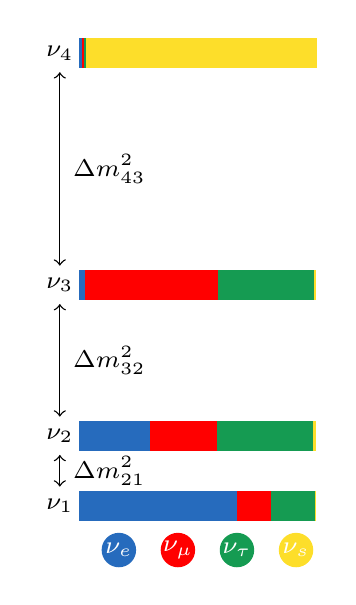}}~
\end{center}
\caption{ Illustration of normal neutrino mass ordering and mixing for the three (left) and four (right) neutrino picture. Note that in the four neutrino picture, $\Delta m^2_{41}=\Delta m^2_{21}+\Delta m^2_{32} + \Delta m^2_{43}$.
\label{threefourbars}}
\end{figure}

In the $\nu$SM we expand the formalism to three neutrino mass eigenstates that are not aligned with three
neutrino flavor eigenstates. The unitary matrix relating these two
bases is parameterized by the PMNS (Pontecorvo-Maki-Nakagawa-Sakata) matrix:
\begin{equation}
\label{PMNS}
\begin{split} 
\begin{pmatrix}
\nu_{e}\\
\nu_{\mu}\\
\nu_{\tau}%
\end{pmatrix}  {=}
& \begin{pmatrix}
\cos\theta_{12} & \sin\theta_{12} & 0\\
-\sin\theta_{12} & \cos\theta_{12} & 0\\
0 & 0 & 1
\end{pmatrix} \\
& \begin{pmatrix}
\cos\theta_{13} & 0 & e^{-i\delta_{CP}}\sin\theta_{13}\\
0 & 1 & 0\\
-e^{-i\delta_{CP}}\sin\theta_{13} & 0 & \cos\theta_{13}%
\end{pmatrix} \\
& \begin{pmatrix}
1 & 0 & 0\\
0 & \cos\theta_{23} & \sin\theta_{23}\\
0 & -\sin\theta_{23} & \cos\theta_{23}%
\end{pmatrix}
\begin{pmatrix}
\nu_{1}\\
\nu_{2}\\
\nu_{3}%
\end{pmatrix}.
\end{split} 
\end{equation}
The three mixing angles in this matrix are probed by different types of experiments. Here,
$\theta_{12}$ is referred to as the solar mixing angle, $\theta_{23}$ as the
atmospheric mixing angle, and $\theta_{13}$ sets the $\nu_e$ disappearance oscillations at reactors and $\nu_e$ appearance for accelerator neutrino beams.
The $\delta_{CP}$ parameter in the matrix is the complex phase associated with $CP$ violation for neutrino oscillations.
With three generations, there are two independent mass-squared
differences associated with the mass eigenstates. These are denoted by $\Delta m_{21}%
^{2}=\Delta m_{Solar}^{2}=m_{2}^{2}-m_{1}^{2}$, and $\Delta m_{31}^{2}=\Delta
m_{Atmospheric}^{2}=m_{3}^{2}-m_{1}^{2}$. 
The resulting formulae for oscillations between the flavors are more complicated, and can be found in Ref.~\cite{threenureview} --- they are not reproduced here.

\begin{table}[tbp] \centering
\begin{tabular}
[c]{cc}\hline
Parameter & Value \\\hline
$\Delta m_{21}^{2}\times10^{-5}\mathrm{eV}^{2}$ & $7.39\pm0.21$\\
$\Delta m_{31}^{2}\times10^{-3}\mathrm{eV}^{2}$ & $\pm2.525\pm 0.033$\\
$\sin^{2}\theta_{12}$ & $0.310\pm0.013$\\
$\sin^{2}\theta_{23}$ & $0.580\pm0.021$\\
$\sin^{2}\theta_{13}$ & $0.02241\pm0.00065$\\\hline
\end{tabular}
\caption{Present values and uncertainties for oscillation parameters determined from
global fits to all data for normal hierarchy~\cite{threenureview}.} \label{OscPar} 
\end{table}%

Experiments have been performed using a wide range of neutrino sources including
solar neutrinos ($\sim$1 to 10~MeV), atmospheric neutrinos ($\sim$0.5 to 20~GeV), reactor
neutrinos ($\sim$2 to 8~MeV), and accelerator neutrino beams with a wide range of
energies from 20~MeV up to 200~GeV. The backgrounds to the experiments are
energy dependent and typically come from natural radioactivity in low energy
experiments and from cosmic ray muon backgrounds at higher energy. Accelerator
neutrinos have the added benefit in that they can use the beam timing to reduce non-beam related backgrounds.

Since the difference between the $\Delta m^{2}$ values for the solar and atmospheric eigenstates is so large, a simple analysis of many of the experimental measurements can be
performed assuming only two neutrino mixing, as discussed in the previous section.
Applying Eq.~\ref{vacosc} -- for
reactor neutrinos of energy $\sim3$~MeV -- the first oscillation maximum
associated with the $\Delta m_{Solar}^{2}$ mass splitting would be at $L=39$~km. For accelerator neutrinos of $\sim1$~GeV, the distance for
$\Delta m_{Atmospheric}^{2}$ would be $L=420$~km.   However, to obtain the most accurate values of the parameters, the full three neutrino mixing formalism must be applied.
Except for $\delta_{CP}$, all of the oscillation parameters have been
determined~\cite{threenureview} through a combination of measurements with the values
and uncertainties listed in Table~\ref{OscPar}.

Fig.~\ref{threefourbars} (left) illustrates the $\nu$SM.   Each bar represents
a mass state.   In this case we show ``normal ordering'' with $\Delta m_{31}^{2} > 0$, as opposed to ``inverted ordering'' with $\Delta m_{31}^{2} < 0$.   The colors within the bar represent the flavor composition of each mass eigenstate.   

Three outstanding questions remain within the $\nu$SM.
The first is the mass ordering---normal versus inverted.   The
accumulated  data favor normal ordering from 2 to 4$\sigma$~\cite{threenureview},
depending on the data sets used in the determination.  The second is whether
a non-zero $CP$-violating parameter, $\delta_{CP}$, appears in the mixing matrix.
Global fits indicate that $\delta_{CP}$ is non-zero, and is large, at
greater than 2$\sigma$~\cite{threenureview}. 
The final question is whether the $\theta_{23}$ mixing angle lies above or below $45\degree$, the answer of which may hold theoretical implications for neutrino masses.   As seen in Table~\ref{OscPar}, the present preferred value is $49.6\degree$, but within 3$\sigma$, both octants are allowed and the question remains open. 
Overall, though, we are rapidly
closing in on a fully consistent three-neutrino picture that fits a
large fraction of the data from neutrino oscillation experiments.

\begin{center}
\begin{table*}
\begin{tabular}{l || c | c | c}
& $\nu_\mu \rightarrow \nu_e$ & $\nu_\mu \rightarrow \nu_\mu$ & $\nu_e \rightarrow \nu_e$ \\
\hline
\hline
\multirow{4}{6em}{Neutrino} & MiniBooNE (BNB) $*$ & SciBooNE/MiniBooNE & KARMEN/LSND Cross Section\\
& MiniBooNE(NuMI) & CCFR & Gallium $*$ \\
& NOMAD & CDHS & \\
& & MINOS & \\
\hline
\multirow{4}{6em}{Antineutrino} 
& LSND $*$ & SciBooNE/MiniBooNE & Bugey \\
& KARMEN & CCFR & NEOS \\
& MiniBooNE (BNB) $*$ & MINOS & DANSS $*$ \\
& & & PROSPECT
\end{tabular}
\caption{The collection of experiments implemented into our global fit analysis, sorted by oscillation types. This noted with a $*$ have $>2\sigma$ signals, and hence exhibit ``anomalies".  We describe these experiments and provide references in Sec.~\ref{experiments}.}
\label{table:experiments}
\end{table*}
\end{center}

\subsection{Deviations from the $\nu$SM Picture}

While most data fits well within the $\nu$SM,  there is a set of data
from short baseline experiments that does not fit well.    These
experiments can be fit within two neutrino oscillation models with
$\Delta m^2 \sim 1$ eV$^2$--much larger than the solar and atmospheric
splitting.    In a global picture, this is equivalent to adding a third independent mass splitting.

Defining a $>2\sigma$ signal as an anomaly, effects are seen in $\nu_\mu \rightarrow \nu_e$ accelerator-based oscillation experiments,
a set of reactor $\bar \nu_e$ disappearance experiments, and source-based experiments that are consistent with $\nu_e$ disappearance.
There is also a large set of data that do not indicate
signals at the 2$\sigma$ level.  These limit the parameter
space of a neutrino oscillation model that seeks to incorporate the
anomalies listed above. Note that in some cases, the experiments with no anomalous signal do
have effects at a lower confidence level.    Within a global fit
to the data, these effects can conspire with the anomalies to enhance
signal regions when those region align, and suppress them when they do not.    
It is particularly striking that no anomaly has been seen in a $\nu_\mu$
disappearance experiment.    This oscillation mode shows only limits, complicating the interpretation of the fit.

The experiments that we will use in the global fits reported here are listed in Table~\ref{table:experiments}.    We explain our choices and describe these experiments in Sec.~\ref{experiments}. The star ($*$) indicates experiments with an anomalous signal.   We note that the measured reactor flux is in disagreement with the first-principles prediction, an effect called the Reactor Antineutrino Anomaly (RAA)~\cite{reactor1}.   However, because of issues with the reactor flux prediction discussed later in this paper, we only employ reactor results that involve ratios of measurements.    We also note that several very recent experimental results are not included in this generation of our global fits, but will be incorporated in the future, as discussed in Sec.~\ref{nearfuture}.

\subsection{3+1:  The Simplest Model Involving Sterile Neutrinos}

The most economical method of adding a third independent mass splitting is
to introduce a single sterile neutrino into the model.
Fig.~\ref{threefourbars} (right) illustrates this idea.  The neutrino is
assumed to be sterile to avoid clashing with the number of active neutrinos measured by the LEP experiment~\cite{Aleph,L3}.     The sterile neutrino
flavor is mixed within the four mass states.  However, three of the mass states
must have very little mixture of sterile neutrino in order to explain
the data contributing to the $\nu$SM.   

The short-baseline anomalies indicate a mass splitting that is $\gtrsim 10$ times larger than the mass splittings between the mostly-active mass-states.  Therefore, we traditionally invoke the ``short baseline approximation'' where we assume $\Delta m^2_{21} \approx \Delta m^2_{32} \approx 0$.   As a result, in a 3+1 model, we typically consider only one splitting between the mostly sterile state and the mostly active states, which we term $\Delta m^2_{41}$, and which is equal to $\Delta m^2_{21}+\Delta m^2_{32} + \Delta m^2_{43}$.

The flavor and mass
states are now connected by a unitary matrix with one extra row and
column.   Writing this in terms of generic matrix elements:
\begin{equation}
\left( 
\begin{array}{l}
\nu _e \\ 
\nu _\mu \\ 
\nu _\tau \\
\nu_s
\end{array}
\right) =\left( 
\begin{array}{llll}
U_{e1} & U_{e2} & U_{e3} & U_{e4} \\ 
U_{\mu 1} & U_{\mu 2} & U_{\mu 3} & U_{\mu 4} \\ 
U_{\tau 1} & U_{\tau 2} & U_{\tau 3} & U_{\tau 4} \\
U_{s 1} & U_{s 2} & U_{s 3} & U_{s 4}
\end{array}
\right) \left( 
\begin{array}{l}
\nu _1 \\ 
\nu _2 \\ 
\nu _3 \\
\nu _4 
\end{array}
\right)~, \label{mx}
\end{equation}
where we have ignored the possible Majorana phases, since they have no observable effect in neutrino oscillation experiments. Our additional heavy-neutrino mass-state can have either Dirac or Majorana mass terms; see Ref.~\cite{Abazajian:2012ys} for a complete discussion. In this review, we assume that neutrinos are described by Dirac mass terms, which implies that constraints from neutrinoless double beta decay are immediately satisfied~\cite{deGouvea:2015euy}; though for MeV to GeV scale sterile neutrinos, kinematic constraints exist in the Dirac scenario~\cite{Bryman:2019ssi}.

The $\nu_\mu
\rightarrow \nu_e$, $\nu_e \rightarrow \nu_e$, and $\nu_\mu \rightarrow
\nu_\mu$ oscillation probabilities are interconnected through these mixing matrix elements:
{\footnotesize
\begin{eqnarray} 
P_{\nu_e \rightarrow \nu_e} &=& 1 - 4 (1-|U_{e4}|^2)|U_{e4}|^2 \sin^2 (1.27 \Delta m_{41}^2 L/E), \label{ee}\\ 
P_{\nu_\mu \rightarrow \nu_\mu} &=& 1 - 4 (1-|U_{\mu4}|^2)|U_{\mu4}|^2 \sin^2 (1.27 \Delta m_{41}^2 L/E), \label{mumu}\\ 
P_{\nu_\mu \rightarrow \nu_e} &=& 4 |U_{\mu 4}|^2 |U_{e 4}|^2 \sin^2 (1.27 \Delta m_{41}^2 L/E), \label{mue}
\end{eqnarray}
}
where $L$ and $E$ are given in kilometers and GeV, or meters and MeV, respectively. Additionally, there are equations for the $\tau$ channel:
{\footnotesize
\begin{eqnarray} 
P_{\nu_\tau \rightarrow \nu_\tau} &=& 1 - 4 (1-|U_{\tau4}|^2)|U_{\tau4}|^2 \sin^2 (1.27 \Delta m_{41}^2 L/E), \label{tautau}\\ 
P_{\nu_\tau \rightarrow \nu_\mu} &=& 4 |U_{\tau 4}|^2 |U_{\mu 4}|^2 \sin^2 (1.27 \Delta m_{41}^2 L/E), \label{taumu}\\ 
P_{\nu_\tau \rightarrow \nu_e} &=& 4 |U_{\tau 4}|^2 |U_{e 4}|^2 \sin^2 (1.27 \Delta m_{41}^2 L/E). \label{taue}
\end{eqnarray}
}
These equations appear similar to the two-neutrino mixing formula in Eq.~\ref{twonu}.  As a result, the matrix element terms are often replaced with effective mixing angles:
\begin{align} 
\sin^2 2\theta_{ee}   &= 4 (1-|U_{e4}|^2)|U_{e4}|^2, \label{mixee}\\
\sin^2 2\theta_{\mu \mu} &= 4 (1-|U_{\mu4}|^2)|U_{\mu4}|^2, \label{mixmumu}\\
\sin^2 2\theta_{\mu e} &= 4|U_{\mu 4}|^2|U_{e 4}|^2, \label{mixmue}
\end{align}
so that the equations appear clearly analogous; and similarly for the $\tau$ sector.

At this point, few experiments sample $U_{\tau 4}$.  Therefore, we will not include oscillations involving $\tau$-flavor in this discussion.   However, we point out that a past analysis that included IceCube matter effects in the global fits did provide a limit on $U_{\tau 4}$.    In that case, fitting world data leads 
a $4\times 4$ mixing matrix of values, using
unitarity to constrain the elements of the final row~\cite{Collin:2016aqd}.
We have reported ranges of allowed values in the past; for example, in late 2016 the 
ranges of values allowed for this $4\times 4$ mixing matrix were~\cite{Collin:2016aqd}:\\

\noindent $|U|=$% \vspace{-0.05in}
\begin{align}    
% = \hspace{-0.2cm}  %\nonumber \\[1mm]
 %\scalemath{0.83}{
 {\scriptsize
 \begin{bmatrix}
0.79 \rightarrow 0.83   \hspace{0.1cm} &0.53 \rightarrow 0.57 \hspace{0.1cm} & 0.14 \rightarrow 0.15 \hspace{0.1cm} & 0.13 \;(0.17) \rightarrow 0.20\:(0.21)\\%[-1mm]
%  &  &  &  {\scriptstyle (0.08 \rightarrow 0.22)} \\[1mm]
   0.25 \rightarrow 0.50 \hspace{0.1cm} & 0.46 \rightarrow 0.66 \hspace{0.1cm} & 0.64 \rightarrow 0.77 \hspace{0.1cm} & 0.09\;(0.10) \rightarrow 0.15\:(0.13)\\%[-1mm]
%  &  &  &  {\scriptstyle (0.08 \rightarrow 0.16) } \\[1mm] 
  0.26 \rightarrow 0.54 \hspace{0.1cm}& 0.48 \rightarrow 0.69 \hspace{0.1cm} & 0.56 \rightarrow 0.75 \hspace{0.1cm} & 0.0\;(0.0)  \rightarrow 0.7\:(0.05)\\%[-1mm]
%  &  &  &  {\scriptstyle (0.00 \rightarrow 0.7) } \\[1mm]
     \ldots \hspace{0.1cm}& \ldots \hspace{0.1cm} & \ldots \hspace{0.1cm} & \ldots %\\[-1mm]
   %{\scriptstyle (\ldots) }  \hspace{0.1cm} &  {\scriptstyle (\ldots) } \hspace{0.1cm} &  {\scriptstyle (\ldots) } \hspace{0.1cm} &  {\scriptstyle (\ldots) } \\[1mm]
 \end{bmatrix}
 }
 %}
 ,
\label{mix_matrix}\,
\end{align}
where the ``\ldots''  indicate parameters constrained by assumed unitarity, the ranges correspond to 90\% confidence level intervals, and the entries in the last column are for $\Delta m_{41}^2 \sim 2~{\rm eV}^2$ ($\Delta m_{41}^2 \sim 6~{\rm eV}^2$).  

Lastly, depending on the experiment, it may be most appropriate to use a rotation of the mixing matrix that is parameterized as $\theta_{14}$, $\theta_{24}$, and $\theta_{34}$ rather than $U_{e4}$, $U_{\mu 4}$, and $U_{\tau 4}$.   The connections between these angles, the ones introduced above, and the matrix elements are given in our handy ``cheatsheet,'' in Table~\ref{cheatsheet}.    
In the case of $\theta_{14}=0$ (or equivalently $|U_{e 4}|^2=0$), note that $\sin^2 2 \theta_{24}$ reduces to $\sin^2 2\theta_{\mu \mu}$.
Therefore they are often used interchangably.   However, we caution that they are not the same, and in global fits that search for $\nu_\mu \rightarrow \nu_e$
and $\nu_e$ disappearance, the assumption that $\theta_{14}=0$ is inconsistent.   This affects the MINOS results that are included in the global fits later in this discussion.

\begin{table*}[tbp] \centering
\begin{tabular}{lllll}\hline
$\sin^2 2 \theta_{ee}$ &=& 
$\sin^2 2 \theta_{14}$ &=& 
$ 4 (1-|U_{e4}|^2)|U_{e4}|^2$\\
$\sin^2 2 \theta_{\mu\mu}$ &=& 
$4 \cos^2 \theta_{14} \sin^2 \theta_{24} (1 - \cos^2 \theta_{14} \sin^2 \theta_{24})$ &=& 
$4 (1-|U_{\mu4}|^2)|U_{\mu4}|^2$ \\
$\sin^2 2 \theta_{\tau\tau}$ &=& 
$4 \cos^2 \theta_{14} \cos^2 \theta_{24} \sin^2 \theta_{34}(1 - \cos^2 \theta_{14} \cos^2 \theta_{24} \sin^2 \theta_{34})$ &=& 
$4 (1-|U_{\tau4}|^2)|U_{\tau4}|^2$ \\
$\sin^2 2 \theta_{\mu e}$ &=& 
$\sin^2 2 \theta_{14} \sin^2 \theta_{24}$ &=& 
$4|U_{\mu 4}|^2 |U_{e 4}|^2$ \\
$\sin^2 2 \theta_{e \tau}$ &=& 
$\sin^2 2 \theta_{14} \cos^2 \theta_{24} \sin^2 \theta_{34}  $&= &
$4|U_{e 4}|^2 |U_{\tau 4}|^2$\\
$\sin^2 2 \theta_{\mu \tau}$ &=& 
$\sin^2 2 \theta_{24} \cos^4 \theta_{14} \sin^2 \theta_{34} $&= &
$4|U_{\mu 4}|^2 |U_{\tau 4}|^2$\\ \hline 
\end{tabular}
\caption{3+1 Sterile Neutrino Mixing Parameter Cheatsheet \label{cheatsheet}}
\end{table*}

\subsection{The Launching Point of This Review \label{launch}}

In this review, we consider the implications of global fits to short-baseline data sets, in order to interpret whether those data indicate the existence of sterile neutrinos.    The metric we use for the 3+1 model fits is the $\Delta \chi^2 =\chi^2_{null}-\chi^2_{3+1}$, discussed further in Sec.~\ref{subsubsec:model_comparison}.  The $\Delta \chi^2$ for a 3+1 fit to the 2016
data described above is 51 for 3 additional parameters~\cite{Collin:2016rao}.    This is an extremely large improvement over a model with no sterile neutrino.   Clearly, the data
strongly favors a correction that behaves like oscillations due to a sterile
neutrino.

Despite this enormous improvement in $\Delta \chi^2$, we are suspicious of this explanation because the fit to a 3+1 model suffers from an observed internal inconsistency.
In principle, if the data sets were divided in half in an arbitrary way,
one should still find that the data subsets will have global-fit solutions
that overlap.    A method to parameterize the agreement between two data subsets is given by the Parameter Goodness of Fit~\cite{MaltoniSchwetz}.  One performs separate fits on the two underlying subsets as well as the full data set, resulting in three $\chi^2$ values; for instance,
$\chi^2_{\text{app}}$, $\chi^2_{\text{dis}}$ and $\chi^2_{\text{glob}}$ for a division along appearance and disappearance data sets.   One then defines an effective $\chi^2$,
\begin{equation}
\chi^2_{\text{PG}}=\chi^2_{\text{glob}}-(\chi^2_{\text{app}}+\chi^2_{\text{dis}}),  \label{chi2pg}
\end{equation}  
with an effective number of degrees of freedom:
\begin{equation}
N_{\text{PG}}=(N_{\text{app}}+N_{\text{dis}}) - N_{\text{glob}},~   \label{npg}  
\end{equation}
where $N_{\text{app}}$, $N_{\text{dis}}$, and $N_{\text{glob}}$ are the number of degrees of freedom for the appearance, disappearance and global data sets, respectively.
These are then interpreted as a $\chi^2$ to obtain a probability.   This ``PG Test'' probability is used to define the underlying ``tension'' between the two data sets.

This scenario is a natural way to divide the data sets in a 3+1 model.   One can see that $|U_{e4}||U_{\mu 4}|$ can be extracted either from measurements of $\nu_\mu \rightarrow \nu_e$ appearance (Eq.~\ref{mue}) or from the combination of electron and muon disappearance data sets (Eqs.~\ref{ee} and \ref{mumu}).   Therefore, it is customary to apply the PG Test to the appearance and disappearance subsets when testing 3+1 global fits.    The state of matters for the past five years is that the PG Test probability is small ($\lesssim 10^{-5}$) for this comparison \cite{Dentler:2018sju}.   We investigate this further in this review.

There are several possible explanations for the tension:
\begin{enumerate}

\item  There are no sterile neutrinos.   In this case, all of the data sets must suffer from biases, and those biases accidentally match the effect of adding a sterile neutrino.  

\item There is one stable sterile neutrino as described in a 3+1 model.  In this case, a few data sets must be suffering from unknown experimental effects.   As we will discuss in this review, in the appearance data sets, the MiniBooNE results are systematics-limited, and those systematic uncertainties may not represent a perfect description.    Also, we note that data sets with limits are generally not examined closely. 

\item There are sterile neutrinos, but the model is more complex than 3+1.   While 3+1 is the simplest model, it also seems highly artificial.  Why would there be only one sterile neutrino contributing?   Why would the sterile neutrino be stable, and not decay?  

\end{enumerate}
At present, all three explanations are in play, and we explore them in the remainder of this review.

\section{Design of Short Baseline Experiments}

Accessing an oscillation signal region requires selection of neutrino sources that
can produce the flavor of interest, and a detector which can observe such a flavor.    The designer must also select the appropriate
$L/E$ for the parameter space of interest,  and this additionally influences the choice of source and detector.
Large distance-of-travel requires intense sources and large detectors.   The
selected energy range affects the choice of source.  This usually 
leads to a limited range of high-rate interaction channels, and in turn, limits the detector design choices.

In this section, we begin by briefly introducing the neutrino
interactions of interest to most oscillation experiments.   Then, we
discuss options for detectors and sources in light of commonly
employed selections of $L$ and $E$.

\subsection{Accessible Flavors and Interaction Modes}

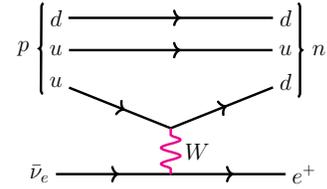
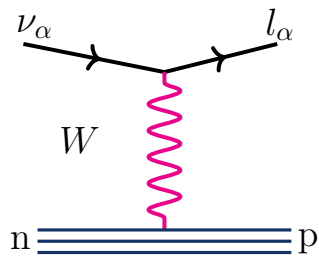
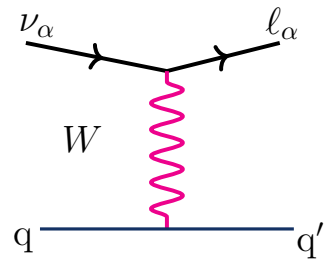
\begin{figure}
\centering
\begin{subfigure}[b]{0.45\columnwidth}
\centering
\includegraphics[width=1.2\textwidth]{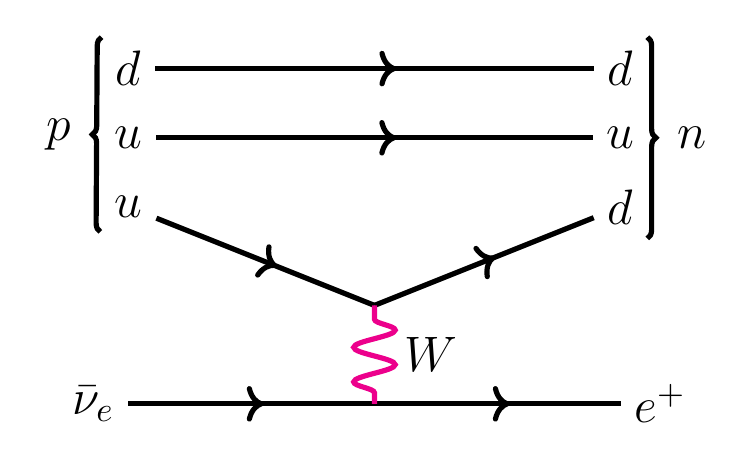}
\caption{Inverse beta decay--electron antineutrino scattering from a free proton.}
\end{subfigure}
\vskip\baselineskip
\begin{subfigure}[bt]{0.45\columnwidth}
\begin{tikzpicture}[line width=1.5 pt, scale=0.75]
		\node at (-0.8, 1.8) {\Large $\nu_\alpha$};
		\draw[fermionout] (-1.0,1.5) -- (1.5,1);
		\node at (3.55, 1.85) {\Large $l_\alpha$};
		\draw[fermionout] (1.5,1) -- (3.5,1.5);
		\draw[gaugeboson] (1.5,1) -- (1.5, -1.8);
		
		\node at (0.0, -0.2) {\Large $W$};
		
		\node at (-1.05, -2.00) {\Large $\mathrm{n}$};
		\node at (4.05, -2.00) {\Large $\mathrm{p}$};
		\draw[fermion] (-0.75,-1.8 ) -- (3.75,-1.8);
		\draw[fermion] (-0.75,-2.0 ) -- (3.75, -2.0);
		\draw[fermion] (-0.75,-2.2 ) -- (3.75,-2.2);
\end{tikzpicture}
\caption{Quasi-elastic neutrino-nucleon scattering from a nuclear target.}
\end{subfigure}
\vskip\baselineskip
\begin{subfigure}[bt]{0.45\columnwidth}
\begin{tikzpicture}[line width=1.5 pt, scale=0.75]
		\node at (-0.8, 1.8) {\Large $\nu_\alpha$};
		\draw[fermionout] (-1.0,1.5) -- (1.5,1);
		\node at (3.55, 1.85) {\Large $\ell_\alpha$};
		\draw[fermionout] (1.5,1) -- (3.5,1.5);
		\draw[gaugeboson] (1.5,1) -- (1.5, -1.8);
		
		\node at (0.0, -0.2) {\Large $W$};
		
%		\node at (-1.05, -2.00) {\Large $\mathrm{N}$};
%		\node at (4.05, -2.00) {\Large $\mathrm{X}$};
		\node at (-1.05, -2.00) {\Large $\mathrm{q}$};
		\node at (4.05, -2.00) {\Large $\mathrm{q^\prime}$};
		\draw[fermion] (-0.75,-1.8 ) -- (3.75,-1.8);
%		\draw[fermion] (-0.75,-2.0 ) -- (3.75, -2.0);
%		\draw[fermion] (-0.75,-2.2 ) -- (3.75,-2.2);
\end{tikzpicture}
\caption{Deep-inelastic scattering diagram, where $q$ and $q^\prime$ are quarks.}
\end{subfigure}
\caption{Some charged-current scattering diagrams relevant for this review.}
\label{feynman}
\end{figure}

Observation of neutrino interactions usually makes use of
charged-current (CC) interactions, which allows observation of the
outgoing lepton flavor.    Figure~\ref{feynman} shows the Feynman
diagrams for CC interactions that will be discussed in this review.  See Ref.~\cite{formaggio-zeller} for
a full review of neutrino interactions from low to high energies.

The lowest energy purely CC interaction that
is commonly employed is inverse beta decay (IBD),  which is electron antineutrino scattering from a free proton,  $\bar \nu_e + p
\rightarrow e^+ + n$, as shown in Fig.~\ref{feynman} (top).   It is called IBD because it is a transformation (a crossing-symmetry diagram) of neutron beta decay,  $n \rightarrow p + e^- + \bar \nu_e$.   
Making use of the relationship to neutron decay, which has a very well-determined lifetime,  
the IBD cross section for this interaction is predicted to $0.2\%$ \cite{IBD1, IBD2}.

Along with the well-determined cross section, there are several other reasons that IBD is a
popular interaction mode for oscillation studies.
It is easy to construct a target of free
protons---one can use water, oil, or plastic, for example.    For a free proton
target,  the energy threshold is very low, at
1.8~MeV, which arises from the mass of the positron, 0.5~MeV, and
the mass difference between the proton and the neutron of 1.3~MeV.
The
experiment can be designed such that the capture of the outgoing free
neutron may be detected.  As a result, the IBD interaction allows a
time-coincidence signal of initial interaction followed by capture,
greatly reducing backgrounds.   

At low energies, neutrino interactions are generally suppressed as Pauli
blocking prevents the conversion of a neutron into a proton.   For
example, in commonly used carbon-base targets, the threshold for 
$\nu_e + \textrm{C} \rightarrow e^-+ \textrm{N}$ is 17~MeV.  An important exception to
this will be gallium, which has a 233.2 keV threshold for 
$\nu_e + \textrm{Ga} \rightarrow e^-+ \textrm{Ge}$. 

As we look to higher energies,  the four momentum transfer can become large
enough to knock the nucleon outside the nucleus.    For argon targets -- used in new, state-of-the-art detectors -- the 
binding energy, that must be overcome
to free the proton in a neutrino interaction, is about 40~MeV.    The
threshold is even higher for carbon, which is a tightly bound
nucleus.   When a single nucleon is knocked out of the target,  
$\nu_e + n \rightarrow e^- + p$,  the interaction is called Charged
Current Quasi-Elastic (CCQE) scattering.   CCQE can also be observed
with high energy antineutrinos, where the target is the protons in
the nucleus.

Muon flavor CCQE scattering, shown in Fig.~\ref{feynman} (middle), is
  employed in several of the experiments we discuss in this 
  review.
  The threshold of around 150~MeV, depending on the target, is
  due to a combination of the binding energy and the muon mass, which
  is 106~MeV.    In principle, this has a very clean signature of
  one outgoing muon and one outgoing proton ($1\mu 1p$). However, this is complicated by interactions between the struck proton and other nucleons during its exit from the nuclear medium.  This leads to a
  cross section that is only known to about 15\%, depending on the
  nuclear target \cite{MBCCQE, MINERVACCQE}.  

A complex set of resonances
contribute to the interactions between around 500~MeV to 20~GeV.  Neutrino cross sections in this region are
difficult to predict and precision measurements are difficult to
obtain.  As a result, the neutrino community is collaborating with the
electron-scattering community at Jefferson Laboratory on a set of experiments
that will better constrain this region.  For example, Ref. \cite{JeffeAr}
reports the first electron-argon differential cross section, at a beam energy of 2.2~GeV.
However, until a full suite of these results are obtained,  oscillation experiments
are generally avoiding the use of this ``Resonance Region'' if possible.

Unfortunately, experiments that are focused on CCQE interactions often
have beam energies that extend into the Resonance Region, and this
produces backgrounds.   A background of particular concern, as we will
discuss later, comes from neutral current (NC) production of the $\Delta$
resonance, rather than CC production of this resonance.  The $\Delta$
can decay to a $\pi^0$ and a proton, and, rarely, to a single photon and
proton.   If the photon is misidentified as an electron, then this
interaction can fake a $\nu_e$ CCQE scatter.

Above 20~GeV, the neutrinos carry sufficient energy to resolve the
quarks within the nucleon ($\nu_\mu + d \rightarrow \mu^- + u$).
This is called Deep Inelastic Scattering or ``DIS,'' shown in  Fig.~\ref{feynman}
  (bottom).    DIS can also be observed using antineutrinos,  where the
  target is the $u$ quarks in the nucleon.     The cross section for
DIS rises linearly with energy and is known to
2\% \cite{PrecisionDIS}.    As a result, this kinematic region represents
an excellent option for precision studies of neutrino oscillations,
assuming that the appropriate matching $L$ is also feasible.  

While we have described interactions of electron-flavor and muon-flavor
neutrinos, we have not discussed tau-flavor.   The reason is that CC interactions
involving the $\tau$  are highly suppressed due to its 1.8~GeV mass.
Even at 100~GeV, the ratio of $\nu_\tau$ interactions to $\nu_\mu$
interactions is only around 80\% due to this mass suppression.   
Therefore, it is relatively rare for an oscillation experiment to
employ $\nu_\tau$ CC interactions as a signature. 

\begin{figure}[t]
\begin{center}
{\includegraphics[width=\columnwidth]{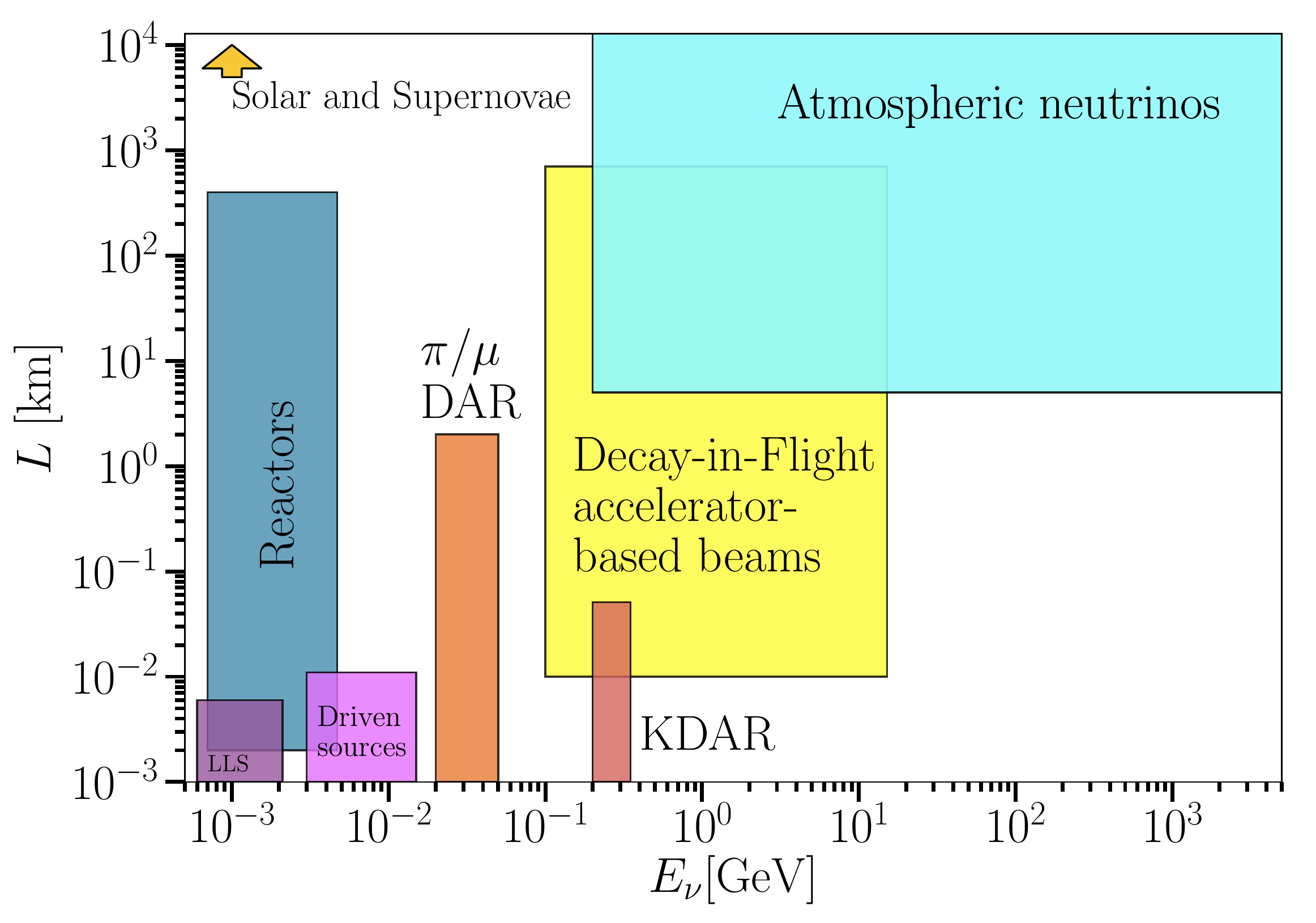}}
\end{center}
\caption{An illustration of the commonly-used neutrino sources for a
  given $L$, distance from neutrino production to detector, and $E$,
  energy of the neutrino. LLS stands for Long-lived sources and KDAR for kaon decay-at-rest.
\label{sourceplane}}
\end{figure}

\subsection{Neutrino Sources \label{sources}}

Because the characteristic experimental parameters of an oscillation
search are $L$ and $E$, let us consider the range of possible neutrino
sources in an $L$ {\it vs.} $E$ plane, illustrated in 
Fig.~\ref{sourceplane}.   The range of $L$ is limited by the flux that
can be directed toward the detector.   In this review, we 
will consider only neutrino fluxes from weak decays of mesons and
baryons, but we note that fusion in the sun and dying stars can also
produce measurable fluxes on earth.    The $E$-range divides into
two regimes: decay-at-rest sources at lower energies and
decay-in-flight sources at higher energies. 

\subsubsection{Low Energy $\nu_e$ and $\bar \nu_e$ Fluxes \\ (Radioactive and Reactor Sources) \label{lowEflux}}

The shortest $L$ and lowest $E$ neutrino fluxes discussed in this
review are from  artificially created megaCurie sources of $^{51}$Cr
and $^{37}$Ar.  These have 27.8~day and 35.0~day half-lives,
respectively.   This makes their use complicated, as the
highly radioactive source must be quickly and safely brought from the reactor site
 -- where it is made -- to the detector.   These isotopes both 
decay exclusively by electron capture with a total decay energy
of $Q_{EC} = 753$ keV ($^{51}$Cr) and $Q_{EC} = 814$ keV ($^{37}$Ar) 
and so they must be paired with a gallium target. 

The short half-life can be addressed through replenishing the isotope
using an accelerator-driven system.      An example of this is the
IsoDAR source which uses an accelerator to produce $^8$Li, which decays
in 841~ms.    While concepts like IsoDAR are in development, such a
source has not been constructed.      We explore this further with a
discussion of next-generation experiments in Sec.~\ref{DAR}.

\begin{figure}
\includegraphics[width=\linewidth]{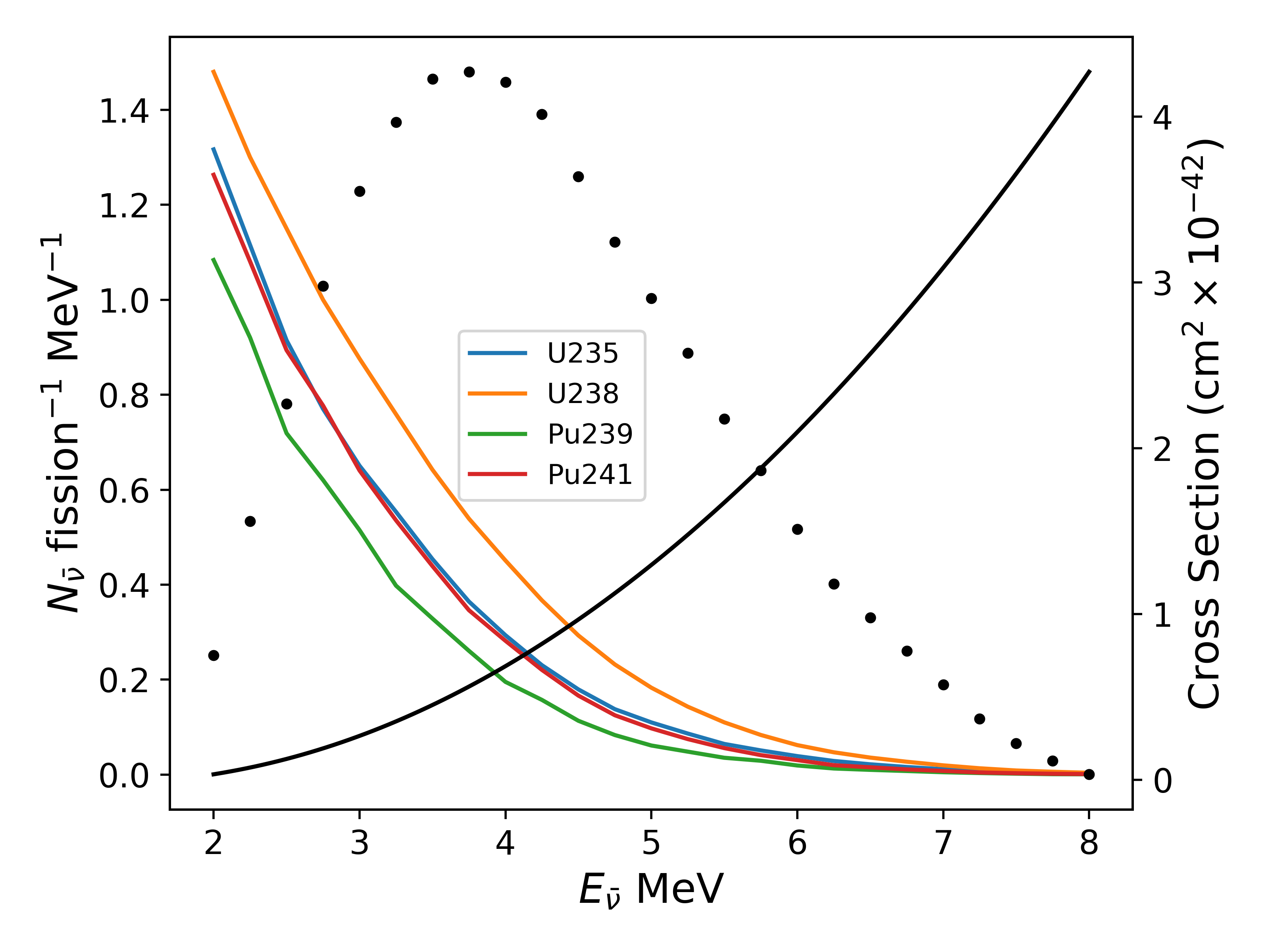}
\caption{Decreasing lines: average number of antineutrinos created per fission per MeV for each fuel component with scale on left y-axis. Black line: IBD cross section versus neutrino energy with scale on right y-axis. The black dots: convolution of the antineutrino flux (assuming equal fuel contributions) and IBD cross section, shown in an arbritrary scale.}
\label{fig:ReactorFlux}
\end{figure}

Reactors are the primary source of low energy antineutrinos for
oscillation experiments today.   Low energy $\bar \nu_e$ are copiously
produced through the $^{235}$U and $^{239}$Pu decay chains.  In fact,
most of the antineutrinos are produced below the IBD threshold, and
so they are not used for the physics discussed here.   When the rising IBD
cross section is combined with the falling reactor flux,  one obtains
an energy spectrum of events that peaks at around 3~MeV, as shown in Fig.~\ref{fig:ReactorFlux}.
Note that while the $x$-axis is neutrino energy, it is often the case that  
reactor experiments present observed spectra as a function of prompt energy, which is $0.8$~MeV lower than the neutrino energy.

Years of effort have gone into predicting the reactor flux.
In 2011, a series of studies \cite{reactor1, Mueller:2011nm, Huber} revisited the
absolute prediction of reactor fluxes, updating 20-year-old cross sections with modern data.   The surprising result was the
Reactor Antineutrino Anomaly (RAA)---a shift in the
predicted reactor flux with respect to measurements that could be
interpreted as a sterile neutrino signal. At this point, the highest precision result comes from a combined analysis of Daya Bay and RENO \cite{GiuntiRAA} which finds an overall rate of data compared to prediction of $0.927 \pm 0.016$ for the weighted averaged of the two experiments and the two isotopes. However, attributing this to sterile neutrinos is complicated
by two further observations.

First, a few percent excess in the reactor visible energy spectrum is observed at 5~MeV.   We discuss the experiments that observe this excess in detail in Sec.~\ref{bump}.    The results of the RENO experiment show that this excess 
scales with both reactor power \cite{RENO5MeV} and with the U-235 content of the core \cite{RENOvenice}. At present, the source of the 5~MeV excess is far from resolved.

\begin{figure}[t]
\begin{center}
{\includegraphics[width=\linewidth]{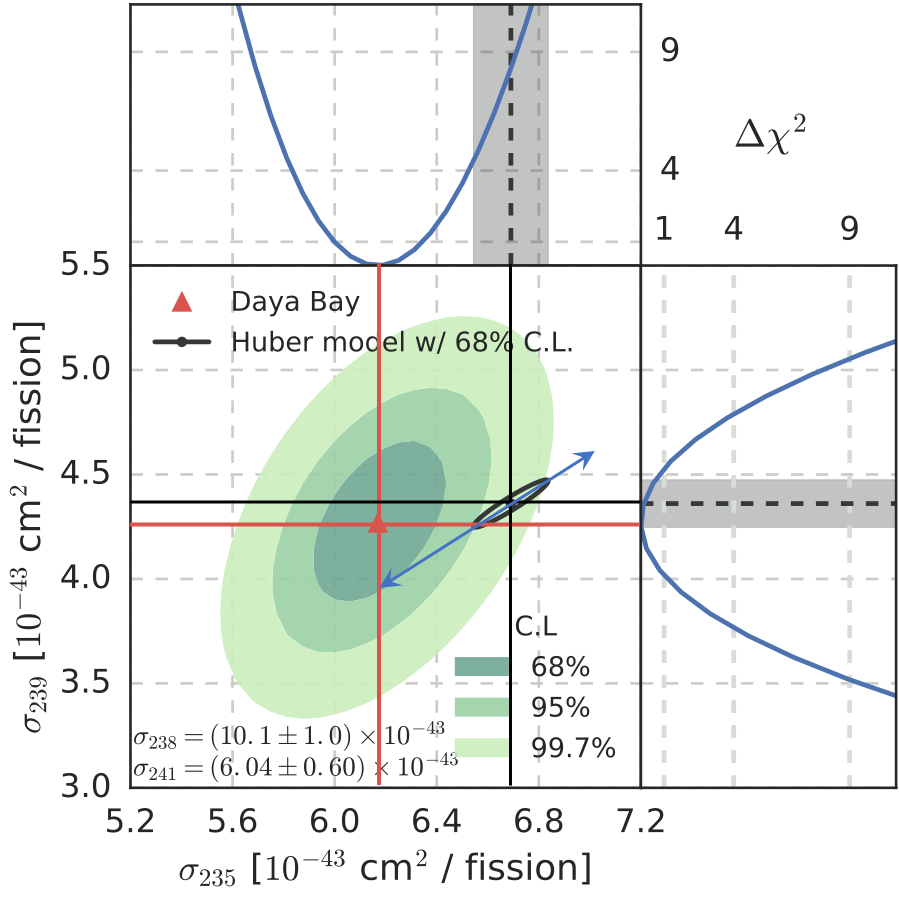}}
\end{center}
\vspace{-0.1in}
\caption{Daya Bay measurement of the $^{235}$U and $^{239}$Pu cross sections averaged over the reactor flux, shown in green.   Huber prediction is shown by black point, with 1$\sigma$ on prediction shown in oval.   Red lines project the central value of the measurement and black lines project the central value of the prediction. The blue line indicates the line along which the prediction moves if sterile neutrinos are included.  Plot modified from Ref.~\cite{DBflux}.
\label{DBUPu}}
\end{figure}

Second, Daya Bay
has shown that an alternative explanation of the RAA is an incorrect prediction of $^{235}$U antineutrino production rate
for power reactors \cite{DBflux}.  This analysis makes use of the
fact that, as a power reactor burns fuel, the relative fission rate of $^{235}$U and $^{239}$Pu changes with time, as well as the fact that 
the $^{235}$U antineutrino flux has a different energy
dependence than the $^{239}$Pu flux. Keeping in mind that the IBD
cross section rises with energy, if one had a source that
was entirely due to antineutrinos from  $^{235}$U, then the average IBD cross
section would be $(6.17\pm 0.17) \times 10^{-43}$ cm$^2$/fission, while it
would be $(4.27 \pm 0.26) \times 10^{-43}$ cm$^2$/fission for $^{239}$Pu.  
Thus, the two
average cross sections can be extracted through a time dependent study that accounts for the relative fission rates.
Fig.~\ref{DBUPu} shows the Daya
Bay result (green), compared to a recent model from Huber, {\it et al}
(black) \cite{Huber}.    We overlay two sets of lines: the black vertical
and horizontal 
lines guide the eye to the central value of the production cross sections by
Huber, while red lines project the central value of the Daya Bay measurement.   One sees that the  $^{239}$Pu crosses the Daya Bay measurement
well within 1$\sigma$, but the  $^{235}$U prediction is outside
the 2$\sigma$ region of the Daya Bay measurement.    On the other hand, the
blue tilted line indicates how a combined  $^{239}$Pu and  $^{235}$U
average cross sections prediction will vary in the presence of sterile
neutrinos.   
This blue line intersects the Daya Bay result at 
1$\sigma$ for an 8\% reduction of both the $^{235}$U and $^{239}$Pu cross section due to a sterile neutrino. Therefore, the two possible explanations--a problematic cross section prediction for $^{235}$U or the existence of sterile neutrinos--cannot be distinguished in this data set.

In response to these issues concerning the interpretation of the absolute-normalization-based RAA as due to sterile
neutrinos, recent reactor experiments have moved to near-far detector
ratios that sample identical fluxes (up to solid angle effects) in the absence
of oscillations. This is a classic approach, which usually employs a single moving detector or two detectors located at different
locations. In the case of reactor experiments, where the $L$ that must be spanned is relatively short,  one can also use a single long detector. In our global fits, we have chosen to include the ratio-based results, and to not include the absolute-normalization-based results. 

\subsubsection{Fluxes from Meson DAR \label{DARsource}}

Moving up in energy, high rates of fluxes in the 20~MeV to 50~MeV
range can be achieved through the pion to muon decay at
rest ($\pi/\mu$ DAR) sequence.   
An accelerator is used to produce $\pi^+$ which comes to a stop in a
target and decays to $\mu^+ + \nu_\mu$.   This is a two-body decay
that produces a monoenergetic, 29~MeV, $\nu_\mu$.   
The muon from the stopped pion decay will also come to a stop and
decays to  $e^+ + \bar \nu_\mu + \nu_e$.   This is a weak decay with a
very well defined energy spectrum for the $ \bar \nu_\mu$ and $
\nu_e$, with an endpoint at 52~MeV.

The $\pi/\mu$ DAR flux produces a negligible amount of $\bar \nu_e$.
Electron-antineutrinos are not produced in the $\pi^+$ initiated
chain and the $\pi^-$ chain is highly suppressed for several reasons.
First, the proton beam energy is typically chosen to be 800~MeV, which
highly suppresses $\pi^-$ production compared to $\pi^+$.   Second,
heavy targets are used, which leads to fast $\pi^-$ and $\mu^-$
capture.   As a result, the typical rate of $\bar \nu_e$ intrinsically
produced in a $\pi/\mu$ DAR flux is 0.01\% of the  $\bar \nu_\mu$ flux.
The combination of this low intrinsic background and the well
understood energy spectrum makes  $\pi/\mu$ DAR ideal
for oscillation studies.   We will refer to experiments using this source often in our
following discussion.

The first step of $\pi/\mu$ DAR offers a source of mono-energetic
neutrinos, which would be ideal for the study oscillations
in a detector that can move, with the exception that 29~MeV $\nu_\mu$
is below CC threshold.   In principle, there is an NC interaction --
coherent neutrino scattering -- which has been observed and can be employed to sample this
flux, but in practice this has never been demonstrated.    In
Sec.~\ref{DAR}, we describe the future Coherent Captain Mills
experiment, which may exploit this signal.   

A second source of mono-energetic neutrinos comes from the two-body K$^+ \rightarrow \nu_\mu + \mu^+$
decay.  Kaon decay at rest (KDAR) produces a 236~MeV 
$\nu_\mu$ flux that was recently observed in CCQE interactions by the
MiniBooNE experiment \cite{KDAR}.   This enables interesting future sterile neutrino searches that we discuss in Sec.~\ref{DAR}.

\subsubsection{Fluxes from Meson DIF \label{DIFflux}}

To reach higher energies, one can use decay-in-flight (DIF) of pions and
kaons.   In most designs, a magnetic ``horn'' is introduced to select
the charge of the meson, so that a relatively pure neutrino or
antineutrino beam will be produced.   As an example, 
in the case of the Booster Neutrino Beam (BNB) at Fermi National Accelerator Laboratory, neutrino running yields 
a $\nu_\mu$ beam content of 93.6\%, a $\bar \nu_\mu$ content of 5.9\%, and a $(\nu_e + \bar \nu_e)$ content of 0.5\% \cite{BNBflux}.   These are fairly typical numbers.   For comparison, the planned DUNE beam will have a 90.6\% $\nu_\mu$, 8.6\% $\bar \nu_\mu$, and 0.8\% $(\nu_e + \bar \nu_e)$ content~\cite{Strait}.
In antineutrino running, these beams tend to be somewhat less clean.  For example, the BNB beam has an 83.7\% $\bar \nu_\mu$, 15.7\% $\nu_\mu$, and 0.6\% $(\nu_e + \bar \nu_e)$ content \cite{BNBflux}.
The energy distribution and the intrinsic $\nu_e$ content of DIF beams is difficult to predict {\it
  ab initio}.    In Sec.~\ref{MBcomment} we describe how the systematic
uncertainty from this can be controlled in a $\nu_\mu \rightarrow
\nu_e$ search through the use of a well-measured $\nu_\mu$ flux as a constraint. 

Atmospheric neutrinos are also produced through DIF.   In this case,
high-energy cosmic rays hit nuclei in the Earth's atmosphere, producing
mainly pions and kaons, which decay to result in a combined neutrino and
antineutrino flux.  The most
famous atmospheric-based oscillation studies, such as the Super Kamiokande results that led to the 2015 Nobel
Prize \cite{SKatmos}, used interactions from 500~MeV to the few
GeV range.    However,  atmospheric neutrino production extends to very high
energies.   We will briefly discuss results from the IceCube
experiment that make use of TeV-energy neutrino interactions.    

\subsection{Neutrino Detectors}

\begin{figure}[t]
\begin{center}
{\includegraphics[width=\columnwidth]{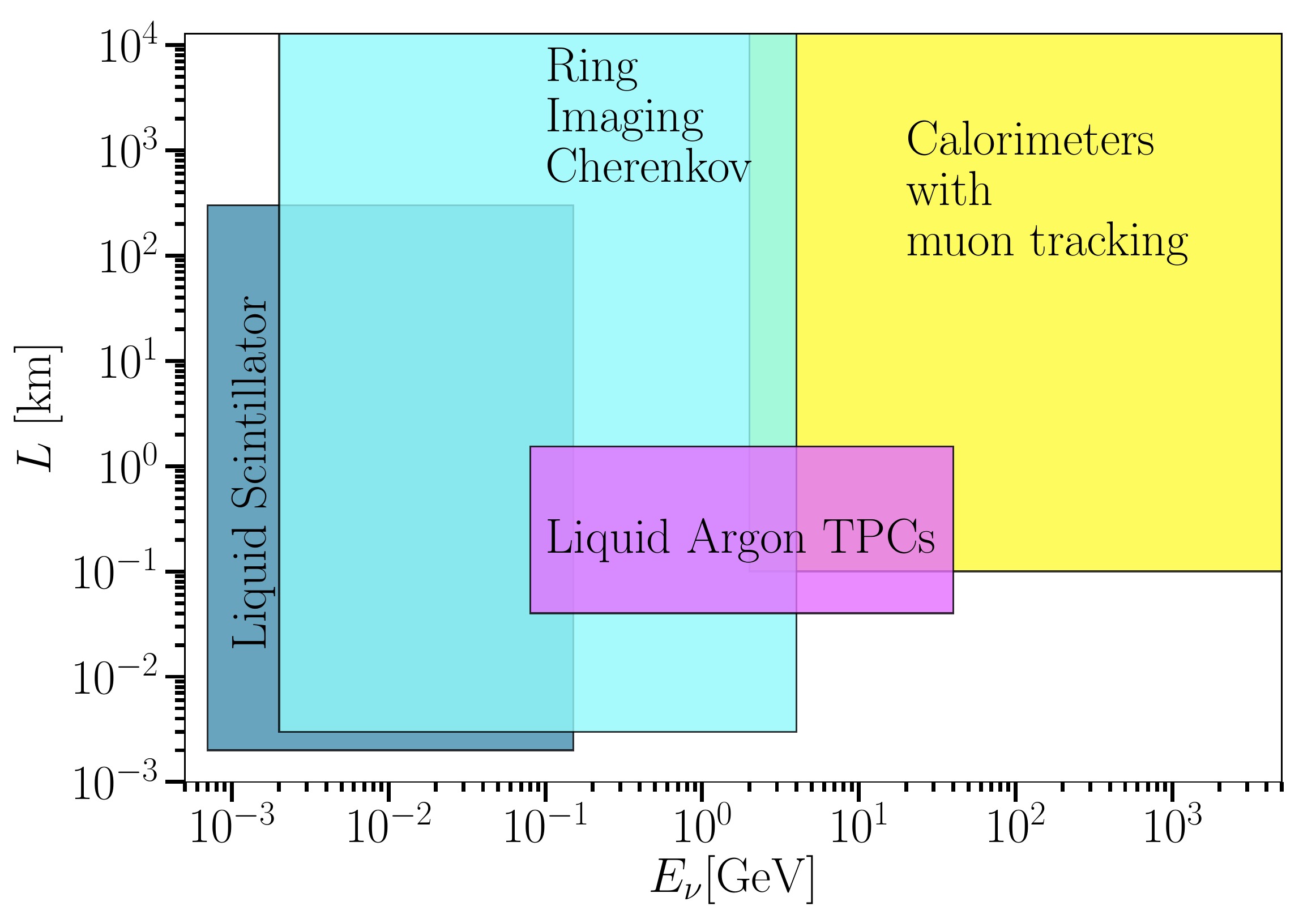}}
\end{center}
\caption{An illustration of the commonly-used neutrino detectors for a
  given $L$, distance from neutrino production to detector, and $E$,
  energy of the neutrino.}
\label{detectplane}
\end{figure}

There are a few common modes for detection of neutrinos in oscillation
searches.    As with any particle physics experiment, there is 
a trade-off between increasing the detector volume and making the detector
more precise.   

\subsubsection{Detectors for Energies below $\sim 20$ MeV}

Detectors at low energies typically make use of the IBD interaction
as the rate is high and the time coincidence of the initial
interaction followed by the neutron capture allows for rejection of
backgrounds.      Hydrocarbon-based scintillator is ideal because it contains many
free-proton targets per ton and it typically produces around
10,000 visible photons per MeV of deposited energy.
The IBD interaction produces a positron, which will stop and
annihilate on an electron in the scintillator.   Thus, the initial signal
is the positron kinetic energy plus the Compton scattering deposited energy of two 511 keV $\gamma$-rays
that are produced by the annihilation.
The protons of the scintillator can also provide a target for the neutron capture, where
a 2.2~MeV $\gamma$ is released, which subsequently Compton scatters to
produce the coincidence signal.

Most often scintillator based detectors are monolithic tanks of scintillator
oil, surrounded by photomultiplier tubes (PMTs), as this is the
cheapest design per ton.   The tanks may have a
buffer region of undoped scintillator between the PMTs and the active
region to prevent radioactive decays in the PMT glass from producing background in the
detector.     Scintillator experiments located close to a source, where
the flux is large, may use detectors that are segmented into
scintillator ``bars.''    This provides better spatial information for
reconstructing event positions,  allowing improved rejection of backgrounds.  
Segmented detectors are often constructed of solid
scintillator, since this avoids leaks,  but in a few cases, such as the KARMEN experiment that
we will discuss later,  the bars are scintillator-oil filled.

Although the neutron capture cross section on hydrogen is relatively
large, certain elements offer much higher neutron capture cross
sections.     If these materials -- such as gadolinium or lithium -- can
be introduced into the scintillator, then the neutron capture time
will be faster, reducing backgrounds from random coincidences.  The
typical neutron capture time of a hydrogen-based detector is $\sim 200$
microseconds, while a detector with a less than 1\% gadolinium
introduced into the scintillator has a
capture time of $\sim 30$ microseconds.   In the early
2000's, a great deal of R\&D was performed to allow gadolinium to be
mixed into the scintillator oil without problematic reactions, such as
the oil turning yellow \cite{BNLGd, DCGd}.    In fact, several early
experiments, including the Chooz experiment, suffered from this
effect.    Modern gadolinium doped scintillators are still regarded as
fragile and must be handled with care.     Lithium has pros and cons
with respect to gadolinium.    As a substantial pro, neutron capture on $^7$Li leads to a
decay involving two alphas that produce light at a well-defined
position; in comparison, gadolinium produces multiple photons, for which the Compton
scatter may occur over many centimeters or exit the detector
entirely.   On the other hand, lithium is more expensive than gadolinium. 

\subsubsection{Detectors for $\sim 10$ MeV to $\sim 1$ GeV}
% GC mark

In the $\sim 10$ MeV to $\sim 1$ GeV range, Cherenkov detectors have a number of attractive features above scintillation detectors.  First, at these energies, protons are generally below Cherenkov threshold, yielding a clean lepton signal for the neutrino interactions.   As we discuss below, the muon and the electron can be distinguished by qualities of the Cherenkov ring.  Also, a Cherenkov detector provides information on the direction of the
outgoing lepton, thus giving the angle with respect to the beam, if the beam direction is known.

Cherenkov detectors are
usually constructed of water, and, in some
cases, pure mineral oil (or
mineral oil very lightly doped with a scintillator).    Oil has a few
advantages for small detectors.    Oil has a slightly larger opening angle for the Cherenkov
ring than water; the PMT high voltage needs no protection because oil
is an insulator; the energy threshold for Cherenkov radiation is lower since oil has a larger index of refraction than water; and a purification system is not needed.   However,
oil costs more per ton, and at about 1 kton, there is a crossing point
where the cheapness of water outweighs the advantages of oil.

Additionally, the flavor
of the outgoing charged lepton can be determined from the topology of the emitted Cherenkov light.     As a result, for higher energy
oscillation experiments where electrons must be distinguished from muons,
Cherenkov detectors are preferred.    Because electrons have a mass
which is 200 times smaller than a muon, they will suffer more multiple
scattering and radiation and produce a ``fuzzy'' ring compared to the
well-defined muon ring.
For the CCQE interaction, given a
well-defined beam direction, one can reconstruct the neutrino energy
from the $\theta$ angle of the track inferred from the Cherenkov ring, 
and the energy, $E_\ell$, of the electron or muon derived from the visible energy seen in the PMTs.
If we define $\ell=e$ or $\mu$, $M_n (M_p)$ as the mass of neutron (proton), and  $B$ as the binding energy of the nucleon, and we define $\Delta=M_n-B$, then the neutrino energy is given by:
\begin{equation}
E_{\nu}^{QE}= 0.5  \dfrac{2 \Delta E_{\ell}-(\Delta^{2}+M_{\ell}^{2}-M_{p}^{2})}{\Delta-E_{\ell}+\sqrt{(E_{\ell}^{2}-M_{\ell}^{2})}\cdot \cos\theta_{\ell}}.
\end{equation}

\subsubsection{Detectors for Energies Beyond $\sim 1$ GeV}

At energies of roughly 1 GeV, it has, historically, been cheapest to develop tracking
calorimeters.    These combine drift chambers and segmented scintillators to
reconstruct outgoing muon tracks and showers from electrons and from
the hadronic vertex.    At higher energies, usually a heavy target is
interspersed with the detectors, such that the detector becomes a
``sampling calorimeter.''     Many of these detectors incorporate
magnets that allow the sign of an outgoing muon to be determined, and
also the momentum to be accurately measured from the radius of
curvature of the track.

The liquid argon time projection chamber (LArTPC) is a new detector that is
being introduced to the field of neutrino oscillation physics.    The
results that we discuss in this review do not, as yet, make use of
this device, but future results will (see Sec.~\ref{nearfuture}).
Therefore, we briefly describe the detector here, and refer the reader
to Ref.~\cite{uBdetector}, for a description of a recently constructed
LArTPC.     The device consists of an electric field cage filled with
liquid argon.    When an interaction occurs, the exiting
charged particles ionize the argon and those electrons drift to one side,
due to the electric field, where they are recorded via wire chambers.
The wire chambers provide information on the event in two dimensions,
while information on the
third dimension is determined by the drift time of the electrons.
The interaction time, which determines the start of the drift, may be
known in two ways.  First, 
liquid argon is an excellent scintillator, and so the time of the
interaction is known through detection of this light.  Second, in the case of
beam-based experiments, it may also be known by the timing of the beam
spill.

\subsection{Short Baseline Experiments Implemented in Global Fits}\label{experiments}

In this section, we provide an overview of the experiments implemented into our global fits.  
Table~\ref{table:experiments} is provided below with a list of these experiments, along with their oscillation type.  To orient the reader to the experimental results we provide fits to a two-neutrino oscillation model for each case in Figs~\ref{appfigs}, \ref{disefigs} and \ref{dismufigs}.   The frequentist confidence regions for the 99\%, 95\%, and 90\% are shown in blue, green, and red, respectively. If a contour includes the lower left edge of the plot, then this is an open contour and a limit is shown. The cases that we refer to as having ``signals'' in this review have closed contours at 95\% confidence level. 

\begin{figure*}[t]
\centering
\includegraphics[width=\textwidth]{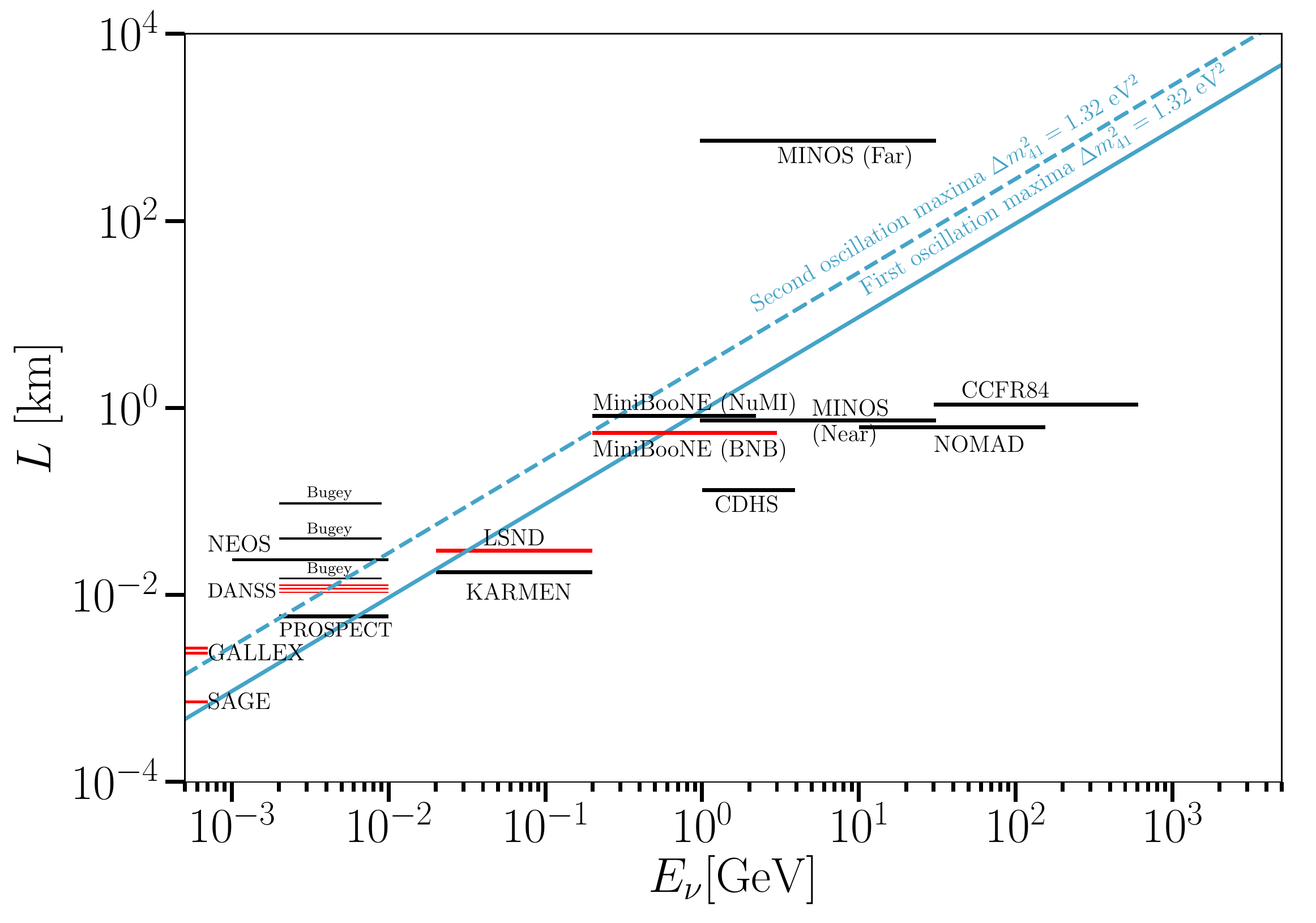}
\caption{The experiments included in this global analysis  shown as a function neutrino energy range and baseline. Red -- those with $>2\sigma$ preference for an additional neutrino state. Blue solid (dashed) line indicates the first (second) oscillation maxima. See text for further discussion.}
\label{fig:all_experiments}
\end{figure*}

These experiments are also organized in Fig.~\ref{fig:all_experiments} where they are shown as a function of the experiment's neutrino energy range and baseline. This figure has been adapted from Ref.~\cite{Diaz:2011ia}, where it was used in the context of Lorentz violation. As marked in Table~\ref{table:experiments}, those with  ``signals'' are indicated  with a red line.    On a $\log$-$\log$ plot of $L$ {\it vs.} $E$, each oscillation maxima for a fixed $\Delta m^2_{41}$ in the 3+1 model will lie on a line.  In solid blue, we indicate the line for the first oscillation maximum of $\Delta m^2=1.32$ eV$^2$, which will be the best fit that we find later in this paper. 
Experiments below this line should see reduced or no oscillation signal, while experiments above the line should see increasingly rapid oscillations.
The second oscillation maximum for the same $\Delta m^2_{41}$ will be offset slightly above, as illustrated by the blue dashed line.   Experiments that do not intercept these lines are not expected to exhibit a signal.   One sees that there is a cohesive picture except for MiniBooNE (NuMI) and PROSPECT.   However, we note that MiniBooNE (NuMI) does exhibit a small excess, as we discuss below, and PROSPECT has not yet garnered sufficient statistics to be sensitive to a potential signal.  Thus, this cartoon presents a coherent picture, overall. 

\subsubsection{Appearance Experiments}
Appearance experiments search for muon flavor neutrinos converting to electron neutrinos.   In the context of two-neutrino global fits, these experiments would be sensitive to the product of the mixing matrix terms $|U_{\mu4}||U_{e4}|$ and $\Delta m_{41}^2$.
In Fig.~\ref{appfigs}, we present two-neutrino ($\nu_\mu \rightarrow \nu_e$) fits to each data set. We fit to the data from the following appearance experiments:

\begin{figure*}[t!]
\centering
\begin{subfigure}{0.44\linewidth}
\centering
\includegraphics[width=\linewidth]{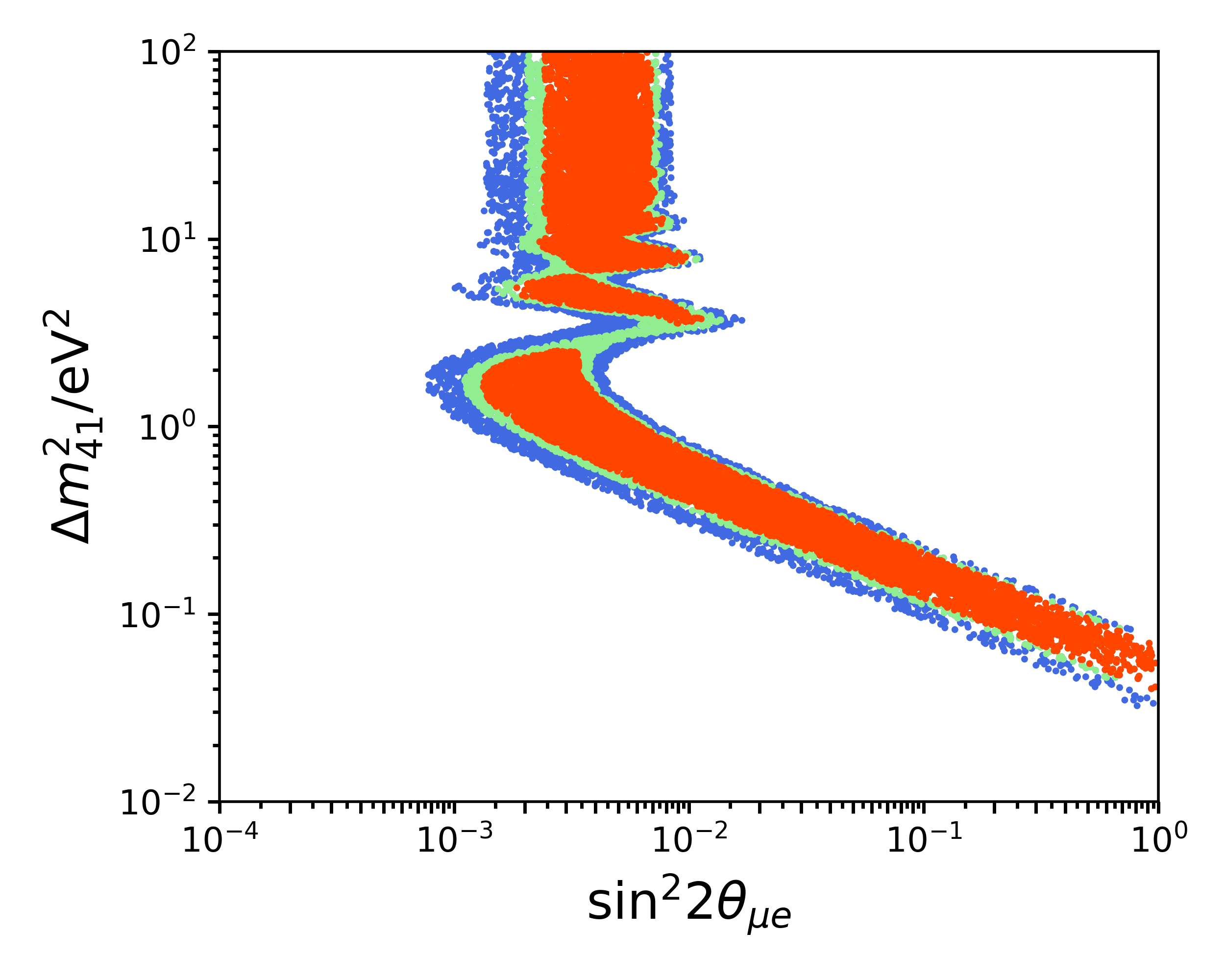}
\caption{LSND}
\end{subfigure}
~ 
\begin{subfigure}{0.44\linewidth}
\centering
\includegraphics[width=\linewidth]{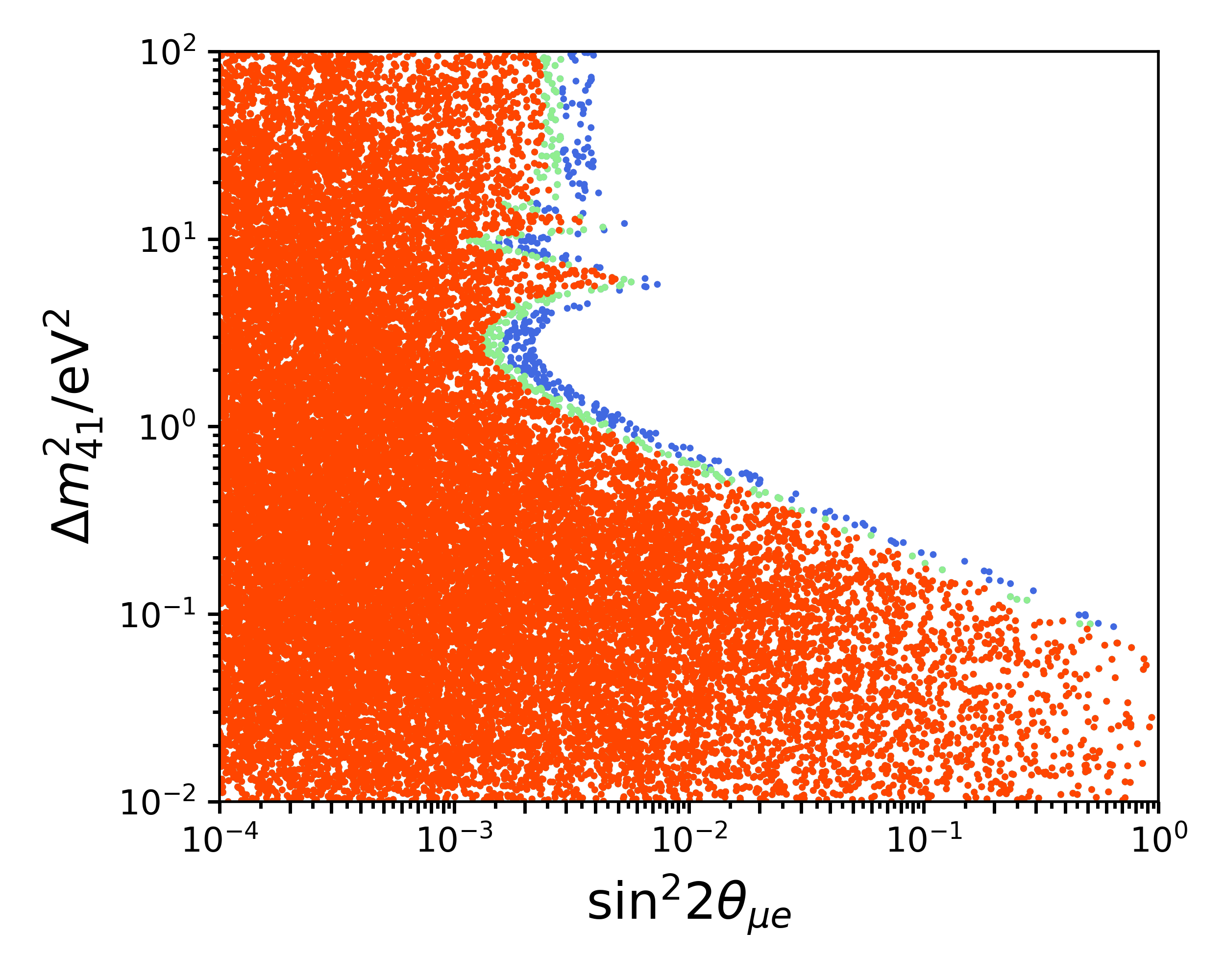}
\caption{KARMEN}
\end{subfigure}

\vspace{\baselineskip}
\begin{subfigure}{0.44\linewidth}
\centering
\includegraphics[width=\linewidth]{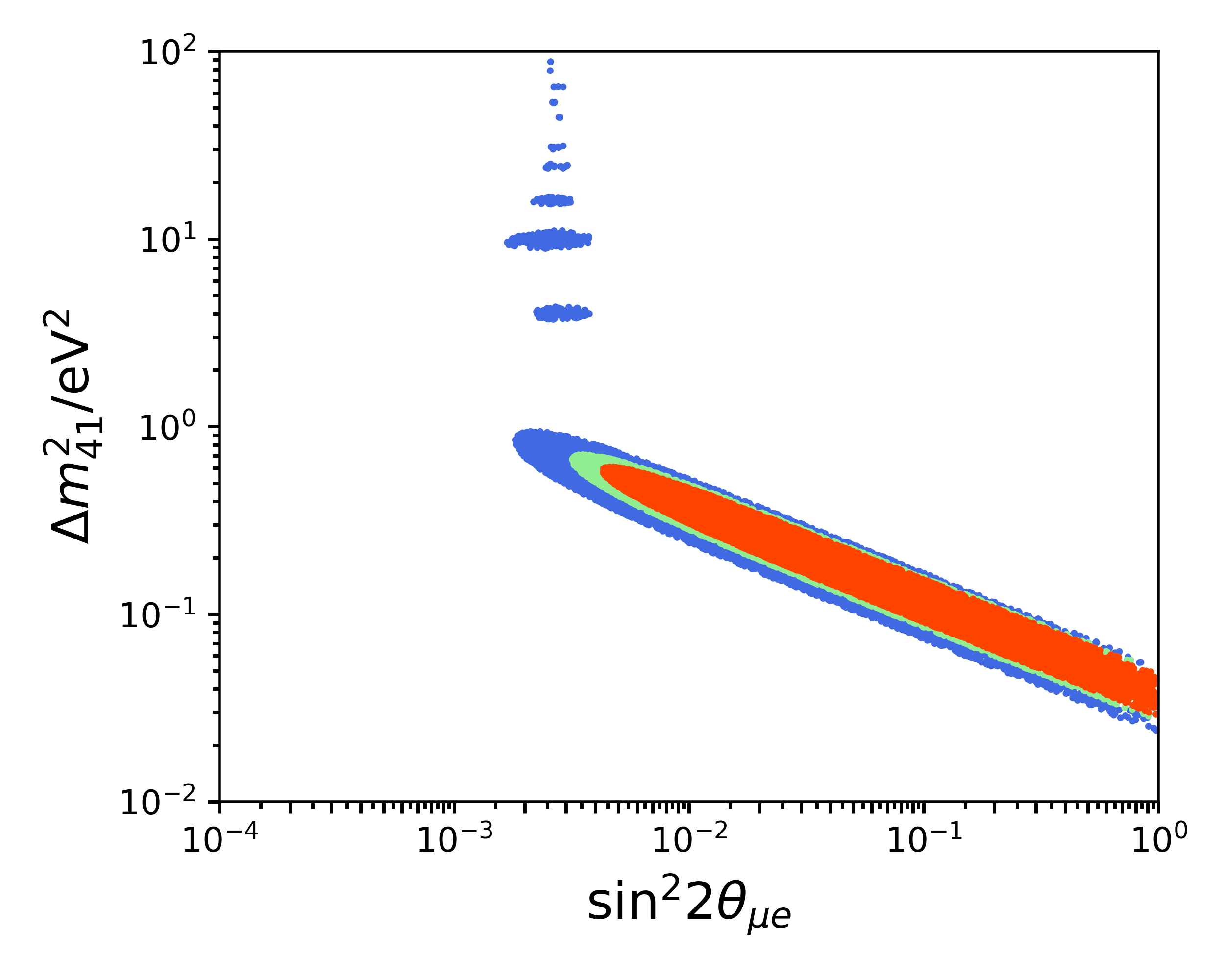}
\caption{MiniBooNE (BNB), $\nu$ and $\bar \nu$}
\end{subfigure}
~ 
\begin{subfigure}{0.44\linewidth}
\includegraphics[width=\linewidth]{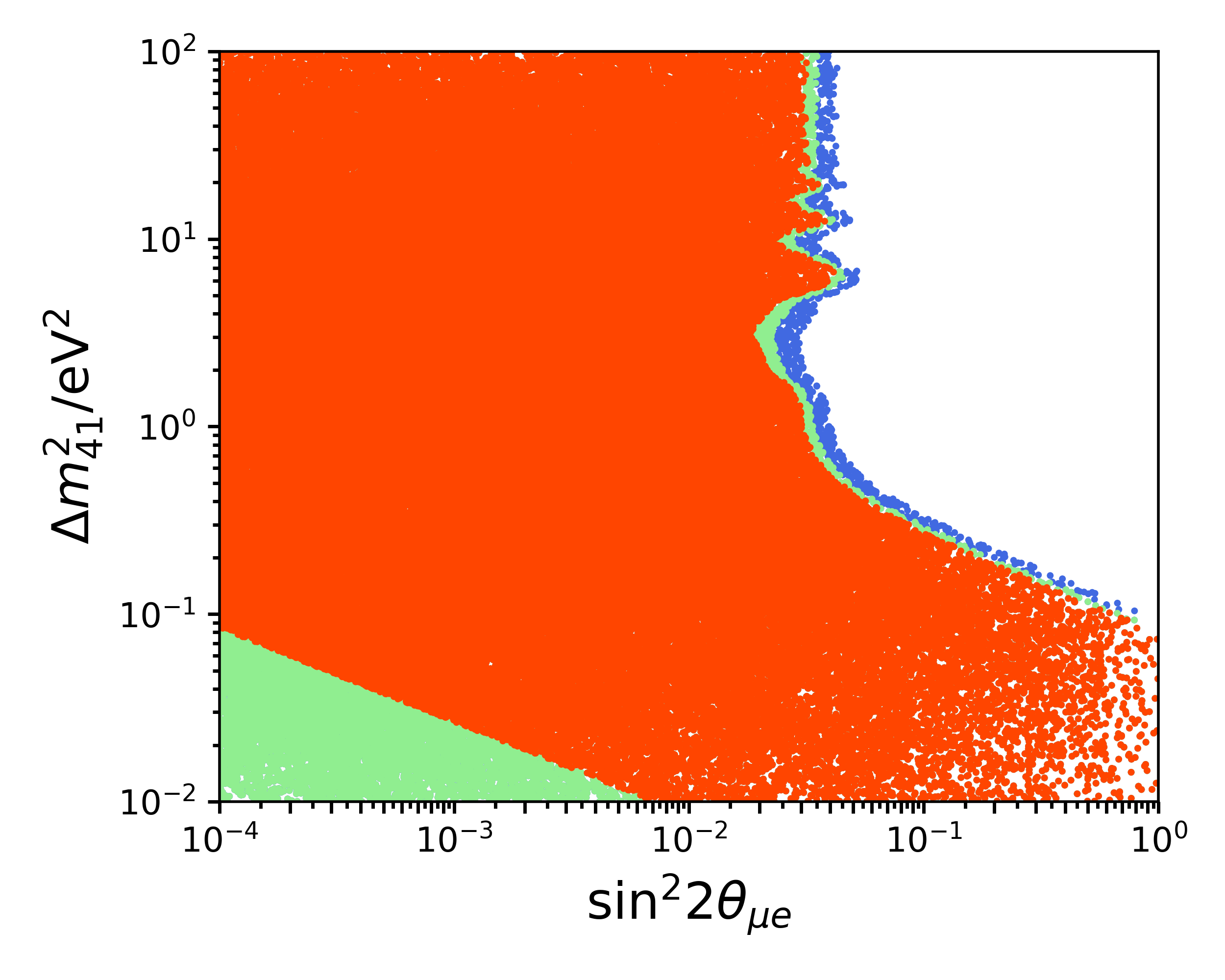}
\caption{MiniBoonE (NuMI)}
\end{subfigure}

\vspace{\baselineskip}
\begin{subfigure}{0.44\linewidth}
\centering
\includegraphics[width=\linewidth]{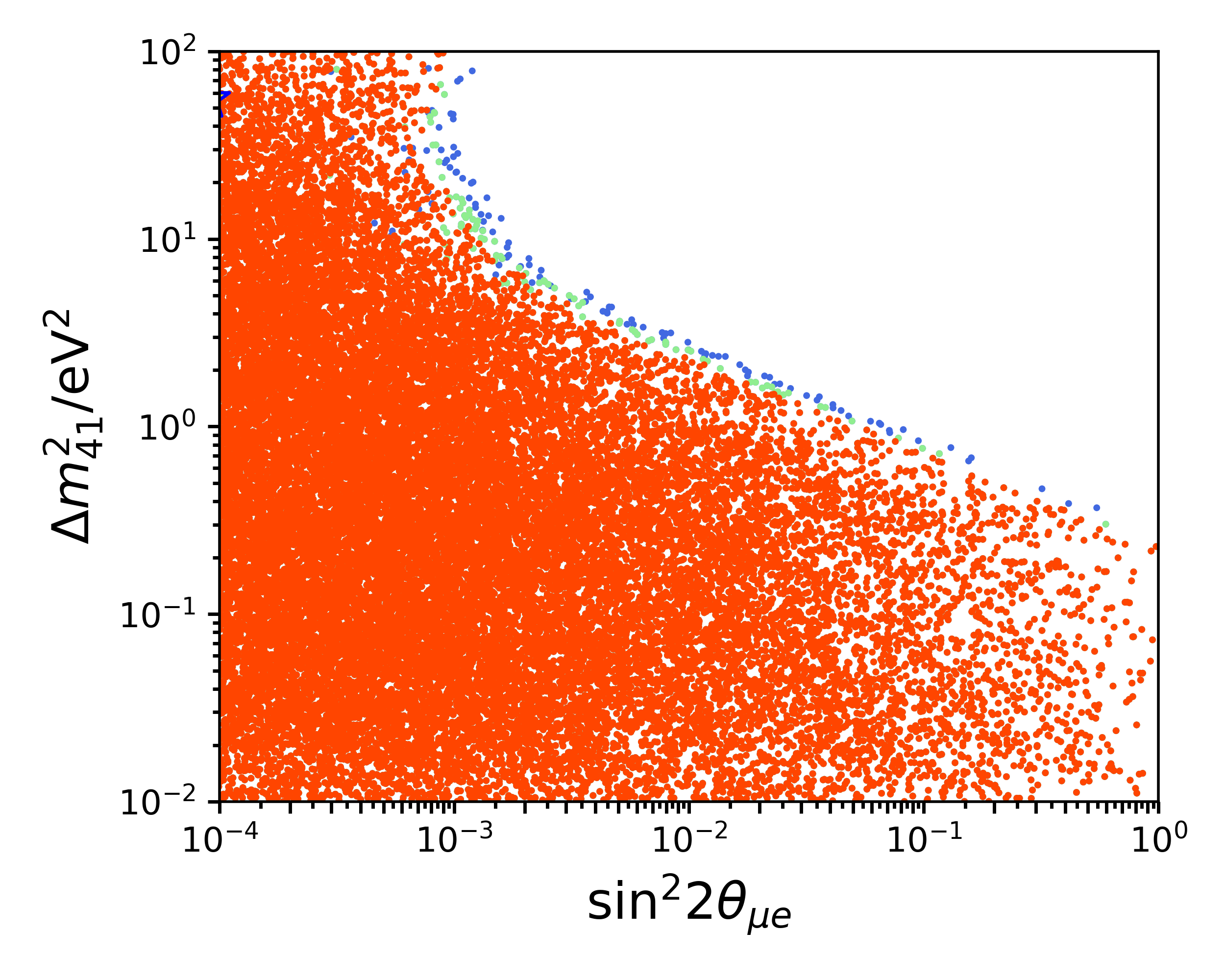}
\caption{NOMAD}
\end{subfigure}

\caption{Fits to $\nu_\mu \rightarrow \nu_e$ to the appearance data sets.  Upper left: LSND; upper right: KARMEN;   middle left:  MiniBooNE (BNB), with neutrino and antineutrino data combined; middle right: MiniBooNE (NUMI); bottom: NOMAD.  The contours for the 99\%, 95\%, and 90\% are shown in blue, green, and red, respectively.}
\label{appfigs}
\end{figure*} 

\paragraph{{\bf LSND} \cite{LSND}: }

The LSND (Liquid Scintillation Neutrino Detector) ran at Los Alamos National
Laboratory in 1993-1998, searching for $\bar{\nu}_\mu \rightarrow \bar{\nu}_e$
appearance using a decay-at-rest (DAR) beam. LSND created its $\bar{\nu}_\mu$ by
 impinging an intense ($\sim$1 mA) beam of 798 MeV protons onto a target. The $\bar{\nu}_\mu$
beam ultimately created extended up to ~55 MeV in energy, with the detector located ~30 m from the
target. The LSND detector was a tank filled with 167 metric tons of liquid
scintillator, surrounded by 1220 8-inch PMTs. LSND observed a $\bar{\nu}_e$ excess of
$87.9 \pm 22.4 \pm 6.0$ events above background, which corresponds to an oscillation
probability of $(0.264\pm0.067\pm0.045)\%$.

%PhysRevD.64.112007

\paragraph{{\bf KARMEN} \cite{KARMEN}: }

The KARMEN (Karlsruhe Rutherford Medium Energy Neutrino) experiment ran at the Rutherford Laboratory in
1997-2001, searching for $\bar{\nu}_\mu \rightarrow \bar{\nu}_e$ appearance using a DAR beam.
This experiment ran in two phases, and we make use of the final KARMEN2 data set.  Similar to
LSND, KARMEN produced its beam by impinging a proton beam on a target and producing $\bar{\nu}_\mu$ by the
same decay chain as LSND. The detector was a segmented liquid scintillation calorimeter, located 17.7 meters from the target at an angle of 100\degree~to the proton beam. KARMEN saw no signal of oscillations, having observed 15 candidate $\bar{\nu}_e$ events with $15.8\pm0.5$ expected from background. KARMEN thus excludes a large area of the LSND signal, but at a lower confidence level because the intensity of the flux was lower than the flux at LSND.

%10.1103/PhysRevD.65.112001

\paragraph{{\bf MiniBooNE (BNB)} \cite{MB2018, MBnubar}: }

The MiniBooNE experiment was commissioned in order to follow up on the LSND anomaly using
different detection techniques and energies while still being sensitive to the same parameter
space. As opposed to a DAR beam, the primary data set for MiniBooNE made use of a decay in flight (DIF) beam by impinging an 8 GeV
proton beam on a berylium target and focusing the charged mesons (primarily pions and kaons)
towards the detectors.   This was produced in the BNB line at Fermi National Accelerator Laboratory, with the fluxes described in Sec.~\ref{DIFflux}.
The target was placed inside of a magnetic focusing horn, allowing the
experiment to focus either positively or negatively charged mesons, which would then decay to
produce either $\nu_\mu$ or $\bar{\nu}_\mu$, respectively. This allowed MiniBooNE to perform both a
$\bar{\nu}_\mu \rightarrow \bar{\nu}_e$ and a $\nu_\mu \rightarrow \nu_e$ search. 

The neutrino beam energy is peaked at $\sim$500 MeV \cite{BNBflux}, and the detector was placed 540 meters
downstream of the target. The MiniBooNE detector is a 450 ton oil Cherenkov detector. In their most
recent result, published in 2018, MiniBooNE reports an oscillation signal in both
neutrino and antineutrino mode and we have combined neutrino and antineutrino data sets in Fig.~\ref{appfigs}. Furthermore, this signal is consistent with the signal seen by
LSND.  Since this signal is associated with an excess at low energy, it has become known as the MiniBooNE ``Low Energy Excess'' (LEE).   In Sec.~\ref{MBcomment}, we discuss in detail how the MiniBooNE constrained backgrounds to the LEE signal using data from the detector.

\paragraph{{\bf MiniBooNE (NuMI)} \cite{MBNumi}: }

The MiniBooNE detector also stands near another neutrino beam, the NuMI beamline. The NuMI beam was directed toward the MINOS detector in Minnesota, and so the beamline was oriented downward and to the north at the Fermilab site.  The neutrinos are created by impinging a 120 GeV proton beam on a carbon target at roughly surface-level, and two magnetic horns focus the positive mesons toward the MINOS detector. The on-surface MiniBooNE detector lies 745 m from the NuMI target, at 6.3\degree off-axis from the NuMI beam.   Thus, MiniBooNE could observe events due to neutrinos produced in the NuMI line.   Unlike the BNB flux,  this spectrum was quite complicated to model. The neutrino energy extended up to $\sim$3 GeV.   In particular, the off-axis NuMI beam had a very high intrinsic electron neutrino content, where 38\% originated from parent mesons produced in non-target material.     This led to large systematic uncertainties on the background in a $\nu_\mu \rightarrow \nu_e$ search, which ran from 2005-2007.   The result yielded a $1.2\sigma$ excess over the background expectation, which is below the level we would note as a signal.   However, this excess is consistent in magnitude and energy range with other anomalies, and so contributes to the overall appearance best fit point.

%10.1103/PhysRevLett.102.211801

\paragraph{{\bf NOMAD} \cite{NOMAD1}: }

NOMAD was an experiment conducted at CERN meant to search for $\nu_\mu \rightarrow \nu_\tau$ oscillations.  The detector was optimized to detect electrons from $\tau^- \rightarrow e^- + \bar{\nu}_e + \nu_\tau$ decays, and so NOMAD could also be used to search for $\nu_\mu \rightarrow \nu_e$ oscillations. The neutrino beam had an average energy of $\sim 20$ GeV, with an average baseline of 625 meters. NOMAD found no signal for oscillation, and excludes the LSND best fit region for $\Delta m^2 \gtrsim 10 \text{ eV}^2$.

%Physics Letters B 570 (2003) 19–31

\subsubsection{Electron Flavor Disappearance Experiments}
\begin{figure*}[t!]
\centering
\begin{subfigure}{0.4\linewidth}
\centering
\includegraphics[width=\linewidth]{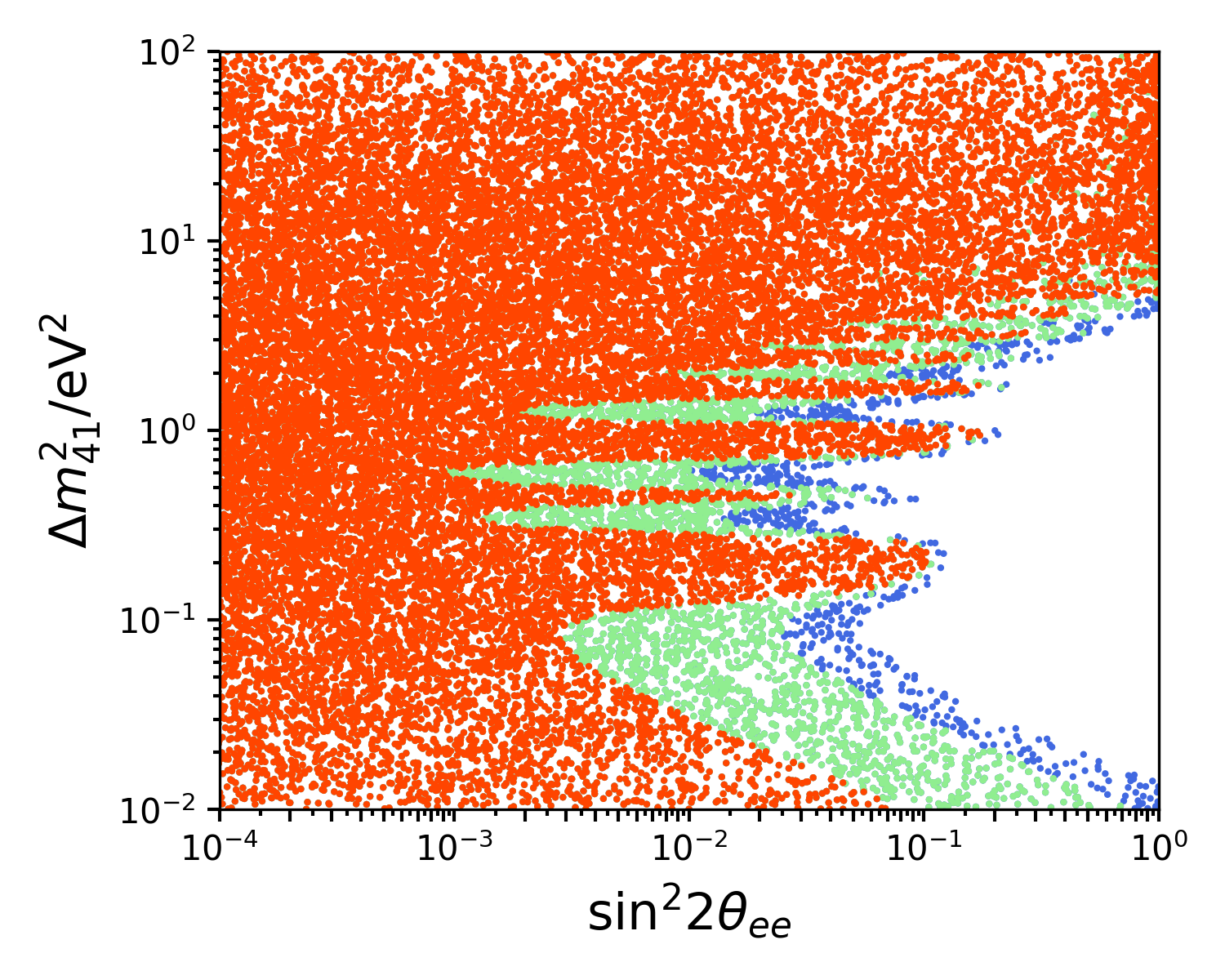}
\caption{Bugey}
\end{subfigure}
~ 
\begin{subfigure}{0.4\linewidth}
\centering
\includegraphics[width=\linewidth]{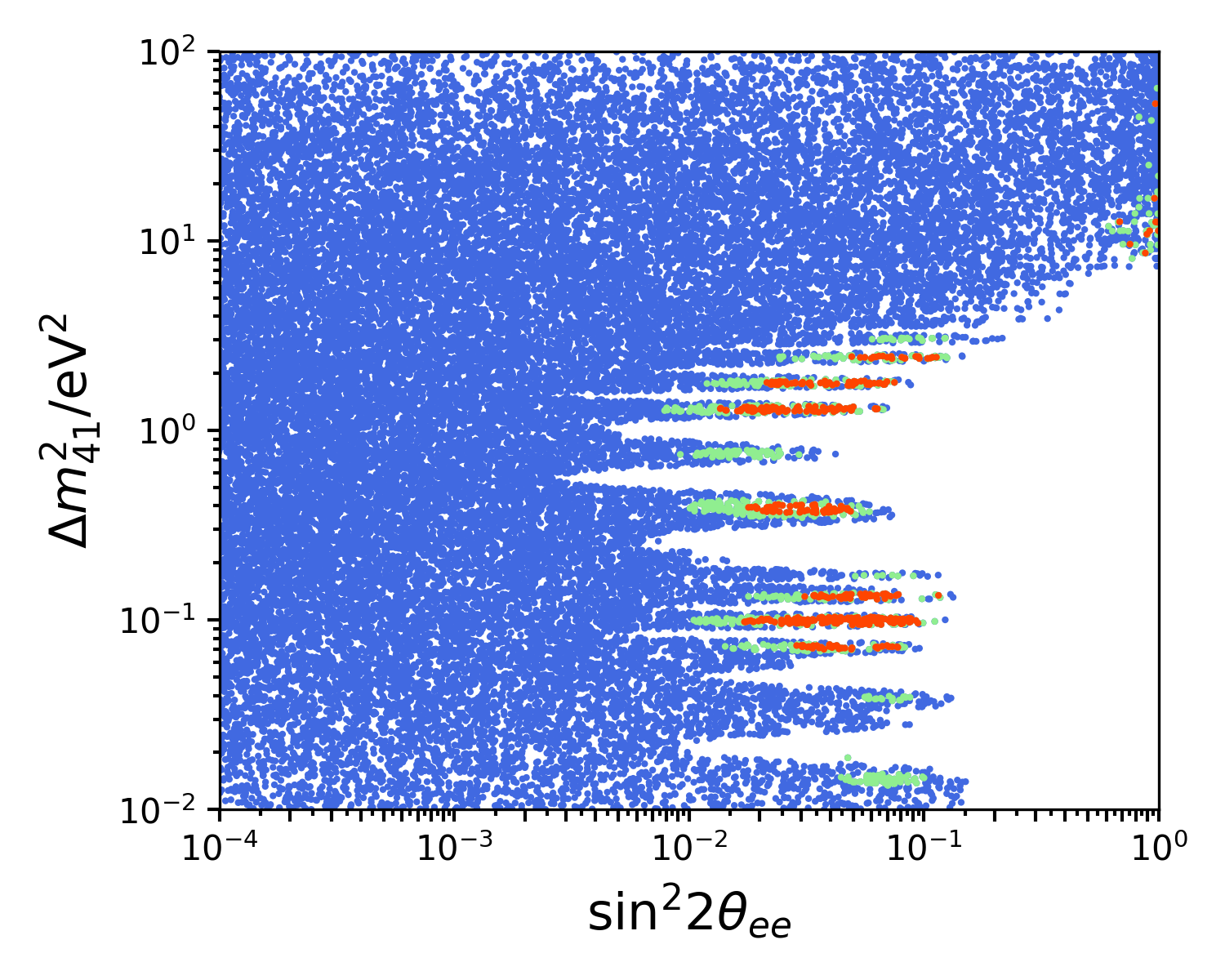}
\caption{NEOS}
\end{subfigure}

\vskip\baselineskip
\begin{subfigure}{0.4\linewidth}
\centering
\includegraphics[width=\linewidth]{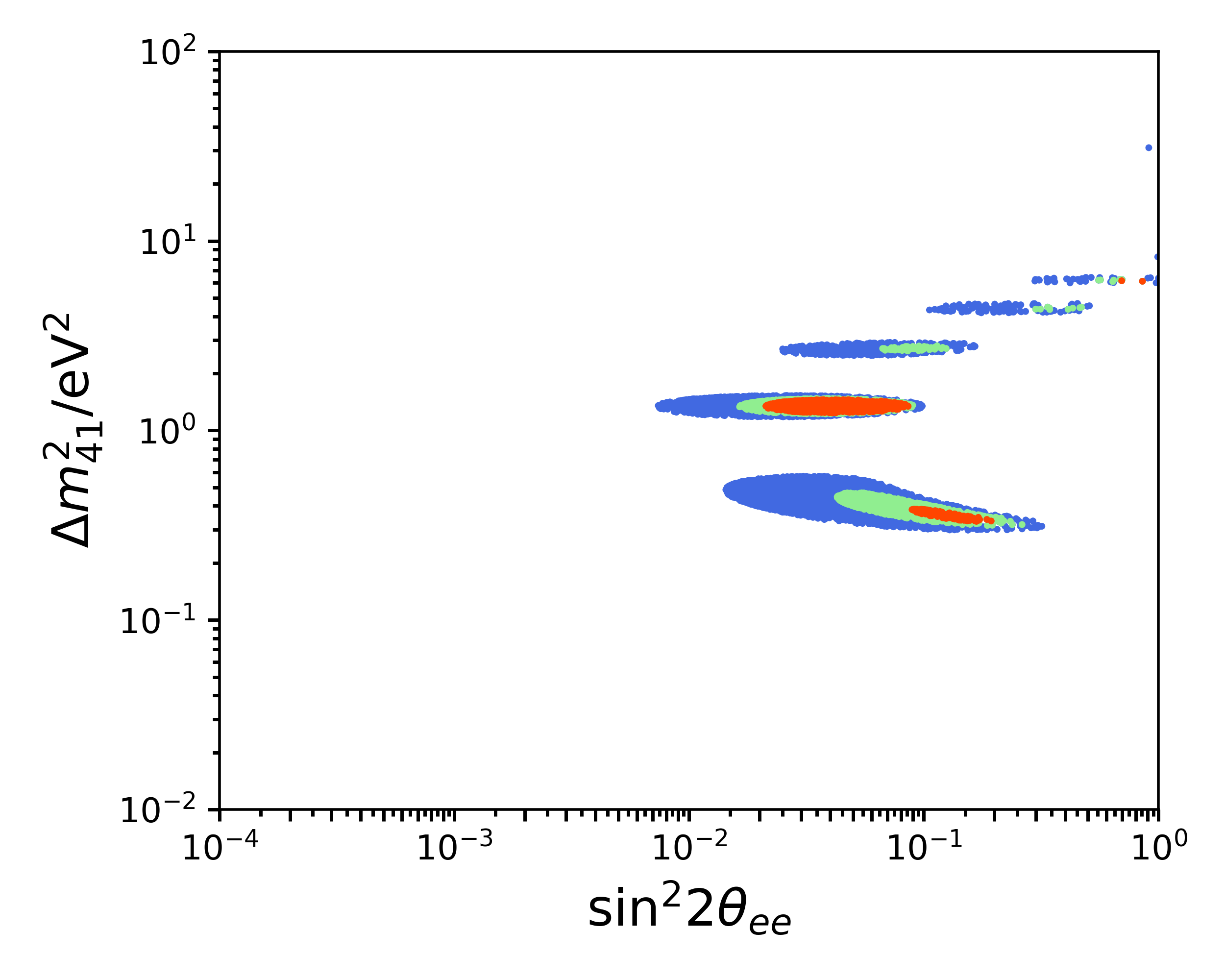}
\caption{DANSS}
\end{subfigure}
~ 
\begin{subfigure}{0.4\linewidth}
\includegraphics[width=\linewidth]{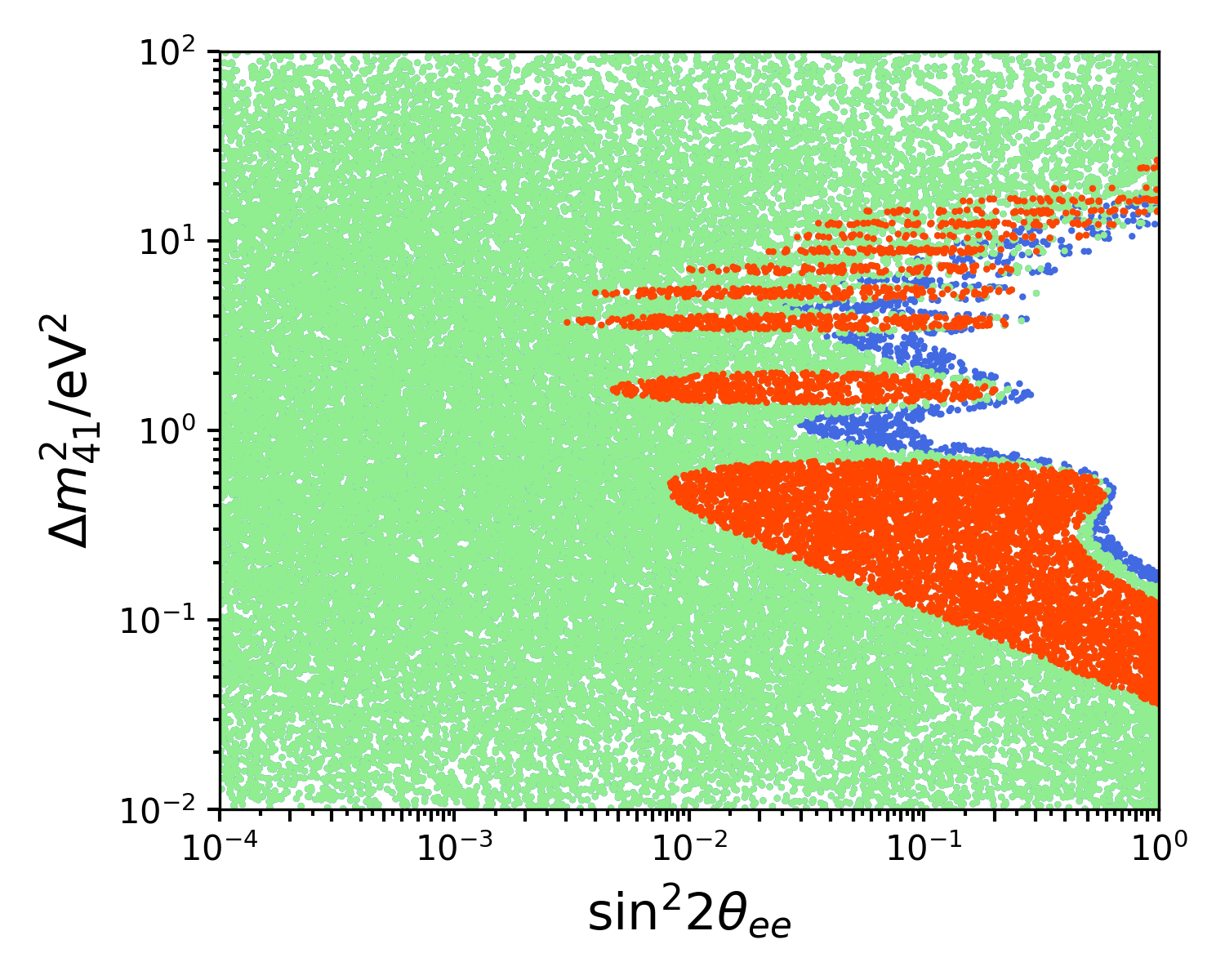}
\caption{PROSPECT}
\end{subfigure}

\vskip\baselineskip
\centering
\begin{subfigure}{0.4\linewidth}
\includegraphics[width=\linewidth]{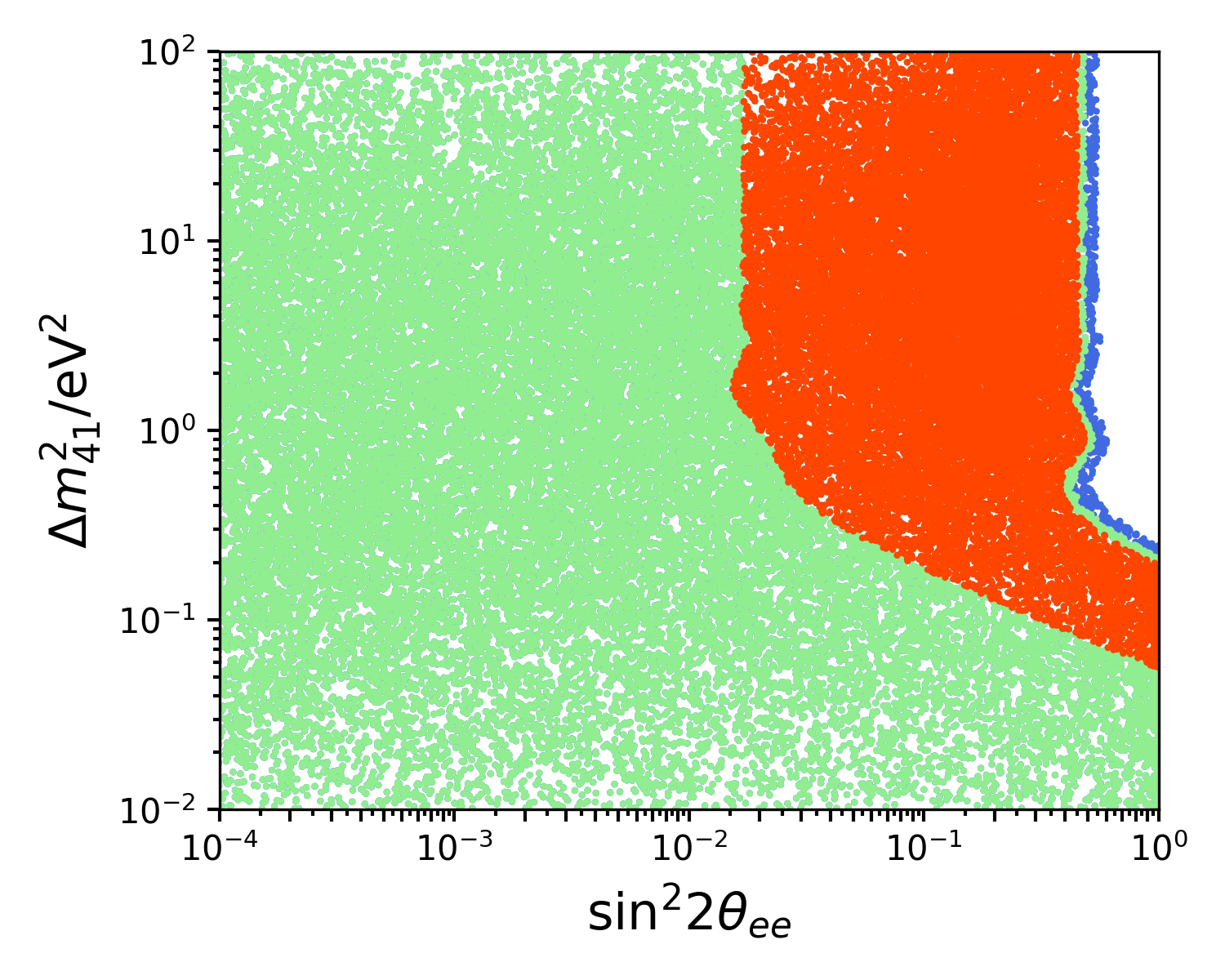}
\caption{SAGE and GALLEX}
\end{subfigure}
~
\begin{subfigure}{0.4\linewidth}
\includegraphics[width=\linewidth]{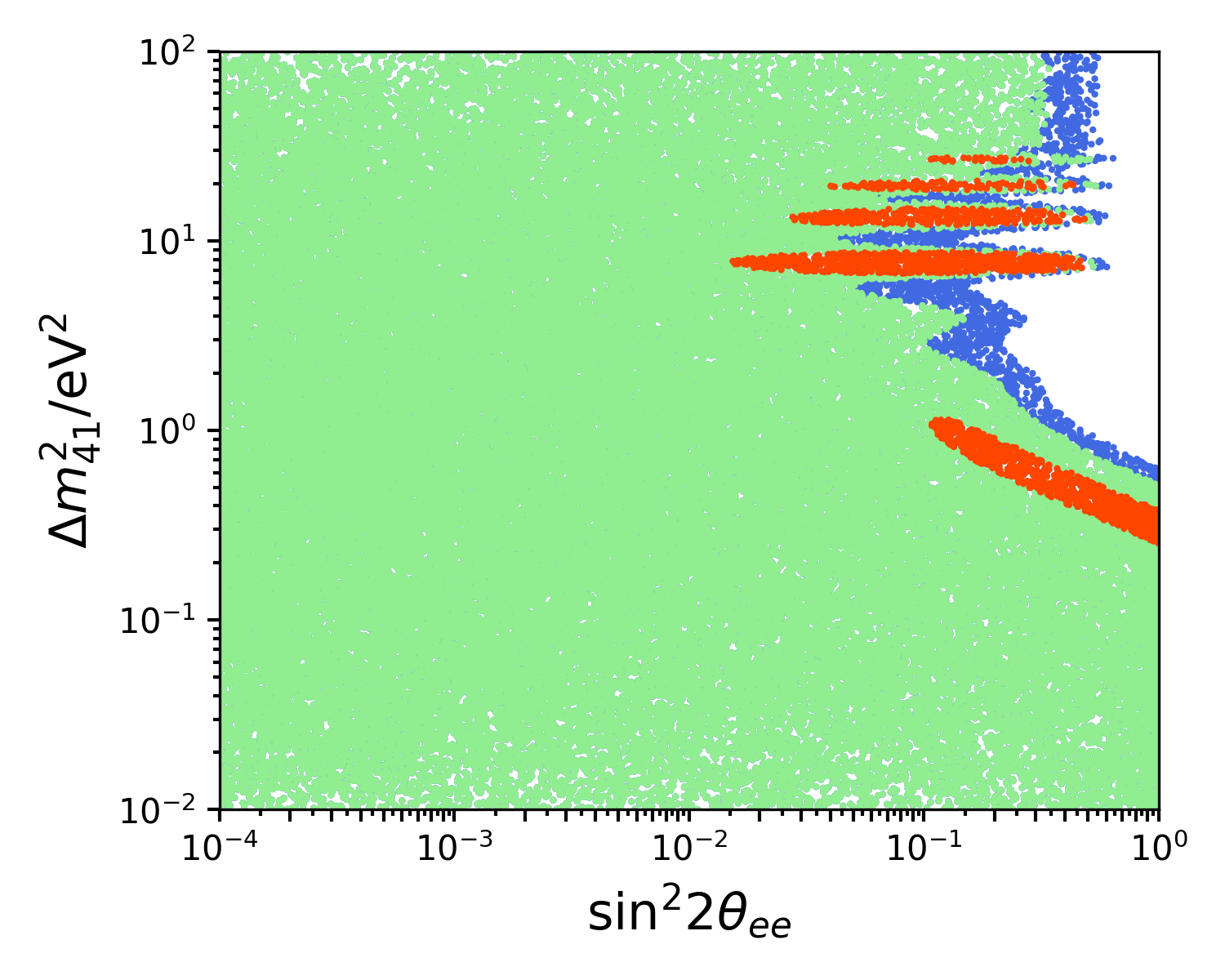}
\caption{KARMEN/LSND Cross Section}
\end{subfigure}

\caption{Fits to $\nu_e \rightarrow \nu_e$ to the electron-flavor data sets.  Upper left:  Bugey; upper right: NEOS;  middle left:  DANSS; middle right:  PROSPECT; bottom left: SAGE and GALLEX, combined; bottom right:  KARMEN and LSND cross section joint fit.}
\label{disefigs}
\end{figure*}

We fit to the data from the following experiments that search for the disappearance of electron flavor flux. In the context of two-neutrino global fits, these experiments would be sensitive to $|U_{e4}|$ and $\Delta m_{41}^2$.   The two neutrino ($\nu_e \rightarrow \nu_e$) fits to these data sets are shown in Fig.~\ref{disefigs}.

\paragraph{{\bf Bugey} \cite{Bugey}: }

Bugey was a $\bar{\nu}_e \rightarrow \bar{\nu}_e$ disappearance reactor experiment. Three detectors were placed at 15, 40, and 95 meters from the reactor. The detectors are each a $\sim600~\text{liter}^2$ tank segmented into 98 segments, filled with $^6\text{Li}$ doped scintillator. The Bugey collaboration conducted two analyses, one where the spectral shapes at each detector was compared to MC prediction, and another where the spectra observed were compared between pairs of detectors. In this analysis, we follow the latter technique. Bugey observed no signal for oscillations at $90\%$ CL.

\paragraph{{\bf NEOS} \cite{NEOS} and {\bf Daya Bay} \cite{An:2016srz}: }

NEOS (Neutrino Experiment for Oscillation at Short baseline) is an ongoing reactor $\bar{\nu}_e \rightarrow \bar{\nu}_e$ disappearance experiment situated in South Korea. The neutrino target of the NEOS detector is 1008 liters of Gd-doped liquid scintillator, positioned 23.7 m from the reactor core center. The detector operated 180 days with the reactor on, and 46 days with the reactor off, averaging 1976 antineutrino events per day. 

In order to compensate for systematic uncertainties in the predicted reactor antineutrino flux, the ratio of the NEOS event rate is taken with the Daya Bay near-detector unfolded spectrum, taking into account differing fuel compositions \cite{Mueller:2011nm}.   Daya Bay, which is located in China, is a high statistics reactor experiment designed initially for a precision oscillation search at an $\sim 1$ km baseline \cite{An:2015qga}.   The two near detector halls are located at 560 m and 600 m, respectively.

In our implementation, both the NEOS and Daya Bay spectra are allowed to oscillate, depending on the oscillation parameters.
No evidence for oscillation is seen in the NEOS/Daya Bay combination, and $\sin^22\theta_{14}$ is excluded up to 0.1 for $\Delta m_{41}^2$ ranging from 0.2 to 2.3 eV${}^{ 2}$ at $90\%$ CL.

%PRL 118, 121802 (2017)

\paragraph{{\bf DANSS} \cite{DANSS}: }

DANSS (Detector of the reactor AntiNeutrino based on Solid Scintillator) is an ongoing reactor
$\bar{\nu}_e \rightarrow \bar{\nu}_e$ disappearance experiment situated in Russia. The DANSS
detector is a 1 m$^3$ volume of highly segmented plastic scintillator strips. To address
systematic uncertainties in the predicted reactor antineutrino flux, the DANSS detector is mobile
and data is taken at three baselines: 10.7, 11.7, and 12.7 m from the reactor core center. The
ratio of the event rates at these different baselines are then taken. The detector averaged 4899
inverse beta decay events per day in the top (10.7 m) position. The most recent
analysis by DANSS, from 2018, only incorporates statistical errors, but systematic uncertainties are expected
to be small due to using the same detector at different baselines. While the DANSS collaboration
has released an exclusion limit, a statistically significant preference for oscillation is found
at $\Delta m^2 = 1.4 \text{ eV}^2$. The collaboration plans to study the significance of this
preference, incorporating systematic uncertainties as more data is collected. 

%arXiv:1804.04046v3

\paragraph{{\bf PROSPECT}  \cite{Prospect}: }

PROSPECT (Precision Reactor OScillation and SPectrum ExperimenT) is an ongoing reactor $\bar{\nu}_e \rightarrow \bar{\nu}_e$ disappearance experiment located at the High Flux Isotope Reactore (HFIR) at Oak Ridge National Laboratory in the US. PROSPECT uses a 4 ton ${}^{6}$Li-doped liquid scintillator detector segmented into 154 optically isolated segments, so that the detector can independently measure the $\bar{\nu}_e$ flux at baselines ranging 7-9 meters, and take their ratios. The HFIR reactor core is compact and composed of highly enriched uranium, with a fission fraction typically $\gtrsim 99\%$ ${}^{235}$U, minimizing baseline uncertainties and fission fragment specific flux uncertainties that commercial reactor experiments face. PROSPECT saw no signal, but the experiment is relatively new;   at the time of inclusions in the fits, this experiment had run for only 33 reactor-on days and 28 reactor-off days in 2018. As a result,  PROSPECT is currently statistically limited. 

%PHYSICAL REVIEW LETTERS 121, 251802 (2018)

\paragraph{{\bf GALLEX} \cite{Gallex} and {\bf SAGE}  \cite{SAGE}: }

A pair of Gallium-based experiments, SAGE and GALLEX, measured the solar neutrino flux by counting the interactions of $\nu_e$ with the gallium in the detectors. Both experiments placed radioactive sources within the detectors for calibration. SAGE conducted two calibrations, once with ${}^{51}\text{Cr}$ and again with ${}^{37}\text{Ar}$. GALLEX  conducted two calibrations, using ${}^{51}\text{Cr}$ both times. The ratio $R$ of the observed interaction rates over the expected due to source strength was taken for each of the four measurements, and the weighted average was found to be $R=0.87\pm0.05$. This $>2\sigma$ can be interpreted as a signal for $\nu_e$ disappearance. 

%10.1103/PhysRevC.80.015807
%doi:10.1016/j.physletb.2010.01.030

\paragraph{{\bf KARMEN/LSND (Cross Section)} \cite{ConradShaevitz}: }

In addition to $\bar{\nu}_e$, both LSND and KARMEN were able to detect $\nu_e$ from the $\nu_e + {}^{12}\text{C} \rightarrow {}^{12}\text{N}_{gs} + e^- $ 
interaction. The  ${}^{12}\text{N}_{gs}$
was identified by the subsequent decay through ${}^{12}\text{N}_{gs} \rightarrow {}^{12}\text{C} + e^+ + \nu_e$. 
These interactions were used to make a $\nu_e$-carbon cross section measurement. Due to
their differing baselines, 17.7 and 29.8 meters for KARMEN and LSND respectively, oscillating
$\nu_e$'s would result in different measured cross sections for the two detectors. No oscillation
signal was found, and the joint analysis excludes a large area of the Gallium confidence levels
while only excluding a modest portion of the Reactor Antineutrino Anomaly confidence level. 

%10.1103/PhysRevD.85.013017

\subsubsection{Muon Flavor Disappearance Experiments}
\begin{figure*}[t!]
\centering
\begin{subfigure}{0.4\linewidth}
\centering
\includegraphics[width=\linewidth]{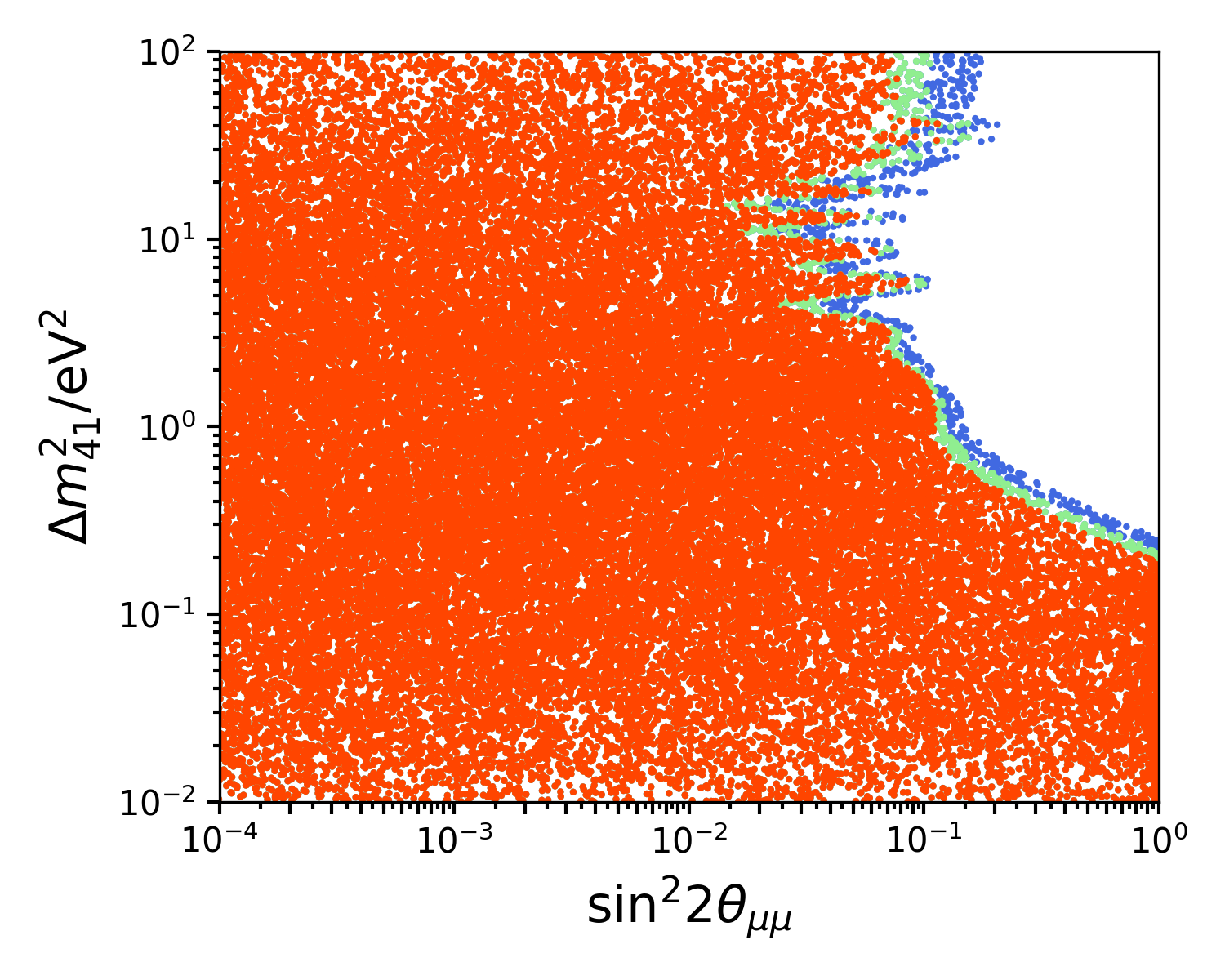}
\caption{MiniBooNE-SciBooNE}
\end{subfigure}
~ 
\begin{subfigure}{0.4\linewidth}
\centering
\includegraphics[width=\linewidth]{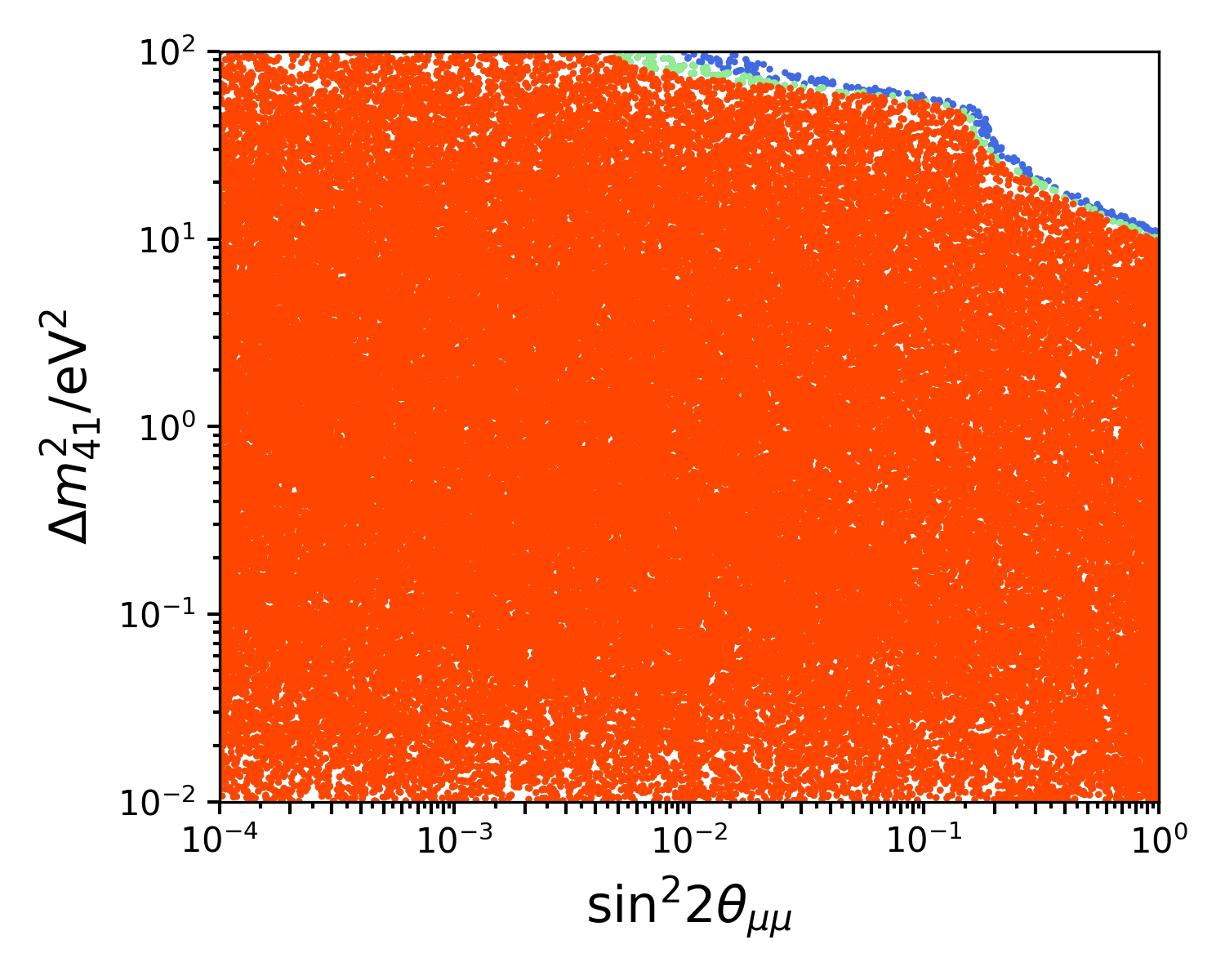}
\caption{CCFR}
\end{subfigure}

\vskip\baselineskip
\begin{subfigure}{0.4\linewidth}
\centering
\includegraphics[width=\linewidth]{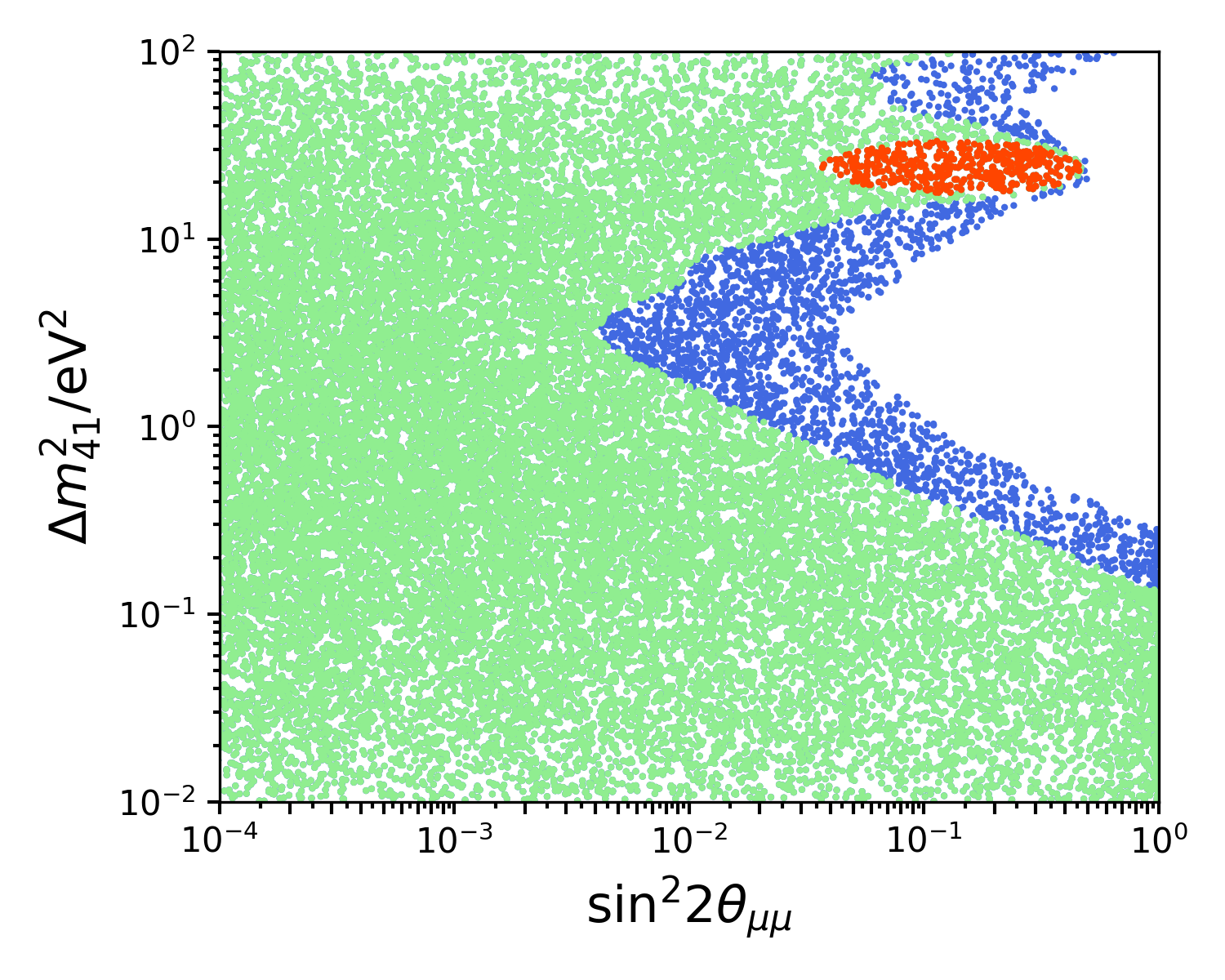}
\caption{CDHS}
\end{subfigure}
~ 
\begin{subfigure}{0.4\linewidth}
\includegraphics[width=\linewidth]{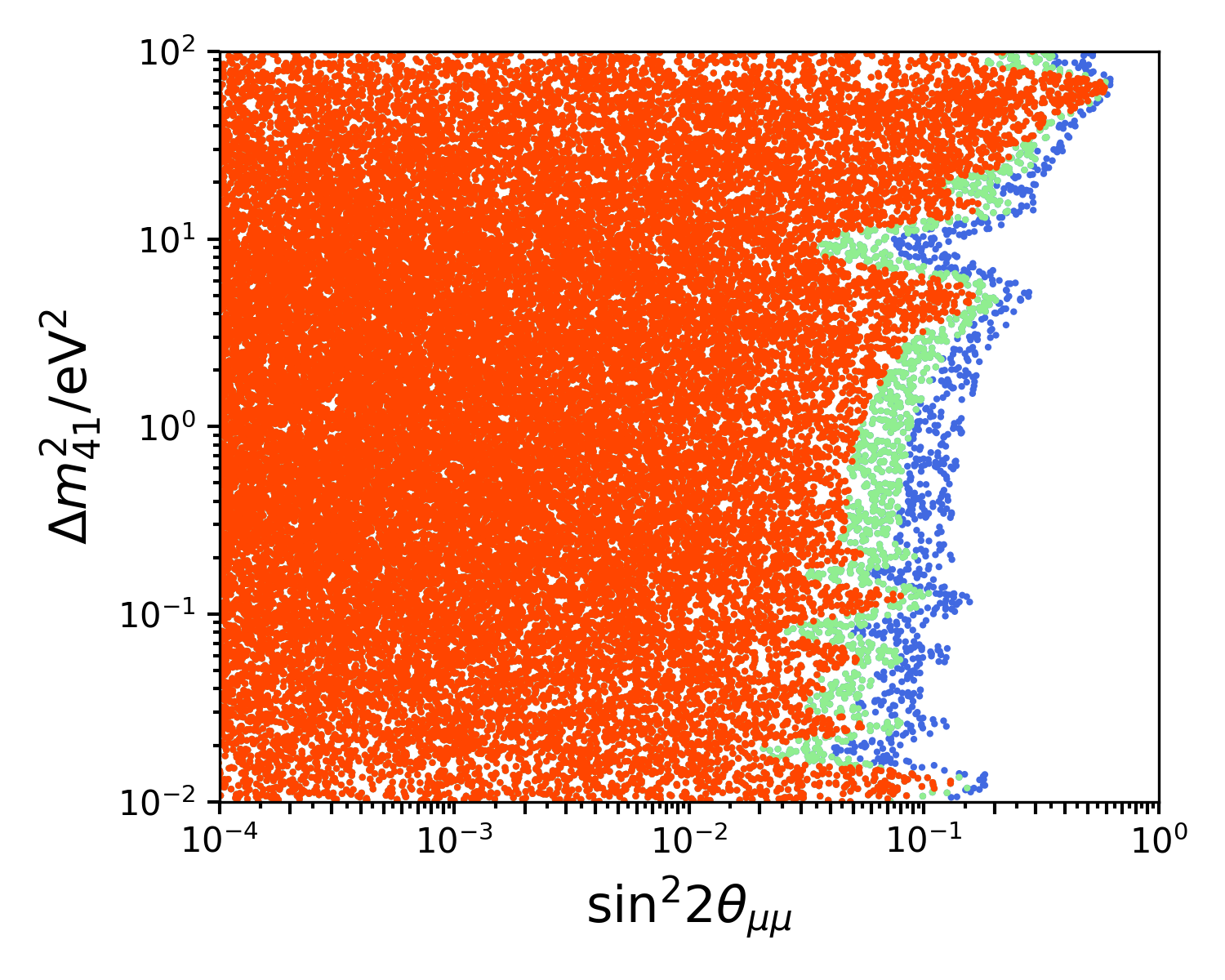}
\caption{MINOS}
\end{subfigure}

\caption{Fits to $\nu_\mu \rightarrow \nu_\mu$ to the muon-flavor data sets.   MiniBooNE-SciBooNE; upper right: CCFR84; lower left:   CDHS; lower right: MINOS-CC data sets, combined.}
\label{dismufigs}
\end{figure*}

We fit to the data from the following experiments that search for the disappearance of muon flavor neutrino flux. In the context of two-neutrino global fits, these experiments would be sensitive to $|U_{\mu4}|$ and $\Delta m_{41}^2$.  Two neutrino fits ($\nu_\mu \rightarrow \nu_\mu$) are shown to these muon-flavor data sets in Fig.~\ref{dismufigs}.

\paragraph{{\bf MiniBooNE/SciBooNE (BNB)} \cite{SBMBnu,SBMBnubar}: }

SciBooNE was a neutrino cross section experiment using the same neutrino beam as MiniBooNE. The SciBooNE detector stood 100 m from the production target, compared to MiniBooNE's 540 m baseline. This allowed a joint analysis between the two detectors to be done, where SciBooNE acted as the near detector and MiniBooNE as the far detector. The joint analysis did separate $\nu_\mu \rightarrow \nu_\mu$  and $\bar{\nu}_\mu \rightarrow \bar{\nu}_\mu$ disappearance searches. In both analyses, the two-detector fit found no signal for $\nu_\mu$ or $\bar{\nu}_\mu$ disappearance.

%10.1103/PhysRevD.85.032007
%10.1103/PhysRevD.86.052009

\paragraph{{\bf CCFR84} \cite{CCFR84}: }

The CCFR collaboration collected data in order to measure $\nu_\mu$ and $\bar{\nu}_\mu$ disappearance using a neutrino beam and a pair of detectors at Fermilab. The narrow band neutrino beam at Fermilab ran on 5 momentum settings for $\pi^+$ and $K^+$ (100, 140, 165, 200, and 250 GeV) and ran at 165 GeV for antineutrino mode. This provided a neutrino energy range between 40 and 230 GeV. The two detectors stood 715 and 1116 meters from the center of the 352 m decay pipe. The CCFR collaboration found no evidence for oscillations in either neutrino or antineutrino mode. The excluded region was approximately $15 < \Delta m^2 < 1000 \text{ eV}^2$ and $\sin(2\theta) > 0.02$. 

%Z. Phys. C - Particles and Fields 27, 53-56 (1985)

\paragraph{{\bf CDHS} \cite{CDHS}: }

The CDHS experiment was designed to study deep inelastic neutrino interactions with iron, using the SPS beam at CERN. A $\nu_\mu \rightarrow \nu_\mu$ oscillation study was done with CDHS by constructing an additional  detector 130 m from the beam target to act as the near detector, while the existing detector acted as the far detector at 885 m from the proton target. The neutrino flux peaked at 1 GeV. Unlike most other neutrino experiments that compare neutrino event rates as a function of reconstructed neutrino energy, CDHS compared rates as a function of the track length of the outgoing $\mu$. CDHS saw no signal for oscillation. 

%Volume 134B, number 3,4 PHYSICS LETTERS 12 January 1984

\paragraph{{\bf MINOS-CC} \cite{MINOSCC2012,MINOSCC2011,MINOS2016}: }

MINOS was a dual detector neutrino experiment based at Fermilab. Utilizing the NuMI beam, MINOS measured $\nu_\mu$ and $\bar{\nu}_\mu$ disappearance using two detectors, at 1.04 km and 735 km from the production target. We consider several data sets, in both neutrino and antineutrino modes, in our analysis. 
For the $\bar{\nu}_\mu$ oscillations, two different data sets from 2011-12  are used. One is from a MINOS analysis where the NuMI beam ran in $\bar{\nu}_\mu$ mode, and $\bar{\nu}_\mu$ disappearance was measured. In the other case, the NuMI beam ran in $\nu_\mu$, and the disappearance of the $7\%$ wrong-signed $\bar{\nu}_\mu$ component of the beam is measured. 
In the case of $\nu_\mu$ oscillations, we use the MINOS 2016 $\nu_\mu$ data set.
 
Because MINOS is a long-baseline experiment, which is affected by the oscillations of the three light neutrinos, assumptions are made in their analyses in order to reduce the number of fit parameters.  For example, in the 2016 data set,  MINOS fit for the ``active-flavor'' mixing parameters $\theta_{23}$ and $\Delta m^2_{23}$, and the sterile parameters $\theta_{24}$, $\theta_{34}$ and $\Delta m^2_{41}$.    They set the remaining active flavor parameters to best fit values as described in Ref.~\cite{MINOS2016}.  Also, they set $\theta_{14}=0$, which is inconsistent with use in a global fit, as it does not allow for $\nu_e$ disappearance or $\nu_\mu \rightarrow \nu_e$ transitions.      We make use of this data set despite this problem, and it is an example of the unfortunate compromises that must be made when performing global fits, as discussed in Sec.~\ref{limits}. 

%PRL 108, 191801 (2012)
%PHYSICAL REVIEW D 84, 071103(R) (2011)

%arXiv:1202.2772v1
%arXiv:1108.1509v2

\subsection{Techniques for Constraining Uncertainties:  The MiniBooNE Example \label{MBcomment}}

\begin{figure}
\begin{center}
{\includegraphics[width=\columnwidth]{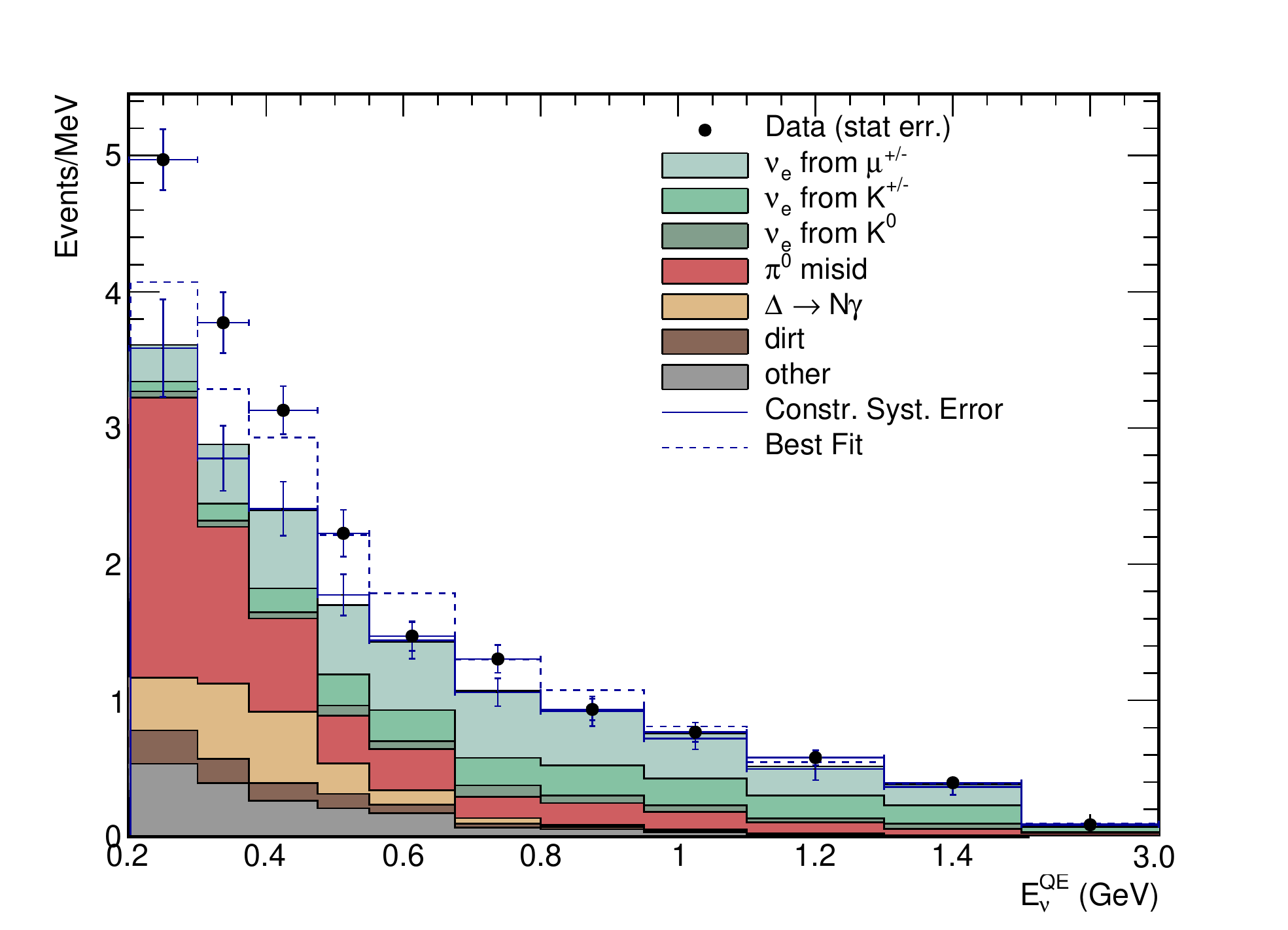}}
\end{center}
\caption{The MiniBooNE $\nu_e$ event sample in neutrino mode, combining data taken from 2002-2007 and 2015-2018.  The stacked plot indicates the Standard Model backgrounds (red/brown/yellow -- misidentification backgrounds; green shades--intrinsic $\nu_e$).}
\label{stackedMB}
\end{figure}

Among the experiments with signals included in our fits, MiniBooNE is unique in that it has high backgrounds.    A stacked plot of backgrounds showing the MiniBooNE excess in neutrino mode is presented in Fig.~\ref{stackedMB}.  
Therefore, as an example, it is worth reviewing MiniBooNE's data-driven techniques for constraining backgrounds in more detail.

The MiniBooNE experiment is a 450 t fiducial volume oil-based
Cherenkov detector running in Fermilab's Booster Neutrino Beamline
(BNB).   The beam energy leads to a signal from charged current
quasi-elastic scattering (CCQE); the detector is searching for an
excess of $\nu_e + n \rightarrow e^- + p$ and $\bar{\nu}_e
+ p \rightarrow e^+ + n$ events in a beam of high $\nu_\mu$ purity (see Sec.~\ref{DIFflux}).

The experiment collected data from 2002-2007 running in neutrino mode~\cite{MBnu},
and then in 2007-2013 in antineutrino mode~\cite{MBnubar}.  The experiment then took a hiatus to search for
dark photon production in the BNB dump~\cite{darkphot}.    When the
MicroBooNE Experiment came online in 2015, the BNB switched back to neutrino
mode running.   MiniBooNE, which is located 70 m upstream of
MicroBooNE and 540 m from the BNB target, continued to take data,
doubling the neutrino data set.   The results were released in
Ref.~\cite{MB2018} in May 2018, and the reader should see this
reference for a full discussion.

\begin{figure*}[t]
\begin{center}
{\includegraphics[width=\textwidth]{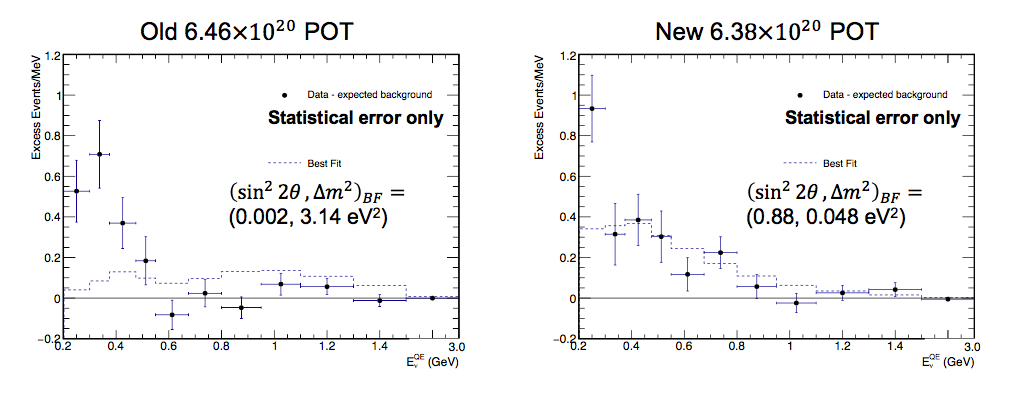}}
\end{center}
\caption{The MiniBooNE excess, after background subtraction, in neutrino mode. The 2002-2007 and 2015-18 data sets are presented separately and compared to the oscillation best fit for each.}
\label{MBoldnew}
\end{figure*}

\begin{figure}[t]
\begin{center}
{\includegraphics[width=\columnwidth]{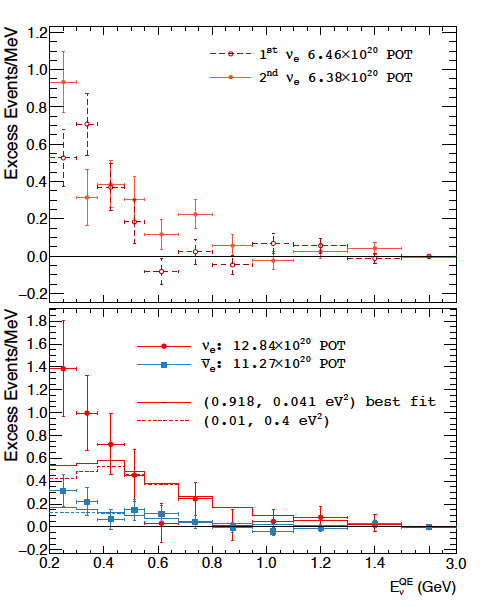}}
\end{center}
\caption{Top:  the background subtracted MiniBooNE excesses in neutrino mode, comparing the 2002-2007 and 2015-2018 data sets.  Bottom:  the combined neutrino mode excess compared to antineutrino mode excess, with two oscillation models for comparison.}
\label{nunubarosc}
\end{figure}

The two MiniBooNE neutrino mode data sets are compared in
Fig.~\ref{MBoldnew}.    The best fit oscillation model is shown in
each case.  
The data sets are an interesting example of
how mis-leading ``$\chi$-by-eye'' can be.       Looking at the
2015-2018 data set, one infers relatively good agreement with oscillations, while
looking at the 2002-2007 data set one infers poor agreement with
oscillations.   In fact, these two neutrino data sets and the
antineutrino data set all agree with one another within statistics, as
discussed in Ref.~\cite{MB2018} and shown in Fig.~\ref{nunubarosc} (top), despite the appearance in Fig.~\ref{MBoldnew}.   One wonders where the sterile neutrino
studies would be today, if the first data set obtained by MiniBooNE
had the form of the 2015-2018 data set.  As shown in Fig.~\ref{nunubarosc} (bottom), where the neutrino and antineutrino data sets are cross compared, and also compared to two oscillation models, one sees that the two results from MiniBooNE are internally compatible although not completely consistent with a $3+1$ oscillation model.

Overall, despite the fact that these data have good fits to $\nu_\mu
\rightarrow \nu_e$ oscillations,  there is a substantial deviation from an oscillation model in the low energy region where backgrounds are large.     Therefore, one might ask if some, or all, of the MiniBooNE signal is coming from background?

The background which peaks at low energy in the MiniBooNE data comes mainly from 
$\pi^0$ decays, where one photon is not detected.    In a Cherenkov
detector, the electromagnetic signature of an $e^+ e^-$ pair from a
converted photon cannot be distinguished from a single $e^-$, hence
a single photon from a $\pi^0$ mimics a signal and is an important background.   The
community has noted that we do not know the cross section for $\pi^0$
production well and have proposed this source of
misidentified background as the probable cause of the entire MiniBooNE
signal.

\begin{figure}[t]
\begin{center}
{\includegraphics[width=\columnwidth]{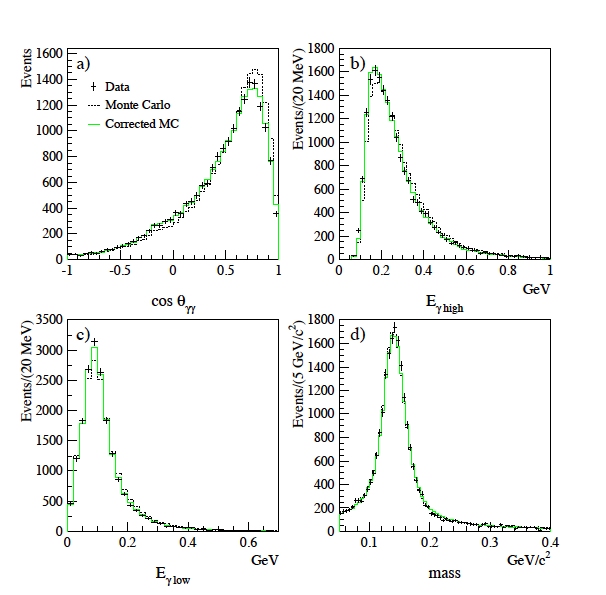}}
\end{center}
\caption{Four kinematic variables associated with neutral current $\pi^0$'s produced in MiniBooNE. Data: black points.  Black dotted histogram:  Monte Carlo prior to re-weighting. Green histogram: Monte Carlo events after re-weighting according to the measured $\pi^0$ momentum. From Ref.~\cite{link}.}
\label{fourpi0plots}
\end{figure}

This view is based on misunderstandings concerning how
MiniBooNE constrains the $\pi^0$ mis-id background.      MiniBooNE does not use
an {\it ab initio} prediction to find the mis-id rate.   Such an
estimation would, indeed, have very large errors and be subject to
suspicion.   Instead MiniBooNE uses the rate of measured two-photon
$\pi^0$ \underline{events} to constrain the two sources of $\pi^0$
mis-id.   The first source, which represents about half of the overall
misidentified events, comes from cases where one photon from the $\pi^0$ exits the
tank.    This is extremely well constrained using the observed event rate,
since it only depends upon understanding the photon conversion length
in oil.   The remaining half of the misidentified events come from
decays of the $\pi^0$ with back-to-back photons, where the decay axis is aligned along the boost
direction for a moving $\pi^0$.      This can lead to an energetic
forward photon and a weak backward photon.   If the weak backward
photon is missed by the reconstruction, then this will be a misidentified
event.     The probability that this will happen is dependent upon the
momentum of the $\pi^0$, which is
measured well using the large sample of reconstructed $\pi^0$ events where the two photons are reconstructed.   This allows
MiniBooNE to correct the production in the simulation as a function of
momentum, leading to a well-constrained prediction for the mis-id background.  
Fig.~\ref{fourpi0plots} shows that, after the simulated events are re-weighted 
according to the measured $\pi^0$ momentum, the simulation agrees well with data
in other kinematic variables associated with the $\pi^0$'s.\cite{link}.

\begin{figure*}[t]
\begin{center}
\includegraphics[width=.3\linewidth]{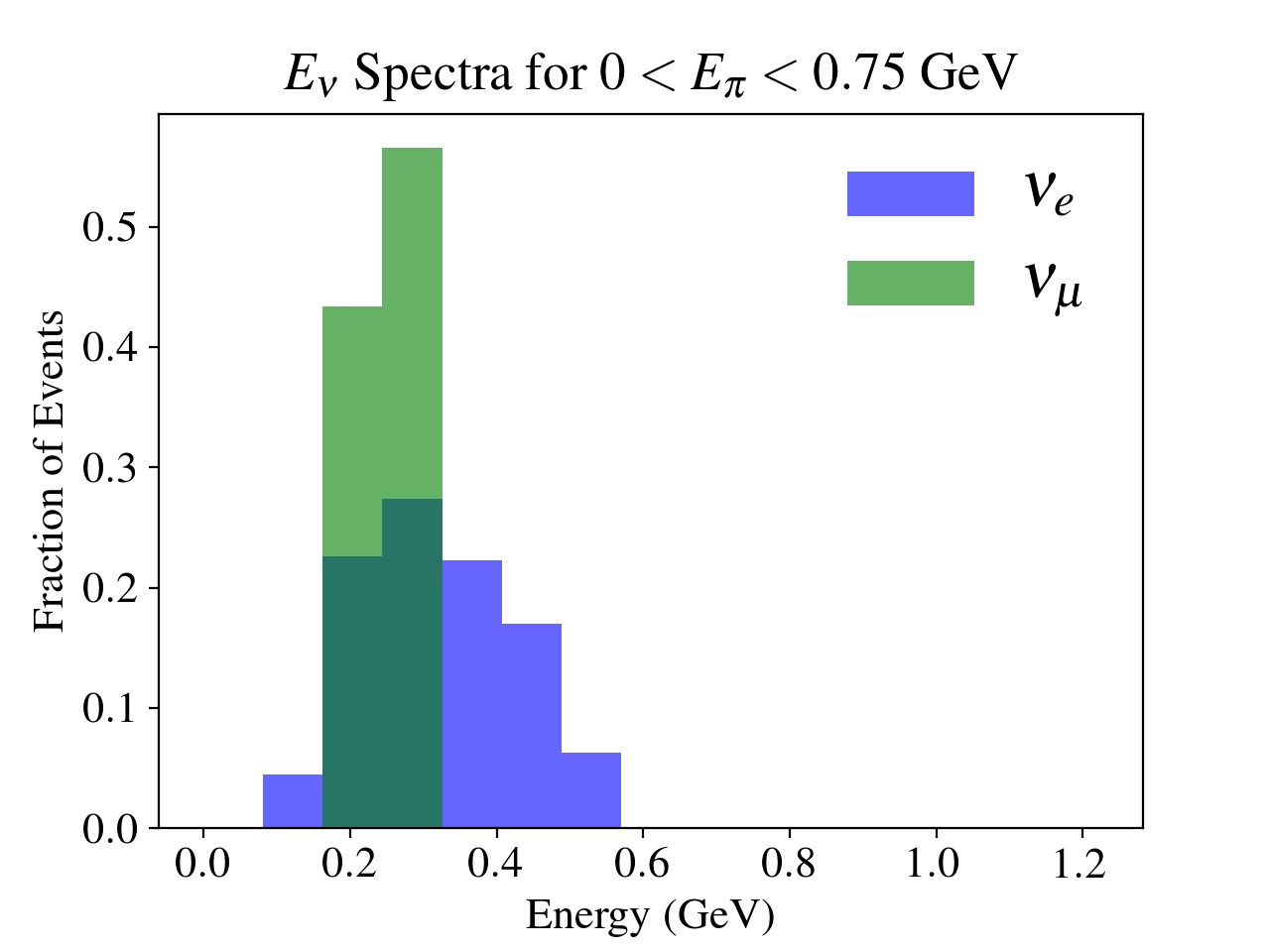}
~
\includegraphics[width=.3\linewidth]{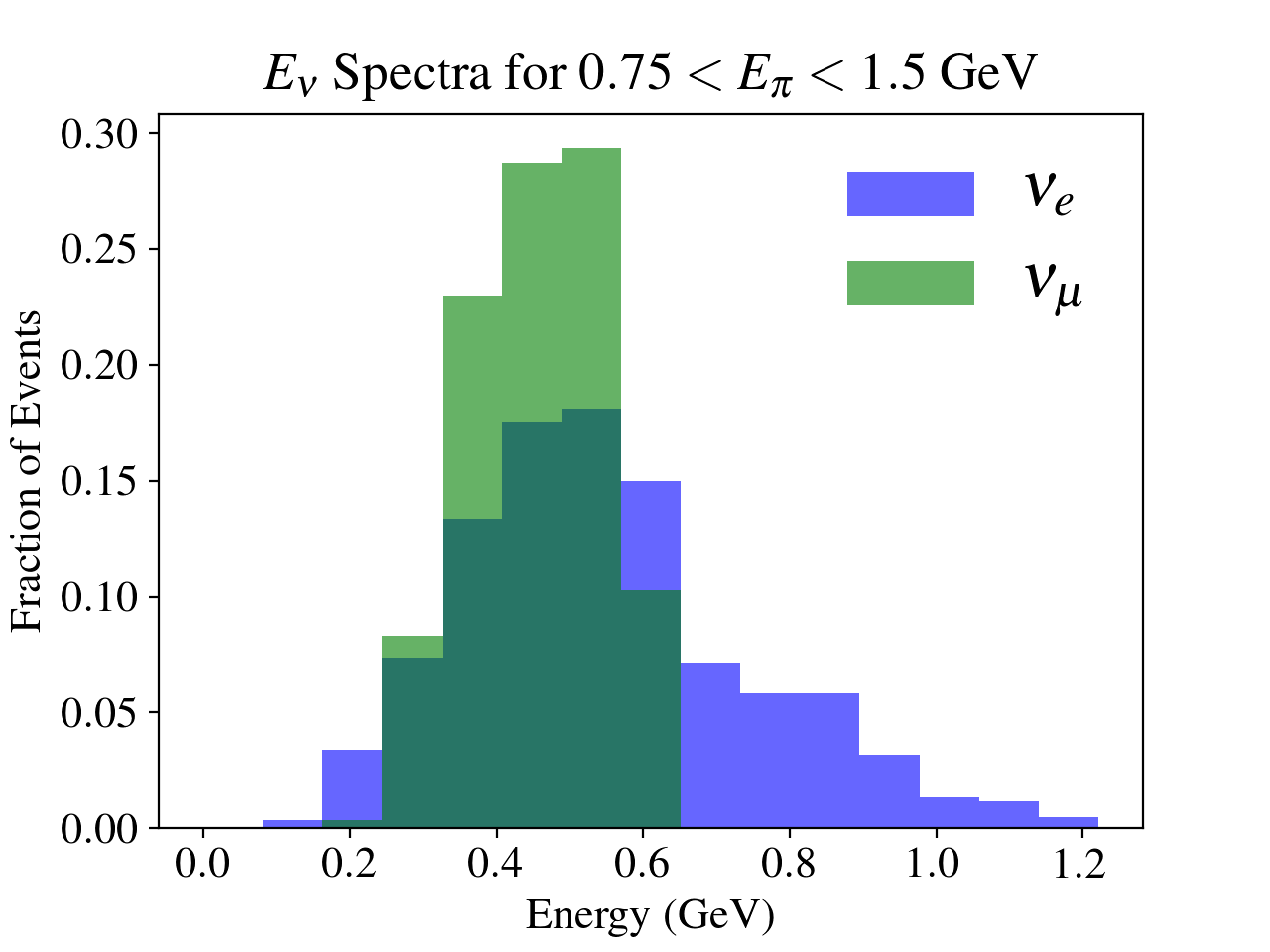}
~
\includegraphics[width=.3\linewidth]{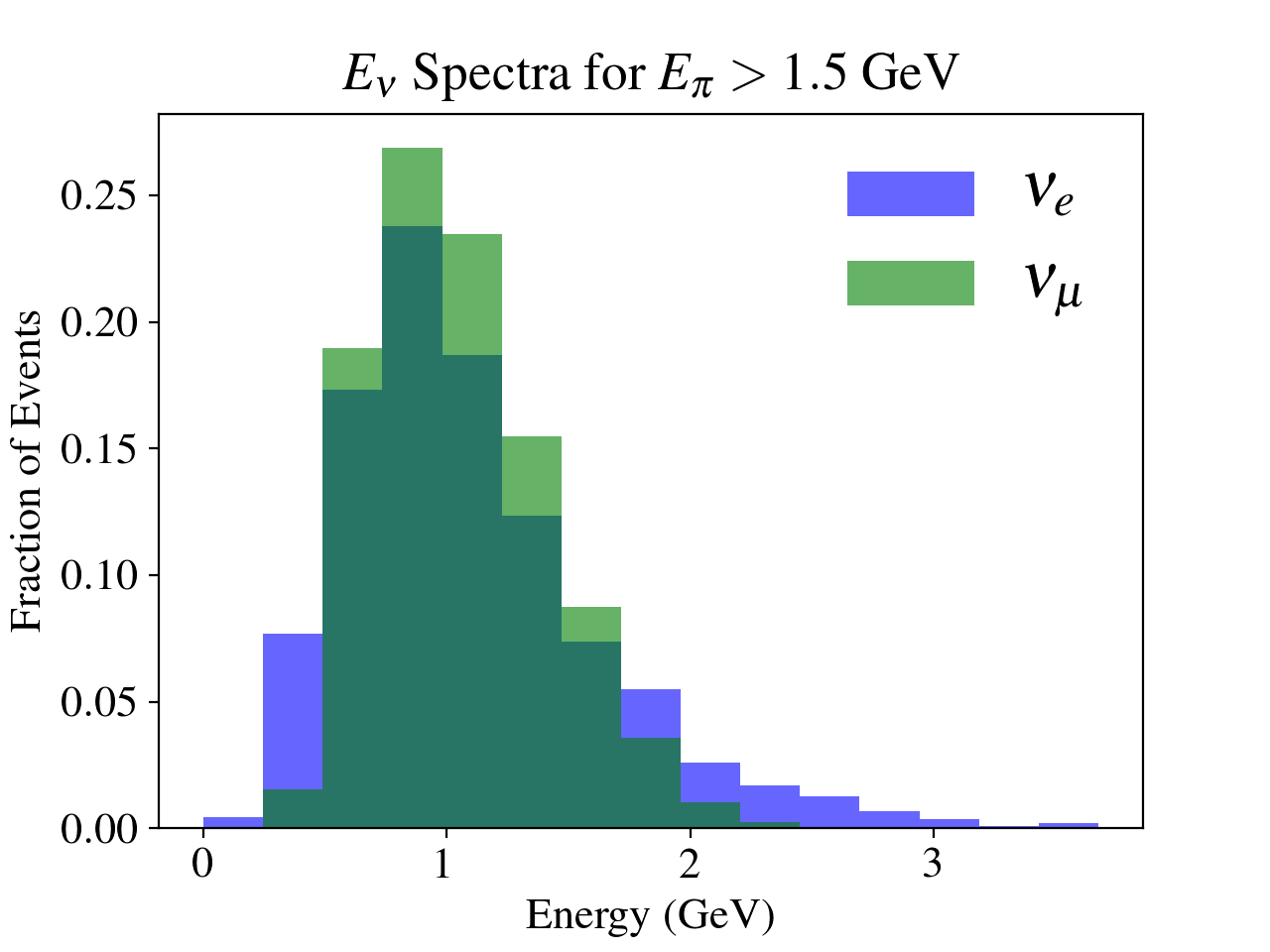}
\end{center}
\caption{For the BNB beam,  three bins in pion energy are displayed.  In each bin, the energy distribution of the $\nu_\mu$ (green) and $\nu_e$ (blue) CCQE events that are related to those pions are shown.  The events in each plot are relatively normalized \cite{lauren}.}
\label{pidk}
\end{figure*}

MiniBooNE was designed so that the uncertainties in the rates of other backgrounds were also constrained by measurements within the detector.  
For example, the $\nu_\mu$ events are used to constrain the flux of
intrinsic $\nu_e$ backgrounds from $\pi$ and $K$ decays in the
beamline using a method developed for MiniBooNE that is also being used in the MicroBooNE analysis.  The method also reduces systematic uncertainties associated with the cross section uncertainties but does rely on the assumption that there is not any sizable $\nu_\mu$ disappearance, which can be shown from other measurements.

In order to understand this $\nu_\mu$ constraint on the intrinsic $\nu_e$'s,
consider the $\pi$ decay chain as an example. This is $\pi^+ \rightarrow \mu^+  \nu_\mu$ followed by $\mu^+ \rightarrow e^+ \nu_e \bar \nu_\mu$.    The first decay is two-body.    This means that, at rest, the muon and the neutrino exit the decay with equal momentum and fixed kinetic energy.     The energy of the muon and the neutrino in the laboratory frame depends upon the magnitude of the boost and the angle of the particle production with respect to the boost.    The kinematics leads to a maximum energy that the $\nu_\mu$ can carry, which is 43\% of the pion energy.    
To calculate this expectation, use 
$\gamma = E_{\pi}^{lab}/m_\pi$; $\beta= p_\pi^{lab}/E_\pi^{lab}$; $\theta$, the angle with respect to the boost in the lab frame; and $E_\nu^{cm} = (m_\pi^2 - m_\mu^2)/(2 m_\pi)$.  Then one can derive:
\begin{eqnarray}
E_\nu^{lab} &=& \gamma E_\nu^{cm}(1+\beta \cos \theta)\\
&=& 0.215 E_\pi^{lab} (1+\beta \cos \theta), \label{pieq}
\end{eqnarray}
which, for $\theta=0$, reduces to $E_{\nu~max}^{lab}=0.43
E_\pi^{lab}$.    This is the maximum energy, but the $\nu_\mu$ associated with a given pion
energy tend to be tightly peaked close to the maximum since the MiniBooNE detector subtends a very small angle, 11 mr.  This is shown
in Fig.~\ref{pidk}, which shows the $\nu_\mu$ energy distribution for
three bins in parent pion energy in blue.  Thus, the measured $\nu_\mu$ energy spectrum constrains the $E_\pi^{lab}$ distribution, which produces it.
The muons from the pion decay also have an energy distribution that is tightly correlated with the $E_\pi^{lab}$ spectrum, although the subsequent three-body decay smears this correlation for
the $\nu_e$.    With that said, this three-body decay is well
understood, and so while the distributions for $\nu_e$ for a given
pion grand-parent, shown in blue in Fig.~\ref{pidk},  are wide, they
are well-predicted since the $E_\pi^{lab}$ spectrum has been effectively measured.    Using these connections, one can use the
measured $\nu_\mu$ events to strongly constrain the intrinsic $\nu_e$
events.

Overall, the MiniBooNE analysis is a good example of how oscillation experiments should exploit in-situ measurements to cross-check backgrounds and reduce systematic uncertainties.

\section{Techniques of Global Fits}

The experimental results discussed previously paint a disparate picture of the sterile neutrino landscape.
If we are to make sense of the available data, they must be combined into a single analysis that considers all results simultaneously: a global fit.
Given a model hypothesis, a global fit computes the likelihood for each experiment and combines them into a global likelihood.

Where experimental results agree, the global likelihood will be reinforced.
This reinforcement reduces the uncertainty on the model parameters, leading to tighter constraints.
In comparison, consider a case where multiple experiments see strong signals in different regions of parameter space.
Each experiment penalizes the others, creating a global likelihood that is highly penalized everywhere.

\subsection{Interpreting the Fit Results\label{interpret}}

\subsubsection{Inference framework}

A $3+1$ sterile neutrino hypothesis both with and without neutrino decay are considered. The $3+1$ model has three parameters that we are sensitive to: the mass splitting $\Delta m^2_{41}$ and the mixing matrix elements $|U_{e 4}|$ and $|U_{\mu 4}|$. The third mixing matrix element, $|U_{\tau 4}|$, is only constrained in regions of high mass, which lie outside our region of interest.

The $3+2$ model has seven parameters: the three parameters already mentioned for the $3+1$ model, an additional mass splitting $\Delta m^2_{51}$, two additional mixing matrix elements $|U_{e 5}|$ and $|U_{\mu 5}|$, and a CP violating phase $\phi_{54}$.

The $3+1$ model with $\nu_4$ decay has four parameters: the three parameters already mentioned for the vanilla $3+1$ model, and an additional decay parameter $\tau$, the lifetime. 

Each of these hypotheses are compared to the null hypothesis, which is formed by setting all the mixing matrix elements to zero. This construction of the null ensures that it forms a nested model.

Scanning over the entire $3+1$ with decay parameter space would require prohibitive computational resources, and so an adaptive sampling approach is used.
A Markov Chain Monte-Carlo (MCMC) is employed to explore only those regions of parameter space that contribute the most to the likelihood. 
The vanilla $3+1$ model is also sampled, to ensure consistency of the analysis with the decay model.

Both frequentist and Bayesian methods are considered.
The MCMC naturally produces samples from the posterior, which can be used to show Bayesian credible regions.
For each sample, a global $\chi^2$ is also calculated from the total of the individual $\chi^2$ values from each experiment.
These are used to show frequentist confidence regions.

\subsubsection{Likelihood function}

The Bayesian analysis makes use of a likelihood function.
In this global fit, all experimental data sets are binned.
Thus, for each experiment indexed by $\rho$, there is a corresponding prediction function $\vec{\Psi}_{\rho}(\vec{\theta})$ that computes the expected number of counts in each bin given the model parameters $\vec{\theta}$.

The likelihood function, $\mathcal{L}_{\rho}(\vec{d}_{\rho}|\vec{\theta})$, gives the probability of the measured data, $\vec{d}_{\rho}$, given the model parameters.
When the measured data is in the high statistics regime, it will be approximately normally distributed, and the likelihood function can take the form of a normal probability density function:
\begin{multline}
	\ln{\mathcal{L}_{\rho}(\vec{d}_{\rho}|\vec{\theta})} = - \frac{1}{2} \left[ \vec{d}_{\rho} - \vec{\Psi}_{\rho}(\theta) \right]^T \mat{\Sigma}_{\rho}(\theta)^{-1} \left[ \vec{d}_{\rho} - \vec{\Psi}_{\rho}(\theta) \right] \\ -
\frac{N_{\rho}}{2} \ln{(2\pi)} - \frac{1}{2}{\left|\mat{\Sigma}_{\rho}(\vec{\theta})\right|},
\end{multline}
where $\mat{\Sigma}_{\rho}(\theta)$ is the covariance matrix of experiment $\rho$, and $N_{\rho}$ is the number of bins of that experiment.

However, in the low statistics regime, the approximation is no longer valid.
Here, the data is assumed to be Poisson distributed, and the Poisson probability mass function is used for the likelihood:
\begin{multline}
	\ln{\mathcal{L}_{\rho}(\vec{d}_{\rho}|\vec{\theta})} = - \sum_{i}^{N_{\rho}} \left( [\vec{\Psi}_{\rho}(\theta)]_i \right.\\\left. - [\vec{d}_{\rho}]_i \ln{\left([ \vec{\Phi}_{\rho}(\theta) ]_i\right)} 	+ \ln{\Gamma(1 + [\vec{d}_{\rho}]_i)} \right),
\end{multline}
where $[\vec{d}_{\rho}]_i$ and $[\vec{\Psi}_{\rho}(\theta)]_i$ are the $i^{\text{th}}$ components of the data vector and prediction function, respectively.

The global likelihood is thus
\begin{align}
	\ln{\mathcal{L}(\vec{d}|\vec{\theta})} = \sum_{\rho}\ln{\mathcal{L}(\vec{d}_{\rho}|\vec{\theta})}.
\end{align}
The global $\chi^2$ is defined as
\begin{align}
	\chi^2(\theta) &= \sum_{\rho} \chi^2_{\rho}(\theta).
\end{align}

For high statistics experiments, the $\chi^2$ uses the standard form:
\begin{align}
	\chi^2_{\rho}(\theta) &= \left[ \vec{d}_{\rho} - \vec{\Psi}_{\rho}(\theta) \right]^T \mat{\Sigma}_{\rho}(\theta)^{-1} \left[ \vec{d}_{\rho} - \vec{\Psi}_{\rho}(\theta) \right].
\end{align} 
For low statistics experiments, the standard $\chi^2$ is no longer appropriate.
Instead, the saturated Poisson function is used:
\begin{align}
	\chi^2_{\rho}(\theta) &= - 2 \left( \ln{\mathcal{L}_{\rho}(\vec{d}_{\rho}|\vec{\theta})} - \ln{P(\vec{d}_{\rho}|\vec{d}_{\rho})} \right),
\end{align}
where $P(\vec{d}|\vec{\lambda})$ is the standard multi-dimensional Poisson distribution with mean $\vec{\lambda}$.

\subsubsection{Bayesian framework}

The likelihood function only specifies the probability of the measured data.
In a Bayesian analysis, the interesting quantity is the probability of the model parameters -- called the posterior distribution -- which can be found via Bayes' rule:
\begin{align}
	p(\vec{\theta} | \vec{d}) &= \frac{\mathcal{L}(\vec{d} | \vec{\theta}) \pi(\vec{\theta})}{ \mathcal{L}(\vec{d}) },
\end{align}
where $\mathcal{L}(\vec{d})$ is called the marginal likelihood, and $\pi(\vec{\theta})$ is called the prior.
As this marginal likelihood typically requires integrating over all model parameters, it can be difficult to compute.
A Markov Chain Monte-Carlo algorithm -- described below -- can avoid this complication, as it draws samples directly from the posterior by comparing ratios of probabilities.

In the Bayesian interpretation, the posterior, $p(\vec{\theta} | \vec{d})$, carries the information known about $\vec{\theta}$ after being updated by -- or conditioned on -- the observed data.
The justification for this interpretation is provided by the information theoretic entropy of the conditional distribution, which is equal to the prior known information of $\vec{\theta}$ plus the mutual information between the observed data and $\vec{\theta}$.
This forms a fundamental theorem of inference; in contrast, frequentist inference is performed in an ad-hoc manner without such a foundation.
However, in practice, frequentist methods can provide a calibration of expected results: if the p-value for the observed data is small, it may suggest that something unexpected -- and interesting -- is happening and warrants further study.
Thus, both Bayesian inference and frequentist methods should be applied to a model fitting problem, as they provide complementary information about the model and the data.

The prior encodes any previously known information on the model parameters.
When no such information is present -- for example in a global fit where all available data is being analyzed -- a choice of prior must be made.
Typically, a wide, high entropy distribution is chosen as the broad range of accepted values reflects our relative ignorance of where the true model parameters lie.

For bounded parameters, uniform, or wide exponential or wide normal distributions are a common choice.
For unbounded parameters, uniform distributions cannot be used as the distribution cannot be normalized.
In this case, a common replacement is a distribution that is uniform inside a certain finite range and exactly zero outside of that range.

In this parameterization, the mass splitting is unbounded by above.
The prior on the mass splittings was chosen to be a log-uniform distribution in the range of $10^{-2}$ eV${}^2$ to $10^{2}$ eV${}^2$.
The decision to use logarithmic coordinates was based on the observation of the fundamental particle masses: they tend to be distributed more uniformly in log-space, compared to linear-space.
The choice of uniform prior was motivated by the need for a hard cut-off at high mass splittings, as the computation time of the likelihood increases with the frequency of the oscillation waves.
Without this requirement, a softer cut-off such as an exponential-family distribution could be appropriate.
The prior on the matrix elements was chosen to also be a log-uniform distribution in the range of $10^{-2}$ to $1$.
The choice follows the presentation of results in logarithmic axes.
In principle, a soft cut-off at low $U$ could be imposed with an exponential prior -- corresponding to a uniform prior in linear-space -- but was not explored for this review.
Finally, the CP-violating phase prior was chosen to be uniform in angle.

Results are most often presented as either a heat-map or histogram of the posterior, or using credible regions.
Credible regions are similar in nomenclature to confidence regions: a $100\alpha$\% credible region is defined as a set $\mathcal{C}(\alpha)$ that satisfies
\begin{align}
    \int_{\mathcal{C}(\alpha)} p(\vec{\theta} | \vec{d}) d\vec{\theta} &= \alpha. \label{eq:credible_region_def}
\end{align}
So for a 95\% region, $\alpha = 0.95$.

Further refinement is needed, as multiple choices of $\mathcal{C}$ can satisfy this requirement.
For multidimensional distributions, the Highest Posterior Density (HPD) credible region is most often used.
This is the unique solution to the requirement that all points in the credible region have a higher probability density than all points outside the region.

The HPD region can be codified by selecting a threshold value $t(\alpha)$ which defines the credible region as
\begin{align}
    \mathcal{C}(\alpha) &= \{ \vec{\theta} : p(\vec{\theta} | \vec{d}) > t(\alpha) \}. \label{eq:hpd_def}
\end{align}
Then, $t(\alpha)$ and $\mathcal{C}(\alpha)$ are the unique solutions to Eqs. \ref{eq:credible_region_def} and \ref{eq:hpd_def}.

\subsubsection{Model comparison}\label{subsubsec:model_comparison}

A frequentist difference of $\chi^2$ metric is used to compare the sterile neutrino model to the null hypothesis.
A Bayes factor is another valid choice, but was not used in this analysis as many experiments included in the global fit use pull terms as nuisance parameters.
The value of the $\chi^2$ or likelihood function for any given specific model parameters is defined to be the minimum over the pull terms.
This minimization procedure will affect the normalization of the likelihood, rendering a meaningless Bayes factor.
To compute the Bayes factor correctly, the pull terms must be promoted to full parameters of the MCMC so that they may be properly marginalized; however, this was considered outside the scope of the current analysis.

A likelihood ratio style model comparison can be performed using a difference between the best-fit $\chi^2$ of the model in question, and the $\chi^2$ of the null hypothesis:
\begin{align}
    \Delta \chi^2 &= \chi^2_{\text{null}} - \chi^2_{\text{min}} \label{dchi2}.
\end{align}
If $\chi^2_{\text{null}}$ and $\chi^2_{\text{min}}$ are $\chi^2$-distributed, then $\Delta \chi^2$ will also be $\chi^2$-distributed. 
The number of degrees of freedom is equal to the number of parameters in the model that are not present in the null hypothesis. 
It should be noted that when fitting appearance only data in a $3+1$ model, the effective number of parameters is only 2, not 3.
In this case, the $|U_{e 4}|$ and $|U_{\mu 4}|$ terms are multiplied together, forming a single free parameter.

The $100(1-\alpha)$\% confidence region can then be defined in terms of this $\Delta \chi^2$ metric. 
Here, the model comparison metric is defined to be between the best-fit and any location in model space.
The confidence region $\mathcal{R}(\alpha)$ is defined as the set of points that do not deviate from the best-fit by more than $\alpha$ significance:
\begin{align}
    \mathcal{R}(\alpha) &= \{ \vec{\theta} : \chi^2(\vec{\theta}) - \chi^2_{\text{min}} < \textrm{CDF}^{-1}_{\chi^2}(k, 1 - \alpha) \}, \label{eq:conf_region_def}
\end{align}
where $\textrm{CDF}^{-1}_{\chi^2}$ is the inverse cumulative distribution function for a $\chi^2$ distribution with $k$ degrees of freedom, and $k$ is the number of effective model parameters.

When the number of model parameters is larger than two, effective presentation of the confidence regions can be difficult.
One approach is to reduce the dimensionality of the parameter space by profiling the $\chi^2$.
For presentation in a two dimensional graphic, two model parameters - here denoted by $\vec{\phi}$ - are chosen.
The $\chi^2$ is then minimized over the remaining model parameters -- here denoted by $\vec{\psi}$:
\begin{align}
    \hat{\chi}^2(\vec{\phi}) &= \min_{\vec{\psi}}{ \chi^2(\vec{\phi}, \vec{\psi}) }.
\end{align}
A two dimensional confidence region can be drawn using $\hat{\chi}^2$, with two degrees of freedom, as all degrees of freedom but two were removed from the $\chi^2$ by the minimization procedure.

\subsection{Markov Chain Monte-Carlo implementation}

The Markov Chain Monte-Carlo algorithm is designed to efficiently draw samples from a probability distribution.
The Markov chain is defined as a history of samples already drawn by the algorithm, with the current sample at the head.
A proposal is then drawn from a proposal distribution, which can be a function of the current sample only.
A common -- if inefficient -- choice of proposal distribution is a normal distribution centered on the current sample.
The probability of this proposal -- defined by the likelihood and prior in our case -- is then compared to the probability of the current sample.
The proposal is accepted as the new head of the chain with probability
\begin{align}
	\alpha_{\text{accept}} &= \min{\left(1, \frac{\mathcal{L}(\vec{d}|\vec{\theta}') \pi(\vec{\theta}') p(\vec{\theta}'\rightarrow\vec{\theta})}{\mathcal{L}(\vec{d}|\vec{\theta}) \pi(\vec{\theta}) p(\vec{\theta}\rightarrow\vec{\theta}')} \right)},
\end{align}
where $\pi(\vec{\theta})$ is the prior, and $p(\vec{\theta}\rightarrow\vec{\theta}')$ is the probability of proposing a move from $\vec{\theta}$ to $\vec{\theta}'$.
Thus, the algorithm always accepts new samples that have a higher probability, and has a chance of accepting samples with lower probability, thus exploring the parameter space of the distribution.

The parallel tempering affine invariant algorithm~\cite{Foreman-Mackey_Hogg_Lang_Goodman_2013} was used to generate proposals for the MCMC.
This algorithm maintains an ensemble of chains, called ``walkers''.
New samples are proposed for each walker by randomly selecting another walker from within the ensemble and moving toward or away from it based on their mutual distance.
In this way, proposals automatically scale to match the current estimate of the posterior distribution.

At iteration $i$, chain $\gamma$ has a chain head $\vec{\theta}_i^\gamma$.
An affine invariant proposal \cite{Goodman_Weare_2010} is then made by first drawing a random chain $\kappa \neq \gamma$, with chain head $\vec{\theta}_i^\kappa$.
Then, the proposal is defined by  
\begin{align}
    {\vec{\theta}_i^{\gamma\prime}} &= \vec{\theta}_i^\kappa + z ( \vec{\theta}_i^\alpha - \vec{\theta}_i^\kappa),
\end{align}
where $z$ is randomly sampled from the distribution
\begin{align}
    h(z) \propto \begin{cases} \frac{1}{\sqrt{z}} & \frac{1}{a} \leq z \leq a \\ 0 & \text{otherwise} \end{cases},
\end{align}
and $a$ is a tunable parameter that is typically set to 2.
This proposal is then accepted with probability
\begin{align}
    \alpha_{\text{accept}} &= \min{\left(1, z^{N-1} \frac{ \mathcal{L}(\vec{d} | {\vec{\theta}_i^{\gamma\prime}}) \pi({\vec{\theta}_i^{\gamma\prime}}) }{ \mathcal{L}(\vec{d} | \vec{\theta}_i^\gamma ) \pi(\vec{\theta}_i^\gamma) } \right)}.
\end{align}

This ensemble of walkers is then organized into a super-ensemble, in which each ensemble operates at a different temperature.
The temperature parameter, $T$, modifies the likelihood surface to increase the probability of accepting proposals:
\begin{align}
    \mathcal{L}(\vec{d} | \vec{\theta}) \pi(\vec{\theta}) &\rightarrow  \exp{\left[- \frac{1}{T} E(\vec{\theta}) \right]},
\end{align}
where the energy, $E$, is defined as
\begin{align}
    E(\vec{\theta}) &= -\ln{\left[ \mathcal{L}(\vec{d} | \vec{\theta}) \pi(\vec{\theta}) \right]}.
\end{align}
This can also be specified in terms of the inverse temperature parameter $\beta=1/T$.

After each affine invariant proposal and update step, a parallel tempering update \cite{Earl_Deem_2005} occurs with a probability of 10\%.
This update -- called replica exchange -- randomly selects a pair of walkers, $\vec{\theta}^\gamma$ and $\vec{\theta}^\kappa$, that belong to two different temperatures --- $\beta_k$ and $\beta_j$ respectively.
The walker then swap positions in parameter space with probability
\begin{align}
    \alpha_{\text{exch}} &= \min{\left(1, \exp{\left[ (\beta_k - \beta_l) (E(\vec{\theta}^\gamma) - E(\vec{\theta}^\kappa)) \right]} \right)}.
\end{align}

\subsection{Presentation of results}

The result of the MCMC is a set of samples from the posterior, $\mathcal{S}$, along with their associated $\chi^2$ values.
Confidence regions are drawn by first selecting the subset of samples whose $\chi^2$ values are less than the critical threshold set by the inverse cumulative distribution function in Eq.~\ref{eq:conf_region_def}:
\begin{align}
    \hat{\mathcal{R}}(\alpha) &= \{ \vec{\theta} \in \mathcal{S} : \chi^2(\vec{\theta}) - \hat{\chi}^2_{\text{min}} < \textrm{CDF}^{-1}_{\chi^2}(k, 1 - \alpha) \},
\end{align}
where $\hat{\chi}^2_{\text{min}}$ is the smallest $\chi^2$ in $\mathcal{S}$.
This subset is then projected into the two dimensional subspace for the desired coordinates, and then plotted.
The plotting is performed in ascending order of $\alpha$, such that the smaller regions with high $\alpha$ overlay the larger regions with low $\alpha$.

An estimate of the posterior must be made to correctly show credible regions in a lower dimensional parameter subspace.
This can be done by either histogramming the samples in this subspace, or using kernel density estimation.
Once this estimate has been generated, Eqs. \ref{eq:credible_region_def} and \ref{eq:hpd_def} can be applied to find the region.
For the results presented in this study, the credible regions were generated with the \texttt{corner.py} library \cite{corner}.

\subsection{Test statistic distributions for ratios}

Even though it has been a standard technique to use the ratio of near-to-far experiments to search for sterile neutrinos, recently reactor neutrino experiments have extended this technique to avoid dependence on the absolute flux normalization and only rely on the shape difference between near and far. In this section, we introduce the ratio test-statistic used and outline its properties.

Consider a detector in two positions or two detectors in two positions. They both measure the distribution of neutrino events in the same set of energy bins. Label the counts in the {\it i}-th energy bin as $N_i$ for the first detector and the counts in the same energy bin in the second detector $\tilde N_i$. We will further assume that the number of expected events per bin is well-described by a normal distribution with parameters $\mu_i$ and $\sigma_i$ for the first detector and similarly tilde parameters for the second detector.

The ratio test statistic is given by
\begin{equation}
R_i = \frac{N_i}{\tilde N_i}.
\end{equation}
This distribution can be used to search for shape and normalization effects.
More over, one can show that this variable can be approximated, under appropriate conditions, as a normal distribution~\cite{doi:10.1287/mnsc.21.11.1338}.
In the case of reactor neutrinos the flux normalization is not well understood; for this reason a new test statistic has been introduced that is agnostic to the observed rate. 
This test statistics is the normalized-ratio
\begin{equation}
{\rm NR}_i = \frac{N_i}{\sum_j N_j} \frac{\sum_j \tilde N_j}{\tilde N_i},
\end{equation}
which is the ratio of the shapes normalized to the observed events counts in each of the detectors. 
Since $NR_i$ depends on the total number of events across all energy bins per detector, this implies that the $NR_i$ are correlated among each other unlike the case of $R_i$.

In the large sample size this variable is well-described by a multidimensional log-normal distribution, with a covariance matrix that is given by
\begin{equation}
\Sigma_{\rm NR} = J \Sigma_{N,\tilde N} J^T,
\end{equation}
where $\Sigma_{N,\tilde N} = {\rm diag}(\sigma_1^2, ..., \sigma_{N_e}^2,\tilde\sigma_1^2, ..., \tilde\sigma_{N_e}^2)$, $N_e$ is the number of energy bins, $\sigma_i$ and $\tilde\sigma_i$ the standard deviations of the normal distributions in the near and far detectors respectively, and $J$ the Jacobian matrix of the function $\vec {\rm NR}(\vec N, \vec {\tilde N})$. This results in the following covariance
\begin{align}
\left(\Sigma_{\rm NR}\right)_{ij} = \sum_k^{N_e}
& \left(\frac{\delta_{ik}}{\mu_i} - \frac{1}{\mu_T}\right)
\left(\frac{\delta_{jk}}{\mu_j} - \frac{1}{\mu_T}\right) \sigma_k^2 \nonumber \\
& +
\left(\frac{\delta_{ik}}{\tilde\mu_i} - \frac{1}{\tilde\mu_T}\right)
\left(\frac{\delta_{jk}}{\tilde\mu_j} - \frac{1}{\tilde\mu_T}\right) \tilde\sigma_k^2,
\end{align}
where $\mu_T = \sum_i^{N_e} \mu_i$ and similarly for the $\tilde\mu_T$ with $\tilde\mu_i$ instead of $\mu_i$. If we assume the $\sigma^2_i = \mu_i$ and similarly for the tilde terms, this simplifies to 
\begin{align}
\left(\Sigma_{\rm NR}\right)_{ij} = \frac{\delta_{ij}}{\mu_i} - \frac{1}{\mu_T} + \frac{\delta_{ij}}{\tilde\mu_i} - \frac{1}{\tilde\mu_T}.
\end{align}

\subsection{Closed form prediction function}

Many experiments use isotropic neutrino sources, which admit closed-form expressions for the predicted number of neutrino events in a bin.
For 3+1 and 3+2, the prediction functions have the general form:
\begin{align}
    \Psi \propto \frac{P(\nu_\alpha \rightarrow \nu_\beta)}{L^2}.
\end{align}
Integrating this expression in length yields a closed-form solution using exponential integrals, for which approximations can be found in many numerical libraries.
In the case of a decay model, the prediction function has an additional exponential term:
\begin{align}
    \Psi \propto e^{-L/L_0} \frac{P(\nu_\alpha \rightarrow \nu_\beta)}{L^2}.
\end{align}
A closed-form solution for the integral of this expression in length is also available in term of complex exponential integral functions.
Although the complex form of these functions are not common in numerical libraries, approximation algorithms exist.
The algorithm of Ref.~\cite{PegoraroJGT11ECVEI} was used for this study.

\section{Models and Global Fit Results}

\subsection{3+1 Model}

In a 3+1 model, we fit for three parameters, $|U_{e4}|$, $|U_{\mu 4}|$, and $\Delta m_{41}$, as introduced in Eqs.~\ref{ee}, \ref{mumu}, and \ref{mue}.

\begin{figure}
\includegraphics[width=\linewidth]{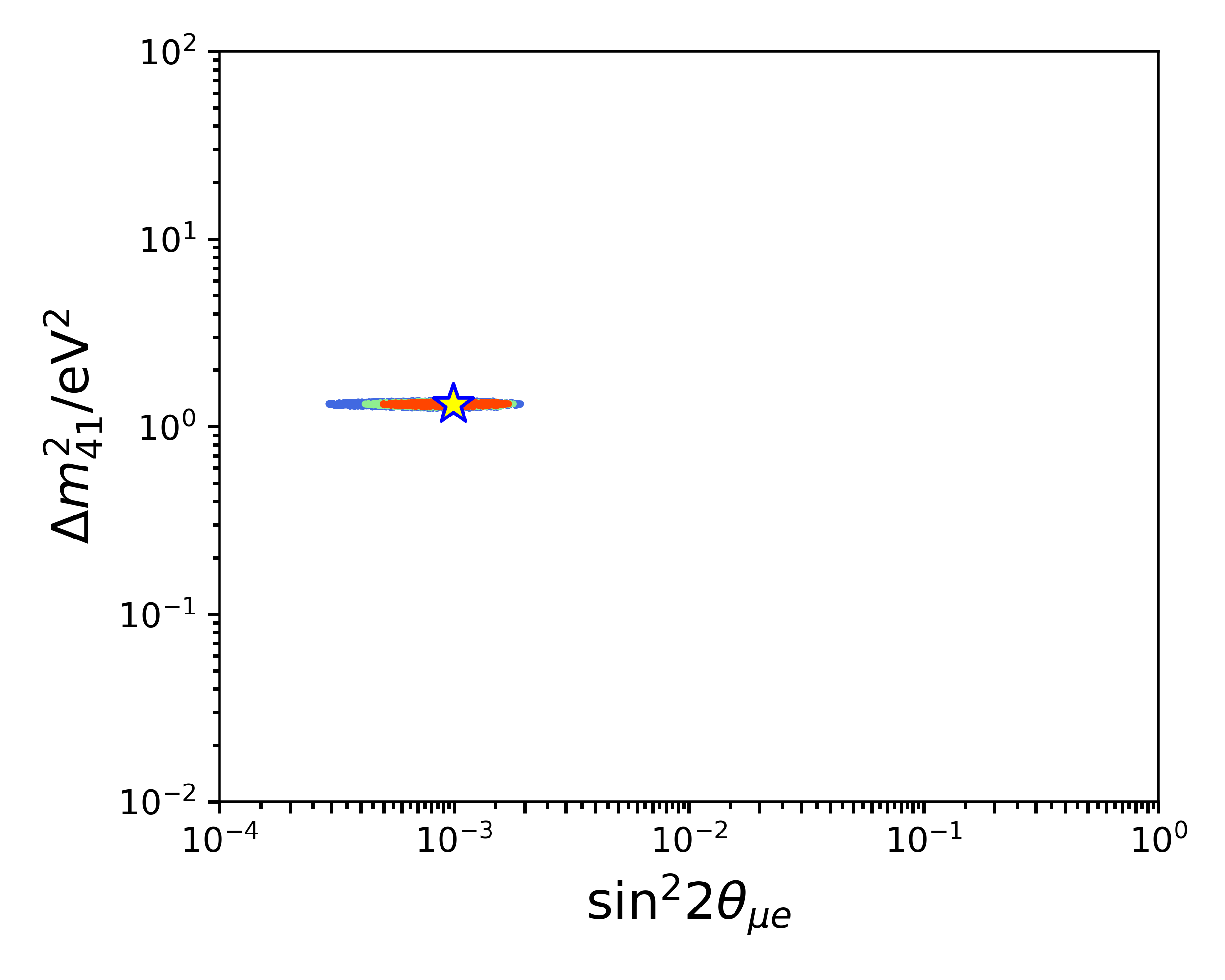}\\
\includegraphics[width=\linewidth]{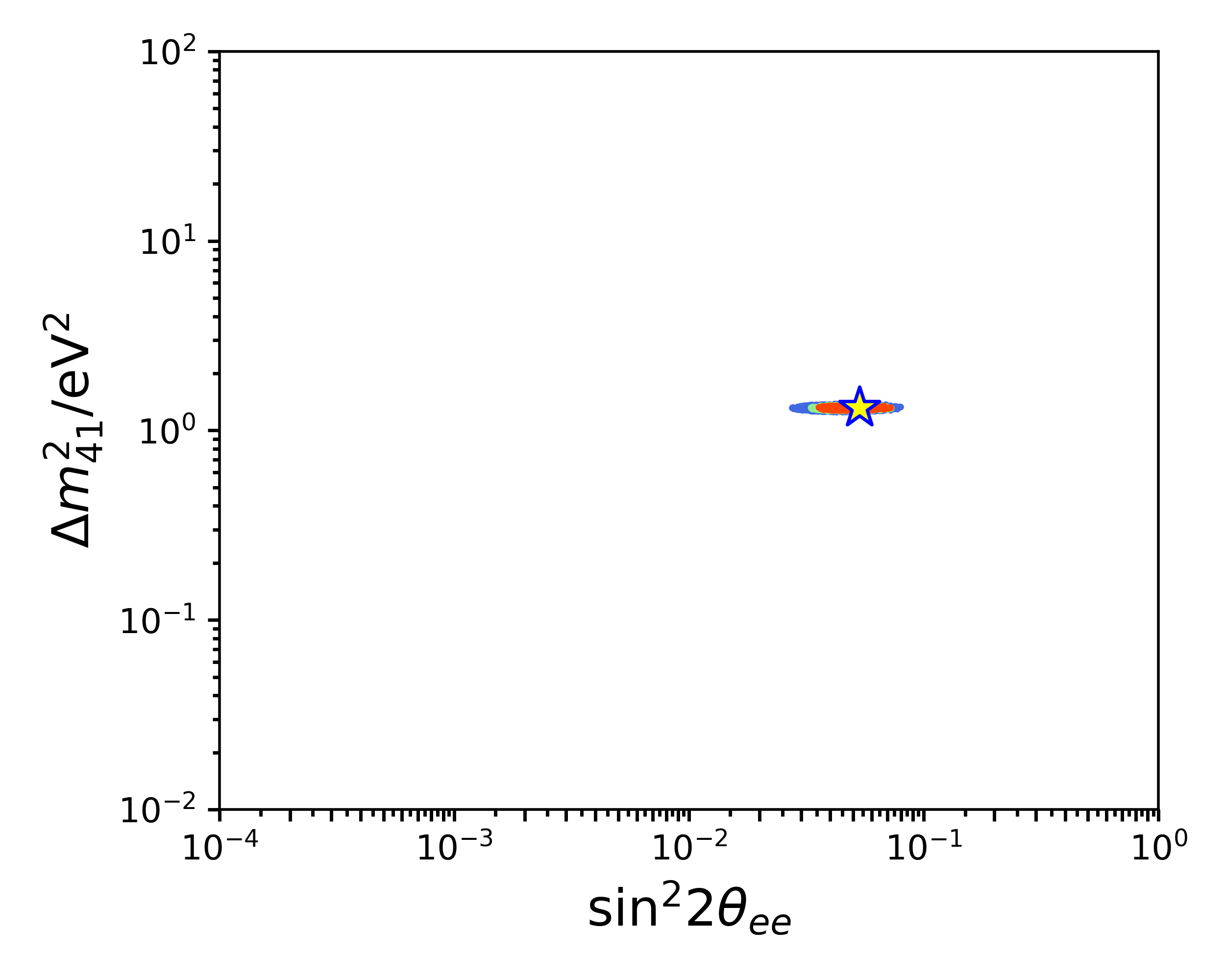}\\
\includegraphics[width=\linewidth]{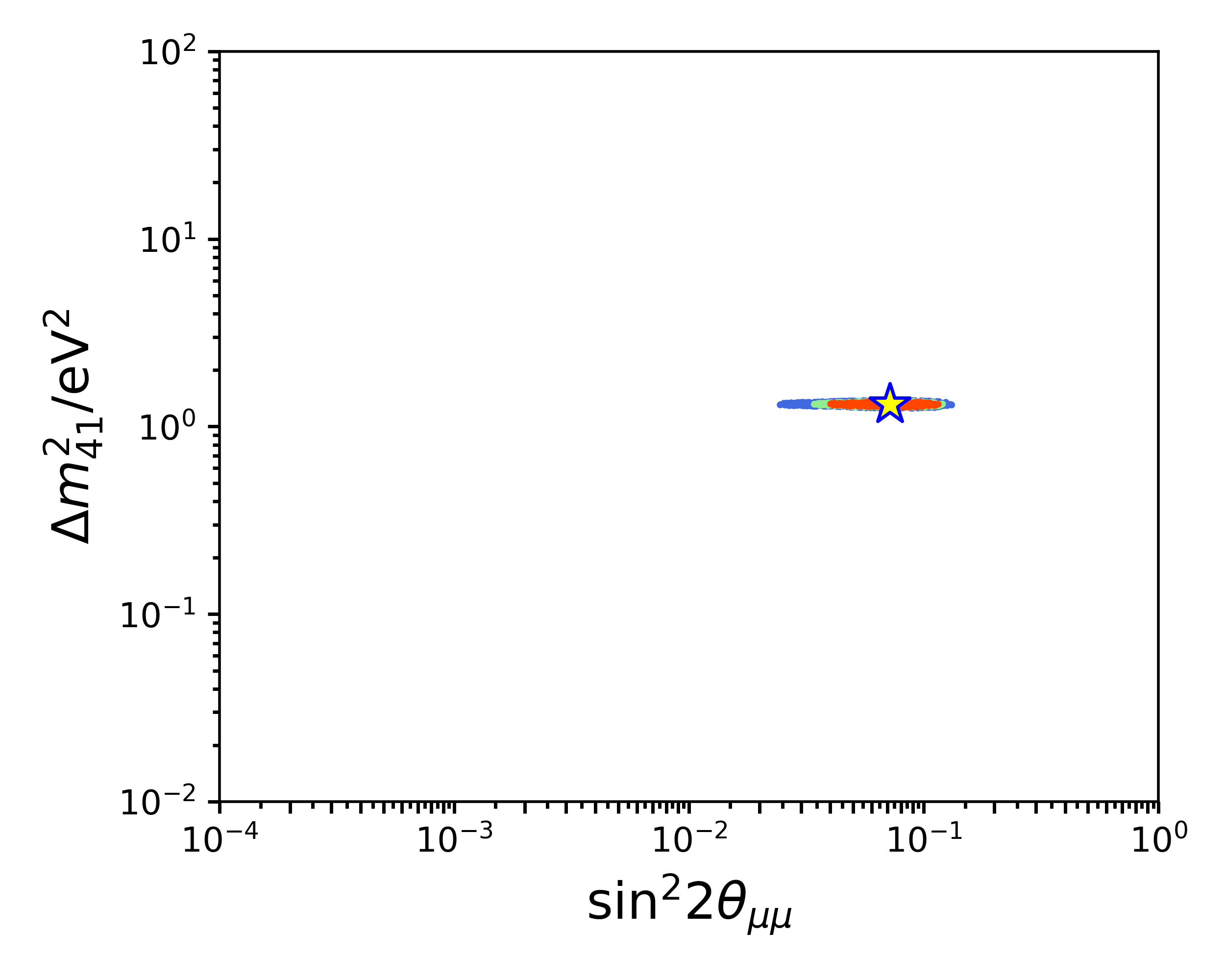}
\caption{Frequentist Confidence Regions for a 3+1 global fit, showing the 99\%, 95\%, and 90\% confidence levels in blue, green, and red, respectively. Top: $\sin^2 2\theta_{\mu e}$ {\it vs.} $\Delta m^2_{41}$; Middle: $\sin^2 2\theta_{e e}$ {\it vs.} $\Delta m^2_{41}$; Bottom: $\sin^2 2\theta_{\mu \mu}$ {\it vs.} $\Delta m^2_{41}$.} 
\label{fig:3plus1}
\end{figure}

\subsubsection{Frequentist method}

Figure~\ref{fig:3plus1} shows the confidence regions for the frequentist fits to the 3+1 model.
These are fits to all experiments described in Sec. \ref{experiments}.
The figures show $\Delta m^2_{41}$ as function of the mixing angles where the three plots correspond to $\sin^2 2\theta_{\mu e}$, $\sin^2 2\theta_{e e}$, and $\sin^2 2\theta_{\mu \mu}$.   Thus, these correspond to predictions for future searches in appearance, $\nu_e$ disappearance, and $\nu_\mu$ disappearance, respectively.    As a reminder, the connections between these mixing angles and the matrix parameters $|U_{\mu4}|$ and $|U_{e4}|$ are given in 
Table~\ref{cheatsheet}.
The regions for the 99\%, 95\%, and 90\% are shown in blue, green, and red, respectively. 
The inclusion of new experiments, particularly the reactor experiments, has diminished the likely parameter space for a sterile neutrino from past global fits \cite{Collin:2016rao,Collin:2016aqd}, leaving only one allowed ``island.''

%\begin{center}
%\begin{table*}
%\begin{tabular}{l | c | c | c | c | c | c | c | c | c }
%&  $|U_{e4}|$ & $|U_{\mu4}|$ & $\sin^2(2\theta_{\mu e})$ & $\Delta m_{41}^2~\text{(eV)}^2$ & $ \tau~\text{(eV)}^{-1}$ & $\chi^2_{\text{Null}} $ & $\chi^2_{\text{Best Fit}}$ & $\Delta \chi^2$ & $\Delta dof$ \\
%\hline
%3 + 1 global & 0.116 & 0.135 & 0.001 & 1.32 & - & 493 & 458 & 35 & 3 \\ 
%\hline
%3 + 1 + Decay global & 0.428 & 0.180 & 0.024 & 0.211 & 1.96 & 493 & 450 & 43& 4\\
%\end{tabular}
%\caption{A summary of the best fit parameters found for each considered model, their best fit $\chi^2$, $\Delta \chi^2=\chi^2_{\text{Null}}-\chi^2_{\text{Best Fit}}$ and $\Delta dof=dof_{\text{Null}}-dof_{\text{Best Fit}}$.}
%\label{table:fitresults}
%\end{table*}
%\end{center}

\begin{center}
\begin{table*}
\begin{tabular}{l | c | c | c | c | c | c | c |c }
Global fit&  $|U_{e4}|$ & $|U_{\mu4}|$ & $|U_{e5}|$ & $|U_{\mu 5}|$ & $\phi_{54}$ (rad) & $\Delta m_{41}^2~\text{(eV)}^2$ &  $\Delta m_{54}^2~\text{(eV)}^2$ & $ \tau~\text{(eV)}^{-1}$ \\
\hline
3 + 1  & 0.116 & 0.135 & - & - &-& 1.32 & -& - \\ 
\hline
3 + 2  & 0.106 & 0.082 & 0.252 & 0.060 & 0.009 & 1.32 &  12.6 & - \\ 
\hline
3 + 1 + Decay & 0.428 & 0.180 & - & - & -& 0.211 & - & 1.96 \\
\end{tabular}
\caption{A summary of the best fit parameters found for each model involving sterile neutrinos.}
\label{table:fitresults}
\end{table*}
\end{center}

\begin{center}
\begin{table*}
\begin{tabular}{l | c | c | c | c  }
Fit type:		&	$3\nu$ (null)	&	3+1	&	3+2	&	3+1+decay	 \\ \hline
Best Fits		&		&		&		&		\\
	$\chi^2$	&	493	&	458	&	449	&	450	\\
	$dof$	&	509	&	506	&	502	&	505	\\
	$p$-value	&	0.687	&	0.938	&	0.957	&	0.962	 \\ \hline
(Null vs Sterile)		&		&		&		&		\\
	$\Delta \chi^2$	&		&	35	&	44	&	43	\\
	$\Delta dof$	&		&	3	&	7	&	4	\\
	$p$-value	&		&	1.2E-07	&	2.1E-07	&	1.0E-08	\\
	$N\sigma$	&		&	5.2	&	5.1	&	5.6	 \\ \hline
(3+1  vs Other)		&		&		&		&		\\
	$\Delta \chi^2$	&		&		&	9	&	8	\\
	$\Delta dof$	&		&		&	4	&	1	\\
	$p$-value	&		&		&	0.0611	&	0.0047	\\
	$N\sigma$	&		&		&	1.5	&	2.6	 \\ \hline
(PG Test)		&		&		&		&		\\
	$\chi^2_{app}$	&		&	77	&	69	&	77	\\
	$N_{app}$	&		&	2	&	5	&	3	\\
	$\chi^2_{dis}$	&		&	356	&	350	&	356	\\
	$N_{dis}$	&		&	3	&	6	&	4	\\
	$\chi^2_{glob}$	&		&	458	&	449	&	450	\\
	$N_{glob}$	&		&	3	&	7	&	4	\\
	$\chi^2_{PG}$	&		&	25	&	30	&	17	\\
	$N_{PG}$	&		&	2	&	4	&	3	\\
	$p$-value	&		&	3.7E-06	&	4.9E-06	&	7.1E-04	\\
	$N\sigma$	&		&	4.5	&	4.4	&	3.2	 \\ \hline
\end{tabular}
\caption{A summary of the quality of the fits.  Columns correspond to the four types of fits.  Top section:  Best fit results for each model; Second section: Comparison of quality of null to each fit including sterile neutrinos; Third section:  Comparison of 3+1 to the extended models; Bottom section:  PG test results for each model, where Eqs.~\ref{chi2pg} and \ref{npg} explain how $\chi^2_{PG}$ and $N_{PG}$ are determined.}
\label{table:fitquality}
\end{table*}
\end{center}

The best fit parameters for the 3+1 model, shown in Table~\ref{table:fitresults}, correspond to $\Delta m_{41}^2 = 1.32~\text{eV}^2$ and $\sin^2(2\theta_{\mu e}) = 0.001$. Compared with our previous result \cite{Collin:2016rao}, the best fit point has shifted to a slightly lower value in both $\sin^2(2\theta_{\mu e})$ and $\Delta m^2_{41}$.   

The quality of the fits are presented in Table~\ref{table:fitquality}.  
In our global fit to the 509 $(L,E)$ bins from all experiments, the 3+1 model has a $\chi^2$ of 458, while the null model has a $\chi^2$ of 493.   Thus, each case has an excellent $\chi^2$/$dof$.  But this occurs because most of $(L,E)$ bins in the fit are not in regions that are sensitive to sterile oscillations.   In order to isolate the $\chi^2$ contribution to the bins with sensitivity, we must use the $\Delta\chi^2$ as described in Eq.~\ref{dchi2} to compare the 3+1 and null models.  
The $\Delta\chi^2$ is found to be 35 with the inclusion of 3 new degrees of freedom--a very strong improvement in the data, which indicates that the 3+1 model is favored over the null model by over 5$\sigma$.  While this does not prove the existence of sterile neutrinos, it indicates that the data strongly prefers a sterile-like signal over the null hypothesis. 

\begin{figure}
\centering
\includegraphics[width=\linewidth]{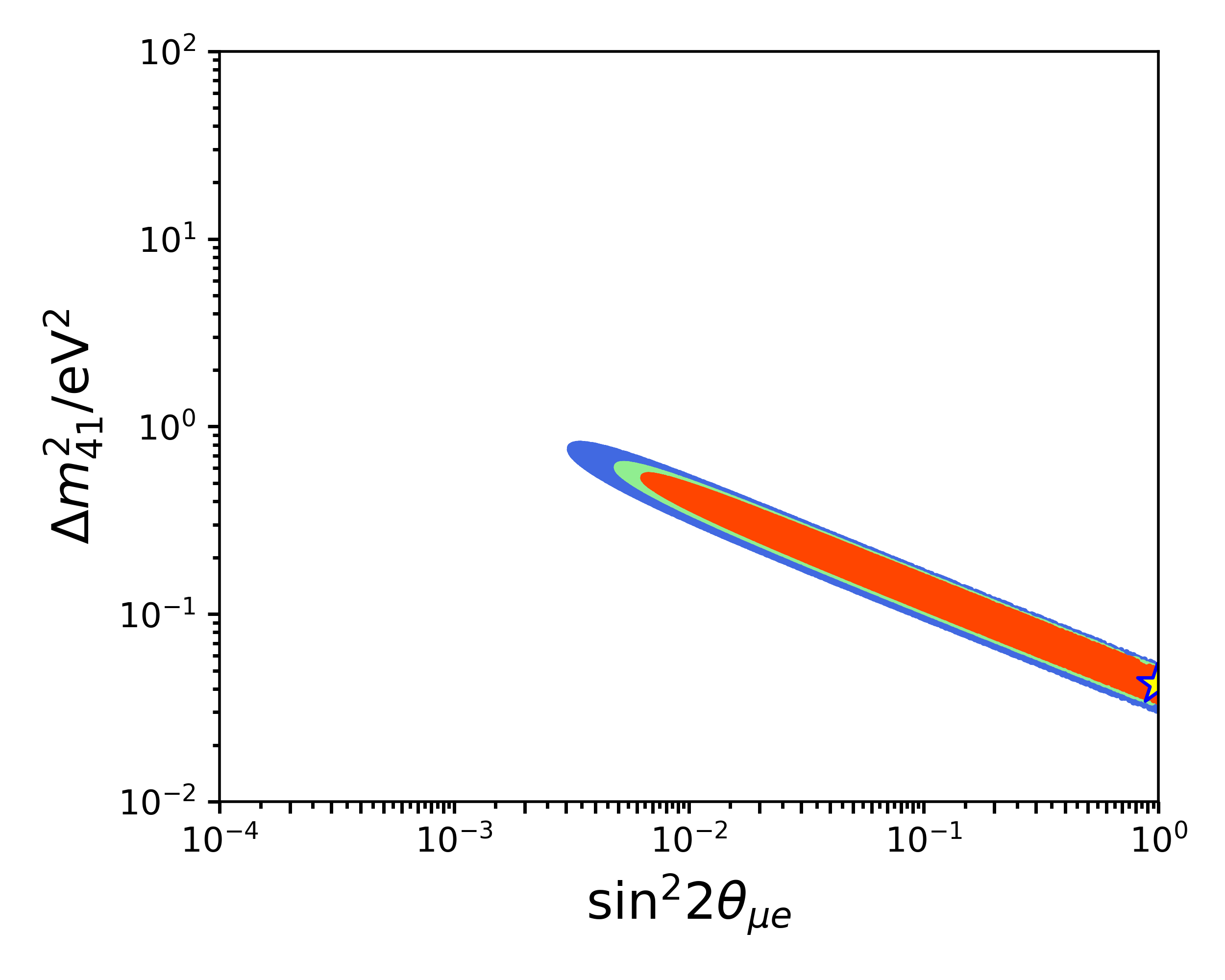}\\
\includegraphics[width=\linewidth]{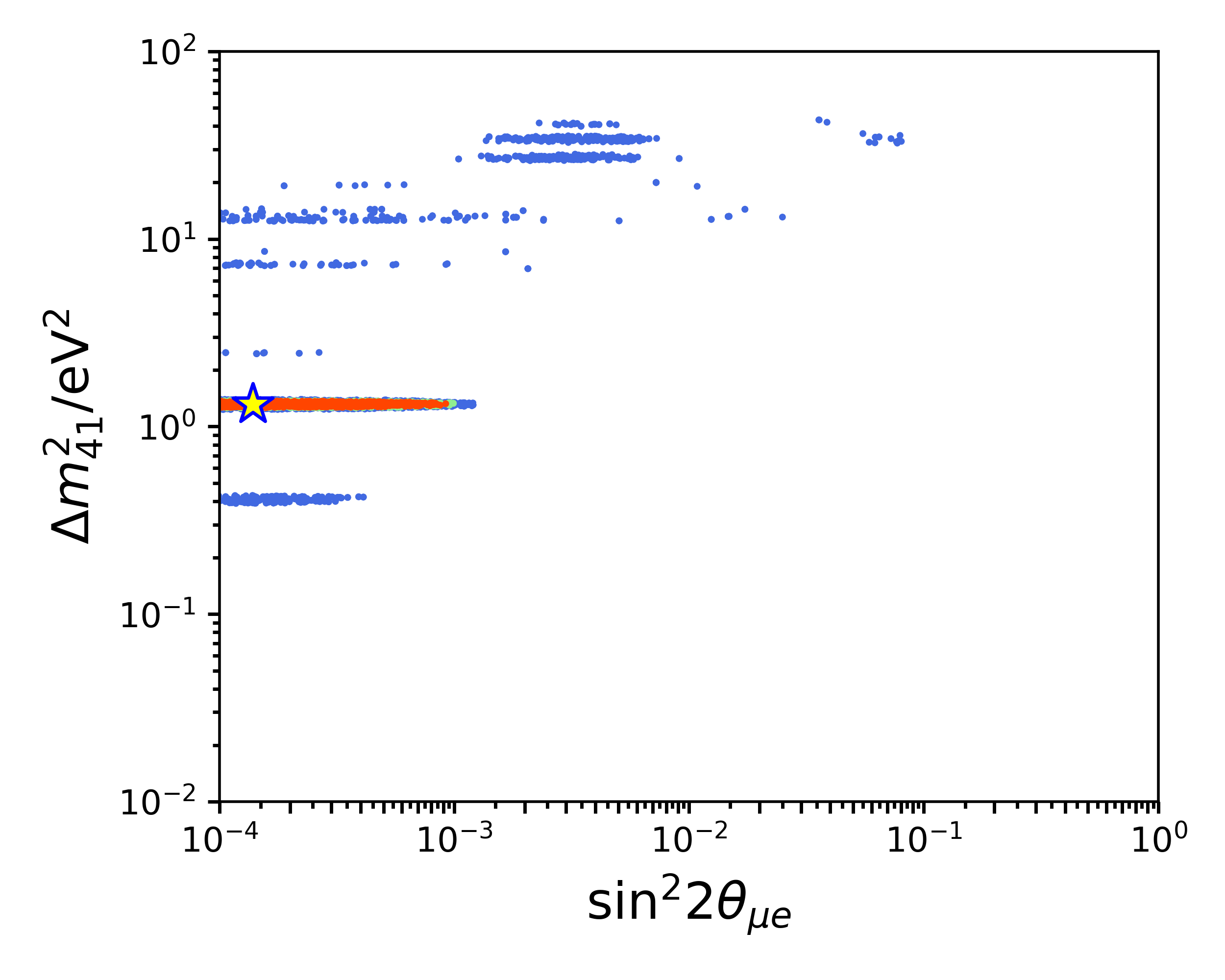}
\caption{Frequentist Confidence Intervals for 3+1 global fits to only appearance (above) data sets and disappearance (below) datasets. These fits demonstrate the tension that is seen within the 3+1 sterile neutrino model.}
\label{fig:tension}
\end{figure}

On the other hand, the 3+1 model has shown tension between the data sets. If one separates appearance experiments (sensitive to the product $|U_{\mu4}||U_{e4}|$) and disappearance experiments (separately sensitive to $|U_{\mu4}|$ and $|U_{e4}|$), a self-consistent model would be expected to show overlapping allowed regions in their respective best fits. It can be seen in Fig.~\ref{fig:tension} that this is not the case and separating the data sets results in differing best allowed regions without any overlap.
The PG test introduced in Sec.~\ref{launch} provides a method for quantifying the tension.  We summarize the inputs to the PG test in Table~\ref{table:fitquality}. 
The $p$-value for this PG test is $3.7\times 10^{-6}$, which indicates that tension between the appearance and disappearance data is at the 4.5$\sigma$ level, if the PG test measure is taken to be a true probability.

\subsubsection{Bayesian interpretation}

\begin{figure}
\includegraphics[width=\linewidth]{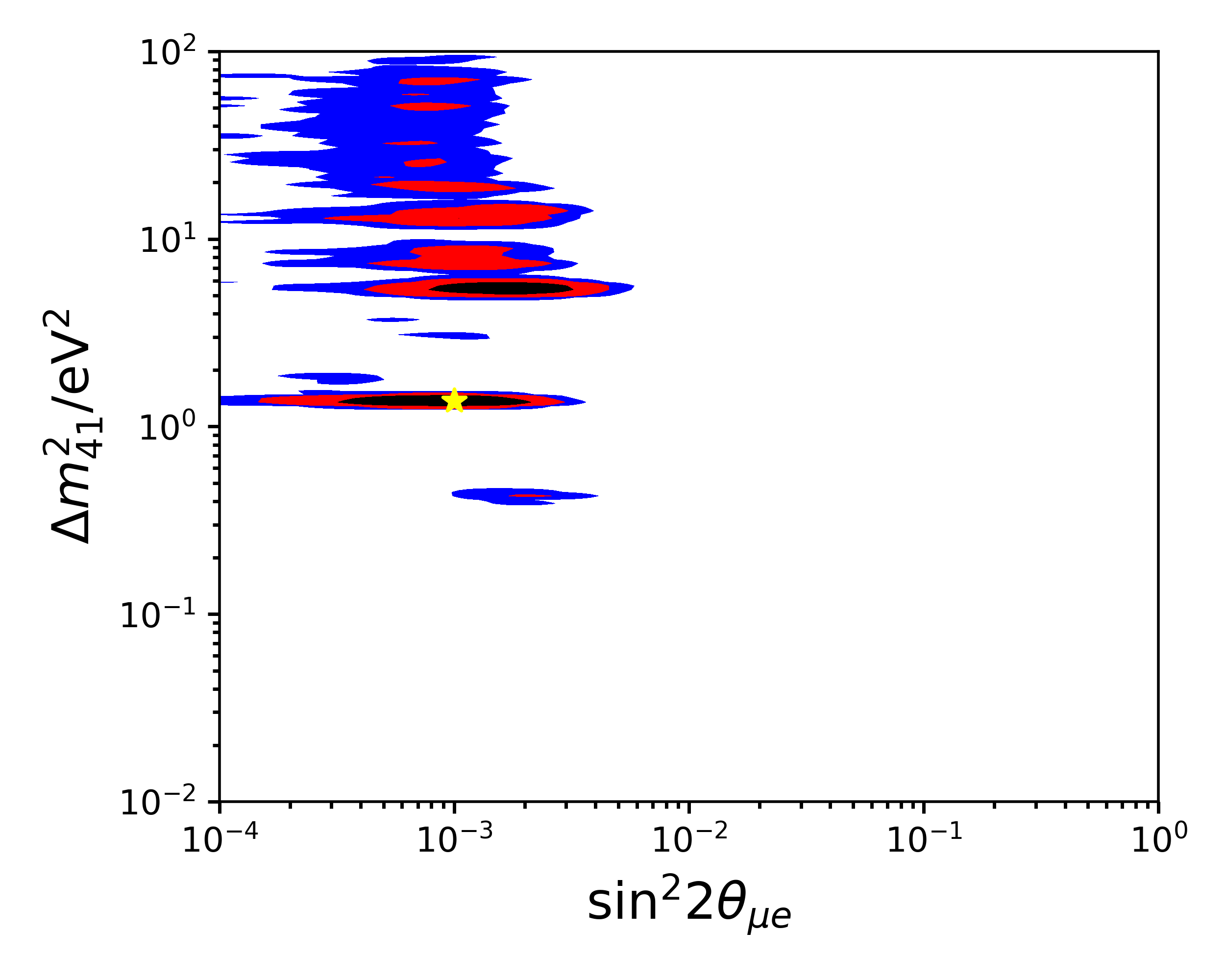}
\caption{Bayesian Credible Regions for a 3+1 global fit, showing the 99\%, 90\%, and 68\% Highest Posterior Density regions in blue, red, and black respectively. The maximum likelihood point is highlighted by the yellow star. } 
\label{fig:3plus1_bayes}
\end{figure}

Bayesian credible regions are shown in Fig.~\ref{fig:3plus1_bayes}.
These results also have one main island at $\Delta m^2_{41} \approx 1.3$ eV$^2$ which contains the best fit point.
However, substantially more parameter space is covered in the high $\Delta m^2_{41}$ region, with multiple islands at each credible level shown.
It should be stressed that Bayesian and frequentist methods address two different questions: Bayesian inference makes statements about probability of model parameters given the observed model, while frequentist methods make statements about the probability of the data given the model.
Thus, it should be no surprise that the regions drawn may differ substantially.
Indeed, confidence and credible regions only agree under special circumstances; for example, in the asymptotic regime where the likelihood function is a single-moded normal distribution with flat priors on the model parameters.

Recall that the confidence regions are themselves random variables.
A 90\% confidence region is defined such that the true model parameters have a 90\% probability of being covered by a randomly realized 90\% confidence region. By construction, this requirement is only met on average. No guarantees are made for any realization of the region\footnote{It is possible for a 90\% confidence region to contain the entire parameter space, or for it to be the empty set. These extremes are not possible for a 90\% credible region.}.
The alternative definition shown in equation~\ref{eq:conf_region_def}, states that the deviation of the best fit from all points in the 90\% region has a p-value of less than 10\%.
That is to say, over repeated observations there will be a 10\% chance or less that a statistical fluctuation will cause the $\Delta \chi^2$ to increase beyond the critical value, causing the point to be drawn outside that particular realization of the 90\% region. 
From this, we can begin to untangle the difference between Fig.~\ref{fig:3plus1} and Fig.~\ref{fig:3plus1_bayes}.
The presence of only one island in the confidence region suggest that there is only one ``stable'' island, in the sense that for any random realisation of the data, this island will lie in approximately the same area of parameter space.
As the higher $\Delta m^2_{41}$ islands seen in the credible region are lacking from the confidence region, this suggest that these islands are not particularly stable to statistical fluctuations.
However, this lack of stability should not be used to discredit these higher $\Delta m^2_{41}$ solutions\footnote{It should be noted that ``allowed region" is common parlance for closed confidence regions that do not include the null, and this terminology is used in this article. However, from the definition of the confidence region -- as a statement about the probability of data -- one cannot infer that a point inside or outside a confidence region is allowed or disallowed by the observed data.}.
Their presence in the credible regions shows that they still contribute significantly to the posterior distribution, and thus there is a high probability that a sterile neutrino may lie at larger $\Delta m^2$.

The observed differences between these methods illustrates why one should perform both Bayesian and frequentist analyses of data sets.
Each provides different information about both the data and the model, complementing each other, and giving a more complete picture of the state of sterile neutrinos.
To translate this information into actions, consider the design of a future experiment.
If the experiment will be sensitive to only a small range of $\Delta m^2$, it should aim for the main island at $\Delta m^2_{41} \approx 1.3$, as this is the least likely island to move around due to statistical fluctuations.
However, the large number of islands in the credible regions suggests that one should build a broad spectrum experiment if at all possible, as there is significant chance that a sterile neutrino will lie in the range of $5 < \Delta m^2_{41} < 100$ eV${}^2$.

As a final concrete example of how credible regions can differ from confidence regions, consider a likelihood function with two modes: one of which contains 60\% of the posterior, and the other 40\%.
Clearly, a 90\% credible region must include both modes by definition; however, one can adjust the likelihood function to ensure that the confidence region only includes one mode.
This can be done by making the 40\% mode narrower, while maintaining the probability that it contains in the posterior.
Such a modification will make the likelihood density at the center of the mode arbitrarily large, and once Wilks' theorem is applied, the confidence region will shrink until it covers only the 40\% mode---despite the fact that it contains less posterior probability than the 60\% mode.
Although this construction is artificial, it describes how these differences in credible and confidence regions can be inevitable under common conditions.

\subsubsection{Summary of where we stand on 3+1}

One should be thoughtful when considering specific global fit 3+1 allowed regions, because these will depend on exactly what question is asked, as demonstrated by our Bayesean versus frequentist comparison. However, we regard the $\Delta \chi^2/\Delta dof$ and the PG test as fair methods for quantifying the frequentist results.   Our conclusion on the 3+1 model is that, although the 3+1 model is favored over the null model at about $5\sigma$, there is a clear problem of internal consistency at the 4.5$\sigma$ level. This leads us to consider other models that go beyond 3+1.

\subsection{3+2 Model}

If one thinks beyond a 3+1 model, an obvious question is:  what if there are additional mostly-sterile states?    In this section we consider the case of adding a second mostly-sterile state, $\nu_5$, in what is called a 3+2 model.  This model will have two large mass splittings,  $\Delta m^2_{54}$ and $\Delta m^2_{41}$.   An additional row and column appear in the mixing matrix.  Because there are two mass splittings of similar magnitude, appearance experiments will be sensitive in this model to a $CP$-violating parameter, $\phi_{54}=\mathrm{arg}(U_{e5}U_{\mu5}^*U_{e4}^*U_{\mu4})$ .  
Therefore, there are seven parameters introduced in a 3+2 model: $\Delta m^2_{41}$,  $\Delta m^2_{51}$, $|U_{\mu4}|$, $|U_{e4}|$, $|U_{\mu5}|$, $|U_{e5}|$, and
$\phi_{54}$.    Note that $\Delta m^2_{54}=\Delta m^2_{51}-\Delta m^2_{41}$.

If we define $\Delta_{ij}=\Delta m^2_{ij}L/E$, then the appearance oscillation probability is given by:
{\footnotesize
\begin{eqnarray}
P(\nu_{\alpha}\rightarrow\nu_{\beta})\simeq 
-4|U_{\alpha5}||U_{\beta5}||U_{\alpha4}||U_{\beta4}|\cos\phi_{54}\sin^2(1.27\Delta_{54}) \nonumber\\
+4(|U_{\alpha4}||U_{\beta4}|+|U_{\alpha5}||U_{\beta5}|\cos\phi_{54})|U_{\alpha4}||U_{\beta4}|\sin^2(1.27\Delta_{41}) \nonumber\\
+4(|U_{\alpha4}||U_{\beta4}|\cos\phi_{54}+|U_{\alpha5}||U_{\beta5}|)|U_{\alpha5}||U_{\beta5}|\sin^2(1.27\Delta_{51}) \nonumber\\
+2|U_{\beta5}||U_{\alpha5}||U_{\beta4}||U_{\alpha4}|\sin\phi_{54}\sin(2.53\Delta_{54}) \nonumber\\
+2(|U_{\alpha5}||U_{\beta5}|\sin\phi_{54})|U_{\alpha4}||U_{\beta4}|\sin(2.53\Delta_{41}) \nonumber\\
+2(-|U_{\alpha4}||U_{\beta4}|\sin\phi_{54})|U_{\alpha5}||U_{\beta5}|\sin(2.53\Delta_{51})~.
\label{2app}
\end{eqnarray}
}
Note that if $\nu$ is replaced by $\bar \nu$,then $\phi \rightarrow -\phi$, so the interference term changes sign.   Thus, unlike the 3+1 model, neutrino and antineutrino data must be considered separately in a 3+2 fit.

Disappearance in a 3+2 model is given by:
\begin{eqnarray}
P(\nu_{\alpha}\rightarrow\nu_{\alpha})\simeq
1-4|U_{\alpha4}|^2|U_{\alpha5}|^2\sin^2(1.27\Delta_{54})\nonumber \\
-4(1-|U_{\alpha4}|^2-|U_{\alpha5}|^2)(|U_{\alpha4}|^2\sin^2(1.27\Delta_{41}) \nonumber \\
+|U_{\alpha5}|^2\sin^2(1.27\Delta_{51}))~. \label{disappeq2}
\end{eqnarray}

\begin{figure}[t]
\includegraphics[width=\linewidth]{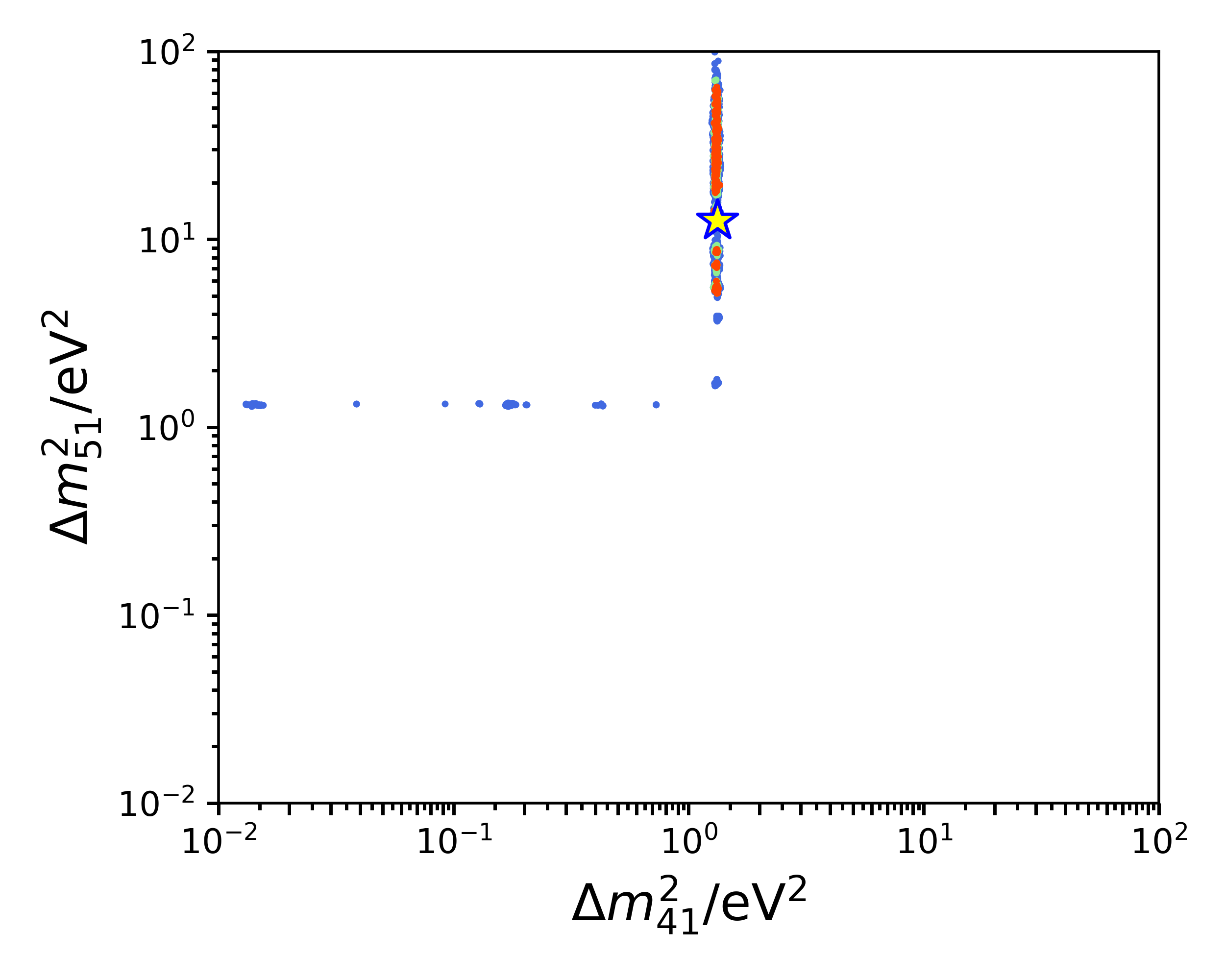}
\caption{3+2 fit allowed regions for the two mass splittings.} 
\label{3plus2}
\end{figure}

We present the parameters of the 3+2 fit in Table~\ref{table:fitresults}.
The first splitting is found at $\Delta m^2_{41}=1.32$ eV$^2$, which is the same as the 3+1 case. This low value of $\Delta m^2$ fits the overall shape well.   The best fit of the second mass splitting is at $\Delta m^2_{54}=12.6$ eV$^2$, but is at a very shallow minimum of the $\chi^2$ distribution that extends across a wide range of higher $\Delta m^2$ values as seen in Fig.~\ref{3plus2}.  Since this is a relatively large value of $\Delta m^2$, this is in a regime where the oscillation signal would vary rapidly for most experiments and average out to a small constant offset.    The best fit $\chi^2$ for 3+2 is 449, thus, compared to the 3+1 model, $\Delta \chi^2 = 9$ for 4 additional parameters, which has a 6\% random probability and indicates about a $1.5\sigma$ improvement for the 3+2 versus the 3+1 fits.

In this 3+2 model, one can also quantify the tension between appearance and disappearance using the PG test.    The parameters for this appear in Table~\ref{table:fitquality}, and, in summary,
$\chi^2_{PG} =30$ with the degrees of freedom, $N_{PG}= 4$, so the $p$-value for this PG test is $4.9\times 10^{-6}$.  This indicates that the tension is at the 4.4$\sigma$ level, which is a very small improvement from the 4.5$\sigma$ value for the 3+1 model.

Our conclusion on the 3+2 model is that there is no compelling improvement beyond the 3+1 model.    One could argue that, if nature follows patterns, a 3+3 model is more likely than a 3+2 model.   But the minimal improvement with the 3+2 case does not encourage us to proceed in this direction.    Instead, we look to other possible improvements to the 3+1 model which could relieve the internal tension.

\subsection{3+1+Decay \label{dk}}

One alternative model we will consider is the 3+1+Decay model, where we allow the fourth neutrino mass state to decay.  This is a more economical model, in that it will introduce only one new parameter beyond the 3+1 case, the lifetime of the $\nu_4$.

In the Standard Model, stable particles must be protected by a symmetry; without this a particle will decay.  Therefore, in principle, neutrinos can decay.   In the Standard Model extended to include neutrino mass, the neutrino can decay.  The lifetimes of the neutrinos are very long~\cite{PhysRevD.25.766,PhysRevD.28.1664} 
\begin{eqnarray}
\nu_i \rightarrow \nu_j + \gamma &\Rightarrow& \tau \simeq 10^{36} (m_i/eV)^{-5} {\rm yr}, \\
\nu_i \rightarrow \nu_j + \gamma + \gamma &\Rightarrow& \tau \simeq 10^{67} (m_i/eV)^{-9} {\rm yr},
\end{eqnarray}
where $i$ corresponds to the more massive neutrino mass state and $j$ a lighter one.

If a fourth neutrino state exists, it may decay.   In that case, the short baseline neutrino experiments may be seeing a combination of 3+1 oscillations and decay. This idea was first suggested as an explanation of LSND in Ref. \cite{PalomaresRuiz:2005vf}. Later, this model was considered in the context of the IceCube experiment~\cite{Moss:2017pur}. The general Lagrangian that governs neutrino decay can be written as~\cite{PalomaresRuiz:2005vf}: 
\begin{equation}
\mathcal{L}=-\sum_{l, h} g_{h l} \overline{\nu}_{l L} \nu_{h R} \phi+\mathrm{h.c.},
\label{eq:decay_lagrangian}
\end{equation}
where the index $l$ runs over the light neutrino mass states and $h$ over the heavy states.  In the case of a $3+1$ model, $l = 1,2,3$ and $h = 4$. The coupling constants $g_{hl}$ are, in general, complex, and control the partial decay width from parent $h$ to daughter $l$. The index $L$ and $R$ refer to the chirality of the field. This is relevant since, in the SM, electroweak interactions couple only left-handed neutrinos and right-handed antineutrinos. In relativistic scenarios--which is the case of neutrino experiments discussed in this review--the helicity and chirality are approximately the same up to order $m/E$. Thus, to this order, we can identify helicity states as neutrinos or antineutrino states. This implies that the Lagrangian in Eq. \ref{eq:decay_lagrangian} allows for chirality-preserving, $\nu_4 \to \bar \nu_l + \phi$, and chirality-flipping, $\nu_4 \to \nu_l + \phi$, processes. For relativistic neutrinos the partial widths for the helicity-preserving and helicity-flipping channels in the lab frame are given by~\cite{Kim:1990km}:  
\begin{equation}
\Gamma_{4 l}=\frac{\left|g_{4 l}\right|^{2} m_{4}^{2}}{32 \pi E_{n_{4}}},
\label{eq:decay_width}
\end{equation}
with the total width given by $\Gamma_{4}=2 \sum_{l} \Gamma_{4 l}$~\cite{Kim:1990km,PalomaresRuiz:2005vf}. In the case of Dirac neutrinos, the decay products of the helicity-flipping channels are invisible, since right-handed neutrinos and left-handed antineutrinos do not participate in SM interactions.

\begin{figure}[t]
    \centering
    \begin{tikzpicture}
        \fill[kyelloworange]   (2cm,0)  circle (0.36cm);
        \begin{feynman}
            \vertex (qi) {$\nu_4$};
            \vertex [right=2cm of qi] (c) {$ig_{4j}$} ;
            \vertex [below right=2cm of c] (qf1) {$\nu_j$};
            \vertex [above right=2cm of c] (pf) {$\phi$};
            \diagram*{
                (qi) -- [fermion] 
                (c) --  [scalar] (pf),
                (c) -- [fermion] (qf1),
                };
        \end{feynman}
    \end{tikzpicture}
    \begin{tikzpicture}
        \fill[kyelloworange]   (2cm,0)  circle (0.36cm);
        \begin{feynman}
            \vertex (qi) {$\nu_4$};
            \vertex [right=2cm of qi] (c) {$ig_{4j}$} ;
            \vertex [below right=2cm of c] (qf1) {$\psi$};
            \vertex [above right=2cm of c] (pf) {$\phi$};
            \diagram*{
                (qi) -- [fermion] 
                (c) --  [scalar] (pf),
                (c) -- [fermion] (qf1),
                };
        \end{feynman}
    \end{tikzpicture}
    \caption{Left: Feynman diagram of visible neutrino decay. Right: Feynman diagram of invisible neutrino decay.}
    \label{decaydiagram}
\end{figure}
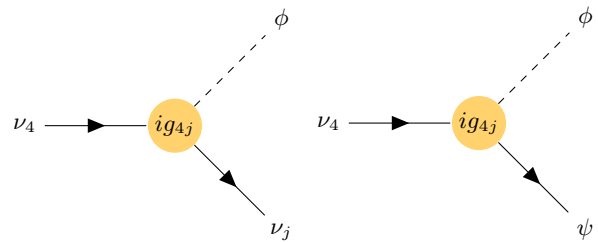

We explore the possibility that the fourth neutrino state decays by one of two cases shown in Fig.~\ref{decaydiagram}.   In the first case (left), the decay produces an active neutrino plus a beyond standard model particle, and hence is called ``visible.''   In the second case (right), the decay produces two beyond standard model particles through a new force (sometimes called a ``secret force''), and hence is ``invisible.''  While both cases were explored for IceCube, in this study we will consider only the invisible decay, which has the property of reducing tension within cosmological models that involve sterile neutrinos, as discussed in Sec.~\ref{cosmo}.

In the 3+1+Decay analysis, the additional parameter included in our fits is the lifetime $\tau_4 = 1/\Gamma_4$, measured in the $\nu_4$ rest frame. Experiments would only be directly sensitive to the lifetime in the lab frame, so it is necessary to know the mass of $\nu_4$ to transform to the lifetime in the $\nu_4$ rest frame. We account for this by taking the approximation that $m_4 \approx \sqrt{\Delta m^2_{41}}$. It is apparent from~\eqref{eq:decay_width} that the maximum width of a particle -- respectively its shortest lifetime -- is bounded by the particle mass and maximum allowed coupling. To stay in the perturbative regime and satisfy unitarity constraints, we only consider couplings such that $g< 4\pi$, which leads to the condition $\tau \ge \tau_{min}(m_4) = \pi/m_4$ with $\hbar = c = 1$. 
We use a the prior on the neutrino decay lifetime that is log-uniform in the range of $\tau_{min}(m_4)$ to $10^{2}$~eV$^{-1}$. We implement this by using a log-uniform prior from $10^{-2}$~eV$^{-1}$ to $10^{2}$~eV$^{-1}$ and then restricting ourselves to the allowed parameter space. Note that interplay between decay and oscillations is relevant when $\tau_4/m_4 \sim 1/m_4^2$, a condition that is more readily satisfied for smaller lifetimes. A log-uniform prior biases towards this scenario more than a linear prior.

\begin{figure*}[t]

\centering
\begin{subfigure}{0.30\linewidth}
\centering
\includegraphics[width=\linewidth]{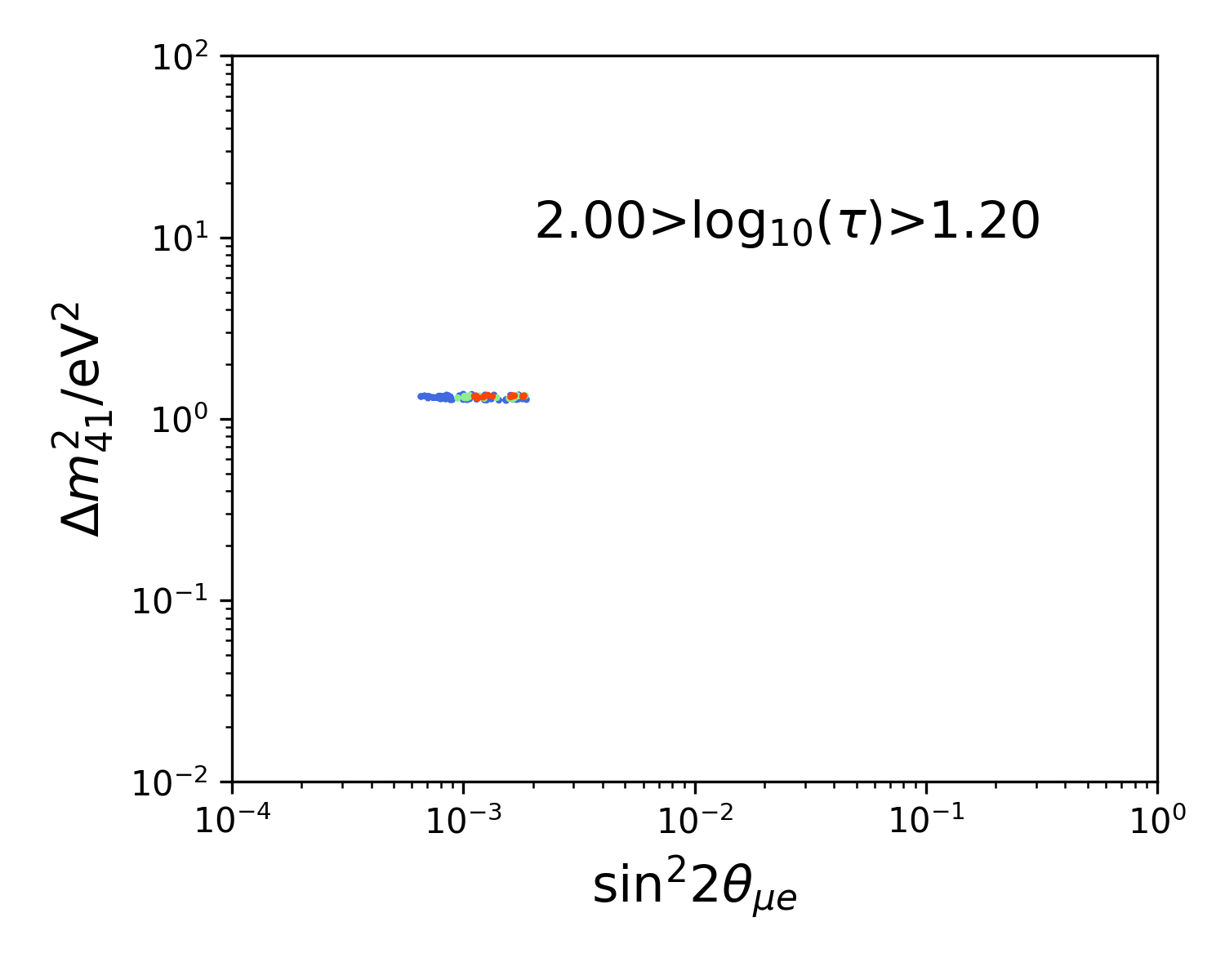}
\caption{Allowed points for $1.2 < \log_{10}(\tau_4/{\rm eV^{-1}}) < 2.0$.}
\end{subfigure}
~
\begin{subfigure}{0.30\linewidth}
\centering
\includegraphics[width=\linewidth]{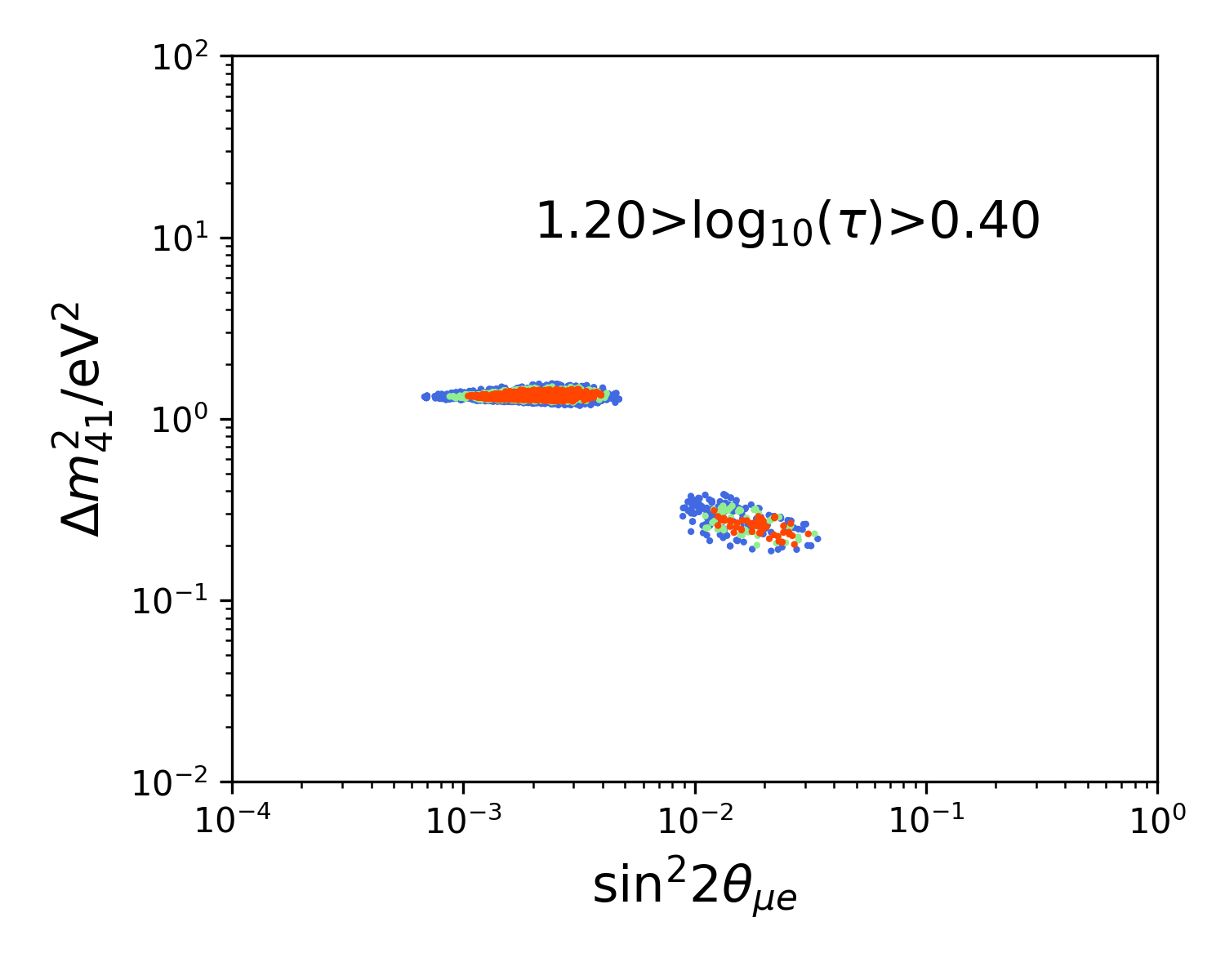}
\caption{Allowed points for $0.4 <\log_{10}(\tau_4/{\rm eV^{-1}}) < 1.2$.}
\end{subfigure}
~
\begin{subfigure}{0.30\linewidth}
\includegraphics[width=\linewidth]{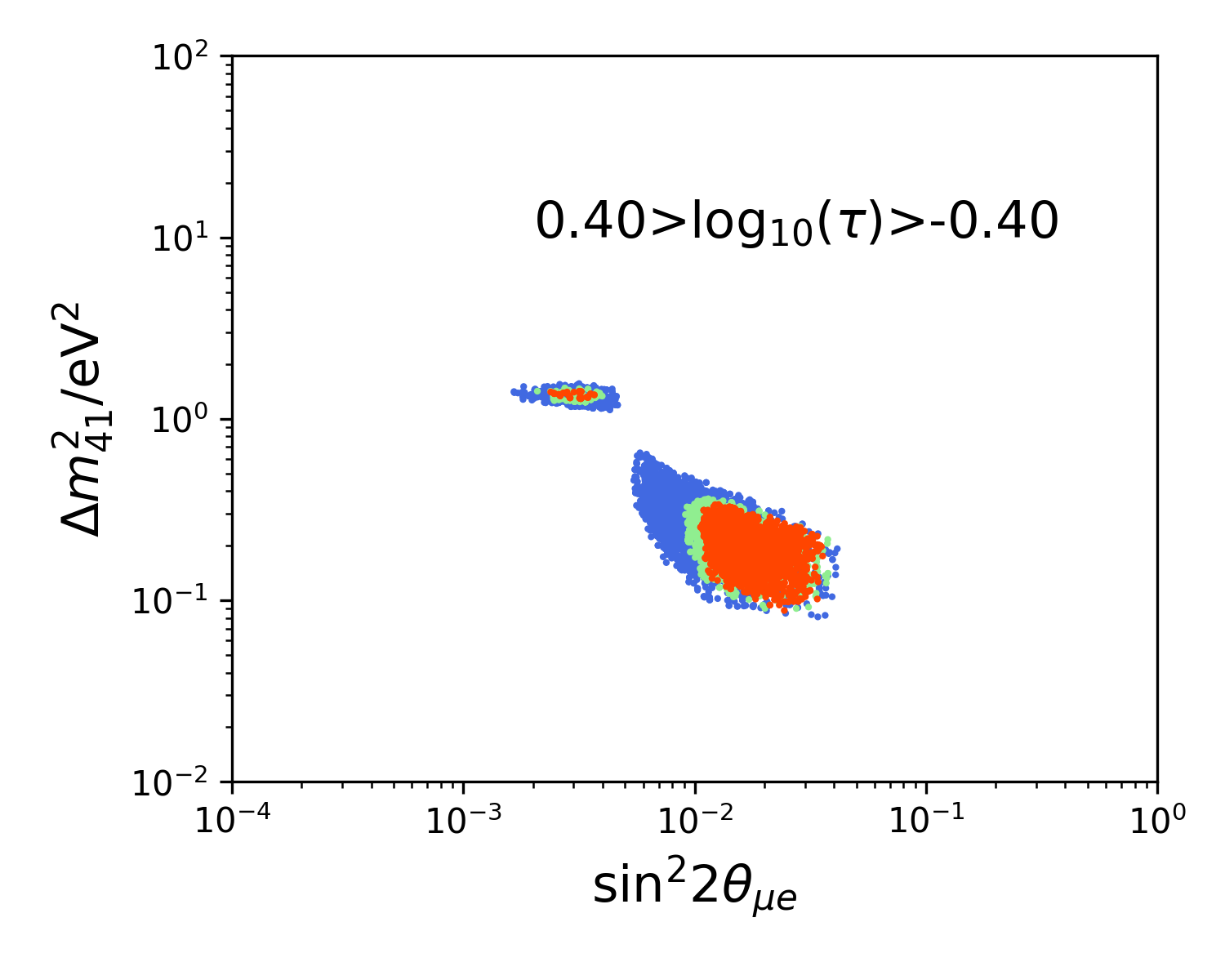}
\caption{Allowed points for $-0.4 <\log_{10}(\tau_4/{\rm eV^{-1}}) < 0.4$.}
\end{subfigure}
\caption{Frequentist confidence intervals for the 3+1+Decay global fit. The different frames show the contours as the lifetime of the $\nu_4$ state decreases.}
\label{fig:decay}
\end{figure*}

In Fig.~\ref{fig:decay}, we show the results of the these fits. We display the allowed regions for lifetimes in the range $-2.0 < \log_{10}(\tau~\text{eV}^{-1}) < 0.4$. The plot shows how the best fit contours change as we decrease the lifetime of the $\nu_4$ state.  

The best fit points are reported in Table~\ref{table:fitresults}.  The best fit mixing and mass splitting  are found to be very different than 3+1, with $\sin^2(2\theta_{\mu e}) = 0.024$ and $\Delta m_{41}^2 = 0.211~\text{eV}^2$, and a  lifetime of $\tau = 1.96~\text{eV}^{-1}$.  The $\chi^2 = 450$ corresponds to an improvement over the null of 43.  Comparing the $\chi^2$ for the 3+1+Decay model to the $\chi^2$ for the 3+1 model leads to an improvement in the fit with a $\Delta \chi^2 =8 $ for only 1 additional parameter and indicates about a $2.6\sigma$ improvement over the 3+1 only model.

\begin{figure*}[t]
\centering
\begin{subfigure}{0.30\linewidth}
\centering
\includegraphics[width=\linewidth]{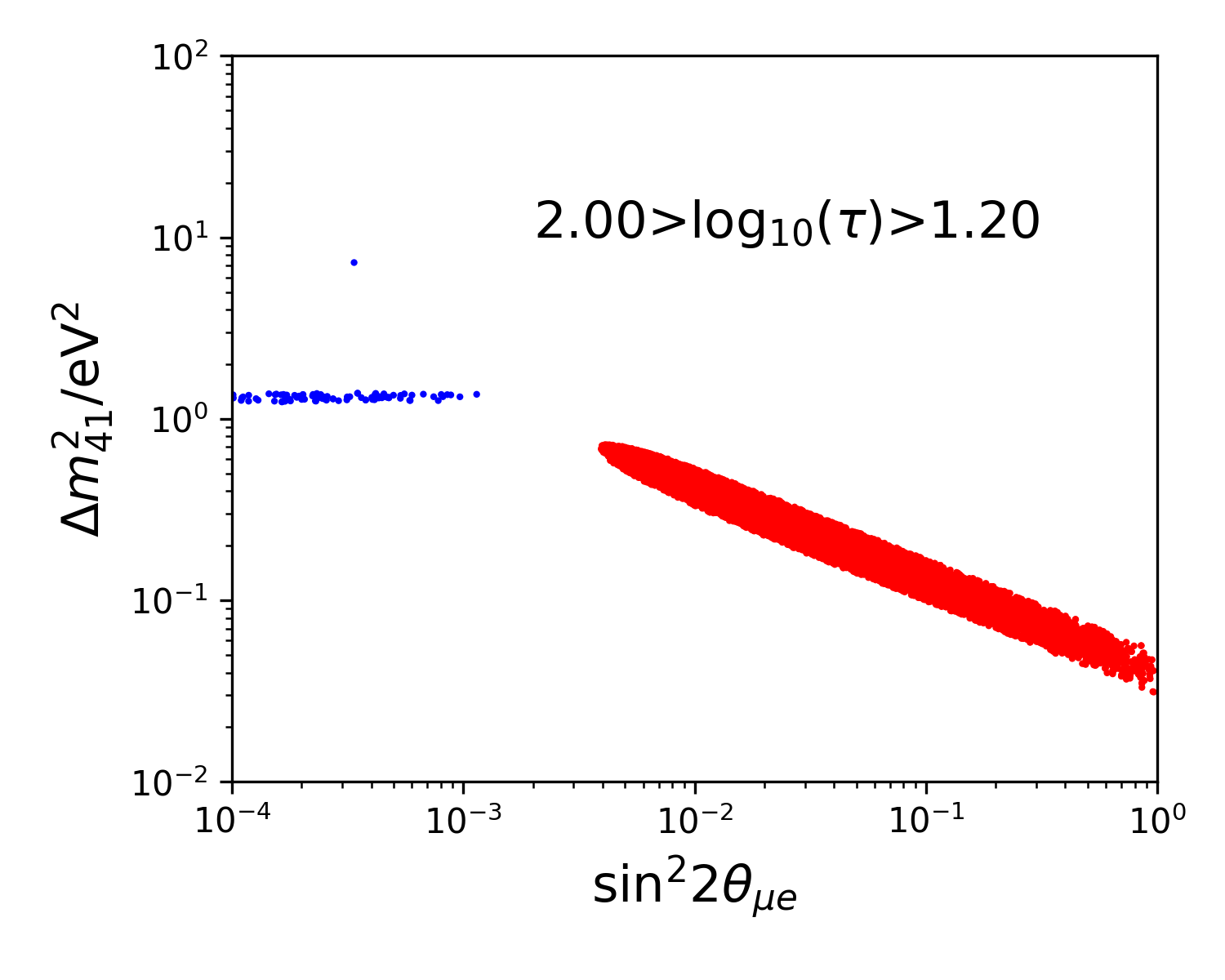}
\caption{Allowed points for $1.2 < \log_{10}(\tau_4/{\rm eV^{-1}}) < 2.0$.}
\end{subfigure}~
\begin{subfigure}{0.30\linewidth}
\includegraphics[width=\linewidth]{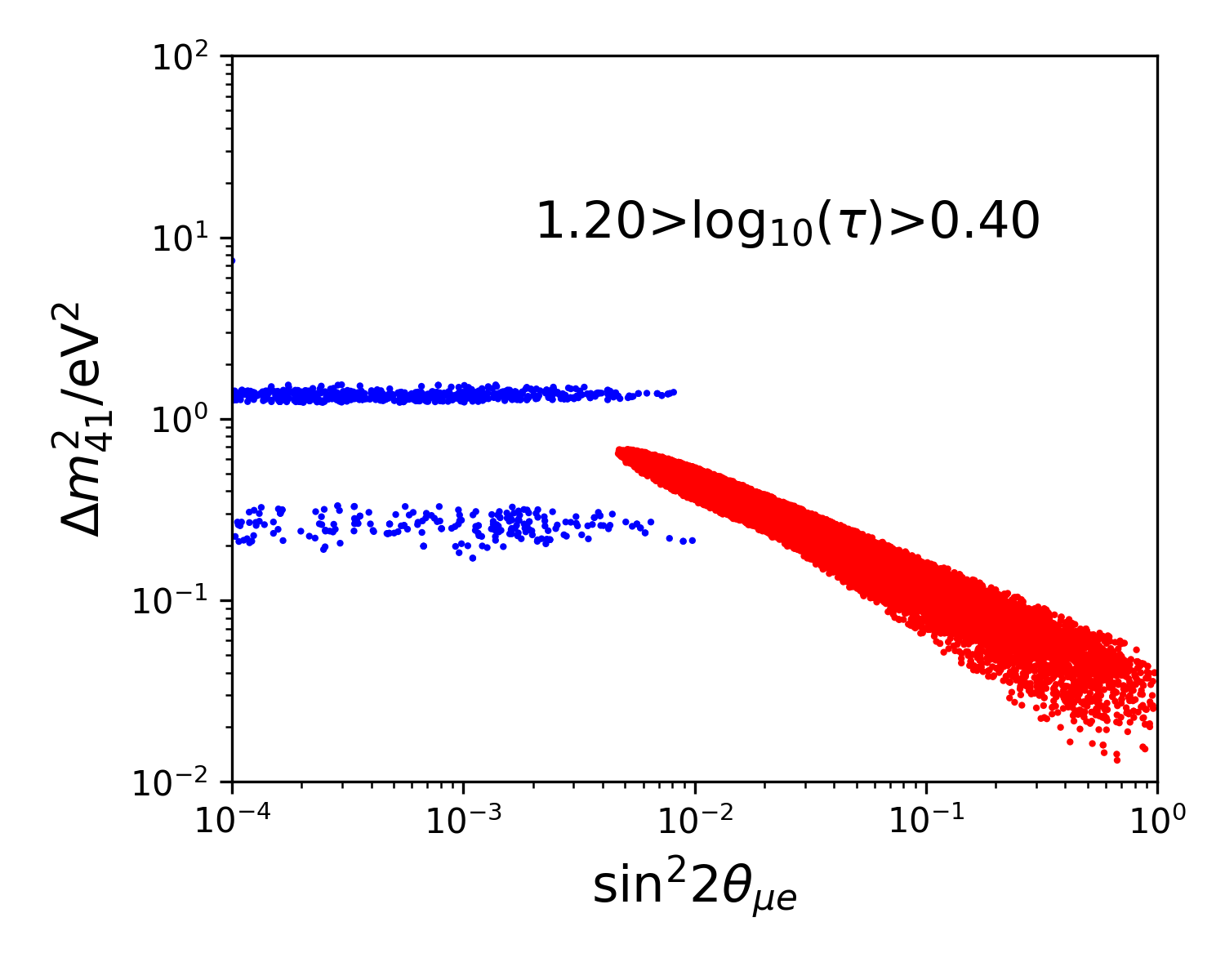}
\caption{Allowed points for $0.4 <\log_{10}(\tau_4/{\rm eV^{-1}}) < 1.2$.}
\end{subfigure}~
\begin{subfigure}{0.30\linewidth}
\includegraphics[width=\linewidth]{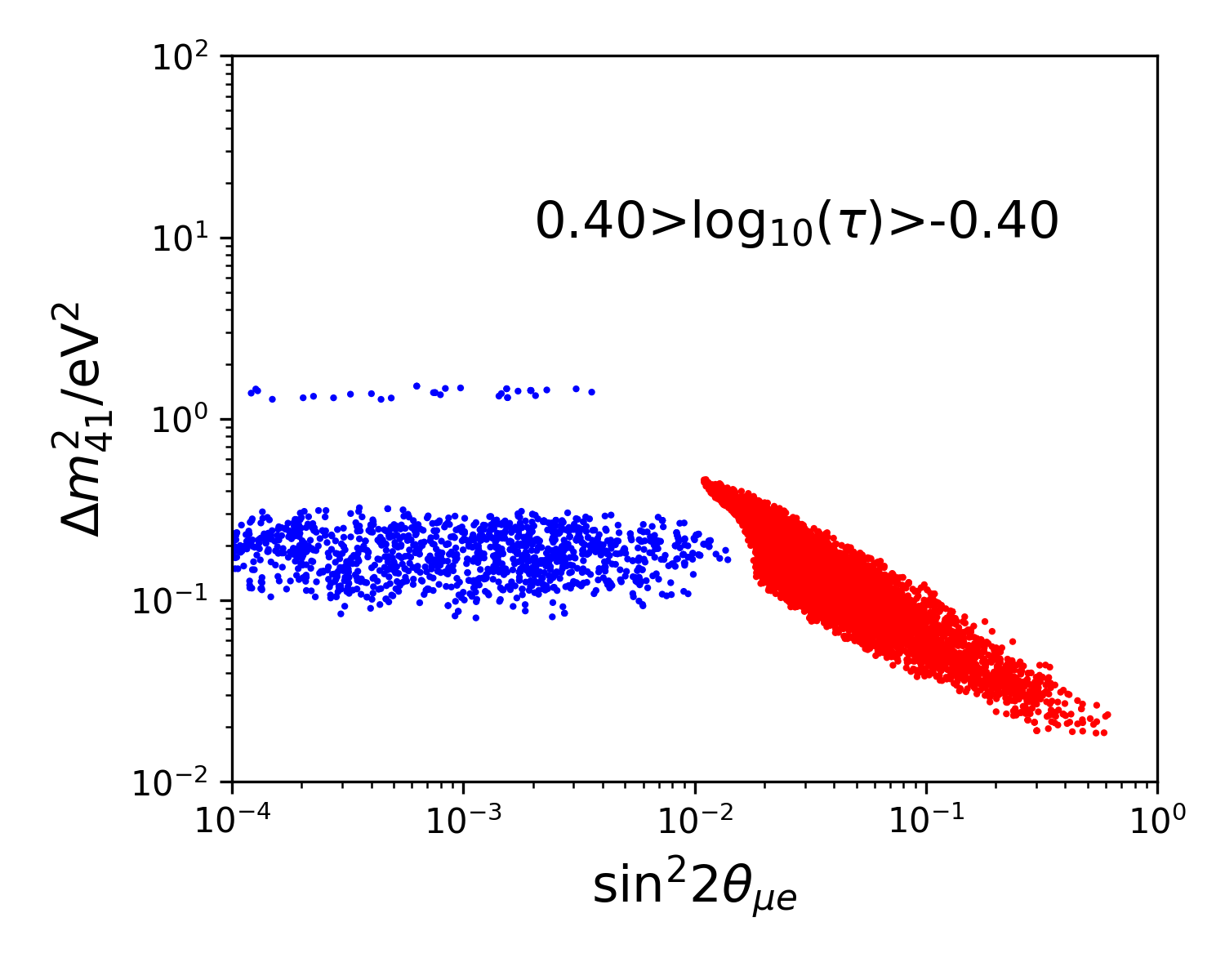}
\caption{Allowed points for $-0.4 <\log_{10}(\tau_4/{\rm eV^{-1}}) < 0.4$.}
\end{subfigure}
\caption{Frequentist confidence intervals for the 3+1+Decay global fit, with appearance experiments in red and disappearance in blue. The contours are each drawn at a confidence level of $95\%$. The different frames show the contours as the lifetime of the $\nu_4$ state decreases.   The proximity of appearance and disappearance for $0.4>\log_{10}(\tau)>-0.4$ indicates decreased tension, as discussed in the text.}
\label{fig:tensiondecay}
\end{figure*}

This model alleviates some of the tension seen between appearance and disappearance experiments, as seen in Table~\ref{table:fitquality}.
Figure~\ref{fig:tensiondecay} compares the appearance and disappearance data sets as a function of $\tau$. Here, we plot only the $95\%$ CL contours, with the appearance data set in red and the disappearance in blue.    From these plots, one sees that inclusion of $\nu_4$ decay improves, but fails to fully relieve, the tension between the separate data sets.   In Fig.~\ref{fig:tensiondecay}, upper right, overlap in $\sin^2 2\theta_{\mu e}$ occurs, but not in $\Delta m^2_{41}$, and in the lower right plot,  overlap in $\Delta m^2_{41}$ occurs, but not in  $\sin^2 2\theta_{\mu e}$.   There is no solution where there is overlap in both parameters.   Measuring the tension using the PG test, one finds that when appearance and disappearance are fit separately, they prefer very long lifetimes (no decay).   Thus, as with the 3+1 fit, the best-fit $\chi^2$'s for the 3+1+Decay are 77 and 356 for appearance and disappearance respectively.   However, because including decay relieves some tension,  the global fit improves by 8 units of $\chi^2$ compared to the 3+1 best fit,  so that the $\chi^2_{PG}= 17$.   The decay parameter adds a degree of freedom to each of the fits so that $N_{PG}= 3$.   Thus, the $p$-value for the PG test for 3+1+Decay is $7.07\times 10^{-4}$, which indicates the tension is reduced to the $3.2\sigma$ level.

In conclusion, introducing decay has yielded   
 a 2.6$\sigma$ improvement in the fit $\chi^2$ and about a factor of 200 increase in the PG-fit $p$-value compared to a 3+1  model.  While the tension probability still remains high,  this indicates an interesting new direction for exploring the source of the appearance/disappearance incompatibility.  The next step is to study visible decay, {\it i.e.} decays to active neutrinos, as opposed to invisible decays.  This scenario replenishes the flux in disappearance experiments, as was shown in the IceCube study~\cite{Moss:2017pur} and, thus, is likely to relieve the tension further.

We end our exploration of improvements to the 3+1 model here.   We simply note that it would not be surprising that a model as simple as 3+1 needs some improvement, and that further development of ideas by the theory community is warranted.

\section{What Can Possibly Go Wrong?}

While global fits can provide general guidance, there are a number of issues that can bias the results and, potentially, contribute to the tension. In this section we examine some of the features of the data that may contribute uncertainty to the global fit results.    Ideally these uncertainties would be quantified, but at present it is not clear how this may be performed.  Therefore, we simply present a qualitative discussion of things that can, possibly, go wrong.

\subsection{The Difficulty of Exactly Reproducing Experimental Results \label{difficulty}}

\subsubsection{Insufficient Data Releases}

Each experiment's implementation in our analysis has two main components: simulating the physics of the experiment, and finding the statistical significance of the data given the hypothesis. To be able to implement these components, we rely on collaborations to release the pertinent information. 

For the former point, simulation of the physics of a particular experiment is necessary to be able to change the predicted observation as a function of the neutrino model. 
To create a minimally acceptable simulation, we ask collaborations to provide expectation of the neutrino flux and a detector response function or matrix that gives the distribution of observed prompt energy in a detector as a function of real neutrino energy. 

\begin{figure}
\centering
\includegraphics[width=\linewidth]{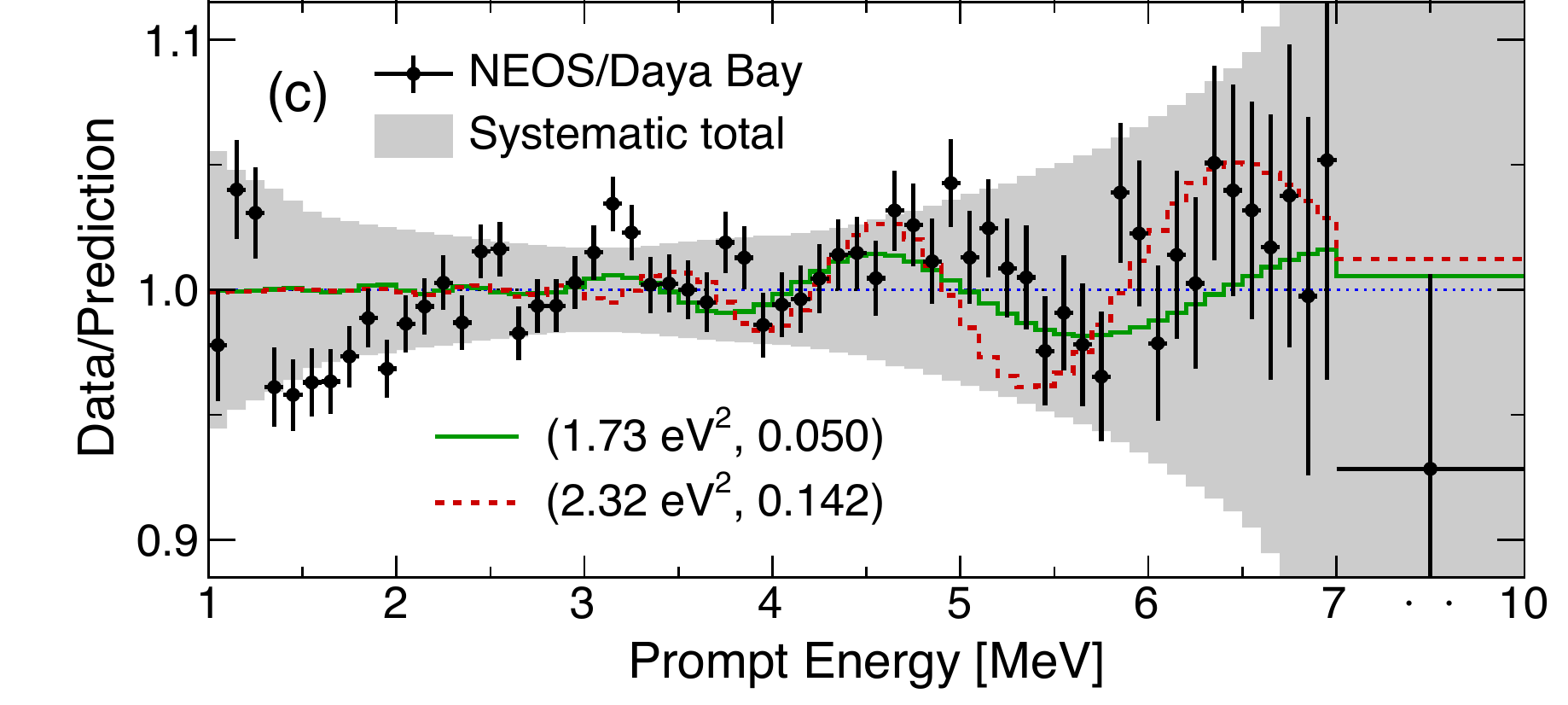}
\caption{Ratio of the observed neutrino spectrum at the NEOS detector over the expected from the Daya Bay results \cite{NEOS}.}
\label{fig:NEOS}
\end{figure}

Regarding the implementation of systematic and statistical uncertainties into the global fits, we find that experiments are especially lacking in providing the necessary information.
Most experiments simply release a data plot that includes the square root of the diagonal elements of the covariance matrix. For example, Fig.~\ref{fig:NEOS} shows how NEOS released their data \cite{NEOS}.
If one were to only consider the error bars shown in the plot, one would find a $\chi^2\approx20$. 
This is a large deviation from their quoted $\chi^2  = 64.0$ and demonstrates the need for a full covariance matrix to be provided. Plenty of hours are invested in trying to reproduce a covariance matrix with the limited information provided in experiments' publications, and these reproduced covariance matrices are undoubtedly inaccurate. 

\begin{figure}
\centering
\includegraphics[width=\linewidth]{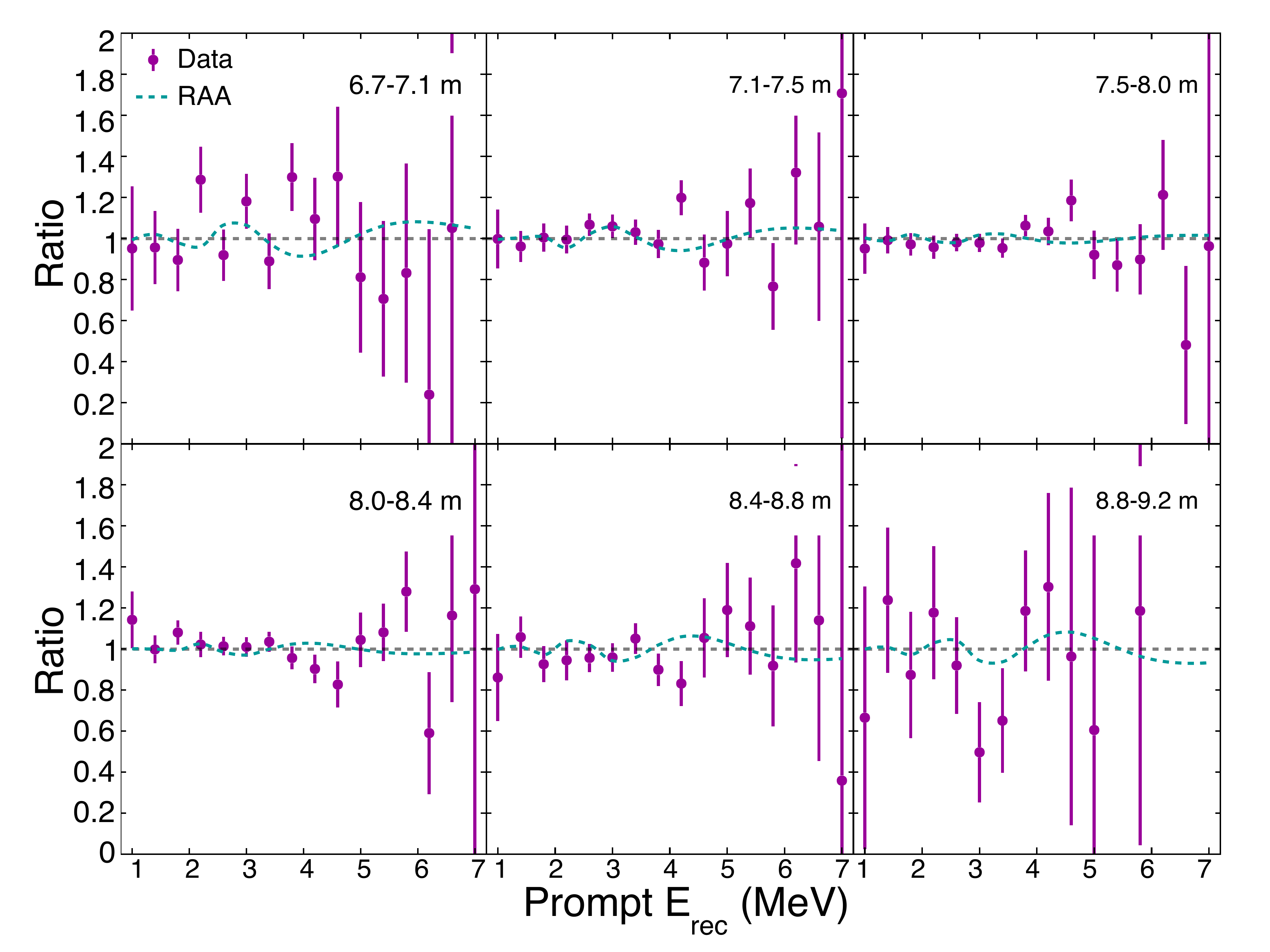}
\caption{Ratio of the observed PROSPECT spectrum over the baseline-integrated spectrum. See \cite{Prospect} for more detail. }
\label{fig:PROSPECT}
\end{figure}

One might expect that experiments which use a near and far detector would not suffer the above problem. Systematic uncertainty would be minimized and one would only have to worry about the statistical errors in an experiment. This would make the diagonal elements of the covariance matrix (i.e. the error bars displayed on plots) enough. Unfortunately, recent reactor experiments normalize the spectra in separate detectors before taking a ratio. This introduces off diagonal statistical errors to the covariance matrix, further complicating the picture. For instance, consider the data release by PROSPECT \cite{Prospect}, displayed in Fig.~\ref{fig:PROSPECT}.  

\begin{figure}
\centering
\includegraphics[width=.5\linewidth]{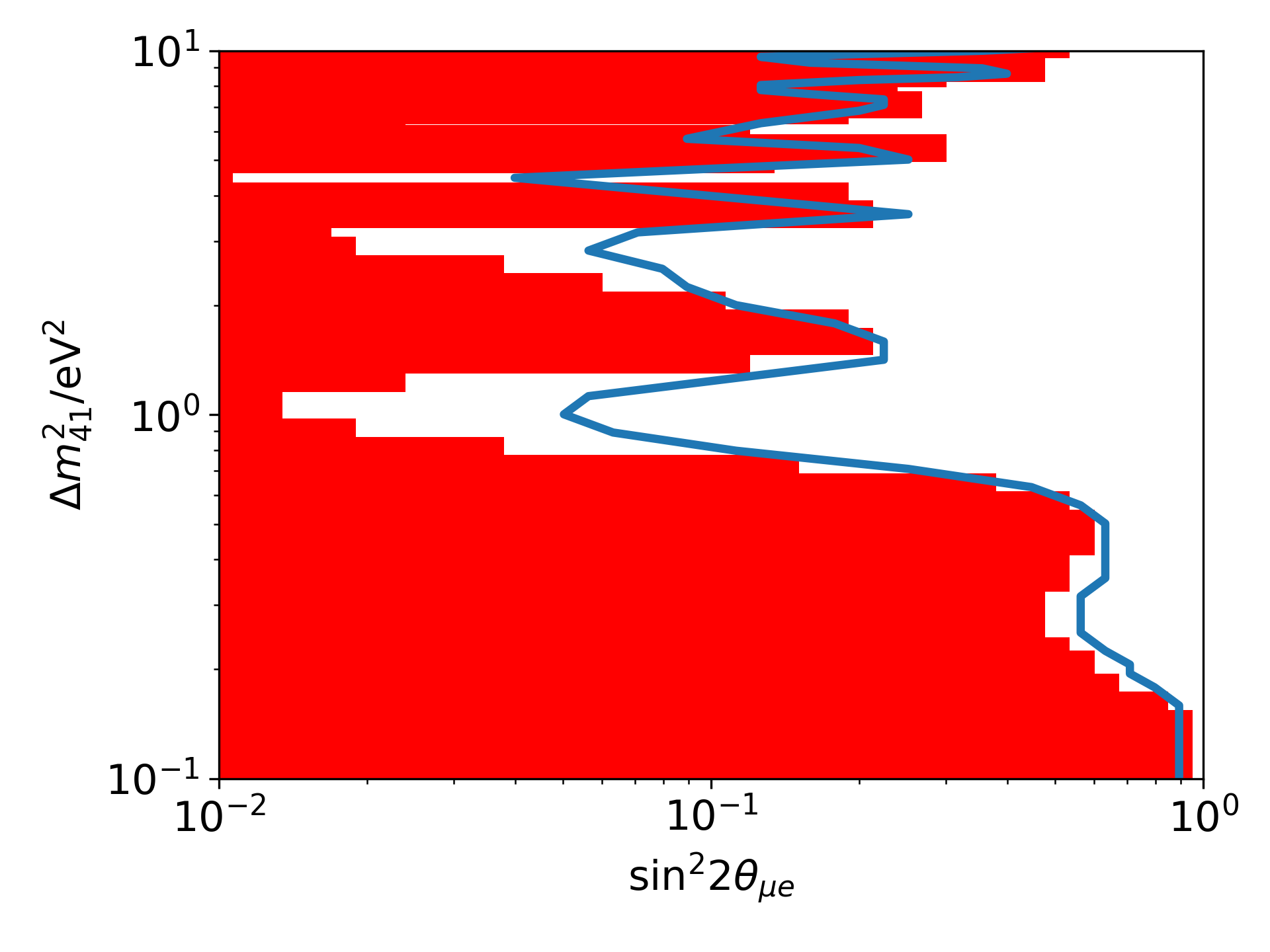}~
\includegraphics[width=.5\linewidth]{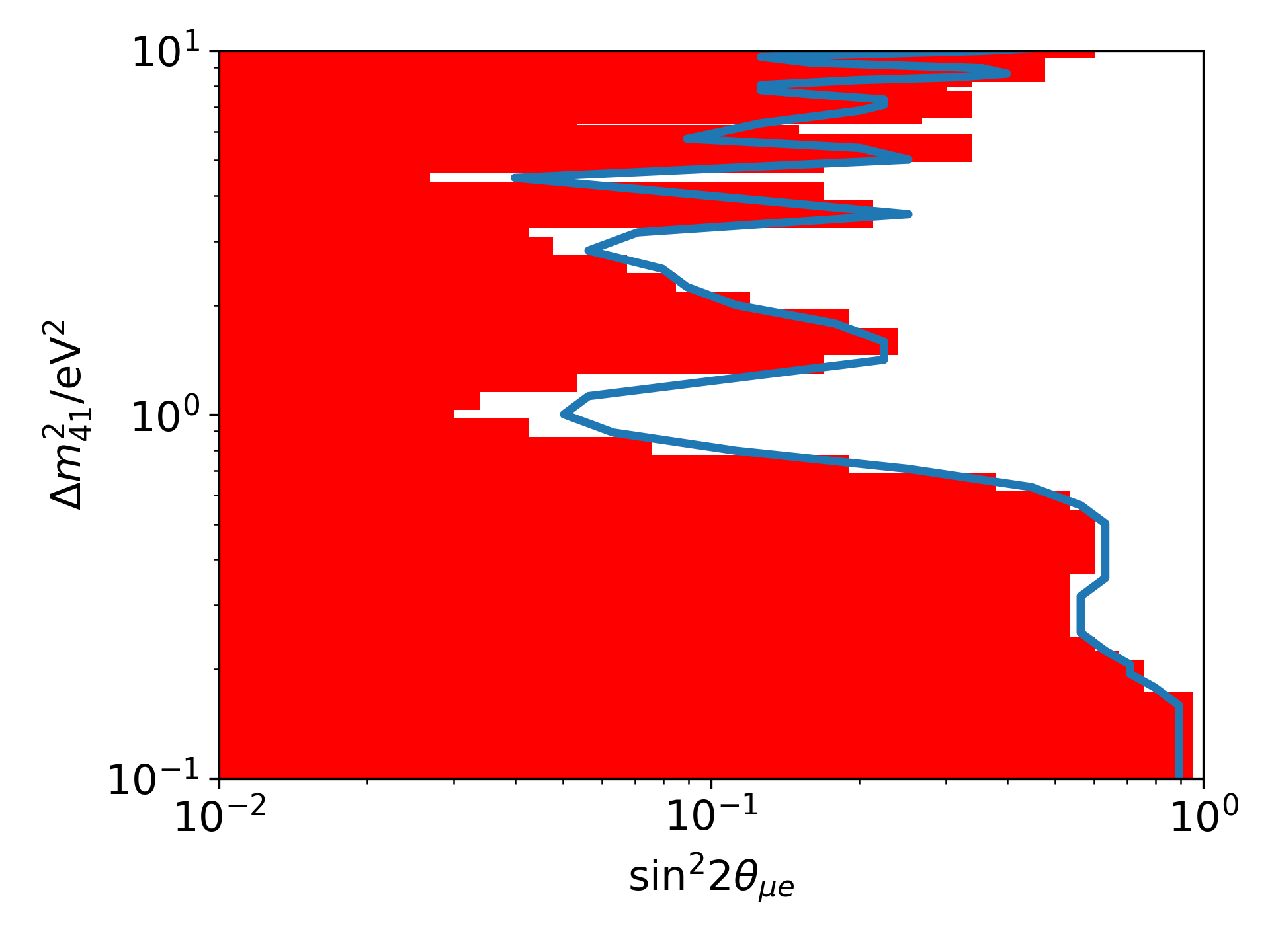}
\caption{Left: Allowed parameter space if one only uses the error provided by PROSPECT, shown in Fig.~\ref{fig:PROSPECT}.  Right: Allowed parameter space after toy studies were done to recreate an approximate full statistical covariance matrix in PROSPECT.\label{Pbeforeafter} }
\end{figure}

If we use only the error bars shown in Fig.~\ref{fig:PROSPECT}, then our 3+1 fits would give us the red allowed region shown in Fig.~\ref{Pbeforeafter}, left. 
The blue line, drawn using a $\chi^2$ map provided by PROSPECT \cite{Prospect}, shows what the exclusion line should be if one assumes that the $\Delta \chi^2$ follows a chi-squared distribution; one finds substantial disagreement.
To approximate a full statistical covariance matrix, we used data from PROSPECT's full spectrum analysis \cite{Prospectflux} to simulate several iterations of an oscillation analysis and recreate a statistical covariance matrix. Using this reconstructed covariance matrix, we find the allowed region shown in Fig.~\ref{Pbeforeafter}, right. Clearly there is a significant improvement.

It should also be noted that these global fits do not take into account correlations between the experiments, which can arise due to common systematics.
To account for these correlations, the experiment analyses would need to present nuisance parameters -- also known as pull parameters -- for basic physical values, such as cross sections.
When performing the global fits, these nuisance parameters would be shared, producing the desired correlations.
However, when not provided, the extraction of these parameters from an already completed analysis is an impossible task.
Thus, they cannot be included in our present fits.
Ideally, future experiments should publish analyses with these parameters included and exposed to allow external modification, so that this effect can be accounted for going forward.

It should be cautioned that it is difficult to guess the effect that these correlations would have on the presented results---they can go either way.
For example, they may make the significance of the best fit point weaker, while simultaneously reducing the tensions in the fits as all muon disappearance limits can become weaker in a correlated fashion.

The authors believe that, for the foreseeable future, no single experiment will be able to fully resolve the sterile neutrino picture. Several experiments will be necessary to probe the various oscillation modes, and global fits need to have the necessary information to combine these results and provide the bigger picture. In order to conduct our global fits, we rely on collaborations to be transparent with their analysis and data to properly implement their results. 

\subsubsection{Assuming $\chi^2$ Statistics}

In our global fits, the frequentist confidence regions are drawn assuming that the $\Delta \chi^2$ statistic is truly $\chi^2$ distributed. This assumption is made because it is computationally very expensive to do fake data studies for all of the experiments. However, the individual experiments have demonstrated that this assumption is not always true, and so they routinely use fake data studies to determine the critical $\Delta \chi^2$ values.   

For example, PROSPECT conducted fake data studies to determine the critical $\Delta\chi^2$ necessary to exclude points of parameter space at $95\%$ confidence level. Using
data provided in the supplementary material of \cite{Prospect}, Fig~\ref{fig:heatmap} shows what this critical $\Delta \chi^2$ is found to be across the parameter space. If one assumes that the $\Delta \chi^2$ test statistic was $\chi^2$ distributed with 2 degrees of freedom, the critical $\Delta \chi ^2$ would be uniformly 6.0. 
Fig~\ref{fig:heatmap} shows that for the majority of the parameter space, and especially in the region of most interest, the critical $\Delta \chi^2$ is actually found to be $\gtrsim9$.
In addition, NEOS demonstrates similar results in their supplementary material for Fig.  \cite{NEOS}. 

\begin{figure}
\centering
\includegraphics[width=\linewidth]{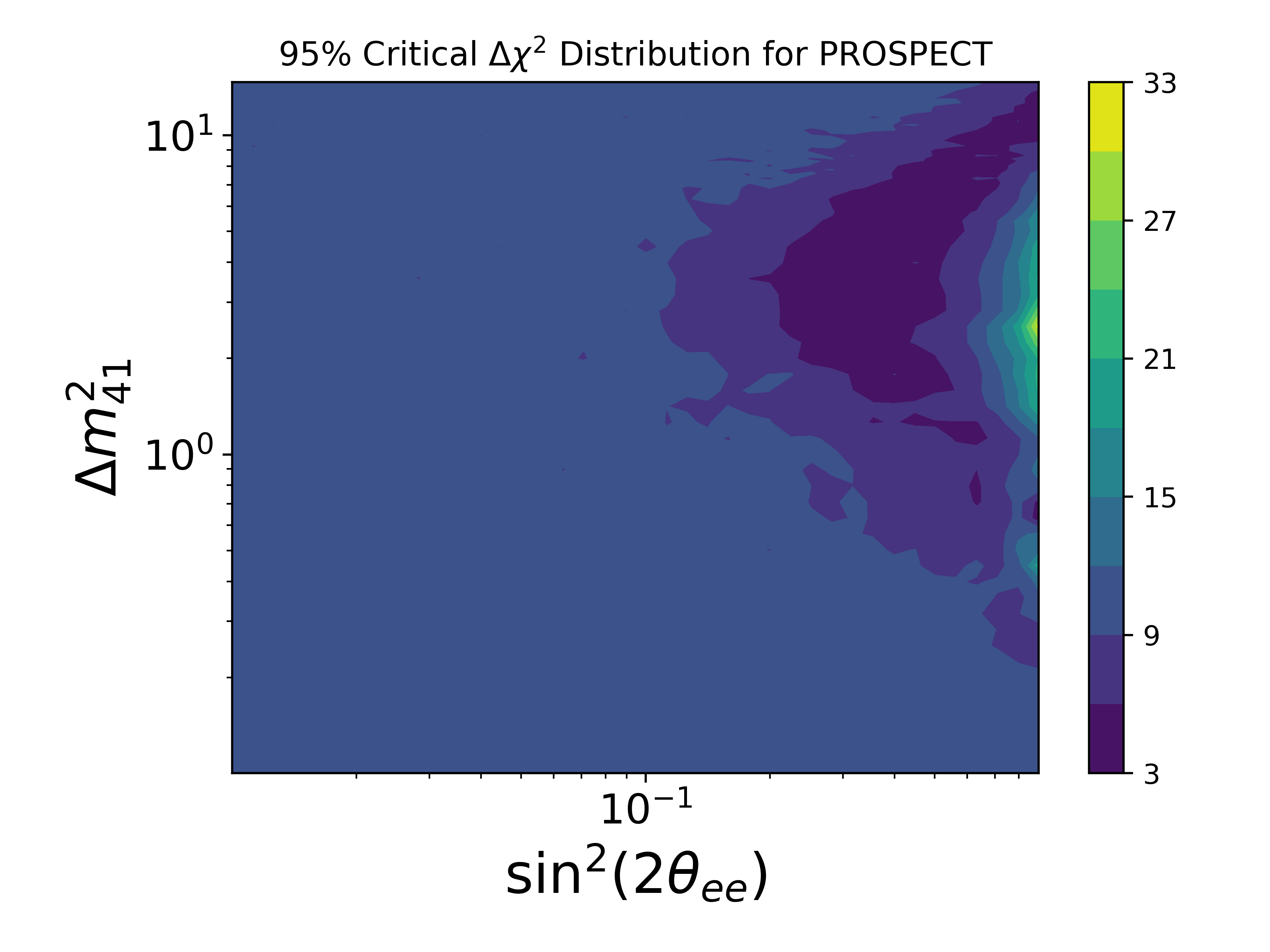}
\caption{Map of the true critical $\Delta \chi^2$ needed to exclude a parameter point at $95\%$, found using fake data studies. Data provided by PROSPECT \cite{Prospect}. A truly $\chi^2$ distributed test statistic would have a critical $\chi^2$ for $95\%$ of 6.}
\label{fig:heatmap}
\end{figure}

Regions of parameter space will be erroneously excluded by this underestimation of the critical $\Delta \chi^2$.
The correct treatment for this issue is numerical estimation of the distribution of test statistics.
At each location in the model parameter space, fake data experiments are thrown to build this estimation---a treatment performed by PROSPECT.
Unfortunately, this is too computational expensive to be conducted for a global fit. Thus, the fits presented above assume that the $\Delta \chi^2$ test statistic is properly distributed. 

\begin{figure}
\centering
\includegraphics[width=\linewidth]{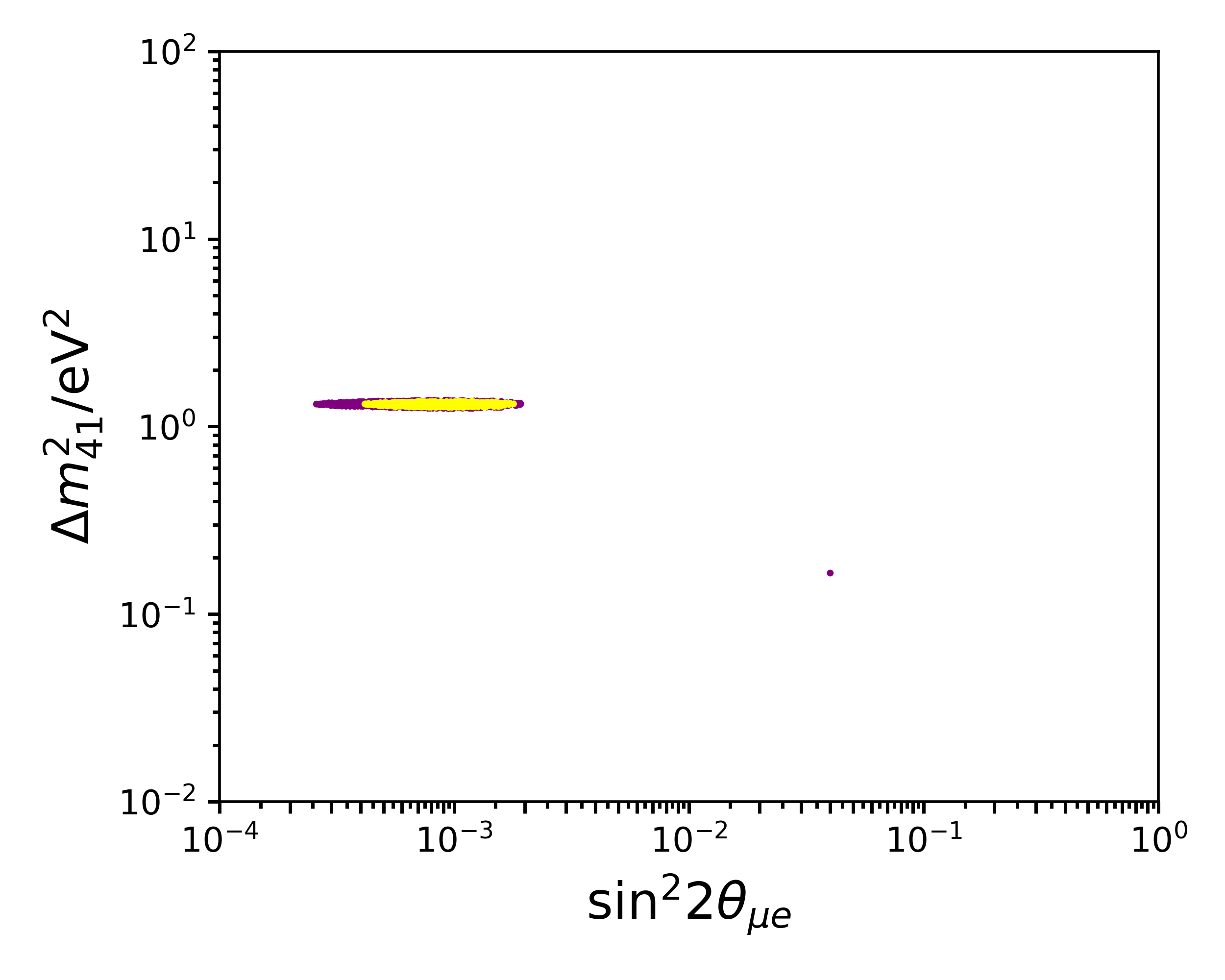}
\caption{$95\%$ confidence interval for 3+1 global fit. Yellow contour assumes that the $\Delta \chi^2$ is properly $\chi^2$ distribution with 2 degrees of freedom. The purple contour is drawing the allowed region if we assume that the critical $\Delta \chi^2$ is the average of that found by PROSPECT. We see that the difference in size is only modest, and negligible in the $\Delta m_{41}^2$ axis.}
\label{fig:biggercritical}
\end{figure}

To estimate an upper bound to the size of this effect on the 3+1 global fits, we perform a comparison where we use the $\Delta \chi^2$ test statistic assuming that it is either properly $\chi^2$ distributed or follows PROSPECT's calculated critical $\Delta \chi^2$. The effect is that, in the former case, points with $\Delta \chi^2 > 6$ are rejected for a $95\%$ confidence region, while in the later case a $\Delta \chi^2 \gtrsim 9$ is required to reject a point. The two results are shown in Fig.~\ref{fig:biggercritical} for the $95\%$ confidence region, where yellow corresponds to the assumption of a $\chi^2$ distributed $\Delta \chi^2$ test statistic, and purple uses PROSPECT's measured critical $\Delta \chi^2$.  We find that even with this conservative inflation of the critical $\Delta \chi^2$, the allowed regions expand only modestly, and almost negligibly in $\Delta m_{41}^2$.  

\subsection{The 5 MeV Excess in the Reactor Flux Can Affect Oscillation Analyses \label{bump}}

In Sec.~\ref{sources}, we briefly noted that many reactor-based experiments have observed an excess of events around an energy of 5~MeV compared to prediction.   
There are, however, a set of reactor experiments that do not observe such an excess.  
In this section, we look at the overall trends in reactor experiments.   Along with the reactor experiments used in our global fits, we consider several others.    Four older experiments, ILL \cite{ILL}, Savannah River \cite{SavannahRiver}, ROVNO \cite{ROVNO}, and Goesgen \cite{gosflux}, are not included in our global fits because their limits have been superseded by more modern experiments.   STEREO has only recently released their data, and so we have not yet had the opportunity to incorporate their results into the global fits. We describe their experiment in more detail in Sec.~\ref{nextreactor}.  However, STEREO has released interesting results with respect to the 5~MeV excess, so we include this experiment in our discussion here.  We omit one reactor experiment, DANSS, as they have provided ratios of rates between detectors at different positions, but not an absolute comparison to simulation of the measured rates.

\begin{figure}[t]
\begin{center}
{\includegraphics[width=\columnwidth]{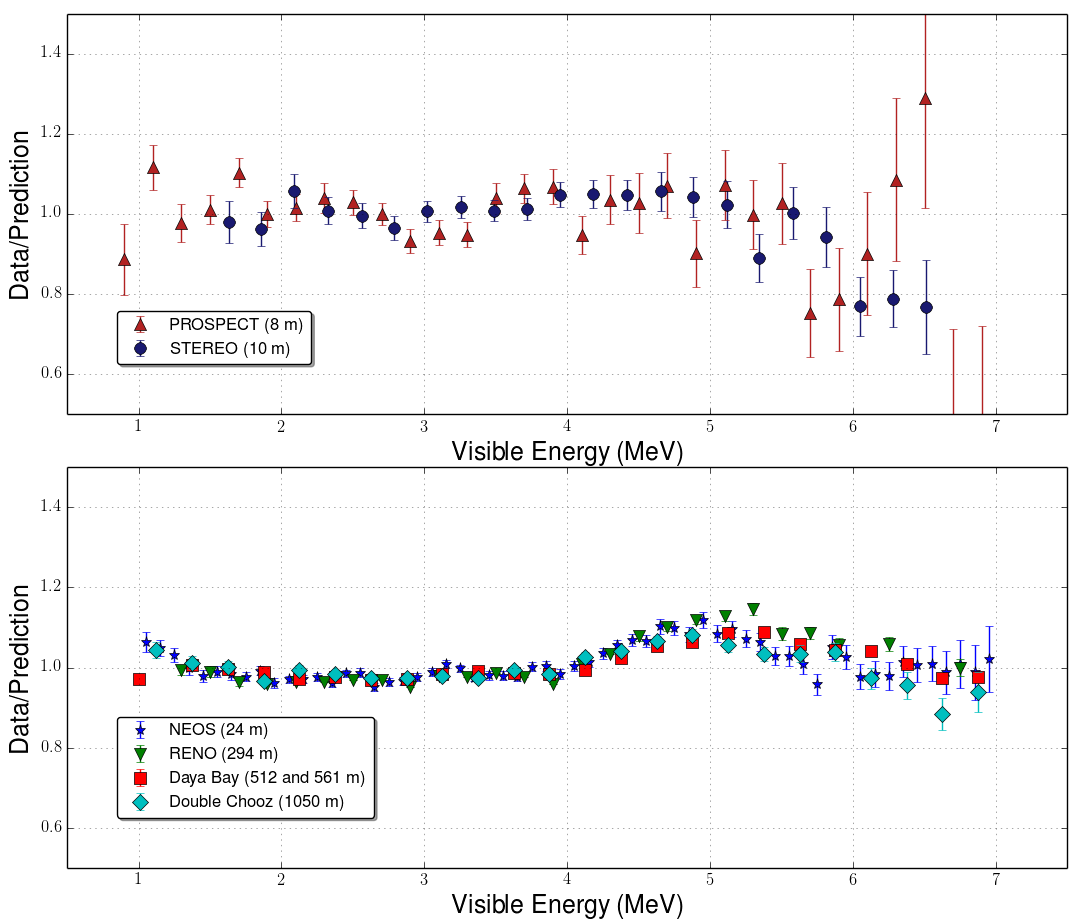}}
\end{center}
\caption{Relatively normalized ratios of data to prediction \cite{Huber}.  Top:  Two recent high-statistics experiments with baseline of $L\sim 10$ m, PROSPECT and STEREO categorized as ``No Excess''.  Bottom:  Four experiments with baseline $>20$ m, NEOS, RENO, Daya Bay and Double Chooz categorized as ``Excess'' \cite{STEREOmoriond}.   See Table~\ref{5MeVcompare} for references for data sets.
\label{5MeVsignals}}
\end{figure}

In Table~\ref{5MeVcompare}, we compare the measured event rate to the simulation provided by the experiment to determine whether the result demonstrates a 5~MeV excess.   The experiments are ordered by distance from the reactor core (column 2).
We summarize our findings in column 3 as ``Excess,'' which indicates a 5~MeV excess of $\sim 10\%$, while  ``No Excess,''  indicates an excess of $\lesssim 5\%$ over the prediction.  We illustrate these categories on  Fig.~\ref{5MeVsignals} which shows examples of ``No Excess'' on top and ``Excess'' on bottom.  In one case, the result is ``Unclear'' as discussed below.   In column 4, we 
indicate the type of core, which may be Highly Enriched Uranium (HEU) in the case of research reactors, or Low Enriched Uranium (LEU) in the case of power reactors.     We categorize the size of each detector as ``$< 1$ t'' of target material,  ``$<10$ t'' but $>1$ t,  and ``$>10$ t'' in column 5.    All detectors are scintillator based, however each experiment has 
augmented neutron capture through interspersing an isotope with high neutron capture cross section in the detector, as indicated in the ``n-capture'' column.

\begin{center}
\begin{table*}[t]
\begin{tabular}{l|c|c|c|c|c|c|}
Experiment & Average $L$ & Observation & Core Type & Detector Size & n-capture & Ref.   \\
 \hline
 PROSPECT & 8 m & No Excess & HEU & $< 10$ t &  Li & \cite{Prospectflux} \\
 ILL & 9 m & No Excess & HEU & $<1$t & He&  \cite{ILL}\\
 STEREO & 10 m & No Excess & HEU & $<10$ t  & Gd & \cite{STEREOmoriond}\\
 Bugey & 15, 40, and 90 m & No Excess at any position & HEU & $<10$ t & Li & \cite{Bugey} \\
 ROVNO & 18 m & Excess & LEU &  $<10$ t & Gd & \cite{ROVNO} \\
 Savannah River & 18 and 24 m & No Excess and Unclear & LEU & $<1$ t & Gd & \cite{SavannahRiver} \\
 NEOS & 24 m & Excess & LEU & $< 1$ t & Gd& \cite{NEOS}\\
 Goesgen & 38, 46 and 65 m & Excess & LEU & $<1$ t & He &\cite{gosflux} \\
 RENO & 294 m & Excess & LEU & $>10$ t & Gd &\cite{RENO5MeV} \\
 Daya Bay & 512 and 561 m & Excess & LEU & $>10$ t & Gd &\cite{DB5MeV} \\
 Double Chooz & 1050 m    & Excess & LEU & $> 10$ t & Gd &\cite{DC5MeV}
\end{tabular}
\caption{Short and long baseline experiments that can potentially observe an excess in the $E_{prompt}\sim 5$ MeV ($E_{positron}\sim 4$ MeV) range in the flux.  ``No Excess''-- agreement with prediction to $<5\%$ in range of interest;   ``Excess''--disagreement at $\sim 10\%$. See text on Savannah River 24 m.  H(L)EU -- Highly (Low) Enriched Uranium core.   Detectors are constructed of liquid or solid scintillator, interspersed with, or mixed with, elements with a  high neutron capture cross section, as noted in the ``n-capture'' column. \label{5MeVcompare}}
\end{table*}
\end{center}

Overall, most of the experiments fall into two categories: those at $\sim 10$~m show no excess, while those at $\gtrsim 20$~m show an excess.   
There are two experiments that deviate from this picture: Savannah River and Bugey.
The Savannah River 18~m data set shows no excess.  The 24~m data set shows an excess, but one that does not have the Gaussian shape seen in the other experiments.  Instead, the deviation monotonically increases from visible energy of $2$ to $\sim 6$~MeV, and then jumps above and below the prediction thereafter. The Bugey 15, 40 and 95~m results show no excess at any of the three positions.   We will set aside these two experiments for the remaining discussion.    

There is a strong correlation between those experiments with excesses and those with low enriched uranium cores (power reactors).   This would lead one to suspect that plutonium-burning, which occurs in LEU and not HEU cores might be the source of the excess.  However, studies of data from RENO \cite{RENOvenice} and NEOS \cite{HuberNEOS} as a function of the burn-cycle contradict this conclusion.   The present RENO result indicates the effect is due to uranium-burning at nearly 3$\sigma$.

Other possible explanations include those that are detector-related.   For example, the absence of an observed peak might be somehow related to the small size of the short-baseline reactor detectors.   This, however, is contradicted by two very small detectors that see the excess: NEOS and Goesgen.   There is also no apparent pattern in the choice of neutron-capture element interspersed or added.

Can the explanation for the short/long baseline disagreement be that the 5~MeV excess is located at the position of an oscillation dip for experiments in the $\sim 10$ to 15 m~range?  In order to study this possibility, we use the measured PROSPECT data \cite{Prospectflux}, and compare it to a prediction that includes various models of the 5~MeV excess.    To produce this prediction, we modify the Huber $^{235}$U flux \cite{Huber} by the excess seen in the Daya Bay measured and unfolded antineutrino spectrum \cite{An:2016srz}. We consider three cases where a) there is no excess with respect to the Huber model, b) each fission fuel component has an equal contribution to the excess and c) the excess originates entirely from the $^{235}$U chain only.  For each of these cases, we modify the Huber model accordingly and do a fit to the PROSPECT data with either no oscillations or with a best fit $3+1$ sterile oscillation model.

\begin{figure*}
\centering
\begin{subfigure}{0.32\linewidth}
\includegraphics[width=\textwidth]{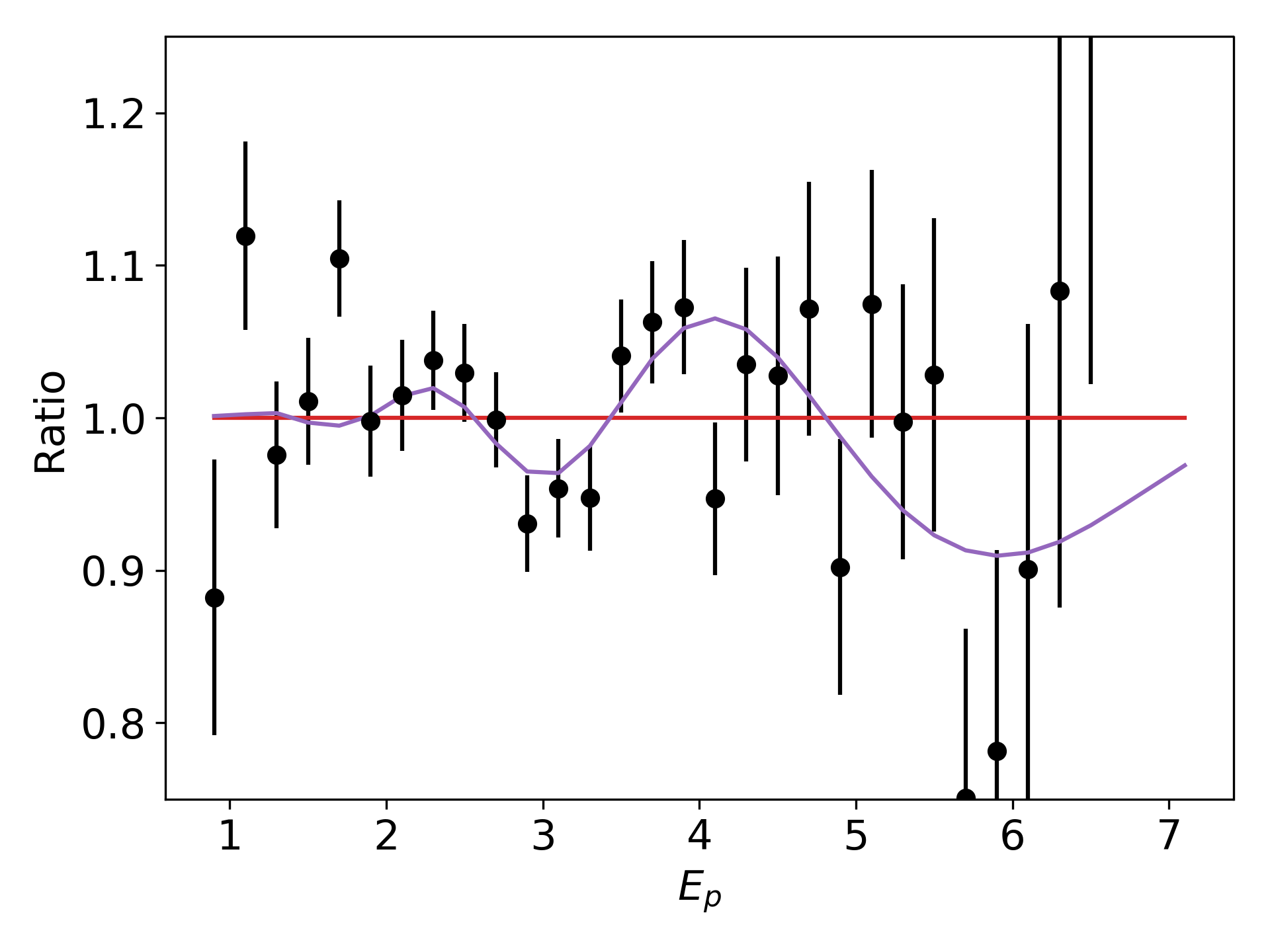}
\caption{No 5 MeV excess flux model.}
\end{subfigure}
\hfill
\begin{subfigure}{0.32\linewidth}
\includegraphics[width=\textwidth]{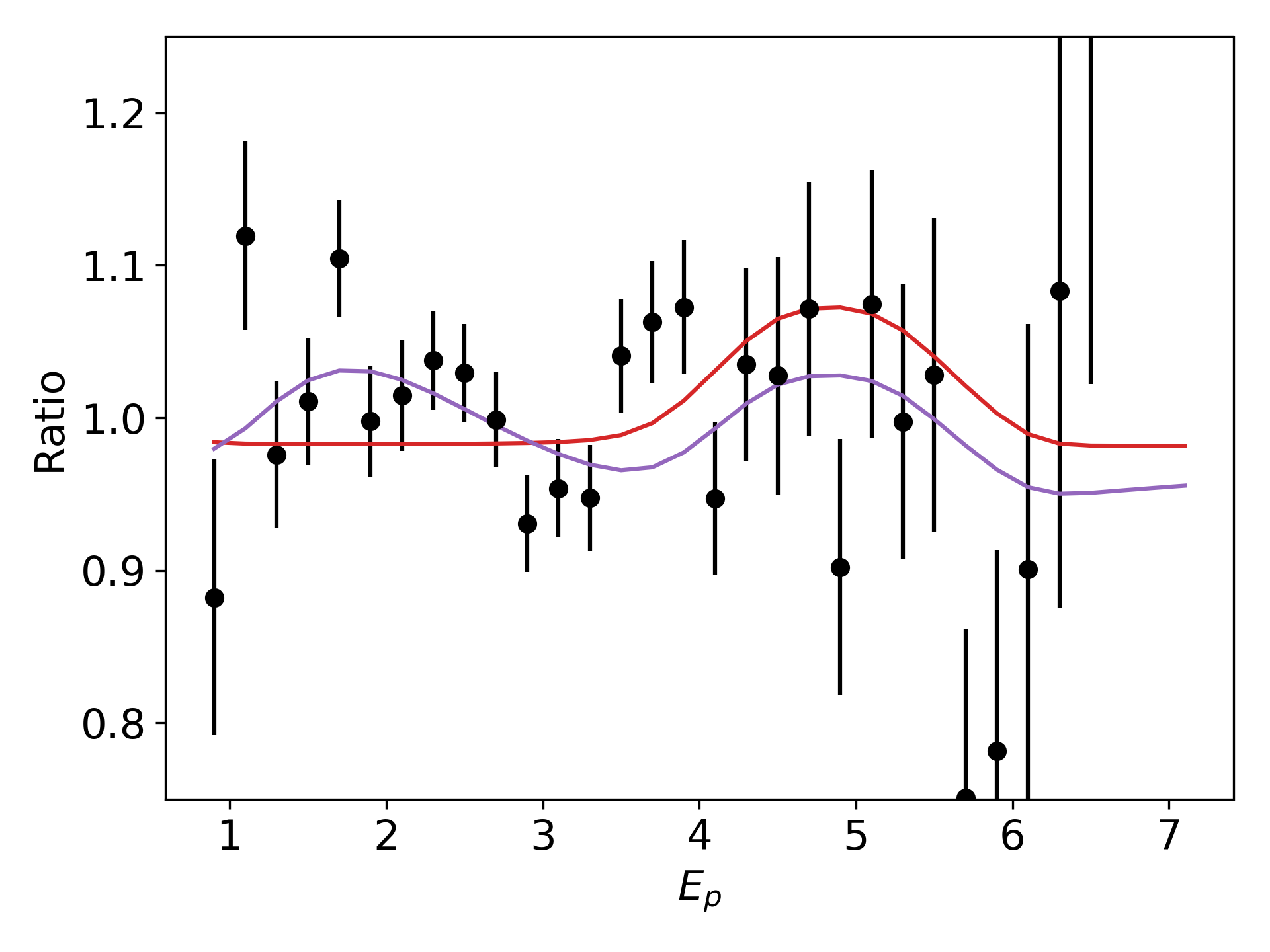}
\caption{An equal 5 MeV excess for all fuel components.}
\end{subfigure}
\hfill
\begin{subfigure}{0.32\linewidth}
\includegraphics[width=\textwidth]{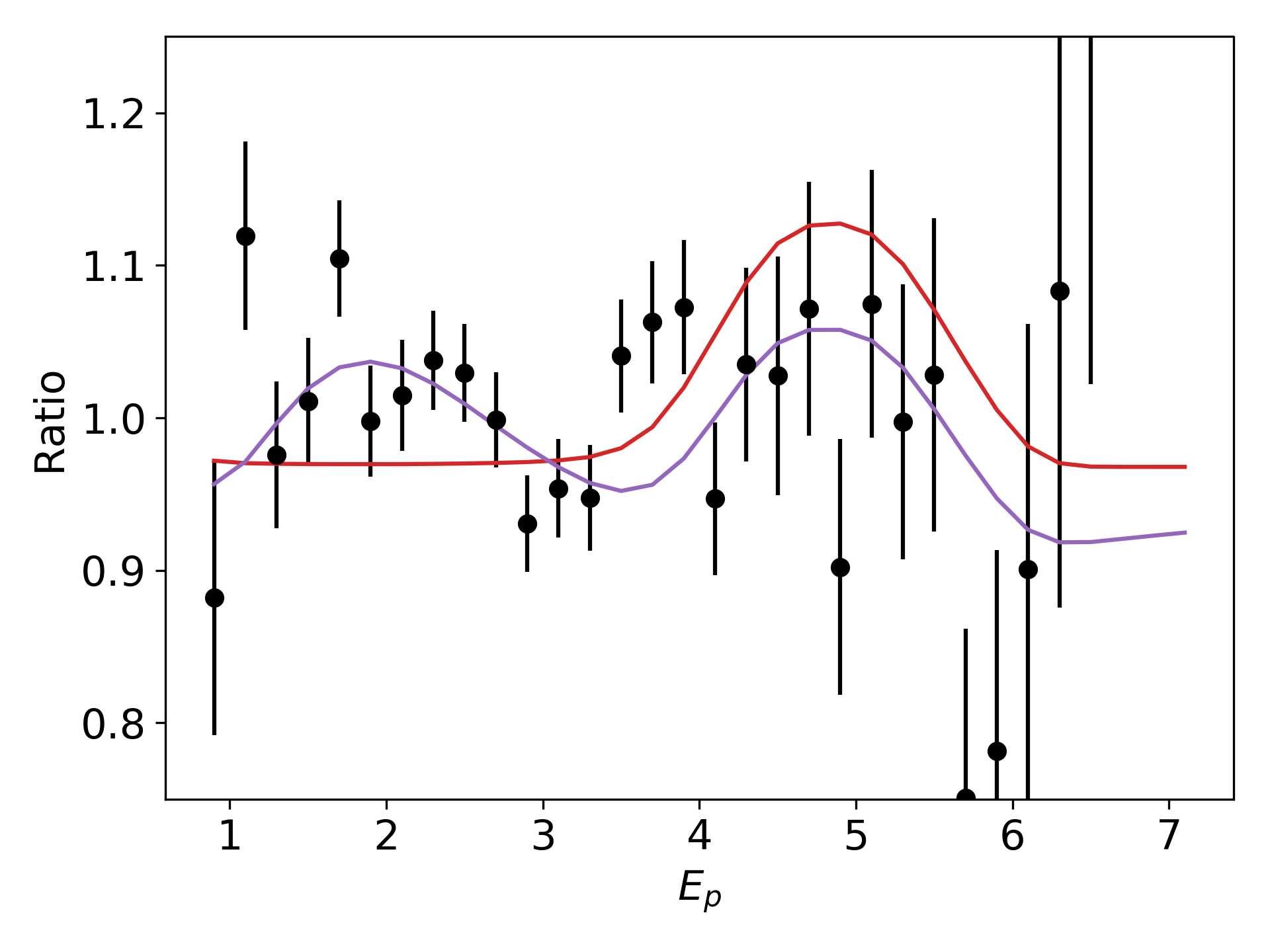}
\caption{A 5 MeV excess for $^{235}$U only.}
\end{subfigure}
\caption{Prospect measured data (black points with error bars) compared to predictions with no oscillation (red curve) or a best fit $3+1$ sterile oscillation model (purple curve).}
\label{bump_plots}
\end{figure*}

The results of these fits are shown in Figure~\ref{bump_plots}.  For cases a), b), and c), the oscillation model reduces the $\chi^2$ from 62 to 50, 69 to 58, and 84 to 61 respectively.  These $\chi^2$ reductions indicate that there is some oscillatory behavior in the data that the fit is picking up both in the excess regions and also around 2 MeV. For the two excess cases, b) and c), the best fit values are similar with values of approximately $\sin^22\theta = 0.14$ and $\Delta m^2 = 0.95$ eV$^2$.  Especially in case c), where there is a substantial change with $\Delta \chi^2 = 23$, we speculate that there could be a 5~MeV excess in the PROSPECT data that is being reduced by a $3+1$ oscillation effect. According to PROSPECT's provided $95\%$ exclusion line \cite{Prospect}, the above point lies directly on the line. 
 
Our conclusion is that both the existence of the 5~MeV excess in most longer-baseline data sets, and the lack of an excess in all data sets at $\sim$10~m needs to be addressed.  One speculative possibility is that there could be $\bar \nu_e$ disappearance at the energy of the 5~MeV flux excess that effectively reduces the bump for experiments with $\sim$10~m baselines.  As more data becomes available, this possibility should be tested as an explanation for the differences among the various reactor experiments.

\subsection{True $E_\nu$, Reconstructed $E_\nu$, and Perils of $L/E$ Plots}

Inverse Beta Decay experiments -- such as the reactor experiments and LSND -- reconstruct $E_\nu$ with excellent resolution; however, most CCQE experiments tend to have resolution on the order of 10\%/$\sqrt{E}$, with asymmetric, long tails at lower reconstructed energies.    The tails are usually due to nuclear effects that remove visible energy from the event.     These include CC single pion production, where the pion is absorbed in the nucleus, leading to the signature of a single lepton and proton.   Hard interactions of the proton within the nuclear environment -- releasing neutrons that are invisible in the detector -- also contribute to this effect. 
In order to address this, experiments with relatively poor energy resolution must provide information on the true energy of simulated reconstructed events.    The oscillation signal prediction must be introduced to the analysis based on this true neutrino energy.    

The oscillation signal must also properly reflect the length the neutrino travels from production to observation.    This can be an important effect for short-baseline experiments using decay-in-flight beams.

\begin{figure}[t]
\begin{center}
{\includegraphics[width=\linewidth]{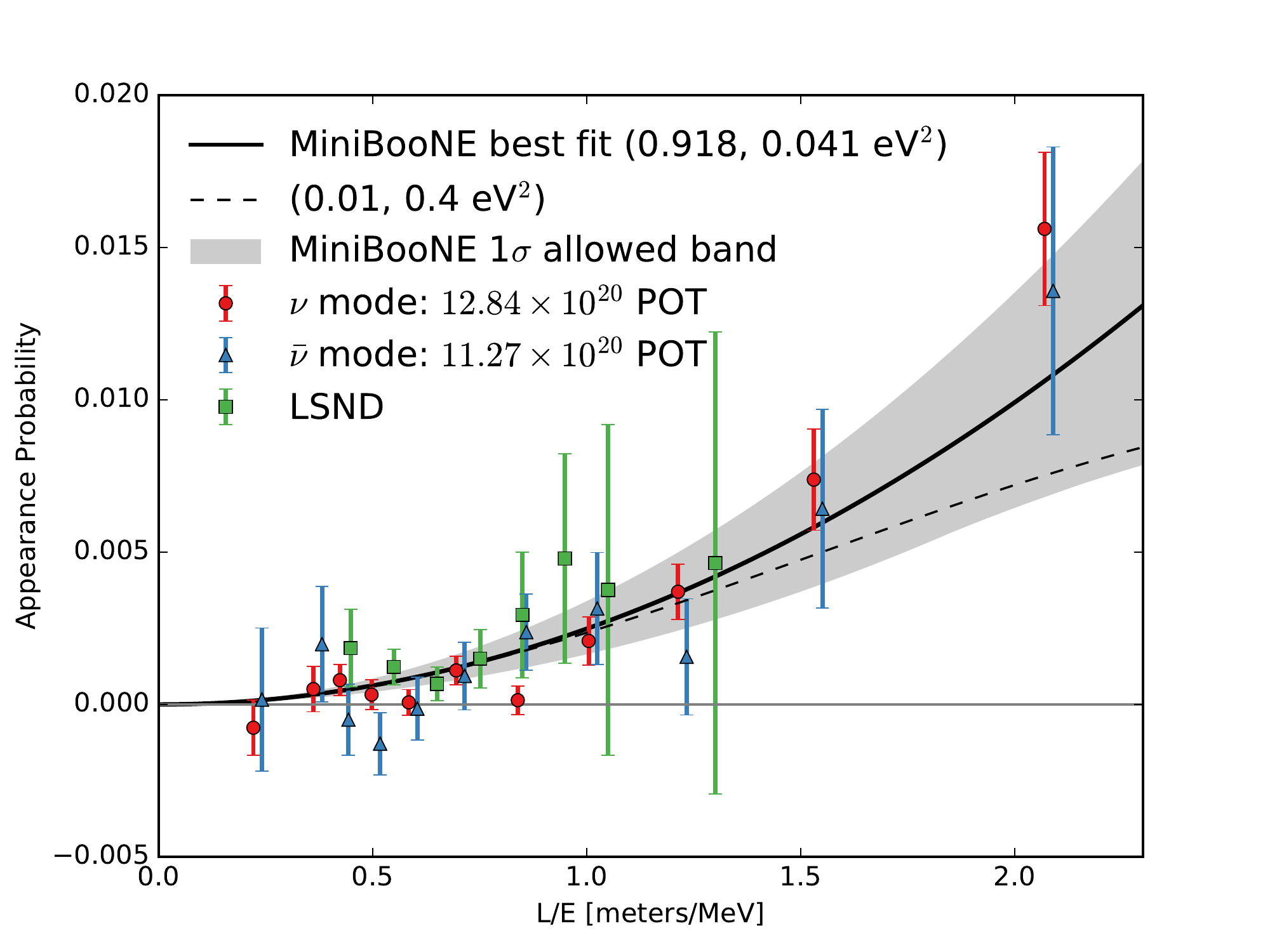}}
\end{center}
\vspace{-0.25in}
\caption{Comparison of $L/E$ distributions from MiniBooNE neutrino running, first and second runs combined, (red); antineutrino running (blue), and LSND (green).  The MiniBooNE 1$\sigma$ allowed range is shown in grey. Plot from Ref. \cite{MB2018}.
\label{loverefig}}
\end{figure}

While this may sound relatively obvious, experiments have, in the past, made incorrect assessments of their results due to leaving out these corrections \cite{ICARUS}.  Quite often this occurs when experimenters make hasty $L/E$ plots \cite{LEpaper}.  For example, the MiniBooNE results will appear to be highly incompatible with LSND if this correction is not included.    However, if handled properly, there is reasonable agreement, as seen in Fig.~\ref{loverefig}. 

An important nuclear effect that is the focus of many recent studies is the case of multi-nucleon interactions.  In this case, the exchange current is interacting with a pair of nucleons.  This will distort the kinematics, since the target is no longer a single nucleon.
This case is not included in the MiniBooNE covariance matrices because this was not a recognized problem at the time of the MiniBooNE first neutrino and antineutrino analyses, and the second MiniBooNE neutrino analysis uses the exact same simulation as the first.  Indeed, the MiniBooNE cross section measurements were central to identifying this possible effect.   This could add as much as 20\% to the overall cross section \cite{Teppeireview}.   Additional rate is not an issue since MiniBooNE constrains the normalization of the $\nu_e$ prediction with observed $\nu_\mu$ interactions, but distortion of the kinematics in the range of muon threshold effects could be an important missing systematic uncertainty in the global fits \cite{Martini}. 

\subsection{``Tension'' and a Problem with the PG Test}

In Sec.~\ref{launch}, we introduced the 
primary issue for sterile neutrino studies today, which is referred to as the ``tension" between disappearance and appearance results.    The tension is parameterized using the PG test, Eq.~\ref{chi2pg}.    As we pointed out in Ref.~\cite{bookchapter}, the PG test relies on the normally-distributed systematic uncertainties.   If the 
number of degrees of freedom is not $N_{PG}$ (see Eq. \ref{npg}), then the
probability of $\chi^2_{PG}$ for $N_{PG}$ is not a valid estimate.    

This is very likely to be the scenario we are in today.  This arises in the cases of low statistics, where Poissonian fluctuations are not properly considered; in the cases of ratio plots, because the error on the ratio is not normally-distributed; and in the cases of experiments dominated by systematic uncertainties, since these are rarely normally-distributed.   This describes the majority of the experiments to which we are applying the test.

Beyond this, the PG test is testing a scenario where no experiment has residual backgrounds after background subtraction.     In fact, the PG Test is relatively insensitive to most types of backgrounds, as we discuss in Ref.~\cite{bookchapter}.  However, there is one case that is highly problematic.     This is the case where a distribution with an underlying signal with parameters ($\Delta m_1^2$, $\theta_{\mu \mu 1}$, $\theta_{e e 1}$, $\theta_{\mu e 1}$) fits well to a signal with a different set of parameters ($\Delta m_2^2$, $\theta_{\mu \mu 2}$, $\theta_{e e 2}$, $\theta_{\mu e 2}$) when an unknown background is added.    This may explain the MiniBooNE case.   In principle, this would not be a major problem for the PG Test if there were many data sets of equal strength in the appearance subset.  However, in the 1 eV$^2$ region, the MiniBooNE data set dominates.

\subsection{Is There Enough Scrutiny Given to Limits? \label{limits}}

When you set a limit at
$N\%$ CL, there is a $(100-N)\%$ chance that it can happen, in principle.   So any
given experiment can just get unlucky.   That should be properly accounted for in the global fits, to the level that the errors are normally-distributed, since limits, as well as signals, are brought in with associated errors.     

However, there are important biases that must be considered with respect to limits: 
\begin{itemize}
\item The most-stringent limits are
limited by systematic uncertainties.   An analysis is more likely
to accidentally neglect to include a systematic effect that weakens the limit, than include systematic effects that are
inflated.   
\item Some of the most stringent limits are not coming from blind analyses.  For example, we have seen the MINOS/MINOS+ limit become rapidly more stringent from 2014 to 2018 in a non-blind analysis.     While most blind analysis discussions focus on the question of whether signals can be manufactured, there is an equal danger of signals being removed.
\item  Our community simply pays less attention to limits than anomalies.     Mistakes can be
made, and if the mistakes go in the expected direction, then people accept
the limit without much scrutiny. 
\end{itemize}

\subsubsection{Recent History Suggests Limits Can Go Wrong}

Before discussing a case-in-point for these global fits, it is worth noting that history bears out the observation that limits can certainly go wrong.  One example is when an signal is observed well within an excluded region from a limit.
A recent example of this comes from a 2-neutrino double
beta decay ($2 \nu \beta \beta$) observation in $^{136}$Xe.  In Ref.~\cite{DAMA},
DAMA liquid xenon sets a limit on the half-life of $2 \nu \beta \beta$ of  $T_{1/2} >
1.0 \times 10^{22}$ years at 90\% CL.     This limit was set in October, 2002.     Then, in
November, 2005,  Baksan reported that $T_{1/2} > 8.5\times 10^{21}$ years at 90\% \cite{baksan},
confirming
the DAMA result.     In November, 2011,  EXO-200 turned on and
measured $T_{1/2} = 2.11\times 10^{21}$ years \cite{exo}.   The bottom line: both of the original limits were incorrect, which may have happened for any of the reasons listed in the introduction to this section.  Therefore, this stands as a cautionary tale--it is worthwhile to question limits as strongly as signals, and to cross check the limits by more than one experiment.

\subsubsection{Comment on the MINOS+ 2018 Limit \label{MINOSsec}}

MINOS/MINOS+ is a two detector experiment.   The near detector is located 1~km from the neutrino target, and the far detector is located 730~km downstream.  The two detectors are
not the same and the near detector intercepts a very different beam than the far detector as the decay pipe (neutrino source) ends only 317~m upstream of the near detector.   Also, because of this proximity, it is only partially instrumented, and yet suffers from a very large dead-time and pile up due to the high beam rate.  These differences, along with the details of the neutrino flux and interactions, are modeled with Monte Carlo simulation.   The systematic uncertainties associated with this modeling are captured in a covariance matrix, which is used to parameterize the neutrino flux, cross section, and detector uncertainties and their correlations between energy bins, neutrino processes, and detectors.

In March 2019, MINOS/MINOS+ published a result for a combined two detector analysis entitled, ``Search for Sterile Neutrinos in MINOS and MINOS+ Using a Two-Detector Fit''\cite{MINOS2019}.  For this paper they performed a disappearance analysis using four data sets by separating out the charged-current $\nu_\mu$ events and neutral-current events for the near and far detector separately.  To include and exploit, the correlations between these four data sets, MINOS built up a combined covariance matrix for these four data sets.  The search for sterile neutrino oscillations was then accomplished using a $\chi^2$ statistic calculated -- with the use of the combined covariance matrix -- for a given oscillation model against these four data sets.

The results presented for the MINOS/MINOS+ 2019 analysis \cite{MINOS2019} are surprising.  Comparing the previous results that were shown at Neutrino~2014 to the results shown at Neutrino~2016 and finally to the latest published results, the $\sin^2 \theta_{24}$ 90\% C.L. limit at large mass-squared-differences ($\Delta m^2 = 1000$ eV$^2$) has improved from 0.226 to 0.1 to 0.023 \footnote{Note that we define $\theta_{24}$ in Table~\ref{cheatsheet} and that MINOS reports $\sin^2 \theta_{24}$, not the more usual $\sin^2 2\theta_{24}$.}.
From the 2014 to the 2016 result, 50\% more data was added for which the energy spectrum was about $\times 1.75$ higher, but the latest published result used the same data as 2016.  From these numbers, it is clearly not possible that these improvements were due to more statistics.  

The MINOS+ collaboration states that these improvements are due to the improved analysis method of fitting to data in the near and far detector directly using the covariance matrix and adding the neutral-current channel to the analysis.  However, the high $\Delta m^2$ limit they claim violates the relation given in Eq.~\ref{eq:hi_dm2_limit}, where the limit is related to the uncertainty in the predicted total event rate $N_{\alpha}$ by $\sin^22\theta_{Limit} = 2(1.28)*\delta N_\alpha/N_\alpha$ for 90\% C.L..  To reach the above $\sin^2 \theta_{24} < 0.023$ (or $\sin^2 2\theta_{24}<0.09$), one would need a normalization uncertainty of $\delta N_\alpha/N_\alpha = 3.5\%$.  MINOS has determined their normalization uncertainty in several ways (see supplement of Ref. \cite{MINOS2019}) and find that $\delta N_\alpha/N_\alpha$ is about 10\%, which would give a $\sin^2 \theta_{24}$ limit of 0.07 that is close to the 2016 result.  In response to this discrepancy, MINOS+ claims that this  type of high $\Delta m^2$ limit is equivalent to a single bin counting analysis that does not give you the proper limit since breaking up the data into smaller bins gives added oscillation sensitivity due to the statistical fluctuations in the many bins.  This does not seem to make sense since statistical fluctuations should just lead to fluctuations in the limit for a given data set and not a trend toward better sensitivity.  If making finer energy bins improves the limit, why not go to hundreds or thousands of bins?

A second issue with the MINOS+ analysis concerns the assumption that $\theta_{14}$ is zero, which ignores the possibility of electron-neutrino appearance. In general, 3+1 models have non-zero values for both $\theta_{14}$ and $\theta_{24}$. Therefore, the analysis should be repeated without using the neutral-current (NC) data sample, which includes electron-neutrino charged-current events, and without assuming that $\theta_{14}$ is zero. The authors claim that the NC data sample has little effect on the oscillation sensitivity but MINOS sees $\sim$6\% more NC events in the far detector than the near detector, which could be due to electron neutrino appearance and/or an unknown systematic uncertainty.  

Since the simple high $\Delta m^2$ test failed, the results seem to be sensitive in an unexpected way to the energy binning, and NC events may affect the disappearance result, we feel that more study is required before we can include this data in our global fits.     Therefore, we use the MINOS/MINOS+ 2016 data set.

%\section{The MiniBooNE Technique of Constraining Backgrounds with $\nu_\mu$ Data \label{MBcomment}}

%\section{Further Comments on the MiniBooNE Data Sets \label{MBcomment}}

%\input{comment_MB}

\section{Results Beyond Vacuum Oscillation Experiments}

Our global fits focus on results from accelerator and reactor sources that can be interpreted using vacuum oscillations.  However, there are other methods of sterile neutrino searches which use signatures beyond vacuum-oscillations.   In this section, we briefly review other approaches using atmospheric, solar, and astrophysical neutrinos.   We also touch on the ongoing controversy concerning sterile neutrinos and cosmology.

\subsection{Atmospheric neutrinos}

Experiments that make use of atmospheric neutrinos for the flux have produced limits on sterile neutrinos. Experiments can set limits through vacuum oscillations and also matter effect  resonances~\cite{Nunokawa:2003ep, Petcov:2016iiu}.   These resonances can produces a large disappearance signal in TeV-energy atmospheric neutrinos traversing the Earth's core for $\Delta m^2 \sim 1 \text{ eV}^ 2$.   

Atmospheric neutrinos are produced in hadronic showers initiated by high-energy cosmic rays impacting the Earth's atmosphere~\cite{Gaisser:2016uoy}. At high energies, pions and kaons are long-lived and lose energy before decaying. Neutrinos from pion, kaon, and muon decay make up what is called the ``conventional atmospheric flux.'' In contrast, charmed mesons and baryons decay before interacting in the atmosphere, resulting in a harder spectrum, {\it i.e.} with higher energy particle content. Due to the immediate decay, this is called the ``prompt component''~\cite{Enberg:2008te}. The muon-neutrino flux is dominated by the conventional component up to approximately 10 TeV~\cite{Fedynitch:2018cbl}.  
This range of the atmospheric neutrino energy and angular distribution is well-understood. 
At higher energies, it becomes a combination of prompt and astrophysical neutrino fluxes~\cite{Aartsen:2013bka,Aartsen:2016xlq}.    These fluxes are not yet well-constrained, and so we will concentrate on results using the conventional flux.

Atmospheric neutrinos allow for the most strict constraint on $|U_{\mu 4}|$ and $|U_{\tau 4}|$ combinations for masses less than $\Delta m^2 < 10~{\rm eV}^2$; at higher masses, constraints from NOMAD~\cite{Astier:2001yj} are relevant. 
When the oscillation length is much smaller than the baseline one has that, in matter, the muon-neutrino two-flavor disappearance probability can be expressed as~\cite{Blennow:2018hto}:
\begin{equation}
P _ { \mu \mu } \simeq 1 - V _ { \mathrm { NC } } ^ { 2 } \left| \alpha _ { \tau \mu } \right| ^ { 2 } L ^ { 2 },
\label{eq:numu_matter_ut4}
\end{equation}
where $L$ is the baseline of propagation, $V_{NV}$ is the neutral current potential in the constant density, and $\alpha_{\tau\mu} = \sum_i^n  U_{\tau i} U_{\mu i}^*$, where $U$ is the extended PMNS matrix to n-flavors.
There is also a constraint by MINOS, by measuring the rate of neutral current events~\cite{Adamson:2011ku}, but the constraints are significantly weaker. 
A search of this nature was performed first by SuperKamiokande~\cite{Abe:2014gda} and later by IceCube using the DeepCore inner array~\cite{Aartsen:2017bap}.
A recent search has also been performed by the ANTARES collaboration~\cite{Salvadori:2017ugh} yielding similar results to the DeepCore bounds.

At TeV neutrino energies neutrinos experience resonance conversion between active and sterile neutrino flavor states~\cite{Nunokawa:2003ep}; see~\cite{Petcov:2016iiu} for a recent review of this effect. For a 3+1 model the resonance happens in antineutrinos, which make about 30\% of the total rate of events in the TeV-energy range due to a diminished flux and cross section. The phenomenology of this effect has been developed in IceCube in~\cite{Esmaili:2012nz,Esmaili:2013fva}. We summarize the results of the IceCube analysis~\cite{TheIceCube:2016oqi} using this technique in Sec.~\ref{sec:icecube}.

These results depends on the standard neutrino matter potential. It was pointed out in~\cite{Liao:2016reh} that introduction of non-standard neutrino interactions can severely modify the results. Adding non-standard interactions~\cite{Liao:2018mbg} shifts the maximum transition probability to lie in the region between the DeepCore and IceCube analysis~\cite{Esmaili:2018qzu}.
These ``secret'' neutrino interactions do not need to involve the active flavors; all of the new interactions can be in the sterile flavor state~\cite{Denton:2018dqq}.
Another way of modifying these sterile bounds is by making the sterile neutrino decay length smaller than the flavor transition scale~\cite{Moss:2017pur}.

In the oscillation-averaged regime, for energies above the resonance behavior, matter enhancement becomes $\Delta m^2$-independent and the fit is performed in the $|U_{\mu 4}|$ and $|U_{\tau 4}|$ plane.
The hint of sterile neutrino observation that was pointed out in~Ref.~\cite{Blennow:2018hto} is in tension with NOMAD measurements~\cite{Astier:2001yj}.
Also, as the resonance energy increases it moves to higher energies where the atmospheric flux is not well modeled.
In fact, the unmeasured charmed contribution is predicted to be approximately 5\% at 20 TeV, as has been pointed out in a recent work studying IceCube public data~\cite{Miranda:2018buo}. 

Further searches will come from KM3Net-ORCA~\cite{Coelho:2017cwp}, as well as INO~\cite{Thakore:2018lgn}, DUNE-atmospherics~\cite{Higuera:2018zjz}, and Hyper-K atmospherics~\cite{Kelly:2017kch}.

\subsection{Solar neutrinos}

The deficit of electron-neutrinos produced in nuclear reactions in the Sun core, which were the first indication of neutrino-flavor morphing, can be modified by the existence of a sterile component.  Therefore, we briefly review matter effects in the Sun, and then consider sterile neutrino effects; see Ref.~\cite{Maltoni:2015kca} for a review of solar neutrino physics.

The flavor conversion in the Sun can only be correctly interpreted if matter effects are properly included because the Solar matter potential is very large. One can write the neutrino propagation Hamiltonian as
\begin{equation}
H = H_{std} + H_{matter},
\label{eq:h_matter}
\end{equation}
where, in the case of three-active neutrinos, $H_{matter}$ only depends on $G_F$ and the electron number density, $N_e$.
The average electron neutrino survival probability is given, in the two-level system, by
\begin{equation}
\bar P_{ee} = \frac{1}{2} + \left(\frac{1}{2} - P_c\right) \cos 2\theta_0 \cos 2\theta_m,
\label{eq:parke}
\end{equation}
where $\cos 2\theta_0$ and $\cos 2\theta_m$ are the vacuum and matter mixing angles.
The term $P_c$ is known as the crossing probability and is given in terms of the adiabaticity parameter, $\gamma$, which is related to the change in matter density by
\begin{equation}
\gamma = \frac{\Delta m^{2} \sin ^{2} 2 \vartheta}{2 E \cos 2 \vartheta\left|\mathrm{d} \ln N_{e} / \mathrm{d} x\right|_{\mathrm{R}}},
\label{eq:parke}
\end{equation}
where $\left|\mathrm{d} \ln N_{e} / \mathrm{d} x\right|_{\mathrm{R}}$ is related to the radial change of electron number density.
This expression provides excellent agreement with exact calculation of neutrino oscillation in matter obtained by solving the neutrino propagation equation~\cite{deHolanda:2004fd,Ioannisian:2004jk,Ioannisian:2008ve}.

The Solar neutrino data, when fit alone, is well-described by the adiabatic conversion, known as MSW effect, with neutrino oscillation parameters given by
$\Delta m_{21}^2 = 4.7\times 10^{-5}~{\rm eV}^2$ and $\sin^2 \theta_{21} = 0.31$, and any model introducing modifications from additional mostly-sterile neutrinos must accommodate this agreement.  Let us consider modifications within three regimes of interest for mass splittings when a fourth state is introduced: 
\begin{itemize}

\item Extremely-small mass splittings, $\Delta m^2_{41} \lesssim 10^{-9} {\rm eV^2} \ll \Delta m^2_{21}$, do not distort the solar matter potential, but introduces oscillations lengths comparable to the Sun-Earth distance.
In fact, this scale of mass splitting was known as the ``Just-so'' solution to the solar neutrino problem, but it is now ruled out as an explanation to solar neutrino flavor morphing.

\item Smaller, but comparable, $\Delta m^2_{41}$ mass-squared splittings to $\Delta m^2_{21}$ are motivated by the absence of the upturn in the Solar neutrino data. Additional sterile neutrino mass states with mass-squared differences of $~0.2 \Delta m^2_{21}$ and mixings of order $\sin^22\theta \sim 10^{-3}$ have been shown to alleviate the tension between solar data and KamLAND~\cite{Maltoni:2015kca}.
Due to the lack of next generation Solar neutrino experiments this tension will remain unsolved for the next years, but DUNE~\cite{Capozzi:2018dat} could address this in the next decade.

\item Larger mass-squared differences, $\Delta m^2_{41} \gg \Delta m^2_{21}$, as motivated by the short-baseline anomalies described in this review, affect the oscillation probability by modifying the high-energy part of the electron neutrino survival probability and causing an overall disappearance of the all-flavor neutrino flux.
Of these effects, the strongest comes from the precise measurement of the all-flavor Solar neutrino flux by SNO yielding a limit of $|U_{e3}|^2  + |U_{e4}|^2 < 0.077$ at 95\% C.L.~\cite{Kopp:2013vaa,Bellini:2013uui}. 
\end{itemize}

Lastly, the reader should note that we have not considered a ``2+2'' model in this review.   These models, where the largest gap is in-between pairs of mass states, are significantly disfavored by the solar neutrino data~\cite{Maltoni:2002xd}.

\subsection{Astrophysics observables}

The presence of sterile neutrinos can also affect the expectations of neutrinos from cosmic beam dumps, which are called ``astrophysical neutrinos.''   These may be galactic or extra-galactic in origin.
At energies above 10 TeV, the IceCube neutrino observatory has observed a component of astrophysical neutrinos that is most likely of extra-galactic origin.    In fact, 
the galactic component is constrained to be $\sim10$\% of the observed astrophysical flux~ \cite{Aartsen:2017ujz}.

The large travel-distance and high energies of astrophysical neutrinos causes the neutrino oscillation phase to be very large. This fact, added to the unknown propagation baseline and finite energy resolution, $\mathcal{O}$(10\%) of the deposited energy in a detector like IceCube,  implies that the flavor transition probabilities have lost all sensitivity to $\Delta m^2$ and are given by:
\begin{equation}
P_{\alpha\beta} = \sum_{i=1}^{n}|{U}_{\alpha i}|^2 |{U}_{\beta i}|^2,
\end{equation}
where $\alpha$ is the initial flavor, $\beta$ is the final flavor, $n$ the number of neutrino species, and $U$ is the extended PMNS matrix.  
Refs~\cite{Brdar:2016thq} and \cite{Rasmussen:2017ert} show that, for astrophysical  neutrinos produced via pion decay, the effects of sterile neutrinos are small on the astrophysical neutrino flavor ratio. 
This is due to the fact that the transition probability involving the mostly-sterile mass states comes in like $|U_{\alpha 4}|^2 |U_{\beta 4}|^2$ where $|U_{\alpha 4}| \sim \mathcal{O}(0.1)$.

However, as noted in~\cite{Brdar:2016thq} a significant change of the astrophysical flavor ratio can be obtained if the initial neutrino state has a dominant sterile neutrino flavor. This can be realized by a mechanism like dark matter decays onto sterile neutrinos via a beyond Standard Model force. In this case, the expected flavor ratio at Earth can be found in regions with large astrophysical tau neutrino fraction, which is forbidden by unitarity of the neutrino evolution given that conventional astrophysical neutrino production mechanisms yield only muon and electron neutrinos at the source~\cite{Arguelles:2015dca, Ahlers:2018yom}. Current uncertainties in the astrophysical flavor composition measured by IceCube cannot yet distinguish this scenario from the standard three-neutrino picture. 

If there were a  high-energy $\nu_4$ flux impacting Earth, it was recently pointed out in~\cite{Cherry:2018rxj} that this provides an explanation of the anomalous ANITA events. The ANITA collaboration has recently reported the observation of very-high-energy neutrino candidates.
These anomalous events are such that the probability that a neutrino traverses the amount of matter corresponding to their emergence angle is at the level of $10^{-9}$~\cite{Fox:2018syq, Reno:2019jtr}. 
One possible explanation of these events is that they are due to an incoming $\nu_4$, whose interaction length is longer than a mostly active neutrino mass state. 

\subsection{Cosmological constraints \label{cosmo}}

Two important quantities are use to synthesize the compatibility of an additional neutrino state with cosmology. These are the number of relativistic neutrino species, $N_{eff}$, and the sum of the neutrino masses, $\Sigma m_{v}$. These two parameters, among other cosmological parameters, can be measured by means of three observables: the cosmic microwave background (CMB), the abundance of light elements from Big Bang Nucleosynthesis (BBN), and the large-scale structures (LSS) in the Universe. All three are used to constrain $N_{eff}$, while the LSS and CMB measurements constrain the sum of neutrino masses. 
Recent Planck results constrain the sum of neutrino masses to $\Sigma m_{v} \lesssim 0.1~\mathrm{eV}$~\cite{Aghanim:2018eyx}. The preferred value of $N_{eff}$ is consistent with the three-neutrino framework. However this quantity is correlated with the value of the Hubble parameter, $H_0$, and can be as large as 3.5 at 95\% C.L. for the larger values of $H_0$ given by CMB measurements~\cite{Bernal:2016gxb}. It is important to note that these values of $H_0$ are in tension with local measurements, which prefer a larger value of $H_0$~\cite{Riess:2016jrr}.

Taking these results at face value, there is severe tension between cosmology and a sterile neutrino of masses of $\mathcal{O}(1)$ eV and mixings of $\mathcal{O}(0.1)$, which are the preferred parameters obtained in this review. This situation is more complex in the 3+2 scenario~\cite{Melchiorri:2008gq,Archidiacono:2013xxa}. This is due to the fact that sterile neutrinos of this mixing and mass are assumed to be in thermal equilibrium with the active neutrinos prior to neutrino decoupling at $T \sim 1~{\rm MeV}$~\cite{Dolgov:2002wy}, which implies that they should modify the observed value of $N_{eff}$ and $\Sigma m_{v}$. This tension can be evaded if equilibrium is avoided.

BBN constraints on sterile neutrinos are dominated by measurements of the abundance of primordial helium-4, $Y_{p}^{^4He}$. Current measurements are obtained by linearly extrapolating the helium mass-fraction measurements of dwarf galaxies, yielding a value of $Y_{p}^{^4He} = 0.2579^{+0.0033}_{-0.0088}$~\cite{2011JCAP...03..043A}. This primordial mass-fraction is related to the equilibrium neutron-proton ratio, which is given by~\cite{Steigman:2012ve}:
\begin{equation}
(n/p)_{eq} = {\rm exp}(-(m_n - m_p)/T),
\end{equation}
where $m_n$ and $m_p$ are the neutron and proton masses respectively. Sterile neutrinos would contribute additional radiation energy density, which would then lead to a higher freeze-out temperature at BBN. This implies a larger number of neutrons, and thus an increased value of $Y_{p}^{^4He}$. This leads to a 95\% credible upper limit on the number of sterile neutrinos, $N_s$, of 1.26~\cite{Hamann:2011ge}. However, this limit depends upon the known value of the neutron lifetime, which has two conflicting values obtained from beam and bottle experiments~\cite{Wietfeldt:2011suo}. Using the upper values of the neutron lifetime~\cite{PDG}, the constraint is slightly strengthened to 1.14~\cite{Hamann:2011ge}. 

With this said, as argued in~\cite{Hamann:2011ge}, the existence of a non-zero chemical potential that is common to the neutrinos~\cite{Kang:1991xa} can significantly weaken the upper limits. This is due to the fact that such chemical potential, $\mu$ modifies the equilibrium neutron-to-proton ratio as
\begin{equation}
(n/p)_{eq} = {\rm exp}(-(m_n - m_p)/T - \mu).
\end{equation}
Thus, a positive value of $\mu$ will reduce the number of available neutrons at freeze-out, cancelling the effect produced by the additional neutrino states on this quantity. Allowing for chemical potentials of $\mathcal{O}(0.1)$ results in an upper bound of $N_s < 2.56$ at 95\% credible level~\cite{Hamann:2011ge}.

The tension between cosmological observables, such as CMB and LSS, and eV-scale neutrinos can be reduced by invoking either non-standard cosmological scenarios~\cite{Gelmini:2004ah,Hamann:2011ge} or introducing new neutrino forces~\cite{Hannestad:2013ana,Dasgupta:2013zpn}. The latter solution, known as ``secret forces,'' has been recently reviewed in Refs.~\cite{Song:2018zyl,Chu:2018gxk}. These secret forces suppress the production of sterile neutrinos in the early Universe prior to neutrino decoupling, but yield $\mathcal{O}(0.1)$ mixing angles at current times. The suppression of the mixing angle before neutrino decoupling avoids the thermalization of the sterile neutrino state and thus avoids the $N_{eff}$ constraints. However, this mechanism implies that, at larger times, sterile neutrinos are in thermal equilibrium with the active neutrinos, as the mixings must return to be $\mathcal{O}(0.1)$, and are efficiently produced via the Dodelson-Widrow mechanism~\cite{Chu:2018gxk}. Even if this recoupling is significantly delayed, thereby avoiding the constraints from $N_{eff}$, the equilibrium between active and sterile content in later times would be in tension with measurements of $\Sigma m_{v}$ via LSS~\cite{Mirizzi:2014ama}. This is due to the fact that, after neutrinos decouple, the sterile neutrino component affects LSS by changing the rate of  free-streaming. As noted in~\cite{Chu:2018gxk} this problem can be avoided if the mostly-sterile neutrino mass state experiences prompt invisible decay. This observation further motivates our study of the 3+1 model with decay discussed in Sec.~\ref{dk}.

The effects that secret forces introduce in order to explain cosmological data must evade the present measurements in other experiments in order to provide an adequate explanation. The effects of these forces in terrestrial experiments has been studied in Ref.~\cite{Kopp:2014fha,Denton:2018dqq}; and the effects of high-energy astrophysical neutrinos is reviewed in Ref.~\cite{Cherry:2016jol}.
Overall, while an interesting approach, secret forces do not appear to be a complete solution to the problem.

In summary, cosmological models are in tension with the vacuum-oscillation-based sterile neutrino results.   This is leading to fruitful investigations of less simplistic cosmological models. While the perfect solution has not been identified yet, there is progress. Should the questions surrounding the Earth-based measurements be resolved by determining the existence of more neutrino states, the path to adapting this into workable cosmological models does not seem to be impossible to find. In particular, studies of cosmology point to a more complex sterile neutrino scenario such as additional forces or decay.

\section{The Immediate Future for Short-Baseline Results \label{nearfuture}}

Returning to our focus on sterile neutrino searches at man-made sources, we emphasize that this is an exciting and fast-growing field.  Within the next two years, a number of the experiments already included in the global fits will provide important updates.   In this section, we review experiments that will provide additional results within the next two years, beyond the experiments already included.  

\subsection{$\nu_e$ Disappearance:  Reactor Experiments \label{nextreactor}}

The immediate future of $\nu_e$ disappearance studies lies with reactor experiments. There are three very interesting new experiments that will produce new data sets for our global fits in the very near future: STEREO, Neutrino-4, and SoLid.

As discussed in Sec.~\ref{bump}, we have not yet included results from the STEREO experiment, which runs at the ILL facility in France using their research reactor with a flux from the $^{235}$U fission chain.   This experiment uses a relatively long, segmented detector filled with Gd-doped liquid scintillator.   Running at a research reactor is advantageous compared to DANSS, which runs at a power reactor, because the reactor core is a factor of three smaller in diameter and uses highly enriched (93\%) $^{235}$U fuel. The detector dimensions are 2.233 m $\times$ 0.889 m $\times$ 1.230 m divided into six cells arranged radially from the reactor core, with the detector center at 10.3m from the core center.  For their first results, the experiment forms ratios of the event rates in cells 2 to 6 to the rate in the first cell~\cite{Stereo66}.  This is less advantageous than a movable detector like DANSS, because it requires careful cell-to-cell calibration, but STEREO is equipped with a sophisticated calibration system to address this concern.   STEREO has published null results for an initial 66 days of running~\cite{Stereo66}.  Recently, STEREO has shown results at conferences for 185 days~\cite{STEREOmoriond} where they have used a new shape only fitting technique with floating normalization parameters for each energy bin and, when published, we will include this data in our fits. 

We have also not yet included results from the Neutrino-4 experiment~\cite{Nu4} in our global fits.    This is another experiment that has reported data but has not made a data release.  This is a Gd-doped segmented liquid scintillator detector that has a 1.4 m$^3$ fiducial volume.  It is unique in that it sits very close to the reactor with $L=6$ m upstream, and $L=12$ m downstream.   Thus, this will be an interesting experiment to include in our 5 MeV excess analysis in the future.    This experiment is already reporting a 2.9$\sigma$ oscillation signal at $\Delta m^2 = 7.34$ eV$^2$ and $\sin^2 2\theta_{ee}\simeq 0.39$.  This is surprising since this large mixing angle is already excluded by other reactor experiments.

The SoLid experiment~\cite{solid} is running at the SCK-CEN BR2 research reactor
in Belgium.  The detector consists of solid scintillator $5 \times 5 \times 5$ cm$^{3}$ cubes 
with $66$LiF:ZnS(Ag) on two faces of each cube.  The set of cubes, with an active mass of 1.6 t, are arranged to cover baselines from 6 to 9 m from the compact reactor core. The detector takes advantage of the positron and neutron position correlation using this highly segmented set of 12,800 detection cells, which can detect and localize both the neutron and electromagnetic signals. The cubes are read out using a 2D grid of 3,200 wavelength shifting fibres coupled to Silicon Photomultipliers. The need to operate the detector on the surface near a reactor, combined with the vast number of detector channels, introduces challenges during data taking, which are being addressed by sophisticated online data reduction techniques that optimize the sensitivity of the experiment whilst achieving a manageable output data-rate.  The experiment is currently taking data in physics mode with the 1.6 ton Phase I detector and expects first physics results in 2019.

\subsection{$\nu_\mu \rightarrow \nu_e$: The SBN Program at Fermilab}

The ongoing Short Baseline Neutrino (SBN) Program at Fermi National Accelerator Laboratory is dedicated to addressing the question of short baseline $\nu_\mu \rightarrow \nu_e$ appearance and $\nu_\mu \rightarrow \nu_\mu$ disappearance signals.    The program has already begun and will extend into the early 2020's.

\subsubsection{MicroBooNE}

The MicroBooNE experiment,  now running
upstream of MiniBooNE at Fermilab, 
was conceived to determine if the MiniBooNE LEE  signal is due
to $\nu_e$ interactions giving an outgoing electron.   
A leading Standard Model hypothesis for
the MiniBooNE LEE anomaly, discussed in Sec.~\ref{MBcomment}, is that the signal could be due, in fact, to an unidentified source
of photons ($\gamma \rightarrow e^+e^-$) coming, for example, from a higher
$\pi^0$ production or a larger branching ratio for the radiative decays of the $\Delta$ baryon.   
The $\gamma \rightarrow e^+e^-$ signature differs from the 
$\nu_e$ CCQE signal in that there is no proton at the vertex, and a
$e^+e^-$ pair is produced rather than a single $e^-$, which leave indistinguishable signals in a Cherenkov detector.

Unlike MiniBooNE, 
which is a Cherenkov detector, the MicroBooNE detector is a Liquid
Argon Time Projection Chamber (LArTPC).   This
has two advantages over a
Cherenkov detector in isolating $\nu_e$ CCQE events from $\gamma$ backgrounds:
1) in a LArTPC, protons above $\sim 25$ MeV kinetic energy are reconstructed, while
in a Cherenkov detector, all protons below 350 MeV are invisible; 2) in a
LArTPC, $\gamma$ conversion to an $e^+e^-$ pair can be distinguished
from a single $e^-$ in about 85\% of the events~\cite{Argoelectronphoton}, while in a Cherenkov
detector, $\gamma$ conversion to $e^+e^-$ cannot be distinguished
from a single $e^-$.

The MicroBooNE detector is installed at 470 m from Fermilab's
BNB beamline target and 90 m upstream of the MiniBooNE detector.
The detector has
a total volume of 170 tons of liquid argon,  with an active region of
$2.3\times 2.5 \times 10$ m$^3$.   The fiducial volume is about 90~tons.    The system comprises two
major subdetectors:  a time projection chamber (TPC) for tracking, and a light
collection system.   
The TPC drifts ionized electrons using a field of 273
V/cm to three wire planes that provide the charge read-out.   The
wire spacing is 3.3 mm, and the shaping time is 2 $\mu$s,
resulting in highly detailed event information that can be exploited in the
analysis.   

\begin{figure}[t]
\begin{center}
{\includegraphics[width=\columnwidth]{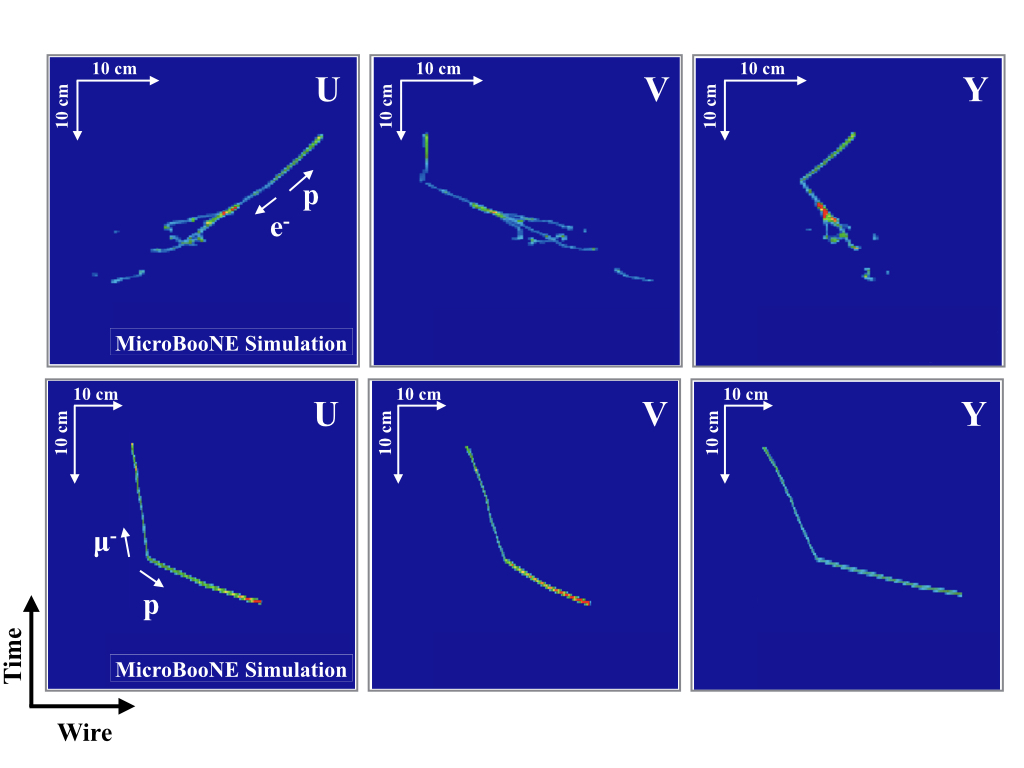}}
\end{center}
\caption{Simulated 1$e$1$p$ and 1$\mu$1$p$ events from MicroBooNE.  Plots show wire number versus drift time for three wire views,  $U$, $V$ and $Y$.   The color indicates the deposited charge measured in ADC count, with a threshold of 10 ADC counts applied.}
\label{ubtopo}
\end{figure}

The key to the MicroBooNE LEE analysis is to utilize events with one
electron and one proton meeting at the vertex (``1$e$1$p$''), as shown in Fig.~\ref{ubtopo}.
Requiring this distinct topology greatly reduces
backgrounds.   In principle, one can develop an analysis that reduces
all $\nu_\mu$ backgrounds, like $\pi^0$ or single photon production,
to a negligible level, leaving  only the intrinsic $\nu_e$
background, which is separated in energy from the LEE signature.
To constrain systematic
errors, MicroBooNE will simultaneously fit to the $\nu_\mu$
interaction counterpart, 1$\mu$1$p$, as described in Sec.~\ref{MBcomment}.    The two track-signal allows accurate neutrino energy reconstruction, which is essential for differentiating the anomalous signal from the intrinsic electron neutrino background.   Analyses that use signatures beyond one-lepton-one proton are also under consideration.

The very high resolution of the LArTPC lends itself to treating the two dimensional plots of wire number versus drift time, such as those shown in Fig.~\ref{ubtopo} as images.   Each cell of the two-dimensional histogram can be thought of as a pixel, where the color is related to the charge, as measured in ADC counts.    This treatment of the data allows MicroBooNE to make use of powerful tools developed for image recognition now routinely used by the artificial intelligence community.   MicroBooNE has been a leader in this area within the neutrino community with a number of published articles \cite{uBpid, uBssnet, MLNature} as important examples.  These techniques can be expected to form the key to the most sensitive LEE analyses from the experiment.

It should be noted that some or all of the MiniBooNE signal may not be due to $\nu_e$ events.  MicroBooNE also has capability to search for beyond Standard Model sources of photons that might produce an LEE-like signal.   However, these are not sterile-neutrino related and, so, are beyond the scope of this review.

The MicroBooNE search has the potential to  be highly impactful.  Should MicroBooNE exclude a LEE electron signal scaled from the measured MiniBooNE rate, no existing sterile neutrino model can accommodate the 
MiniBooNE LEE signal as oscillations because of the proximity of the two detectors on the same beamline.

\subsubsection{MicroBooNE, ICARUS, and SBND:  The Full SBN Program }

The step beyond MicroBooNE is the multi-LAr-detector configuration called the Short Baseline Neutrino (SBN) program \cite{Machado_SBN_AnnRev}.   Ideally, in a multi-detector set-up, one uses two identical detectors that sample identical beams.   Because of the proximity of near-detector sites to the beamline,  it is simply not possible for the near detector to sample an identical beam flux.   And for various practical reasons, it was not possible to build the near detector, SBND, in an identical form  to MicroBooNE.    However,  the proposed SBN program makes up for the systematic uncertainty detector differences might introduce by adding a second far detector, ICARUS.   This three detector combination greatly reduces sensitivity to detector differences.

ICARUS will begin running along with MicroBooNE in 2019. It is located at 600 m from the BNB target.  The detector consists of 500 t of active volume of LAr, which is $5.5$ times larger than MicroBooNE.   Initially, ICARUS can perform a stand-alone LEE search using the same methods as MicroBooNE.  The two experiments can also perform a joint search, since the 130 m separation between the two will constrain the possible ranges for oscillations.    The two experiments can also perform a $\nu_\mu$ disappearance search.

SBND will join MicroBooNE and ICARUS in 2021, rounding out the triad.  This detector is located at 110 m from the production target.   Because it is so close, it can be considerably smaller, with a 112 t active volume, and still have $>6$ times the event rate of ICARUS.   The purpose of SBND is to measure the flux prior to potential oscillation.    The SBND design also serves as a prototype for the DUNE detector.

The use of these detectors for $\nu_\mu \rightarrow \nu_e$ appearance is an interesting test case for the DUNE search algorithms.     They will provide accurate cross comparison of LEE results in two different ways.  First, one can use the proposed method for DUNE of a near-far detector analysis.   Second, one can compare this to the method of constraining the $\nu_e$ with the $\nu_\mu$.   The results should be in agreement.   If they are, then this represent a clear cross check of the MiniBooNE method and raises the possibility of DUNE, and even on-going long baseline $\nu_\mu \rightarrow \nu_e$ appearance searches, of using this constraint method.

\begin{figure}[t]
\begin{center}
\vspace{-0.3in}
\includegraphics[width=2.95in]{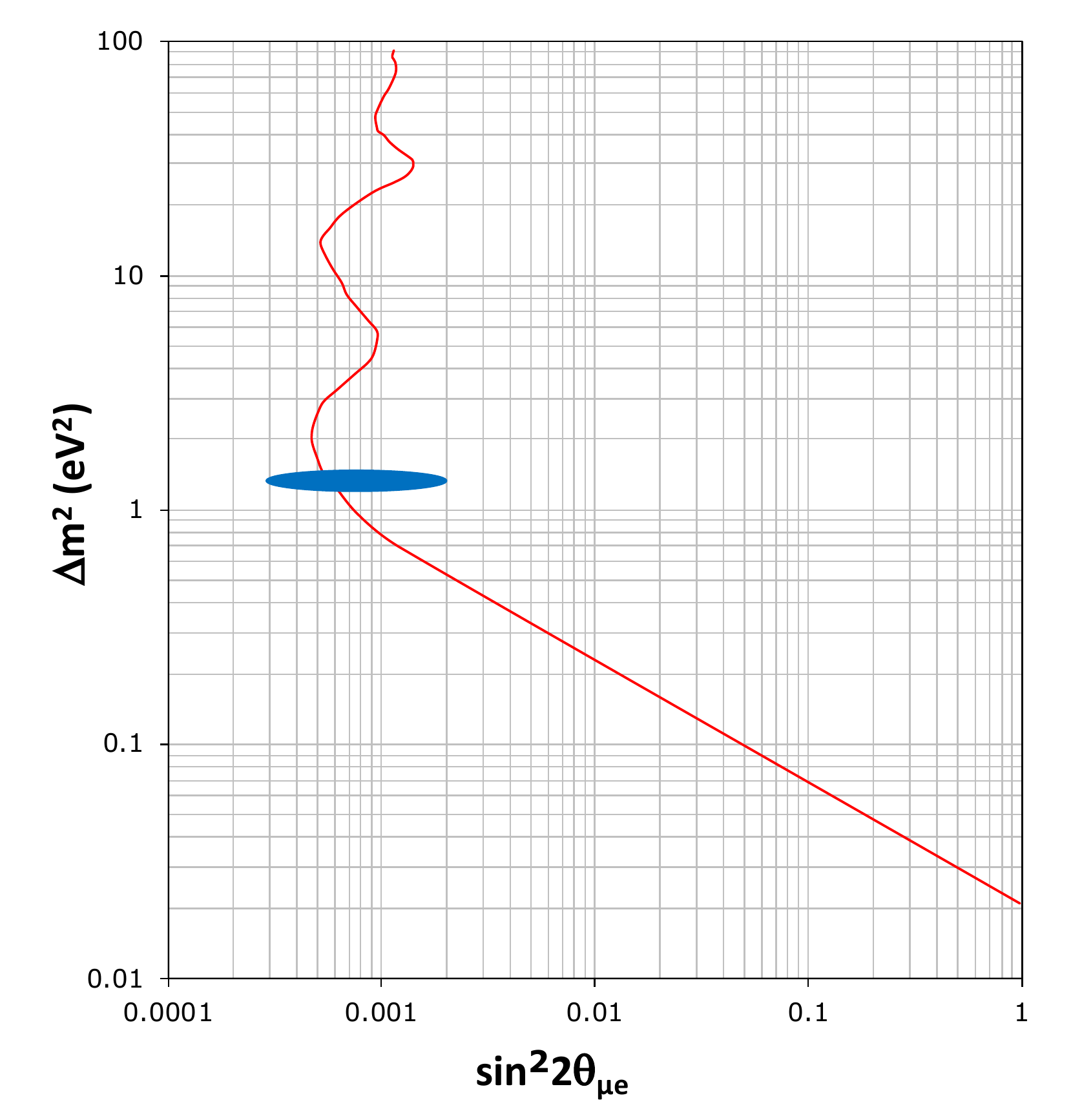}
\end{center} \vspace{-0.2in}
\caption{ \label{SBN_nue_app} 
Red line: SBN sensitivity\cite{Machado_SBN_AnnRev} at 3$\sigma$ for $\nu_\mu \rightarrow \nu_e$ appearance with a 6.6E20 POT data run. Blue area: Global fit 99\% C.L. allowed region from Fig.~\ref{fig:3plus1}.}
\end{figure}

\begin{figure}[t]
\begin{center}
\vspace{-0.15in}
\includegraphics[width=2.95in]{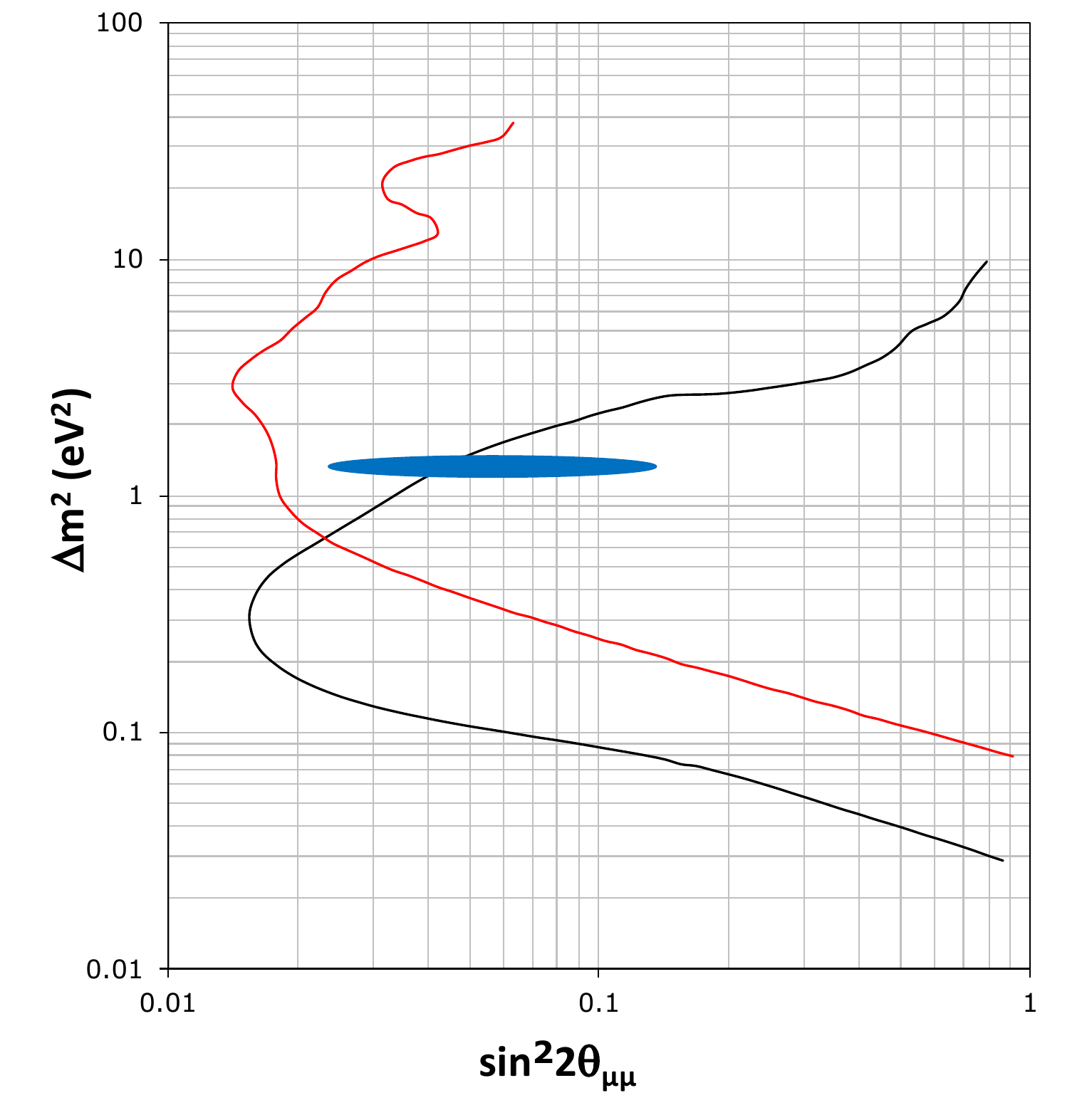}
\end{center} \vspace{-0.2in}
\caption{ \label{numudislim} 
Black line: IceCube current limit at 90\% C.L. for $\nu_\mu$ disappearance. Red line: SBN sensitivity\cite{Machado_SBN_AnnRev} at 3$\sigma$ for $\nu_\mu$ disappearance with a 6.6e20 POT data run. Blue area: Global fit 99\% C.L. allowed region from Fig.~\ref{fig:3plus1}.}
\end{figure}

The combination of the three SBN detectors will have improved sensitivity to $\nu_\mu \rightarrow \nu_e$ appearance and also $\nu_\mu$ disappearance.  This is especially true for $\nu_\mu$ disappearance searches where the systematic uncertainties in the neutrino flux dominates.  The neutrino flux uncertainty can be significantly reduced by using the SBND near detector to measure the flux at short $L$-values before oscillation effects occur and propagate it to the far detector flux using simulations.  Fig.~\ref{SBN_nue_app} and Fig.~\ref{numudislim} show the expected sensitivity of the SBN program for a 6.6E20 POT data run for $\nu_\mu \rightarrow \nu_e$ appearance and $\nu_\mu$ disappearance oscillation respectively, as compared to our global fit allowed region.

\subsection{$\nu_\mu$ Disappearance: IceCube\label{sec:icecube}}

The IceCube neutrino observatory is a one gigaton Cherenkov detector
that consists 
of 5160 digital optical modules (DOMs)~\cite{Aartsen_2017} light detectors arranged on
86 strings, located 1450 m below the top of the Antarctic ice.  Most of the
detector has sparse string spacing (17 m between DOMs and $\sim$125 m
between strings), with an energy threshold of $\sim 100$ GeV.  An
8-string region (7 m between DOMs, $\sim 50$ m between strings)
with a $\sim 10$ GeV threshold, called DeepCore, has also been installed.

The IceCube plan exploits the detector's one-of-a-kind capability to observe a resonance signature from sterile-induced
matter-effects in upward-going antineutrinos.
The resonance
is mass-hierarchy-dependent and most models favor its appearance in the
antineutrino flux.
The signal is resonant depletion of up-going $\bar \nu_\mu$
propagating through the Earth, producing a deficit at a specific energy
and zenith angle.   Because the deficit is well-localized and large 
compared to vacuum oscillations, IceCube has deep reach in $\sin^2 2\theta_{\mu\mu}$.

No signal for $\bar \nu_\mu$ disappearance was
observed in one year of analyzed data.  The limit assumed $|U_{\tau 4}|^2=0$ to
permit cross comparison to the SBL results.  The current IceCube limit\cite{TheIceCube:2016oqi} for $\nu_\mu$ disappearance is shown in Fig.~\ref{numudislim} where it is compared to the global fit allowed region from Fig.~\ref{fig:3plus1}.  
IceCube has collected about $13$ times the data set used for this previous publication and plans to publish new results soon.

\section{The Next Generation: What Will Resolve the Sterile Neutrino Picture? \label{DAR}}

The past approach for addressing sterile neutrino anomalies has been to develop new experiments that are ``good enough'' under the best of conditions to provide some new information.     This strategy will likely continue to result in leaving the field in a confusing situation since most of the new experiments cannot provide decisive, highly significant results.  As a comparison,
the sterile neutrino situation now is similar to the three-neutrino oscillation results available in the late 1980's and mid-1990's.    The key at that time to resolving the question of the resilient, but confusing, anomalies  was for the community to invest in definitive experiments--Super K and SNO.  In the case of SNO, an entirely new approach was applied to the problem.   
We are going to need to approach today's resilient, but confusing, sterile anomalies with a similar strategy.     We need truly decisive and definitive experiments--ones that cover the anomalies at 5$\sigma$ with conservative assumptions--designed to address the specific questions that are arising from the anomalies.        Employing new strategies and techniques, as happened with SNO, is also important.   We see single-particle decay-at-rest (DAR) sources as a bold new approach.    These sources can be produced with high fluxes by decays of a single isotope, such as $^8$Li; specific mesons, {\it i.e.} pions and kaons; and muons.
The advantages of decay-at-rest sources are that the flavor content and energy distribution are defined by nature
and, in all of these cases, quite well understood.

Resolving the anomalies with these purpose-built experiments is, in fact, an excellent investment for our field.    These experiments are at a much lower cost scale than Super-K or SNO.  While some technological advances are needed,  these experiments represent smaller steps than that which were required for the success of SNO.    And, like Super-K and SNO, these experiments provide much needed, new infrastructure to the field, that can do physics beyond the program of addressing the sterile neutrino anomalies.

\subsection{$\nu_e$ Disappearance: IsoDAR} 

\begin{figure}[t]
\begin{center}
\includegraphics[width=.3\textwidth]{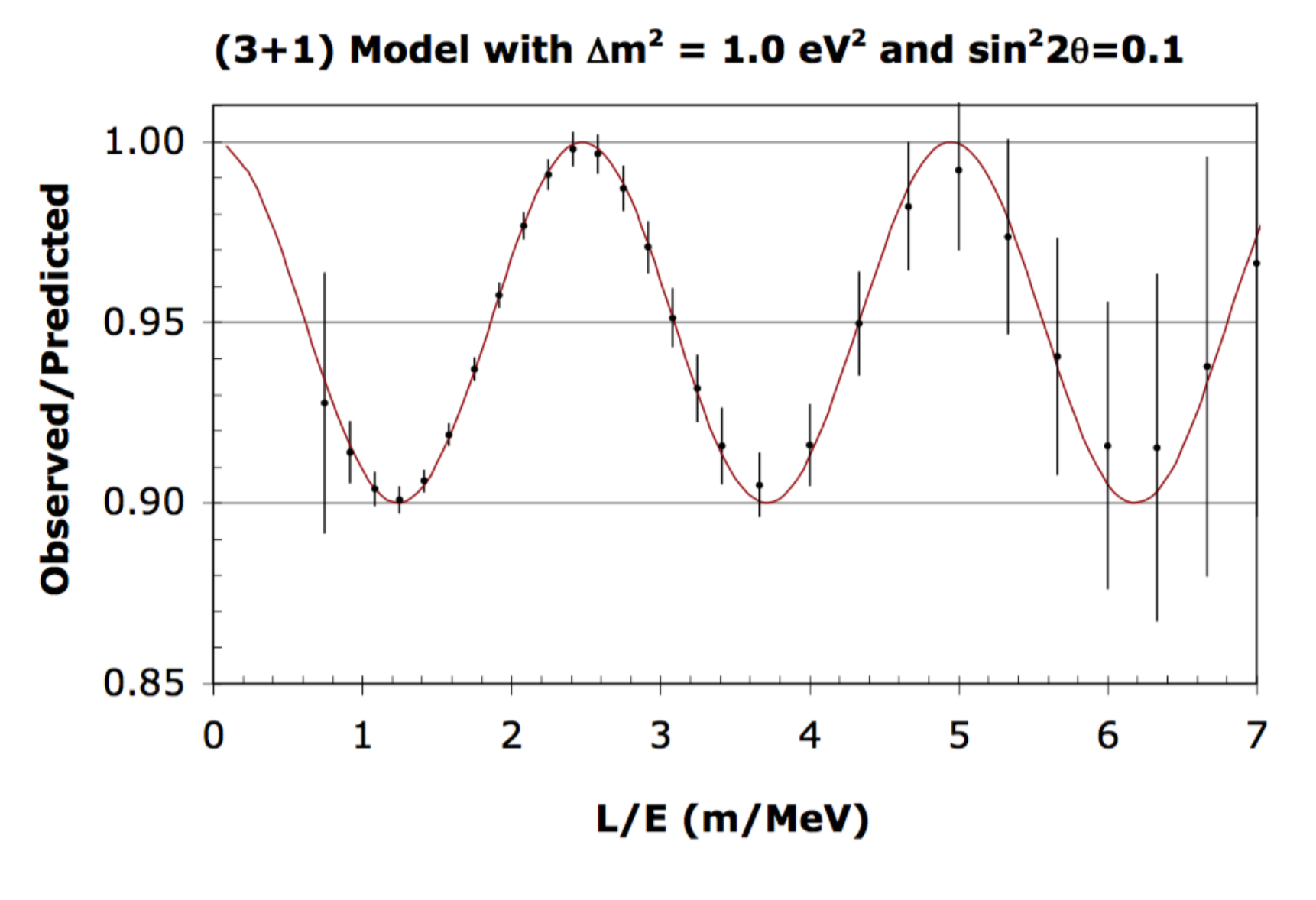}\\\includegraphics[width=.3\textwidth]{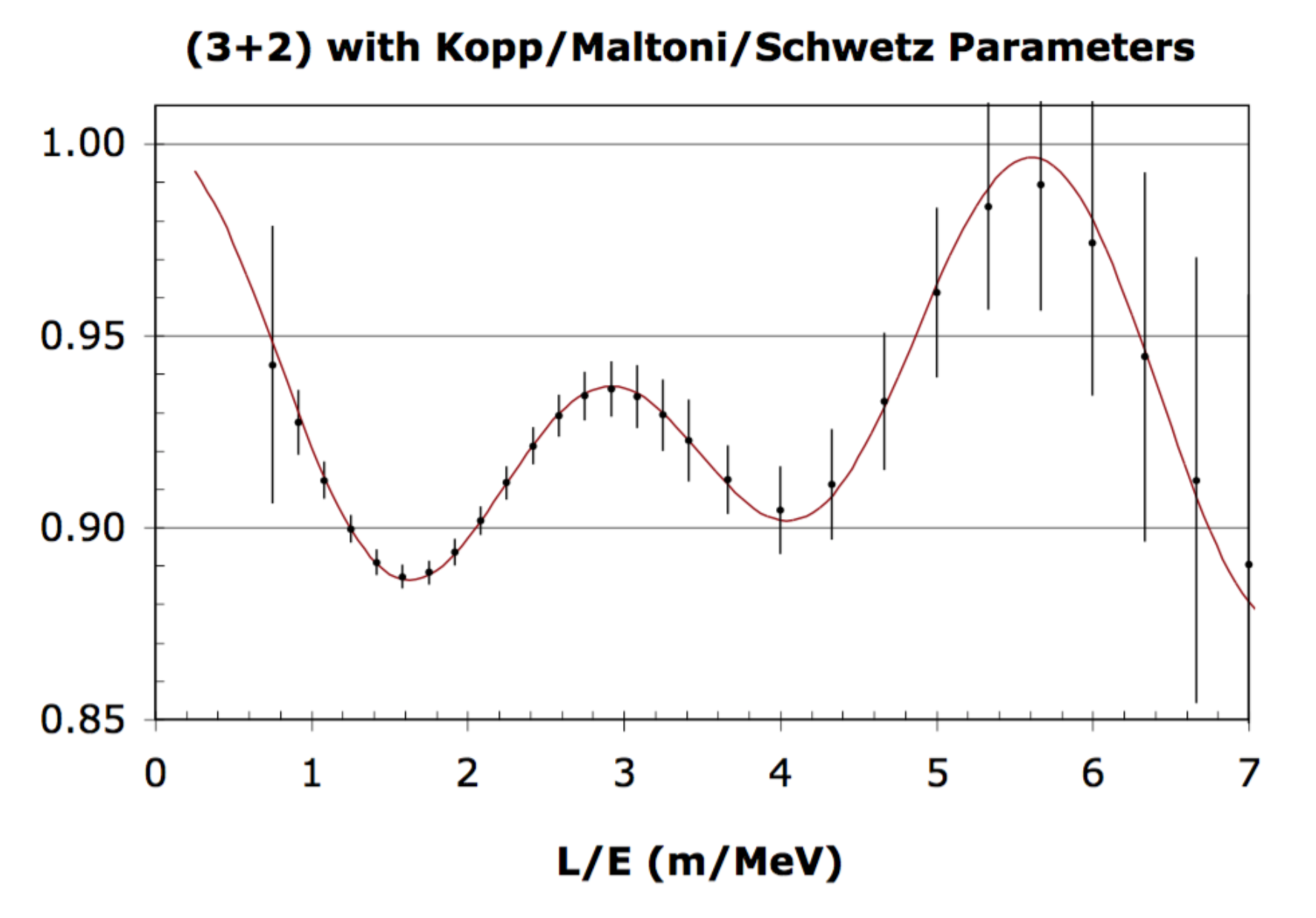}\\\includegraphics[width=.3\textwidth]{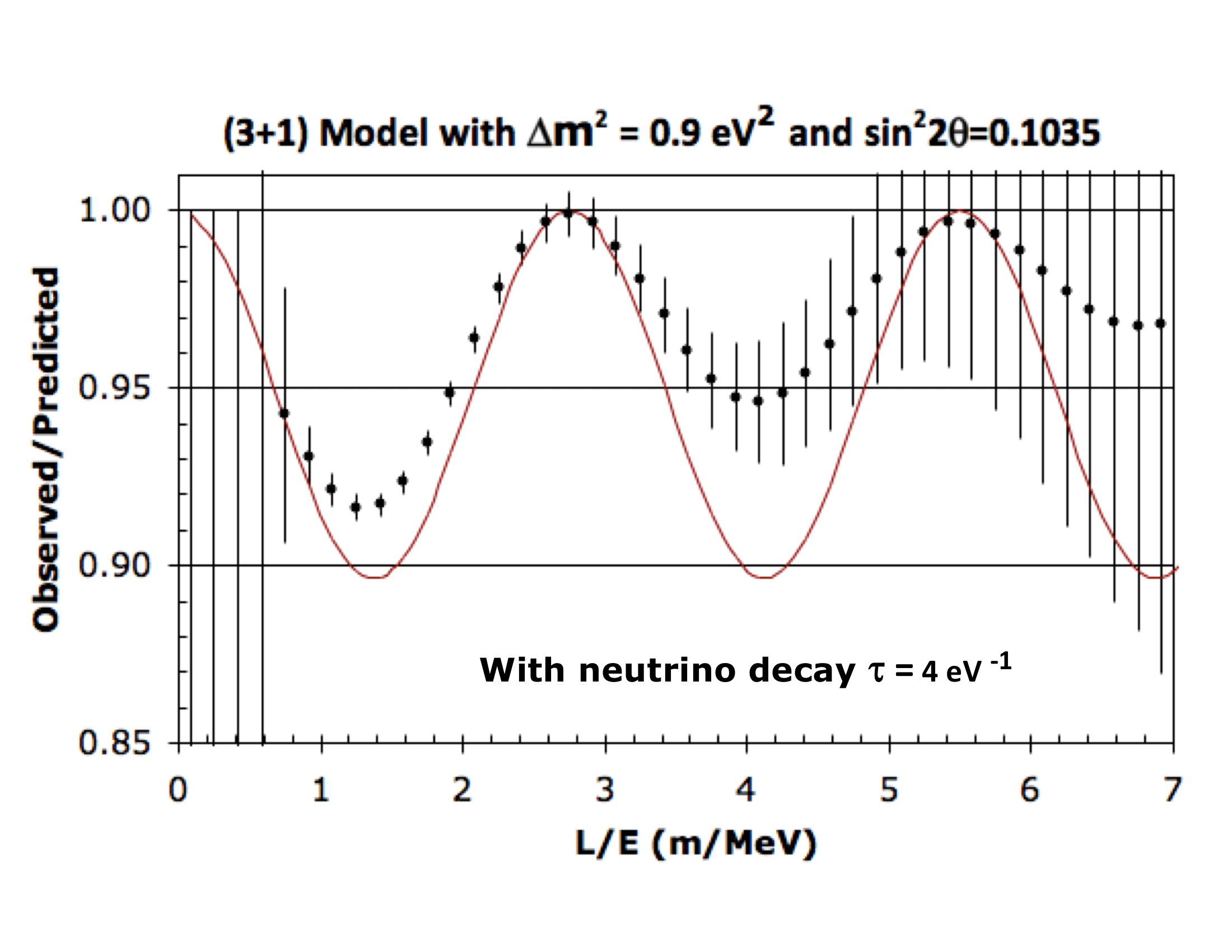}
\end{center}
\vspace{-.5cm}
\caption{ IsoDAR$@$KamLAND $L/E$ dependence, 5 years of running,
  for 3+1 (top) and 3+2
  (middle) sterile neutrinos, and 3+1+Decay (bottom).   Solid
  curve shows no smearing in the reconstructed position and energy
  and no decay for the bottom plot.  Data points with error bars include smearing.  
\label{osc3N}}
\end{figure}

IsoDAR offers $>$5$\sigma$ coverage of the $\bar \nu_e$ disappearance results.   The experiment 
will use a novel, high-intensity,
single-isotope $\bar \nu_e$ source, paired with a $\sim 1$ kt scintillator
detector.    Pairing this source with KamLAND allows for 5$\sigma$ coverage in 5 years with conservative assumptions.    Another alternative is a special purpose, segmented neutrino detector such as the  CHANDLER design \cite{Chandler}, that can deliver similar capability but with less mass because it can be optimally arranged in a vee shape near to the source.
In either case, IsoDAR's strength comes from
high-precision reconstruction of the $L/E$ dependence of the neutrino
disappearance  ``oscillation wave.'' Fig.~\ref{osc3N}
illustrates the ability to distinguish potential signals that may describe the anomalies.

IsoDAR is unique in providing a single-isotope flux 
with endpoint of $\sim$13 MeV.  To produce this, 
a high-intensity H$_2^+$ ion source feeds a 
60 MeV/amu cyclotron via an RFQ \cite{IsoDARNIM}.  The extracted beam is 
transported to a novel 
$^9$Be target with boiling-water cooling \cite{Adriana} where the full intensity of beam is used to produce many neutrons.  The neutrons enter 
a $\ge$99.99\% isotopically pure $^7$Li sleeve, where 
capture results in $^8$Li.  The  $^8$Li undergoes
$\beta$ decay, producing an isotropic, pure
$\bar \nu_e$ flux \cite{Adriana}.
Pairing this very high-intensity $\bar \nu_e$ source with a
hydrogen-based detector 
allows for the inverse beta decay and  $\bar \nu_e$-$e^-$ elastic scattering.

\begin{figure}[t]
\begin{center}
\vspace{-0.3in}
\includegraphics[width=2.55in]{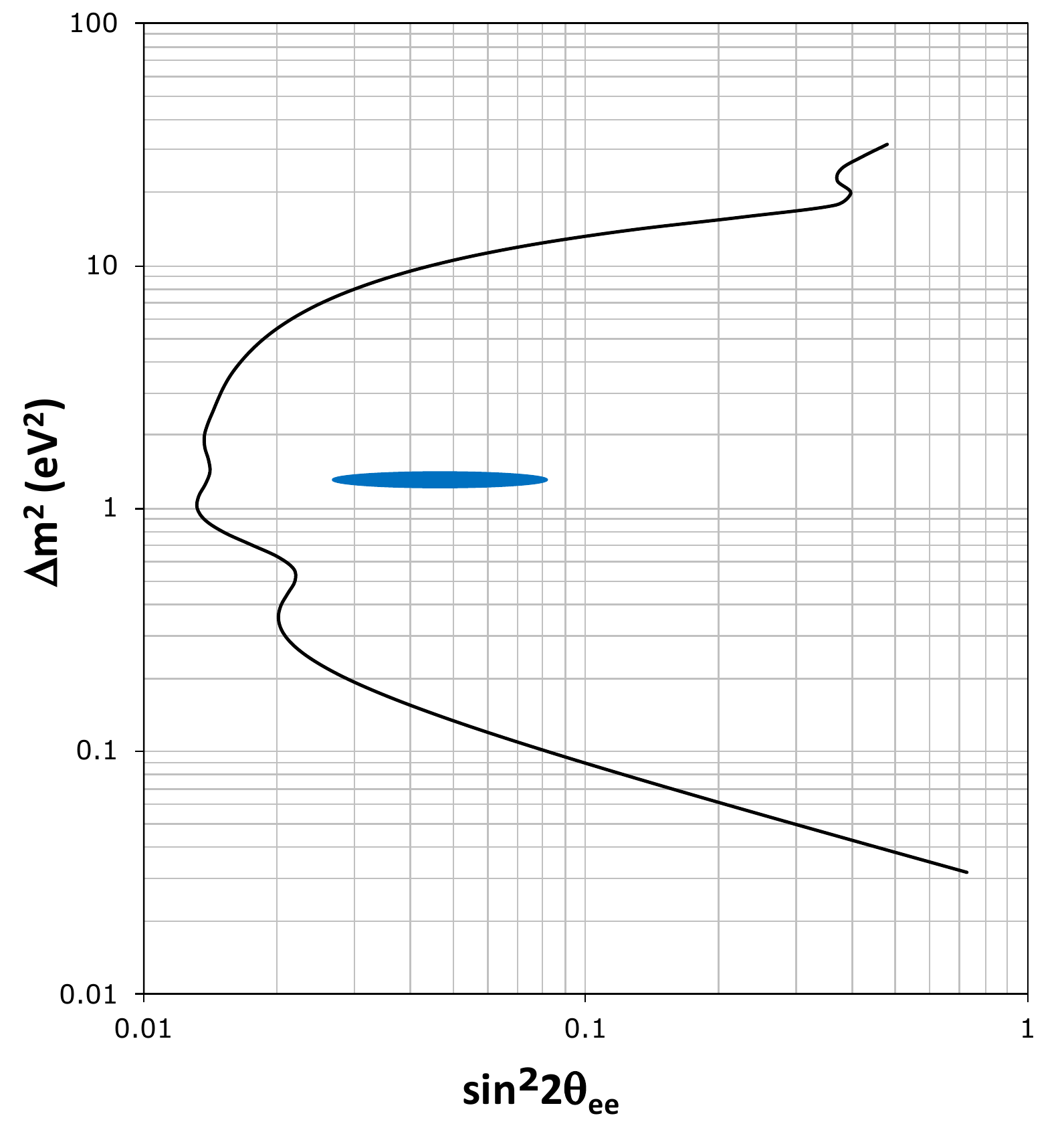}
\end{center} \vspace{-0.2in}
\caption{ \label{isonew} 
Black line: IsoDAR$@$KamLAND $\nu_e$ disappearance sensitivity at 5$\sigma$ for a 5yr run. Blue area: Global fit 99\% C.L. allowed region from Fig.~\ref{fig:3plus1}.}
\end{figure}

The IsoDAR design requires new technology, and more than five years have been invested in proving this neutrino source can be feasibly constructed.     IsoDAR is an accelerator-driven source that makes use of a cyclotron that is an order of magnitude higher in intensity than on-market proton accelerators used for medical isotope production purposes.  The most significant challenge to raising the intensity is accelerating the current without unacceptable beam losses.   These mostly arise due to the coulomb repulsion of the accelerated ions, which increases the size of the beam bunches.   To solve the problems that arise from this, a number of novel approaches have been introduced, including accelerating H$_2^+$ rather than protons; highly efficient bunching via an RFQ; and harnessing the space charge effects to induce vortex motion.  A review of the accelerator system for IsoDAR is described in Ref.~\cite{IsoDARNIM}.

IsoDAR$@$KamLAND covers the allowed region
with 5$\sigma$ sensitivity in 5 years (Fig.~\ref{isonew}, black)
with a higher-energy,   better-controlled $\bar \nu_e$ source than that of reactors.   In particular, the single isotope spectrum of $^8$Li is well predicted.
The $L/E$ dependence (Fig.~\ref{osc3N}) allows models to be clearly
differentiated.   Discriminating power is a trade-off between excellent resolution on $L/E$ and high statistics.  Thus, surprisingly, the 1 kt KamLAND scintillator detector and the 25 kt Super-K Cherenkov detector have very similar sensitivity.    KamLAND sensitivity is extended if the proposed upgrade to the light collection, leading to energy resolution of $3\%/\sqrt{E}$, is introduced.  IsoDAR's ability to reconstruct the $L/E$ dependence of the oscillation wave extends to
$\sin^22\theta_{ee}\sim 0.01$ at KamLAND. 
Thus IsoDAR decisively addresses the anomalies.

\subsection{$\nu_\mu \rightarrow \nu_e$:  A Next-generation JSNS$^2$ is Needed}

The LSND result has been so resilient against technical criticisms because it is very hard to go wrong when pairing a high flux $\pi/\mu$ DAR source with a detector that relies on IBD events.   The field desperately needs a high-statistics follow-up to LSND with this design, with a few features that would further improve the result.     The first is for the experiment to be located at $>90 ^\circ$ from the direction of the incoming proton beam, removing any potential decay-in-flight component.   The second is to run at a very high intensity beam dump that delivers the protons in few nanosecond pulses, rather than the relatively long spill used at LANSCE, where LSND took data.    The result is that, through beam timing, one can separate the $\nu_\mu$ from pion decay and any decay-in-flight, which represent relatively prompt flux, from the $\bar \nu_\mu$ and $\nu_e$ from muon decay, which is relatively delayed.   The third is to use oil that contains Gd, which reduces the time delay between the prompt light and the neutron capture, thus reducing accidental coincidences.    The fourth is to build the detector with an enhanced veto and a $\gamma$-catcher region.   Ability to move the detector would also be very valuable.

Two such experiments have been proposed:  OscSNS \cite{oscsns} at the Spalation Neutron Source (SNS) in the US and JSNS$^2$ \cite{jsns2} at the Materials and Life Science Experimental Facility (MLF) in Japan.    The primary purpose of both facilities is to produce neutrons, but $\pi/\mu$ DAR also occurs.    The JSNS$^2$ experiment is approved to run, but the 17 t detector is too small to cover the LSND range at 5$\sigma$.    
%For scale, this is about the same size as a Daya Bay near detector.    
This experiment will begin to take data in 2020.   The OscSNS detector was proposed to be 1 kton, which would have provided more than adequate coverage, but this experiment was not approved.

We advocate for an upgraded JSNS$^2$ experiment to at least 100 t. 
An important point is that all of the technology for this experiment exists. This detector is not a technical stretch.  The beam source exists and is being run with funding from the spallation neutron experiments.   Such an experiment could easily be mounted within a few years.   This experiment should be done since it would be a key and definitive high statistics test of the LSND result.

\subsection{$\nu_\mu$ Disappearance:  KPipe and CCM}

An important difference between the SNS and MLF is that the former uses 1 GeV protons on target while the latter employs 3 GeV protons on target.    As a result,  kaon decay-at-rest (KDAR) is produced at MLF.    As discussed in Sec.~\ref{DARsource}, this provides a monoenergetic flux of $\nu_\mu$ at 236 MeV, which is above the threshold of CCQE interactions.

The premise of KPipe \cite{kpipe} is to make use of the KDAR flux at MLF.   The detector vessel is proposed to be 3~m in diameter and 120~m long, extending radially at a distance of 32~m to 152~m from the MLF beam dump.   This is filled with liquid scintillator and instrumented with hoops of silicon photomultipliers (SiPMs).     The signal is a coincidence between the light from the initial CCQE interaction and the light produced by the Michel electron from the decay of the muon exiting the interaction which stops after traveling a very short distance in the scintillator oil.   This coincidence greatly reduces background.    This is a very robust search for sterile neutrino oscillation and decay because it relies only on the measured rate of detected events as a function of distance, with no required knowledge of the neutrino interaction cross section or the initial isotropic neutrino flux, which falls as (1/distance)$^2$.   The liquid scintillator does not require Gd or Li doping, since the second signal of the coincidence is not from a neutron, greatly reducing cost.      There is only very modest technological development required for this experiment.

An alternative approach that does require substantial technological development will use the 29 MeV monoenergetic $\nu_\mu$ from pion decay in a disappearance experiment based on coherent neutrino scattering, which is only sensitive to active neutrino scattering.  This is the premise behind the Coherent Captain Mills (CCM)~ \cite{CCM_VanDeWater} experiment, which is a liquid-argon-based detector running at the LUJAN spallation neutron facility at Los Alamos.  The LUJAN facility has a high instantaneous power and a low duty factor, which should have good background rejection.  The detector is a 7 t fiducial volume LAr detector with photomultiplier readout giving good energy and timing resolution plus an energy detection threshold of 10-20 keV, appropriate for detecting coherent neutrino scattering.  The experiment expects to detect about 2700 (680) prompt mono-energetic 29 MeV $\nu_\mu$ coherent elastic scattering events per year with the detector located at 20 m (40 m) from the source. Background mitigation is crucial using beam and detector timing along with instrumented vetoes and shielding.  With this experimental setup, the CCM experiment estimates a sensitivity at the 90\% C.L. level for $\nu_\mu$ disappearance that will cover the global fit region shown in Fig.~\ref{fig:3plus1}.  CCM is expected to start data taking in the summer of 2020 with commissioning runs before that.

\section{Conclusions}

In conclusion, this paper has provided a snapshot of where we are at in exploring the question of the existence of light sterile neutrinos, especially through accelerator and reactor experiments.  The picture is far from clear.    

Anomalies have been observed in a set of short-baseline experiments.  Introducing an additional, mostly sterile mass state to explain this provides an improvement of $>5\sigma$,
%$\Delta \chi^2$/$\Delta dof$ of 35/3, 
which is highly improbable as an accidental improvement.     We find that adding additional sterile neutrinos to the model yields only a modest $1.6\sigma$ improvement over the 3+1 model.   On the other hand, introducing decay of the fourth mass state, which can reduce tension with cosmological measurements, leads to a larger $2.6\sigma$ improvement with respect to a 3+1 model.
%$\Delta \chi^2$/$\Delta dof$ of 8/1.   
We note that the decay model tested only has decays to new invisible particles, and that decays from the fourth state to active neutrinos is likely to produce further improvement.

The data are clearly indicating that something is missing from the model, and that what is missing takes a form similar to a model of 3+1+Decay.     However, any model that explains the results must be self-consistent.    The consistency of the 3+1 and 3+1+Decay models can be tested by dividing the data into appearance and disappearance data sets.     First, one observes that the global fit allowed regions for appearance and disappearance do not overlap at 95\% CL in either model.   Second, one can apply the PG test to quantify the disagreement.   One finds that the tension within 3+1 is at the $4.5\sigma$ level.  The disagreement is improved in a 3+1+Decay model, but is still at the $\sim 3\sigma$ level. 

We have discussed the value of considering Bayesian credible regions as well as frequentist allowed regions.   We show that for our 3+1 fits, 
substantially more parameter space is allowed in the high $\Delta m^2_{41}$ region in the Bayesean study.   The difference arises because
Bayesian and frequentist methods address different questions about the data.   Bayesian inference makes statements about probability of model parameters given the proposed model, while frequentist methods make statements about the probability of the data given the model.  Thus, when interpreting results of global fits, one must carefully consider exactly what question one wants to ask.

With either fitting approach, we urge caution in interpreting global fit results for many reasons.  For example, some fraction of the tension may arise from problems with handling the experimental data.     We have shown that, given the assumptions we must make in global fitting, we end up with imprecise representations of the published results.  This could be mitigated with improved data releases from experiments.   We also note that we are not thoroughly exploring all of the anomalous features in these experiments that may be relevant.   The 5 MeV excess in the reactor experiments, that seems to be greatly reduced for experiments at distances less than 10 m, is an example where systematic flux uncertainties may be interacting with the sterile neutrino oscillation phenomenology.     There are also inherent problems with experiments with large systematic errors.  We assume that these errors are Gaussian, when most probably they are not.  Also, the nature of systematic uncertainty is to quantify the unknown, which is extremely difficult for an experiment to do accurately.

We are excited by the opportunity presented by new experiments coming online soon.  While these will resolve some of the issues, we urge the community to think toward a program of high statistics, low systematic uncertainty decay-at-rest experiments in the 5-year future that we believe will finally be decisive.

The conclusion is that the picture is unclear, but is very thought-provoking.   We are in a situation similar to where we were with three neutrino oscillations in the early 1990's.   Anomalies are observed, but they do not fit comfortably with observed limits.   This could be due to some combination of an incomplete model and unknown systematic effects.  As then, the results call for further exploration.   No matter what we find, the results will move the field significantly forward, but if new physics is the culprit, then this has the potential to revolutionize particle physics.   

\section*{Acknowledgements}

AD, CA and JMC are supported by NSF grant PHY-1801996.   MHS is supported by NSF grant PHY-1707971.   GHC is supported by Institute for Data, Systems, and Society at MIT.  We thank Roger Barlow, Paul Grannis, Adrien Hourlier, Patrick Huber, Keng Lin, Bryce Littlejohn, William Louis, Pedro Machado,  Sergio Palomares-Ruiz, Jordi Salvado,  Robert Shrock, and Lindley Winslow for valuable discussions.  We thank MicroBooNE for the approved event display appearing in Fig.~\ref{ubtopo}.  We thank the STEREO experiment for permission to use their data in Sec.~\ref{bump}.

\bibliographystyle{unsrt}
\bibliography{rmp}

\end{document}